\documentclass[10pt,a4paper]{article}
\pdfoutput=1
\usepackage{jstyle}
\usepackage{amsmath, amsfonts, amssymb,subcaption}
\usepackage{graphicx, hyperref, color, verbatim}
\usepackage{diagbox}

\usepackage{xcolor}
\usepackage{tikz}
\usepackage{amsmath}

\usepackage{tikz-cd}
\usetikzlibrary{matrix, calc, arrows}

\newcommand\zb{\bar{z}}

\newcommand{\bea}{\begin{eqnarray}}
\newcommand{\eea}{\end{eqnarray}}
\newcommand\restr[2]{{\left.\kern-\nulldelimiterspace#1\vphantom{\big|}\right|_{#2}}}

\newcommand{\ext}{\mathtt{ext}}

\newcommand{\sX}{\mathfrak{X}}
\newcommand{\sXt}{\tilde{\mathfrak{X}}}
\newcommand{\sXs}{\mathfrak{X}_{\star}}
\newcommand{\Li}{\mathrm{Li}}

\newcommand{\D}{\Delta}
\newcommand{\h}{h}
\newcommand{\JJh}{j^2_{\h}}
\newcommand{\JJ}{j^2_{\D}}
\newcommand{\hExt}{\h_{\mathtt{ext}}}
\newcommand{\DExt}{\D_{\mathtt{ext}}}
\newcommand{\DOdd}{\D_{\mathtt{odd}}}
\newcommand{\braid}{\big(\tfrac{\chi}{\chi-1}\big)}
\newcommand{\g}{g}

\title{Unmixing the Wilson line defect CFT. \\
Part II: analytic bootstrap}

\author[\varphi]{Pietro Ferrero}
\author[\Psi,f]{Carlo Meneghelli} 
\affiliation[\varphi]{Simons Center for Geometry and Physics, SUNY, Stony Brook, NY 11794, USA}
\affiliation[\Psi]{Dipartimento SMFI, Universit\`{a} di Parma, Viale G.P.
Usberti 7/A, 43100, Parma, Italy}
\affiliation[f]{INFN Gruppo Collegato di Parma}

\emailAdd{pferrero@scgp.stonybrook.edu, \\ \hskip 38pt carlo.meneghelli@unipr.it}
\abstract{In this second installment of a series of two papers on the $\tfrac{1}{2}$-BPS Wilson line defect CFT in $\mathcal{N}=4$ super Yang-Mills, we focus on dynamical aspects of the theory, in particular studying four-point functions with analytic bootstrap methods. Relying on the results of \cite{Ferrero:2023znz} for the kinematics and strong coupling spectrum, we consider various four-point functions in the planar limit, in an expansion for large 't~Hooft coupling. Our ultimate goal is to provide a detailed derivation of the four-point function of the displacement supermultiplet at three loops, first presented in \cite{Ferrero:2021bsb}. Along the way, we present a large amount of new results including four-point functions with zero, one or two long external supermultiplets. The last two represent a novelty in the analytic bootstrap literature and are instrumental in addressing the problem of operators degeneracy. Such phenomenon leads to the necessity of resolving a mixing problem that is more complicated than those usually encountered in the study of holographic correlators, thus leading us to the development of a new approach that we believe will have a wider range of applicability. Related to this issue, we analyze in some detail the structure of the dilatation operator in this model. Some of the ingredients that we use apply more generally to holographic theories, although a thorough investigation of these aspects is missing, to the best of our knowledge, in most interesting cases.}

\begin{document}
\maketitle
\tableofcontents

\newpage

\section{Introduction}\label{sec:intro}

Despite several attempts over the years, solving an interacting quantum field theory (QFT) still remains a formidable challenge. In four dimensions, the most promising candidate is 4d $\mathcal{N}=4$ super Yang-Mills (SYM) theory, due to the simultaneous applicability of most of the existing tools to study strongly coupled QFTs: integrability \cite{Minahan:2002ve}, supersymmetric localization \cite{Pestun:2007rz}, holography \cite{Maldacena:1997re} and the conformal bootstrap \cite{Rattazzi:2008pe}. Yet, even in the planar limit, we are still far from claiming that we have fully solved the theory, mainly due to the difficulty in computing the operator product expansion (OPE) coefficients. It could then be fruitful to focus on an even simpler model, where said techniques are still applicable, to develop new ideas that could be applied to the full-fledged four-dimensional gauge theories.

An interesting proposal is that of focusing on the one-dimensional superconformal field theory (SCFT) defined by operator insertions along a one-half BPS Wilson line in planar 4d $\mathcal{N}=4$ SYM \cite{Maldacena:1998im}. In the simplest case of Wilson lines in the fundamental representation of the gauge group $SU(N)$, the defect CFT on the line has been discussed for small 't~Hooft coupling $\lambda$ in terms of deformations of the circular Wilson loop \cite{Cooke:2017qgm,Cooke:2018obg}, while at strong coupling the dynamics of the theory is captured by the fluctuations of a semiclassical string with AdS$_2$ worldsheet in AdS$_5\times S^5$ \cite{Giombi:2017cqn}. This allows for a perturbative approach both at weak and strong coupling. Moreover, it was already observed by various authors that, much like its four-dimensional parent, this defect CFT can be studied with integrability \cite{Kiryu:2018phb,Grabner:2020nis,Cavaglia:2021bnz,Cavaglia:2022qpg} supersymmetric localization \cite{Giombi:2018qox,Giombi:2018hsx,Giombi:2020amn}, holography \cite{Drukker:2005kx,Gomis:2006sb,Gomis:2006im,Giombi:2017cqn,Giombi:2020kvo}, and the conformal bootstrap \cite{Liendo:2016ymz,Liendo:2018ukf,Ferrero:2021bsb}. In particular, we would like to highlight two recent developments. On the one hand, the powerful information on the spectrum of the theory obtained (numerically) with the quantum spectral curve (QSC) \cite{Gromov:2009tv,Gromov:2009bc,Gromov:2013pga} has been combined with numerical bootstrap techniques to put strong constraints on the OPE coefficients, in a framework that was termed {\it bootstrability} in \cite{Cavaglia:2021bnz,Cavaglia:2022qpg}. Recently, interesting an interesting approach that combines bootstrap, reinforcement learning and the integrability input has been also proposed \cite{AlessandroThesis,Niarchos:2023lot}. On the other hand, analytic bootstrap methods have been proved particularly fruitful at strong coupling, where correlation functions were computed at unprecedented high order and new techniques to address mixing problems were developed in previous work by the authors \cite{Ferrero:2021bsb}.

In this paper, which alongside \cite{Ferrero:2023znz} expands on the results presented in \cite{Ferrero:2021bsb}, we consider an analytic bootstrap approach to the half-BPS Wilson line defect CFT in the fundamental representation, in planar $SU(N)$ 4d $\mathcal{N}=4$ SYM, in the regime of large 't~Hooft coupling $\lambda$. Our main focus is the four-point function of the super-displacement multiplet, $\mathcal{D}_1$, which we compute analytically at fourth perturbative order in the large $\lambda$ expansion. To this end, we employ the techniques developed in \cite{Liendo:2018ukf,Ferrero:2019luz} for one-dimensional CFTs (that we review in detail) that use an ansatz with harmonic polylogarithms (HPLs) and demonstrate their power and applicability, in principle, to each order in $1/\sqrt{\lambda}$, once all the relevant CFT data at previous orders are known. The main obstacle to this approach is the degeneracy of states at $\lambda=\infty$, which makes it impossible to extract all the CFT data at each order from a single correlator. To address this issue, in \cite{Ferrero:2021bsb} we considered various families of four-point functions. Besides correlators between half-BPS operators of arbitrary weight, it proved crucial in this case to consider external unprotected supermultiplets, a case which has not yet been considered in detail in the bootstrap literature, despite its great potential. In the companion paper \cite{Ferrero:2023znz}, we develop the kinematics and superconformal blocks for correlators involving long multiplets and perform an analysis of the spectrum of the theory at strong coupling, allowing for the implementation of the bootstrap machinery in this paper.

We note that our setup is akin to the analytic bootstrap of holographic four-point functions between half-BPS operators, that has been developed at tree level for theories with maximal \cite{Rastelli:2016nze,Rastelli:2017udc,Alday:2020lbp,Alday:2020dtb} and half-maximal \cite{Alday:2021odx} supersymmetry and pushed to one loop for some cases \cite{Aprile:2017bgs,Alday:2017xua,Alday:2020tgi,Alday:2021ajh,Behan:2022uqr,Alday:2022rly} and even up to two loops in AdS$_5\times S^5$ \cite{Huang:2021xws,Drummond:2022dxw}. It is then interesting to draw a parallel with that program. First, we emphasize that in terms of Witten diagrams our result presented in \cite{Ferrero:2021bsb} and reviewed here corresponds to a three-loop computation, which is hardly approachable in $d>1$ with the current technology. Moreover, while for the resolution of mixing between double trace operators (at first order) it has proven sufficient in most known holographic setups to focus on four-point functions between half-BPS operators of arbitrary weight, this is not sufficient in our case for reasons that we will explain. Rather, our mixing problem is probably more akin to that faced at higher orders, when higher-trace operators enter the game: for this reason we believe that the techniques developed here and in \cite{Ferrero:2021bsb,Ferrero:2023znz} could be relevant for the bootstrap of holographic correlators in higher dimensions as well. From this perspective, key aspect of our analysis is a thorough investigation of the spectrum of the theory and of the structure of the dilatation operator at strong coupling, in the spirit of \cite{Beisert:2004ry}. Thanks to the simplicity of this model we are able to obtain the complete expression of the dilatation operator at first order, for arbitrary long multiplets, and moreover we provide an understanding of some of it structures at all orders. 

It is also worth emphasizing that we work in position space, while in the literature on holographic correlators both position space and Mellin space \cite{Penedones:2010ue,Fitzpatrick:2011ia} are employed. On the one hand, experimenting with HPLs in 1d CFT is simpler than in higher dimensions and our setup can prove fruitful for a better understanding of the basis of functions that is relevant for perturbative computations. On the other hand, a naive application to Mellin space to one-dimensional theories fails due to the presence of a unique cross ratio in one dimension and the half-BPS Wilson line defect is an ideal arena to test proposals for a 1d Mellin space, such as that of \cite{Bianchi:2021piu}.

\subsection{Summary and results}

Let us now briefly outline the bootstrap problem that we consider, highlighting in particular the main assumptions and results. The goal of this paper is to compute the four-point function $\langle \mathcal{D}_1\,\mathcal{D}_1\,\mathcal{D}_1\,\mathcal{D}_1\rangle$ of the displacement operator $\mathcal{D}_1$ in the CFT defined by the half-BPS fundamental Wilson line in planar $\mathcal{N}=4$ SYM, in a perturbative expansion for large 't~Hooft coupling $\lambda$. The string theory description of the theory at strong coupling \cite{Giombi:2017cqn} offers, in principle, a weakly coupled description through a QFT in AdS$_2$, but the computation of loop Witten diagrams quickly becomes challenging and it is therefore convenient to adopt a bootstrap approach, as already done in \cite{Liendo:2018ukf,Ferrero:2021bsb} for the same theory\footnote{See also \cite{Gimenez-Grau:2019hez,Bianchi:2020hsz} for bootstrap approaches to other defect theories.}. We shall employ the techniques originally developed in \cite{FernandoNotes,  Liendo:2018ukf,Ferrero:2019luz} for 1d CFTs, which postulate an ansatz for a perturbative four-point function in terms of a certain class of HPLs (see Appendix \ref{app:polylogs} for more details):
\begin{align}\label{genericansatz_intro}
G^{(\ell)}(\chi)=\sum_i r_i(\chi)\, \mathcal{T}_i(\chi)\,,
\end{align}
where $G^{(\ell)}(\chi)$ is the $\ell$-th order four-point function in terms of the cross-ratio $\chi$, $r_i$ are rational functions and $\mathcal{T}_i$ are the aforementioned polylogarithms. This is our first and main assumption, which has been already successfully tested in various contexts \cite{Liendo:2018ukf,Gimenez-Grau:2019hez,Ferrero:2019luz,Bianchi:2020hsz,Ferrero:2021bsb,Abl:2021mxo} and for which we shall provide more evidence in Section \ref{sec:bootstrap}.

Once the ansatz is established, the name of the game is to constrain the rational functions appearing in \eqref{genericansatz_intro}, and we have various types of constraints. First, we have functional constraints between the rational functions $r_i$ dictated by the crossing equations of $G^{(\ell)}(\chi)$. There are two such conditions: one coming from the invariance of the correlators under cyclic permutations of the operator insertions, which holds in general for 1d CFTs that can be equivalently defined on a line or on a circle, the second arising from swapping the positions of two neighboring operators, which on the other hand is not a genuine symmetry of 1d CFTs due to the absence of a continuous rotation group in one dimension. We will comment on how this symmetry, which we shall refer to as braiding, can be still employed in a holographic setting under certain assumptions. The third main assumption is related to the Regge limit, which for 1d CFTs was studied in \cite{Mazac:2018mdx,Mazac:2018ycv,Ferrero:2019luz} and is related to unitarity of the theory. This might be broken perturbatively, but we shall assume that at each order the behavior of correlators in the Regge limit is as mild as possible, and following \cite{Ferrero:2019luz} we characterize such growth in terms of the behavior of the anomalous dimensions $\gamma^{(\ell)}_{\Delta}$ developed at $\ell$-th order by operators of free theory dimension $\Delta$ for large $\Delta$. We find, empirically, that the mildest asymptotic behavior for this model corresponds to
\begin{align}
\gamma^{(\ell)}_{\Delta}\sim \Delta^{\ell+1}\,,\quad (\Delta \to \infty)\,,
\end{align}
at perturbative order $\ell$. As we shall discuss, this is well-understood for tree level results ($\ell=1$), but it constitutes an assumption for loop level results ($\ell>1$), although its origin is clearly related to the existence of solutions to the bootstrap equations corresponding to higher-derivative terms in the AdS$_2$ Lagrangian. Finally, a constraint comes from the requirement that the expansion of the correlators in the various OPE limits is compatible with the conformal blocks expansion.

Given these assumptions, as well as some additional external input (for example, three-point functions between one-half BPS operators known from localization \cite{Pestun:2007rz,Giombi:2009ds,Giombi:2018qox}), the method proceeds as follows. At each order, one can compute certain terms in $G^{(\ell)}(\chi)$ once the relevant CFT data at previous orders are known: such terms, which we shall refer to as ``highest logarithmic singularities'', are those that contain singularities $\log^k\chi$ with $k\ge 2$ in the small $\chi$ expansion. They are akin to terms generating double discontinuities in the language of the Lorentzian inversion formula \cite{Caron-Huot:2017vep}, and the idea is completely analogous to the AdS unitarity method of \cite{Aharony:2016dwx}. Once such terms are known, the functional constraints between the rational functions $r_i(\chi)$ arising from cyclicity and braiding fix the four-point functions unambiguously, up to terms which have the same structure (and obey the same crossing equations) as tree-level correlators, therefore containing at most $\log$ singularities, but no higher powers of $\log$. This is where the constraint on the Regge behavior enters the game, selecting one among the infinite solutions to the tree-level bootstrap problem. 

The method outlined above applies, in principle, at each perturbative order, provided that the relevant CFT data are known to compute the highest logarithmic singularities.  Here is where we encounter one of the main themes of this paper: the mixing problem, which we now briefly describe. Recall that we are focusing mainly on the $\langle \mathcal{D}_1\,\mathcal{D}_1\,\mathcal{D}_1\,\mathcal{D}_1\rangle$ four-point function. For this observable, the simplest instance of mixing occurs at one loop ($\ell=2$), where one has a $\log^2\chi$ logarithmic singularity arising in the $\chi\to 0$ OPE limit, with coefficient
\begin{align}\label{mixing_gamma1^2}
\sum_{\D}\langle a^{(0)}\,(\gamma^{(1)})^2\rangle_{\D}g_{\D}(\chi)\,,
\end{align}
where $\D$ are the free-theory dimensions of the exchanged operators, $\gamma^{(1)}$ their first-order anomalous dimensions and $a^{(0)}$ their squared OPE coefficients with pairs of external operators in the free theory, while $g_{\D}(\chi)$ are conformal blocks. The obstacle to the computation of this sum is the presence of degeneracy between operators in the free theory: for each value of $\D$, one has more than one operator, which is why we have used the average symbol $\langle \cdot\rangle$ in \eqref{mixing_gamma1^2}: for each $\D$, the quantity in the brackets is actually given by a sum of terms
\begin{align}\label{a0gamma1^2_intro}
\langle a^{(0)}\,(\gamma^{(1)})^2\rangle_{\D}:=\sum_{\mathcal{O}|\h^{(0)}_{\mathcal{O}}=\D}(\mu_{\mathcal{O}}^{(0)})^2\,(\gamma^{(1)}_{\mathcal{O}})^2\,,
\end{align}
where $\mathcal{O}$ are eigenstates of the dilatation operator and $\mu_{\mathcal{O}}$ their OPE coefficients with pairs of external operators\footnote{Our notation for OPE coefficients will be the following. We reserve the symbol $\mathtt{C}$ for OPE coefficients between three half-BPS operators, while we use $\mu$ when three-point functions involve at least one long multiplet.}. The issue is that if one were to compute only a single four-point function at tree level, the only information that could be obtained on the first order anomalous dimensions would be
\begin{align}\label{a0gamma1_intro}
\langle a^{(0)}\,\gamma^{(1)}\rangle_{\D}:=\sum_{\mathcal{O}|\h^{(0)}_{\mathcal{O}}=\D}(\mu_{\mathcal{O}}^{(0)})^2\,\gamma^{(1)}_{\mathcal{O}}\,,
\end{align}
for each $\Delta$, where the average is over the specific OPE coefficients associated to the specific four-point function that we considered. This way one would then only obtain one specific linear combination of the anomalous dimensions $\gamma^{(1)}_{\mathcal{O}}$, which is not enough to compute the average \eqref{a0gamma1^2_intro}. Instead, one should compute independently averages of the type \eqref{a0gamma1_intro} for a high enough number of correlators such as to allow to extract all $\gamma^{(1)}_{\mathcal{O}}$ for a given $\Delta$ individually: only then, it would be possible to compute \eqref{a0gamma1^2_intro}.

This is a common problem in perturbative CFTs and it is well known that a way to address it is to consider multiple correlators and to extract the quantity $\langle a^{(0)}\,\gamma^{(1)}\rangle_{\D}$ from each of them. If the OPE coefficients $\mu^{(0)}_{\mathcal{O}}$ appearing in \eqref{a0gamma1_intro} are independent in the various correlators, by considering a high enough number of them it is possible to obtain enough inequivalent averages $\langle a^{(0)}\,\gamma^{(1)}\rangle_{\D}$ that one can compute the actual anomalous dimensions $\gamma^{(1)}_{\mathcal{O}}$, thus allowing to compute the quantity in \eqref{a0gamma1^2_intro} and therefore giving access to the coefficient of the logarithmic singularity given in \eqref{mixing_gamma1^2}. 

A similar problem has already been discussed in various contexts, ranging from the $\epsilon$-expansion \cite{Alday:2017zzv,Carmi:2020ekr} to supersymmetric holographic setups \cite{Aprile:2017bgs,Alday:2017xua,Alday:2020tgi,Alday:2021ajh,Behan:2022uqr,Alday:2022rly}. The latter cases might seem, at least {\it a priori}, somewhat analogous to the one considered here, and the solution discussed by various authors is to consider four-point functions between half-BPS operators of arbitrary weight to ``unmix'' the anomalous dimensions and OPE coefficients. However, we have at least two differences compared to those cases. The first is that, as we shall discuss in more detail, the free theory degeneracy is not lifted at first order, so that operators with the same dimension $\Delta$ in the free theory all share the same anomalous dimension $\gamma^{(1)}_{\Delta}$ at tree level: the problem discussed above is therefore shifted to fourth perturbative order and in particular to the computation of $\langle a^{(0)}\,(\gamma^{(2)})^2\rangle_{\Delta}$, which is the actual challenge that we face in our computations. The second difference is that, as opposed to more standard holographic setups where the half-BPS operators are given by Kaluza-Klein (KK) modes that are all independent of each other, here higher-rank half-BPS operators $\mathcal{D}_p$ are all given by powers of the displacement, $\mathcal{D}_p\sim(\mathcal{D}_1)^p$, in the free theory at $\lambda=\infty$\footnote{Note that there is only one $\mathcal{D}_p$ for each $p$, see \cite{Ferrero:2023znz}.}. This implies that the OPE coefficients $\langle \mathcal{D}_p\mathcal{D}_p\mathcal{O}\rangle$ and $\langle \mathcal{D}_q\mathcal{D}_q\mathcal{O}\rangle$ can be thought of as parallel vectors in the degeneracy space of operators $\mathcal{O}$ with the same quantum numbers, even if $p\neq q$. Hence, the averages \eqref{a0gamma1_intro} computed on $\langle\mathcal{D}_p\mathcal{D}_p\mathcal{D}_q\mathcal{D}_q\rangle$ correlators are not enough to explore all $\gamma^{(1)}_{\mathcal{O}}$, since by increasing $p$ and $q$ one simply obtains the same equations over and over. We are then led to introduce a new method to address the mixing problem, that involves studying correlators with external unprotected operators (which we denote with $\mathcal{L}$ and will also refer to as long multiplets): $\langle\mathcal{D}_1\mathcal{D}_1\mathcal{D}_2\mathcal{L}\rangle$ and $\langle\mathcal{D}_1\mathcal{D}_1\mathcal{L}\mathcal{L}\rangle$. 

Another technical complication of our mixing problem, when compared to those that arise at first order in holography, is related to the dimension of the degeneracy spaces. In particular, so far in the literature the analysis has been limited to the mixing between double trace operators, and in that case the number of degenerate operators grows {\it linearly} with their conformal dimension in the generalized free theory. By contrast, in our case the relevant degenerate operators are such that their degeneracy grows {\it quadratically} with their dimension, so that the number of correlators that one has to consider in order to resolve the degeneracy must ``follow'' this quadratic growth. From this perspective, the family $\langle\mathcal{D}_1\mathcal{D}_1\mathcal{D}_2\mathcal{L}\rangle$ of correlators is particularly interesting: by taking the external operators $\mathcal{L}$ to be of the same type of the exchanged operators, we are guaranteed to have a family of correlators whose size is big enough as to assure that we can resolve the degeneracy for (in principle) arbitrary values of the conformal dimension, provided we consider enough independent choices for $\mathcal{L}$. The quadratic degeneracy is relevant to extract the relevant data to use as input for $\langle \mathcal{D}_1 \mathcal{D}_1 \mathcal{D}_1 \mathcal{D}_1\rangle$ at $\ell=4$, while at higher orders the degeneracy spaces will be even larger, but the same strategy applies to these more complicated cases as well. This fact also makes our approach more flexible than those used so far to study mixing problems in perturbative CFTs, and we expect that it should be useful in other contexts such as the mixing between triple trace operators in holography.

We are now ready to outline our main results, although we shall do so in an order which does not necessarily match the order of appearance in the paper (for that, see the next subsection). First, without dealing with any sort of mixing problems one can use our bootstrap strategy to compute $\langle\mathcal{D}_p\mathcal{D}_p\mathcal{D}_q\mathcal{D}_q\rangle$ for all $p,q$ up to one loop ($\ell=2$). The tree-level result is a key ingredient for the proof that the free theory degeneracy between operators is not lifted at first order, and that in particular the tree level anomalous dimensions of all operators are proportional to the quadratic superconformal Casimir eigenvalue $\mathfrak{c}_{\mathfrak{osp}(4^*|4)}$ corresponding to their representation:
\begin{align}\label{gamma1casimir_intro}
\gamma^{(1)}_{\mathcal{O}}=-\frac{1}{2}\mathfrak{c}_{\mathfrak{osp}(4^*|4)}(\omega_{\mathcal{O}})\,,
\end{align}
where we have stressed that the result only depends on the weights $\omega_{\mathcal{O}}$ associated with the representation under which $\mathcal{O}$ transforms in the free theory, not on the specific choice of $\mathcal{O}$ for given weight $\omega$. This allows to compute the one loop results with almost no conceptual effort, and the latter provide some partial input for the mixing problem. An analogous relation was observed to hold for the exchanged operators in the displacement four-point function on the half-BPS Wilson line in ABJM at strong coupling \cite{Bianchi:2020hsz}, and it would be interesting to investigate if it persists in the case of half-BPS surface defects in the 6d $\mathcal{N}=(2,0)$ CFTs \cite{Drukker:2020swu}.

The result \eqref{gamma1casimir_intro} can be then used as an input for the study of $\langle\mathcal{D}_1\mathcal{D}_1\mathcal{D}_2\mathcal{L}\rangle$, which we determine at tree level and one loop up to some unfixed parameters that are not relevant for our purposes, as we shall discuss. We remark here that this is, as far as we are aware, the first instance of analytic bootstrap involving long multiplets. By considering different choices for the external long operator $\mathcal{L}$ one is able to compute enough averages $\langle a^{(0)}\gamma^{(2)}\rangle_{\D}$, for some $\D\le \D_{\text{max}}$, that the quantity $\langle a^{(0)}(\gamma^{(2)})^2\rangle_{\D}$ can also be computed for $\D\le \D_{\text{max}}$. Note that this assumes that the free theory OPE coefficients are known: while in principle they are simply computed using Wick contractions, the explicit determination of superconformal primary operators in terms of fundamental fields in the AdS$_2$ Lagrangian description is a non-trivial problem which we solved in \cite{Ferrero:2023znz}, allowing to simply borrow those results here.

To quote here the simplest instance of the mixing problem, let us focus on long multiplets that are singlets under the R-symmetry and transverse spin subalgebras of $\mathfrak{osp}(4^*|4)$. From the analysis of \cite{Ferrero:2023znz}, it is clear that in the sector with $\D=4$ at strong coupling there are only two such operators. At $\lambda=\infty$ they are exactly degenerate, so one can choose any basis to describe them and we find it convenient to use a basis where each operator has fixed ``length'' in terms of the AdS$_2$ fields. Their schematic form, in terms of the scalar fields $\varphi^a$ which are the superconformal primaries of the $\mathcal{D}_1$ multiplet, is\footnote{In the rest of the paper, we are going to distinguish between two different types of basis for degeneracy spaces of the free theory. Operators that diagonalize the dilatation operator are going to be denote with $\mathcal{O}$, while a basis of operators that have well-defined length, but are not necessarily eigenstates of the dilatation operator, are going to be denoted with $\widehat{\mathcal{O}}$. The distinction will be particularly important in Section \ref{sec:mixing}.}
\begin{align}
\widehat{\mathcal{O}}^{(\D=4)}_{L=2}\sim \varphi\partial^2\varphi\,,\qquad\widehat{\mathcal{O}}^{(\D=4)}_{L=4}\sim \varphi^4\,.
\end{align}
Since the dilatation operator is proportional to the identity at $\lambda=\infty$ as well as at first order in $1/\sqrt{\lambda}$, any linear combination of these two operators is a good eigenstate. However, this structure is broken by perturbations at second order and only specific linear combinations of such two operators are actual eigenstates of the dilatation operator. Let us refer to these as ${\mathcal{O}}^{(\D=4)}_{1}$ and ${\mathcal{O}}^{(\D=4)}_{2}$: then, by also computing $\langle \mathcal{D}_1\mathcal{D}_1\mathcal{D}_q\mathcal{D}_q\rangle$ and $\langle \mathcal{D}_2\mathcal{D}_2\mathcal{D}_2\mathcal{D}_2\rangle$ at third order, we are able to diagonalize the dilatation operator and find, for their dimensions,
\begin{align}
\begin{split}
h[{\mathcal{O}}^{(\D=4)}_{1}]&=4-\frac{14}{\lambda^{1/2}}+\frac{7\,(1033-\sqrt{62569})}{120\,\lambda}-\frac{7\,(673805561-1581637 \sqrt{62569})}{22524840\,\lambda^{3/2}}+\ldots\,,\\
h[{\mathcal{O}}^{(\D=4)}_{2}]&=4-\frac{14}{\lambda^{1/2}}+\frac{7\,(1033+\sqrt{62569})}{120\,\lambda}-\frac{7\,(673805561+1581637 \sqrt{62569})}{22524840\,\lambda^{3/2}}+\ldots\,,\\
\end{split}
\end{align}
while for their squared OPE coefficients with half-BPS operators, which we are able to resolve up to first order\footnote{As we will stress in the main text, due to the fact that the degeneracy is unlifted at first order, the resolution of the mixing problem for $\gamma^{(\ell)}$ is tied to the OPE coefficients at order $\ell-2$, rather than $\ell-1$ as more commonly found.}, we find
\begin{align}\label{OPEs_Delta=4sector_intro}
\begin{split}
\mu^2_{pp{\mathcal{O}}^{(\D=4)}_{1}}=&\frac{(43+120p-\sqrt{62569})^2\,p^2}{875966-2282 \sqrt{62569}}+p^2\Big[\frac{389-249p+219p^2-108p^3}{630}\\
&-\frac{510526661+5856154953 p-5169773793 p^2+1101464676 p^3}{39418470 \sqrt{62569}}\Big]\frac{1}{\lambda^{1/2}}+\ldots\,,\\
\mu^2_{pp{\mathcal{O}}^{(\D=4)}_{2}}=&\frac{(43+120p+\sqrt{62569})^2\,p^2}{875966+2282 \sqrt{62569}}+p^2\Big[\frac{389-249p+219p^2-108p^3}{630}\\
&+\frac{510526661+5856154953 p-5169773793 p^2+1101464676 p^3}{39418470 \sqrt{62569}}\Big]\frac{1}{\lambda^{1/2}}+\ldots\,.
\end{split}
\end{align}
These results are found to agree perfectly with the numerical analysis of \cite{Cavaglia:2021bnz,Cavaglia:2022qpg} after a numerical fit at strong coupling\footnote{We thank the authors of those papers for sharing the relevant data with us.}.

Finally, our main result and the reason for considering such a complicated mixing problem, is the computation of the four-point function $\langle\mathcal{D}_1\mathcal{D}_1\mathcal{D}_1\mathcal{D}_1\rangle$ at three loops, or fourth perturbative order ($\ell=4$). While the explicit result, which we give in \eqref{final1111_3loop}, may look complicated at first sight, we would like to emphasize that it is in fact much simpler than it might be {\it a priori}. One first simplification is that the degree of transcendentality of four-point functions in this model only increases by one unit at each perturbative order. In particular, the three-loop results only have transcendentality four, which is the same as for one loop results in a toy model with no supersymmetry \cite{Ferrero:2019luz}. Moreover, while a basis of the HPLs we are interested in for transcendentality four contains 31 functions, while we find it possible to express our final result only in terms of products of $\log$ and a weight-three function called $L_3$ (defined in \eqref{L3_firsttime}), evaluated at arguments $\chi$ and $1-\chi$. This form of the result highlights the fact that the functions $\Li_2$ and $\Li_4$, while in principle allowed by transcendentality, actually drop out of the final result. We will make some comments and speculations on the properties of the functions appearing in the three-loop result in the main text. 

Finally, let us mention that while the resolution of mixing at two and three loops is beyond the scope of our work, we can still obtain exact results for the conformal dimension and OPE coefficient associated with the only singlet multiplet that is not degenerate in the free theory, which we denote schematically with $\varphi^2$:
\begin{align}
\varphi^2\,\,\, \longleftrightarrow\,\,\, \mathcal{O}_{L=2}^{(\D=2)}\,.
\end{align}
The result, already anticipated in \cite{Ferrero:2021bsb}, is\footnote{Note that we also know the OPE coefficient between two $\mathcal{D}_p$ operators and $\varphi^2$ up to two loops:
\begin{align}
\begin{split}
\mu^2_{pp\varphi^2}=&p^2\Big[\frac{2}{5}-\frac{7+36p}{30\,\lambda^{1/2}}+\frac{-163+123p+90p^2}{60\,\lambda}\\
&+\left(\frac{42115+51876p-32400p^2-5616p^3}{8640}-2(-4+p+p^2)\zeta(3)\right)\frac{1}{\lambda^{3/2}}+\ldots\Big]\,.
\end{split}
\end{align}}
\begin{align}
\begin{split}
h[\varphi^2]&=2-\frac{5}{\sqrt{\lambda}}+\frac{295}{24} \frac{1}{\lambda}-\frac{305}{16} \frac{1}{\lambda^{3 / 2}}+\left(\frac{351845}{13824}-\frac{75}{2} \zeta(3)\right) \frac{1}{\lambda^2}+\ldots\,,\\
\mu^2_{11\varphi^2}&=\frac{2}{5}-\frac{43}{30 \sqrt{\lambda}}+\frac{5}{6 \lambda}+\left(\frac{11195}{1728}+4 \zeta(3)\right) \frac{1}{\lambda^{3 / 2}}-\left(\frac{1705}{96}+\frac{1613}{24} \zeta(3)\right) \frac{1}{\lambda^2}+\ldots\,.
\end{split}
\end{align}
Both of these results match beautifully the numerical predictions of \cite{Grabner:2020nis,Cavaglia:2021bnz,Cavaglia:2022qpg} at large $\lambda$.

\subsection{Guide to the reader}

Let us conclude this introduction by presenting a plan of the paper, in the form of a guide to the reader that can help navigate through both this work and \cite{Ferrero:2023znz}. We have divided the material in such a way that \cite{Ferrero:2023znz} contains all the information about the $\mathfrak{osp}(4^*|4)$ superconformal algebra and its representation theory that is relevant for this paper, as well as the study of the superconformal kinematics and superconformal blocks that we will use to bootstrap four-point functions. The reader is invited to refer to that paper for all such derivations. Moreover, in addressing the mixing problem an essential piece of information comes from the understanding of the spectrum of the defect CFT at $\lambda=\infty$, which is made possible by the countings performed in \cite{Ferrero:2023znz}. Once again, that paper contains all derivations while here we shall limit to refer to certain results.

In Section \ref{sec:setup} we review the holographic description of the half-BPS defect CFT at strong coupling, focusing on the implications of the latter on the structure of the dilatation operator in perturbation theory, whose understanding is crucial for the mixing problem. Then in Section \ref{sec:bootstrap} we will discuss some features of 1d CFTs and review the bootstrap method developed in \cite{FernandoNotes,Liendo:2018ukf,Ferrero:2019luz}, with the aim of giving a self-contained presentation where all the assumptions are spelled out in detail. We will also make some general comments about mixing. We shall then apply this machinery to the bootstrap of the displacement four-point function $\langle \mathcal{D}_1\mathcal{D}_1\mathcal{D}_1\mathcal{D}_1\rangle$ up to fourth order in Section \ref{sec:1111}. The computation is straightforward provided we assume two results: the fact that the first-order perturbation does not lift the free theory degeneracy and the fact that we are able to ``solve'' the mixing problem at second order, at least in the sense that we can compute all quantities that are necessary for the bootstrap of $\langle \mathcal{D}_1\mathcal{D}_1\mathcal{D}_1\mathcal{D}_1\rangle$ at fourth order. We then deal with these two assumptions separately. In Section \ref{sec:ppqq} we consider the bootstrap of more general $\langle \mathcal{D}_p\mathcal{D}_p\mathcal{D}_q\mathcal{D}_q\rangle$ correlators up to second order: besides revealing some interesting structures and results, the analysis proves our claims on the structure of the dilatation operator at first order. Finally, in Section \ref{sec:mixing} we address the mixing problem at second order and describe a method which uses $\langle \mathcal{D}_1\mathcal{D}_1\mathcal{D}_2\mathcal{L}\rangle$ correlators to access certain information about the second order dilatation operator, in particular allowing to compute the quantity $\langle a^{(0)}\,(\gamma^{(2)})^2\rangle_{\Delta}$ which is necessary for the bootstrap of $\langle \mathcal{D}_1\mathcal{D}_1\mathcal{D}_1\mathcal{D}_1\rangle$ at fourth order. We end with some discussion and outlook in Section \ref{sec:discussion}. Our presentation is complemented with various appendices. In Appendix \ref{app:polylogs} we give a general review of HPLs as well as a discussion of the specific functions appearing in this paper; in Appendix \ref{app:perturbativeOPE} we give some technical details on the structure of the OPE for perturbative 1d CFTs, with particular emphasis on perturbations around points where the theory admits braiding symmetry; in Appendix \ref{app:bootstrapresults} we summarize some of the results of our bootstrap analysis for various correlators and CFT data on the Wilson line at strong coupling.

\section{Setup}\label{sec:setup}

Let us begin by giving more details about the problem at hand. We will start by defining the 1d CFT that we are interested in terms of a line operator in 4d $\mathcal{N}=4$ SYM and then focus on the regime of parameters of interest for our work: the planar limit ($N\to \infty$) in an expansion for large 't~Hooft coupling $\lambda$, where the model admits an holographic Lagrangian description in AdS$_2$. To end the section, we will work out the implications of the structure of such Lagrangian on the dilatation operator of the theory at strong coupling, making some observations that will be instrumental for the analysis of the mixing problem carried out in Section \ref{sec:mixing}.

\subsection{The half-BPS Wilson line}

We consider the one-dimensional CFT defined by the insertion of local operators along the one-half BPS Maldacena Wilson line \cite{Maldacena:1998im} in 4d $\mathcal{N}=4$ SYM, defined as 
\begin{align}
\mathcal{W}_{\mathcal{C}}=\frac{1}{N}\text{tr}\,\text{P}\,\text{exp}\int_{\mathcal{C}}\text{d}t\,
\left(i\,A_{\mu}\,\dot{x}^{\mu}+\Phi^6\,|\dot{x}|\right)\,,
\end{align}
where the contour $\mathcal{C}$ parametrized by $x^{\mu}(t)$ is either a circle or a straight line, the trace is taken in the fundamental representation of the gauge group $SU(N)$ and $\text{P}$ denotes the path-ordering. Here $A_{\mu}$ is the SYM gauge connection while $\Phi^6$ is one of the six fundamental scalars of the theory: the arbitrary choice of coupling one out of six scalars to the line manifestly breaks the R-symmetry from $SO(6)$ to $SO(5)$. Moreover, a rotational symmetry $SO(3)$ around the line is preserved, as well as a 1d conformal symmetry $SL(2)$ and sixteen supercharges: all in all, this line defect is known to be invariant under the superconformal algebra $\mathfrak{osp}(4^*|4)$ \cite{Drukker:2005af}. 

Correlation functions in the defect CFT are defined by \cite{Drukker:2006xg}
\begin{align}
\begin{split}
\langle\langle \mathcal{O}_1(t_1)\dots \mathcal{O}_n(t_n)\rangle\rangle&=
\frac{\langle\text{tr}\,\text{P}\, \mathcal{O}_1(t_1)\,\mathcal{W}_{t_1,t_2}\,\mathcal{O}_2(t_2)\ldots \mathcal{O}_n(t_n)\,\mathcal{W}_{t_n,t_1}\rangle}{\langle \mathcal{W}_{\mathcal{C}}\rangle}\\
&=\frac{\langle\text{tr}\,\text{P}\, \mathcal{O}_1(t_1)\,\mathcal{O}_2(t_2)\ldots \mathcal{O}_n(t_n)\,\mathcal{W}_{\mathcal{C}}\rangle}{\langle \mathcal{W}_{\mathcal{C}}\rangle}\,,
\end{split}
\end{align}
where $\mathcal{O}_i$ are composite operators transforming in the adjoint representation of the gauge group $SU(N)$ (see \cite{Ferrero:2023znz} for more details), while $\mathcal{W}_{t_i,t_j}$ are segment of the Wilson loop $\mathcal{W}_{\mathcal{C}}$ connecting the position $t_i$ and $t_j$ along the line. In this paper we are particularly interested in the case of planar $\mathcal{N}=4$ SYM
\begin{align}
N\to \infty\,, \qquad g^2_{YM}\,N\equiv \lambda\,\,\,\text{fixed}\,,
\end{align}
and in the perturbative expansion for large 't~Hooft coupling $\lambda$. In this limit, the theory has an holographic description in terms of semiclassical strings in AdS$_5\times S^5$, which we review in the next subsection.

The relevant details of the superalgebra $\mathfrak{osp}(4^*|4)$ and its representation theory are collected in the companion paper \cite{Ferrero:2023znz}, where we also study the spectrum of the defect theory at weak ($\lambda=0$) and strong ($\lambda=\infty$) coupling. Here we briefly review some important results of that analysis. We label operators according to their $\mathfrak{osp}(4^*|4)$ weights $\omega=\{\h,s,[a,b]\}$, where $\h$ is the $\mathfrak{sl}(2)$ conformal dimension, $s$ labels the $\mathfrak{su}(2)$ transverse spin and $[a,b]$ are $\mathfrak{sp}(4)$ Dynkin labels. We want to emphasize here that at strong coupling only two types of supermultiplets are relevant, namely absolutely protected half-BPS multiplets
\begin{align}
\mathcal{D}_k\,,\qquad \omega(\mathcal{D}_k)=\{k,0,[0,k]\}\,,
\end{align}
and generic long multiplets
\begin{align}
\mathcal{L}^{\h}_{s,[a,b]}\,,\qquad \omega(\mathcal{L}^{\h}_{s,[a,b]})=\{\h,s,[a,b]\}\,,\qquad \h\ge a+b+1+\tfrac{1}{2}s\,,
\end{align}
whose dimension is not protected against perturbative corrections. Supermultiplets of other type do not appear in the strong coupling perturbation theory, and in particular this is true for those semi-short multiplets that could potentially undergo recombination: the latter phenomenon is therefore absent at strong coupling.

\subsection{The holographic description}

The half-BPS Wilson loop in the fundamental representation has a long history, dating back to \cite{Maldacena:1998im}, where it was first observed that such Wilson loop in planar 4d $\mathcal{N}=4$ SYM with strong 't~Hooft coupling $\lambda$ is dual to a semiclassical string in AdS$_5\times S^5$, whose worldsheet is a minimal surface AdS$_2\subset$ AdS$_5$, ending on the contour defining the loop at the boundary. The holographic description of Wilson loops in other representations of the gauge group is more complicated and involves D3 and D5-branes, see \cite{Drukker:2005kx,Gomis:2006sb,Gomis:2006im}.

In this paper we are interested in correlation functions between local operators located along the contour of the Wilson line, or in other words we will only focus on defect observables and not in bulk ones. The holographic description of the defect degrees of freedom was found in \cite{Giombi:2017cqn}, which we now briefly review. Focusing for simplicity on the bosonic sector, the embedding of the superstring dual to the Wilson loop in AdS$_5\times S^5$ is described by three coordinates $x^i$, $i=1,2,3$, parametrizing fluctuations of the string inside AdS$_5$, as well as by five coordinates $y^a$, $a=1,\ldots 5$, accounting for fluctuations along the internal $S^5$. Note that in terms of the components of the displacement supermultiplet\footnote{We denote by $\h$ the conformal dimension of operators in the CFT and the subscripts on the fields denote the dimension of the corresponding representations under the compact bosonic subalgebra $\mathfrak{su}(2)\oplus \mathfrak{sp}(4)\subset\mathfrak{osp}(4^*|4)$.}
\begin{align}
\mathcal{D}_1: \quad \varphi_{(\mathbf{1}, \mathbf{5})}^{\h=1} \longrightarrow \Psi_{(\mathbf{2}, \mathbf{4})}^{\h=3 / 2} \longrightarrow f_{(\mathbf{3}, \mathbf{1})}^{\h=2}\,,
\end{align}
the bosonic coordinates of the superstring are identified with the two most important operators of the defect CFT: the displacement $f$, dual to $x^i$ and related to the breaking of 4d conformal invariance, and the tilt $\varphi$, dual to $y^a$ and related to the breaking of the $SO(6)$ R-symmetry of $\mathcal{N}=4$ SYM.

The local dynamics of the excitations $x^i$ and $y^a$ (as well as their fermionic partner in $\mathcal{D}_1$, via supersymmetry) is captured by the Nambu-Goto action of type IIB superstring theory, where the tension $T$ of the string is related to the 't~Hooft coupling $\lambda$ by the usual relation $T=\frac{\sqrt{\lambda}}{2\pi}$. Following \cite{Giombi:2017cqn} and expanding the action for small fluctuations, one obtains an AdS$_2$ Lagrangian that begins at quadratic order in the fields and contains an infinite number of vertices with higher and higher number of derivatives and fields, whose coupling constant is $1/\sqrt{\lambda}$. This allowed the authors of \cite{Giombi:2017cqn} to compute the first-order correction to $\langle\mathcal{D}_1\mathcal{D}_1\mathcal{D}_1\mathcal{D}_1\rangle$ at large $\lambda$ using tree-level Witten diagrams, a result that was later reproduced in \cite{Liendo:2018ukf} using the bootstrap methods that we also use in this paper.

Since here we will also adopt a bootstrap approach, we will not be interested in the details of the Lagrangian, but only in certain qualitative features that we now discuss (but see \cite{Giombi:2017cqn} for more details). Moreover, since (at least in principle) the Lagrangian is fully fixed by $\mathfrak{osp}(4^*|4)$ invariance once the terms involving fields $y^a$ are known, we shall focus on the interactions between the $y^a$'s in order to describe the features of the Lagrangian that we are interested in. We then schematically write
\begin{align}\label{AdSlagrangian}
\mathcal{L}_{\text{AdS}_2}\sim \sum_{\ell=0}^{\infty}\frac{1}{\lambda^{\ell/2}}(\partial y^a)^{2(\ell+1)}\,,
\end{align}
where $\ell=0$ gives the kinetic term for a massless field in AdS$_2$ and $\ell>0$ corresponds to an infinite tower of interactions between an even number of fields (so we have a symmetry $y\to -y$), which are such that contact terms involving $2(\ell+1)$ fields contain a number of derivatives ranging from 0 to $2(\ell+1)$ (note that in \eqref{AdSlagrangian} we have only emphasized the contribution of the terms with the highest number of derivatives). In particular, $\ell=1$ corresponds to quartic contact terms that contain at most four derivatives. Due to this structure, at $\ell$-th perturbative order all correlation functions in the theory can be extracted from the $(2\ell+2)$-point function
\begin{align}\label{npointPhi}
\langle\Phi(1)\ldots \Phi(2\ell+2)\rangle^{(\ell)}\,,
\end{align}
where $\Phi$ is the superfield representing the $\mathcal{D}_1$ multiplet in superspace. This is due to the structure of the spectrum of the theory at $\lambda=\infty$ \cite{Ferrero:2023znz}, which we recall below, and to the fact that $\langle\Phi(1)\ldots \Phi(n)\rangle^{(\ell)}$ is given by a sum of disconnected Witten diagrams for all $n>2\ell+2$, as a consequence of the Lagrangian \eqref{AdSlagrangian}.

As discussed in \cite{Ferrero:2021bsb}, the fact that the dynamics of the theory is fully captured holographically by a Lagrangian which only involved fields transforming as components of the $\mathcal{D}_1$ multiplet implies that, at least at strong coupling $\lambda=\infty$, all states can be built as graded-symmetrized tensor products of states in $\mathcal{D}_1$, where we are identifying operators and states by the usual state-operator correspondence. More precisely, associated with $\mathcal{D}_1$ is a vector space $\mathbb{V}_{\Phi}$ and the Hilbert space of the theory at strong coupling is built as
\begin{align}\label{Hilbertspacestrong}
\mathcal{H}^{\text {strong }}=\bigoplus_L \mathcal{H}_L^{\text {strong }}, \quad \mathcal{H}_L^{\text {strong }}=(\underbrace{\mathbb{V}_{\Phi} \otimes \cdots \otimes \mathbb{V}_{\Phi}}_{L \text { times }})^{S_L}\,,
\end{align}
where $S_L$ denotes graded symmetrization, and we have introduced a quantum number $L$ which denotes the {\it length} of states/operators. Note that while to each state belongs to a certain representation $\mathcal{R}$ of $\mathfrak{osp}(4^*|4)$, this identification is not one-to-one, as two inequivalent states might belong to the same representation. We will express this fact by saying that there is a degeneracy of operators in the free theory at $\lambda=\infty$, which is generically broken by perturbations (and in particular is expected to be fully broken in the interacting theory). Accordingly, we can write the Hilbert spaces of operators with fixed length as
\begin{align}\label{degeneracyspaces}
 \mathcal{H}_L^{\text {strong }}=\bigoplus_\mathcal{R}\mathtt{d}_L^{\text {strong }}(\mathcal{R})\otimes \mathcal{R}\,,
\end{align}
where $\mathtt{d}_L^{\text {strong }}(\mathcal{R})$ are multiplicity spaces, whose dimension we study in \cite{Ferrero:2023znz} for certain representations. Following \cite{Ferrero:2021bsb}, we specify representations with their weight $\omega=\{\h,s,[a,b]\}$ under the bosonic subalgebra, where $\h$ is the conformal dimension, $s$ the transverse spin and $[a,b]$ are $\mathfrak{sp}(4)$ Dynkin labels. See \cite{Ferrero:2023znz} for a more detailed account of our conventions and a description of the superconformal algebra $\mathfrak{osp}(4^*|4)$ and its representations. Focusing on singlet supermultiplets, with $\omega=\{\h,0,[0,0]\}$, let us repeat here a result that is derived in \cite{Ferrero:2023znz}, namely the dimension of the degeneracy spaces for such multiplets. For general $L$, we have the following scaling at large $\h$:
\begin{align}
\text{dim}\left(\mathtt{d}_L^{\text {strong }}(\mathcal{L}^{\h}_{0,[0,0]})\right)\sim \h^{L-2}\,,\qquad (\h\to \infty)\,,
\end{align}
and in particular 
\begin{align}\label{degeneracyfreeL23}
\text{dim}\left(\mathtt{d}_{L=2}^{\text {strong }}(\mathcal{L}^{\h}_{0,[0,0]})\right)=1\,,\qquad 
\text{dim}\left(\mathtt{d}_{L=3}^{\text {strong }}(\mathcal{L}^{\h}_{0,[0,0]})\right)=0\,,
\end{align}
meaning that there is a unique singlet long supermultiplet of length two for each $\h$, while there is none of length three. At length four, we find
\begin{align}\label{degeneracyfreeL4}
\text{dim}\left(\mathtt{d}_{L=4}^{\text {strong }}(\mathcal{L}^{\h}_{0,[0,0]})\right)=
\begin{cases}
\lfloor (\tfrac{h}{4})^2\rfloor\,,\qquad h\ge 4\,\,\,\text{even}\,,\\
\lfloor (\tfrac{h-3}{4})^2\rfloor\,,\quad h\ge 3\,\,\,\text{odd}\,.
\end{cases}
\end{align}
The explicit construction of singlet supermultiplets with length $L=2,4$ is discussed in \cite{Ferrero:2023znz}.

The setup is reminiscent of the study of holographic correlators in large $N$ CFTs with AdS duals, where one has a weakly coupled description of the field theory in the regime of large central charge, which is a Lagrangian field theory in AdS. However, in all controllable examples of the AdS/CFT correspondence one has an internal space with size comparable to the AdS radius, so that the Lagrangian contains an infinite amount of fields, arising from a Kaluza-Klein (KK) reduction on the internal space. Moreover, the fundamental fields in the AdS Lagrangian description are dual to ``single-trace'' or ``single-particle'' operators in the CFT. The OPE between two single-trace operators then contains both single and double-trace operators. Here the situation is quite different: there is no internal space\footnote{There would be an internal $S^2$ if we were to study Wilson lines in other representations of the gauge group, which would make the setup closer to the standard one for holographic correlators.}, hence no KK modes and only a single multiplet in the AdS Lagrangian. Moreover, as we discuss in \cite{Ferrero:2023znz}, in the gauge theory description of this model there are only {\it open-trace} operators at large $N$, which form a closed subsector under the OPE. This makes the theory a well-defined 1d CFT at $N=\infty$, where the only perturbative parameter is $\lambda$. Studying perturbative $1/N$ corrections is of course an interesting problem, but it is not necessary in this sense.

\subsection{The dilatation operator}\label{sec:dilatationoperator}

While in the rest of this paper we will never make use of the detailed expression for the Lagrangian \eqref{AdSlagrangian}, we can use its rough structure to draw some important conclusions on the structure of the dilatation operator in the strong coupling perturbation theory, which complements and extends the discussion of \cite{Ferrero:2021bsb}. As we discussed at length in \cite{Ferrero:2023znz} and summarized in the previous subsection, a convenient way to arrange the spectrum of the theory at $\lambda=\infty$ is to consider an additional quantum number on top of the $\mathfrak{osp}(4^*|4)$ weights of each supermultiplet, namely the length $L$ of states. At each perturbative order, correlation functions between such operators can be then computed (at least in principle) using Witten diagrams: even though length is a conserved quantum number only in the free theory, it still makes sense to consider correlation functions between operators of well-defined length. Moreover, when degeneracy is present, one is free to choose any arbitrary basis for each degeneracy space, and again we will find it convenient to work in a basis that does not mix operators of different lengths. Such a basis does not necessarily correspond to a basis of eigenstates of the dilatation operator, hence two-point functions will not necessarily be diagonal in the length basis. In particular, associated with each degeneracy space $\mathtt{d}(\D)$, we can define a matrix that we will refer to as the {\it anomalous dimensions matrix} $\Gamma_{\D}$\footnote{Note that we are suppressing the labels of $\mathfrak{su}(2)\oplus \mathfrak{sp}(4)$ since they will not play a role in this discussion: operators with different quantum numbers cannot possibly mix.}, whose entries at each perturbative order can be read off from the two-point functions between the basis elements of $\mathtt{d}(\D)$. To be more precise, let $\mathcal{O}_1$ and $\mathcal{O}_2$ be two eigenstates of the dilatation operator at the non-perturbative level, with dimension $\h_1(\lambda)$ and $\h_2(\lambda)$ respectively, sharing the same dimension $\D$ in the free theory: $\h_1(\infty)=\h_2(\infty)=\D$. Then the two-point functions read
\begin{align}
\langle \mathcal{O}_{\alpha}(1)\mathcal{O}_{\beta}(2)\rangle=\begin{pmatrix}
\frac{1}{t_{12}^{2\h_1(\lambda)}} & 0\\
0 & \frac{1}{t_{12}^{2\h_2(\lambda)}}
\end{pmatrix}\,,
\end{align}
for $\alpha,\beta=1,2$. Expanding perturbatively, we find
\begin{align}
\begin{split}
\langle \mathcal{O}_{\alpha}(1)\mathcal{O}_{\beta}(2)\rangle&=\frac{1}{t_{12}^{2\D}}\left[\delta_{\alpha\beta}-\,\begin{pmatrix}
\tfrac{1}{\lambda^{1/2}}\gamma^{(1)}_1+\tfrac{1}{\lambda}\gamma^{(2)}_1+\ldots & 0\\
0 & \tfrac{1}{\lambda^{1/2}}\gamma^{(1)}_2+\tfrac{1}{\lambda}\gamma^{(2)}_2+\ldots
\end{pmatrix}_{\alpha\beta}\,\log t^2_{12}
+\mathcal{O}(\log^2 t^2_{12})
\right]\,,\\
&:=\frac{1}{t_{12}^{2\D}}\left[\delta_{\alpha\beta}-\,(\Gamma_{\D})_{\alpha\beta}\,\log t^2_{12}
+\mathcal{O}(\log^2 t^2_{12})
\right]\,,\\
\end{split}
\end{align}
where we are neglecting powers of $\log t^2_{12}$ higher than one and we have defined the matrix $\Gamma_{\alpha\beta}$, which is the anomalous dimensions matrix in the basis of eigenstates. Now, if instead of a basis of eigenstates of the dilatations we chose an arbitrary basis via
\begin{align}
\widehat{\mathcal{O}}_{\alpha}=M_{\alpha}^{\,\,\,\beta}\mathcal{O}_{\beta}\,,
\end{align}
then we would obtain the two-point functions 
\begin{align}
\langle \widehat{\mathcal{O}}_{\alpha}(1)\widehat{\mathcal{O}}_{\beta}(2)\rangle=\frac{1}{t_{12}^{2\D}}\left[(MM^T)_{\alpha\beta}+(M\,\Gamma_{\D} M^T)_{\alpha\beta}\,\log t_{12}^2+\mathcal{O}(\log^2t_{12}^2)\right]\,,
\end{align}
which allows us to defined the anomalous dimensions matrix
\begin{align}
\widehat{\Gamma}_{\D}=M\,\Gamma_{\D} M^T\,,
\end{align}
in the new basis chosen for the degeneracy space, which is no longer diagonal due to the fact that $\widehat{\mathcal{O}}_{\alpha}$ are no longer eigenstates of dilatations. In the rest of this paper, we will choose a basis where operators have well-defined length and we will use the symbol $\widehat{\mathcal{O}}_{\alpha}$ to stress that we are working in such basis. Clearly then, in this basis $\widehat{\Gamma}_{\D}$ admits a decomposition in blocks that act on degeneracy spaces of operators with fixed length\footnote{Since $\widehat{\Gamma}_{\D}$ is a symmetric matrix we have that $ \widehat{\Gamma}_{\D,L_1\to L_2}=\left( \widehat{\Gamma}_{\D,L_2\to L_1}\right)^T$.}:
\begin{align}\label{defGammaDelta}
\widehat{\Gamma}_{\D}=\sum_{L_1,L_2} \widehat{\Gamma}_{\D,L_1\to L_2}\,,\qquad \widehat{\Gamma}_{\D,L_1\to L_2}:\,\,\,\mathtt{d}_{L_1}(\D)\to\mathtt{d}_{L_2}(\D)\,.
\end{align}
It then makes sense to ask the following question: which of the $\widehat{\Gamma}_{\D,L_1\to L_2}$ are non-zero at each perturbative order? 

\begin{figure}[h!]
\begin{center}
        \includegraphics[width=0.3\textwidth]{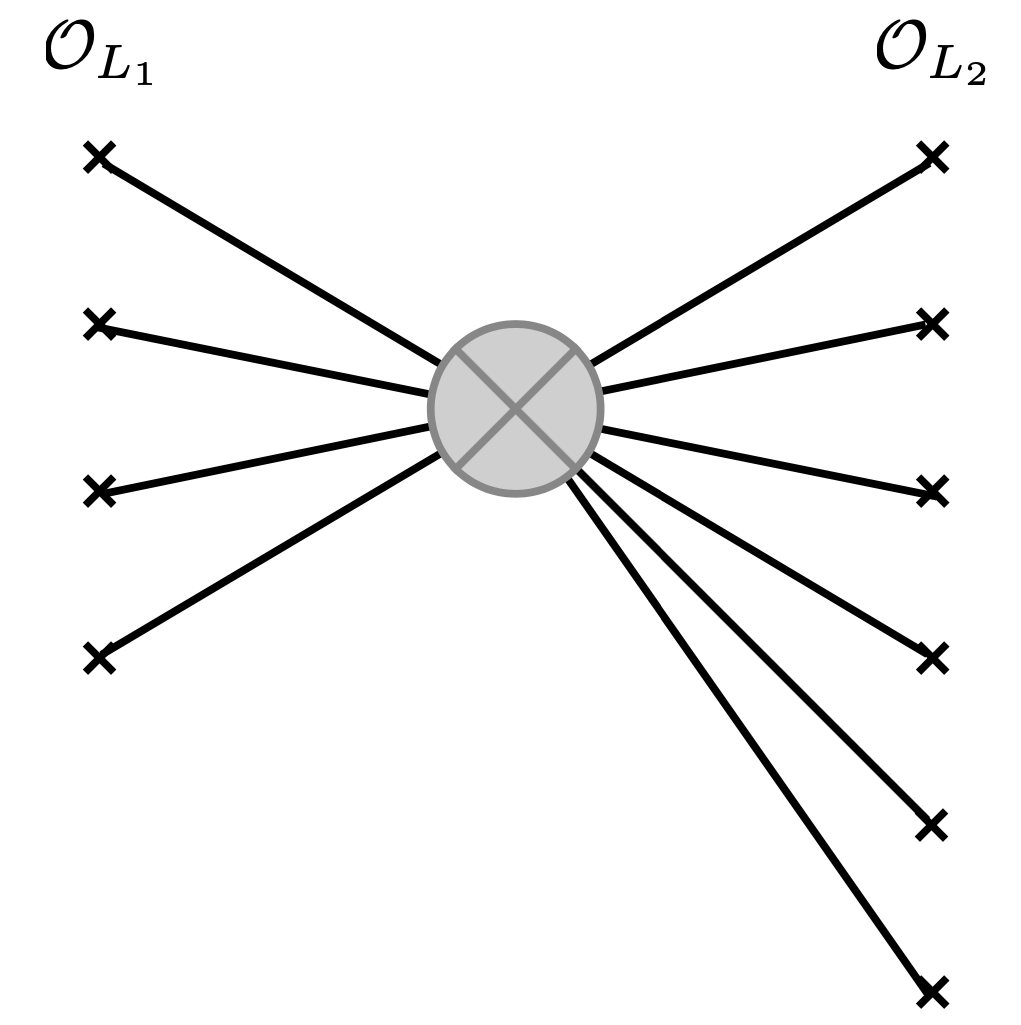}
\caption{Schematic representation of the two-point function between operators ${\mathcal{O}}_{L_1}$ and ${\mathcal{O}}_{L_2}$ of length $L_1$ and $L_2$. The crosses represent the fundamental letters $\mathcal{D}_1$ used to build the operators according to \eqref{Hilbertspacestrong}.}
\label{fig:OO_general}
\end{center}
\end{figure}

To address this, let us consider two operators ${\mathcal{O}}_{L_1}$ and ${\mathcal{O}}_{L_2}$ of length $L_1$ and $L_2$ respectively, both belonging to the same degeneracy space $\mathtt{d}(\D)$. The question above can be then rephrased by asking: what is the relation between $L_1$ and $L_2$ such that the two-point function
\begin{align}\label{2pointLL'}
\langle {\mathcal{O}}_{L_1}{\mathcal{O}}_{L_2}\rangle\,,
\end{align}
is non-zero? To answer it, recall that an operator of length $L$ is defined as in \eqref{Hilbertspacestrong} from the tensor product of $L$ copies of the fundamental letters contained in the multiplet $\mathcal{D}_1$, see \cite{Ferrero:2023znz} for more details. Then, schematically, one is computing an object of the type depicted in Figure \ref{fig:OO_general}.
The relation between $L_1$ and $L_2$ is then dictated, at each order, by the type of diagrams that one is allowed to draw, which in turn is fixed by the structure of the Lagrangian \eqref{AdSlagrangian}. The rules of the game are simple: one should use the vertices in \eqref{AdSlagrangian} to saturate the power of the coupling at each order, and then connect all the remaining crosses in Figure \ref{fig:OO_general} from the left to the right. It is important to stress that self-contractions of fields defined at the same point are forbidden: such diagrams (which are naively infinite) give zero once an appropriate prescription to define composite operators is chosen, such as point-splitting or dimensional regularization. We will then neglect their contribution in the following, see also \cite{Barrat:2021tpn,Barrat:2022eim} for another instance of this in the same theory at weak coupling.

\begin{figure}[h!]
\begin{center}
        \includegraphics[width=0.3\textwidth]{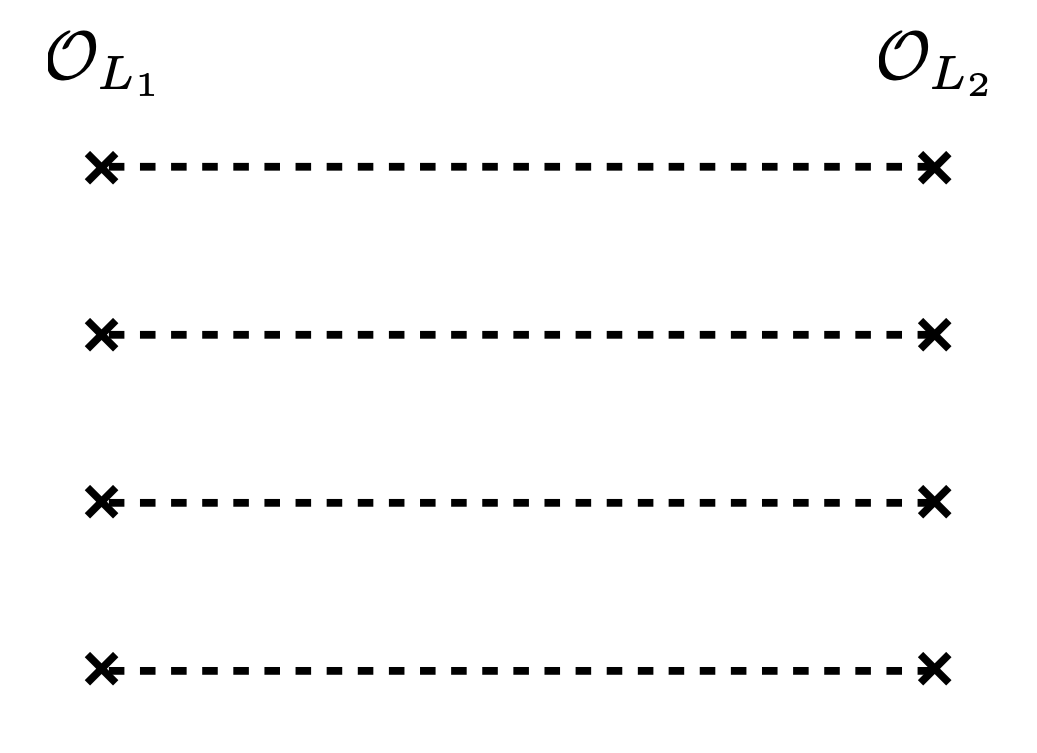}
\caption{In the free theory, $\langle {\mathcal{O}}_{L_1}{\mathcal{O}}_{L_2}\rangle$ only receives contributions from free propagators, represented with dashed lines. This requires that $L_1=L_2$.}
\label{fig:OO_free}
\end{center}
\end{figure}

\begin{figure}[h!]
    \centering
    \begin{subfigure}[b]{0.20\textwidth}
        \includegraphics[width=\textwidth]{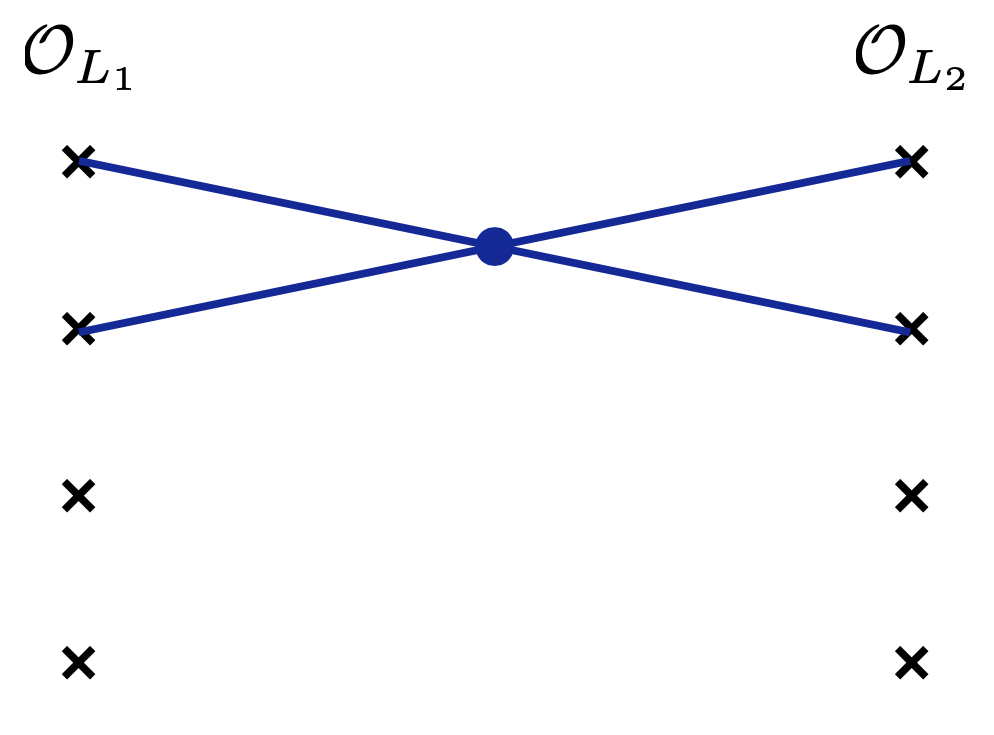}
        \caption{$L_1=L_2$}
        \label{fig:OO_4pt_delta0}
    \end{subfigure}\quad\quad
    \begin{subfigure}[b]{0.20\textwidth}
        \includegraphics[width=\textwidth]{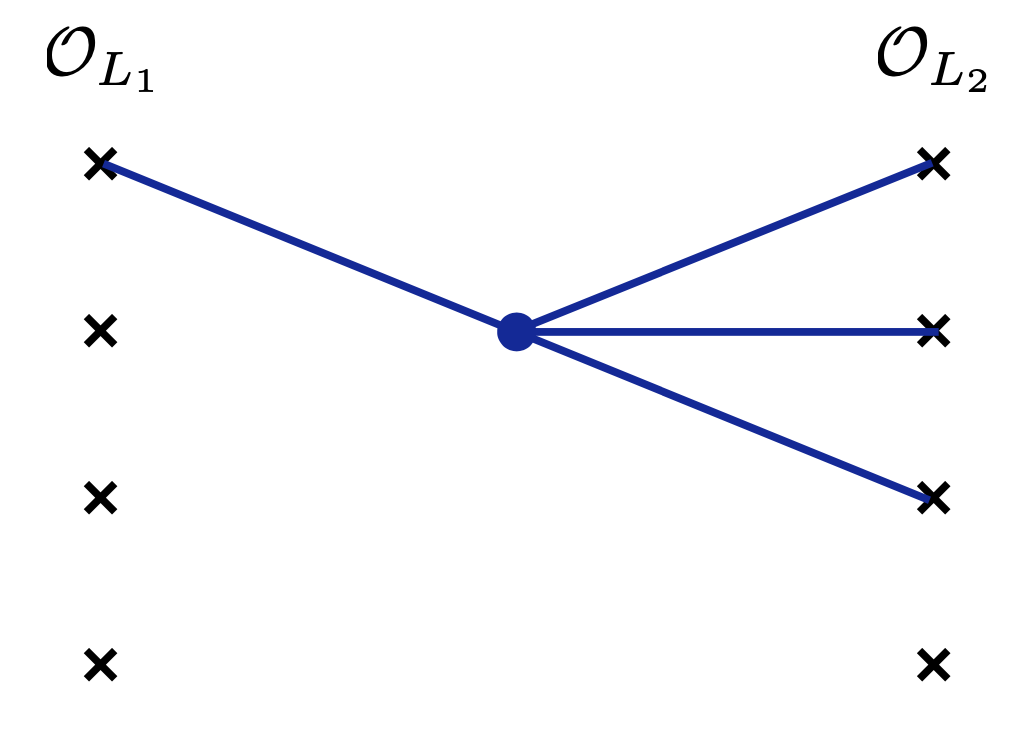}
         \caption{$L_1=L_2\pm 2$}
        \label{fig:OO_4pt_delta2}
    \end{subfigure}\quad\quad
\caption{Contributions to $\langle\mathcal{O}_{L_1}\mathcal{O}_{L_2}\rangle^{(1)}$.}
\label{fig:OO_tree}
\end{figure}

Let us start from the free theory, where the answer is extremely easy: one can only draw free theory propagators to connect each $\mathcal{D}_1$ in $\mathcal{O}_{L_1}$ to a $\mathcal{D}_1$ in $\mathcal{O}_{L_2}$, which implies that necessarily $L_1=L_2$. The situation is represented schematically in Figure \ref{fig:OO_free}.
At first order, only the quartic vertices in \eqref{AdSlagrangian} play a role and, neglecting self-contractions as discussed above, the two-point function \eqref{2pointLL'} receives contributions from the two diagrams in Figure \ref{fig:OO_tree}.
Note that we are using blue lines to denote the use of a quartic vertex and that we are no longer drawing free propagators connecting the remaining points: we simply assume that all other points are linked in a trivial way. The difference between the number of points connected to a vertex on the two sides then gives us a selection rule for the allowed $L_1$ and $L_2$. In the case at hand, the diagrams of the type \ref{fig:OO_4pt_delta0} cannot change the length of operators from the left to the right, hence they lead to $L_1=L_2$. On the other hand, diagrams of the type \ref{fig:OO_4pt_delta2} lead to the selection rule $L_1=L_2\pm 2$. However, this is not quite the end of the story: indeed, all diagrams with the structure of \ref{fig:OO_4pt_delta2} factorize in a free theory contribution and the first order contribution to a two-point function of the type $\langle \mathcal{D}_1 X_{L=3}\rangle$, where $X_{L=3}$ is some operator of length three: roughly speaking,  $\ref{fig:OO_4pt_delta2} \sim \langle \mathcal{D}_1 X_{L=3}\rangle^{(1)}\langle \dots \rangle^{(0)}$, where the second is a free-theory correlator computable by Wick contractions. Regardless of the details of the theory under inspection, by $\mathfrak{osp}(4^*|4)$ invariance such two-point function can only be non-zero if $X_{L=3}$ is a $\mathcal{D}_1$-type multiplet itself: in that case, the two-point function is protected by supersymmetry and does not generate logarithms in the perturbative expansion, hence it does not contribute to the anomalous dimensions matrix. In the particular case at hand, moreover, there is only one multiplet of the type $\mathcal{D}_1$ and it has length one, so the two-point function $\langle \mathcal{D}_1 X_{L=3}\rangle$ vanishes identically. We then conclude that, due to supersymmetry, the tree-level diagram changing the length by two is actually suppressed. Therefore we can write
\begin{align}
\langle \mathcal{O}_{L_1}\,\mathcal{O}_{L_2}\rangle^{(1)} \propto \delta_{L_1,L_2}\,.
\end{align}  
We anticipate here that, moving to higher orders, changes of length will be allowed, but similar arguments always suppress the maximal change of length naively allowed at each order.

\begin{figure}[hbt!]
    \centering
    \begin{subfigure}[b]{0.20\textwidth}
        \includegraphics[width=\textwidth]{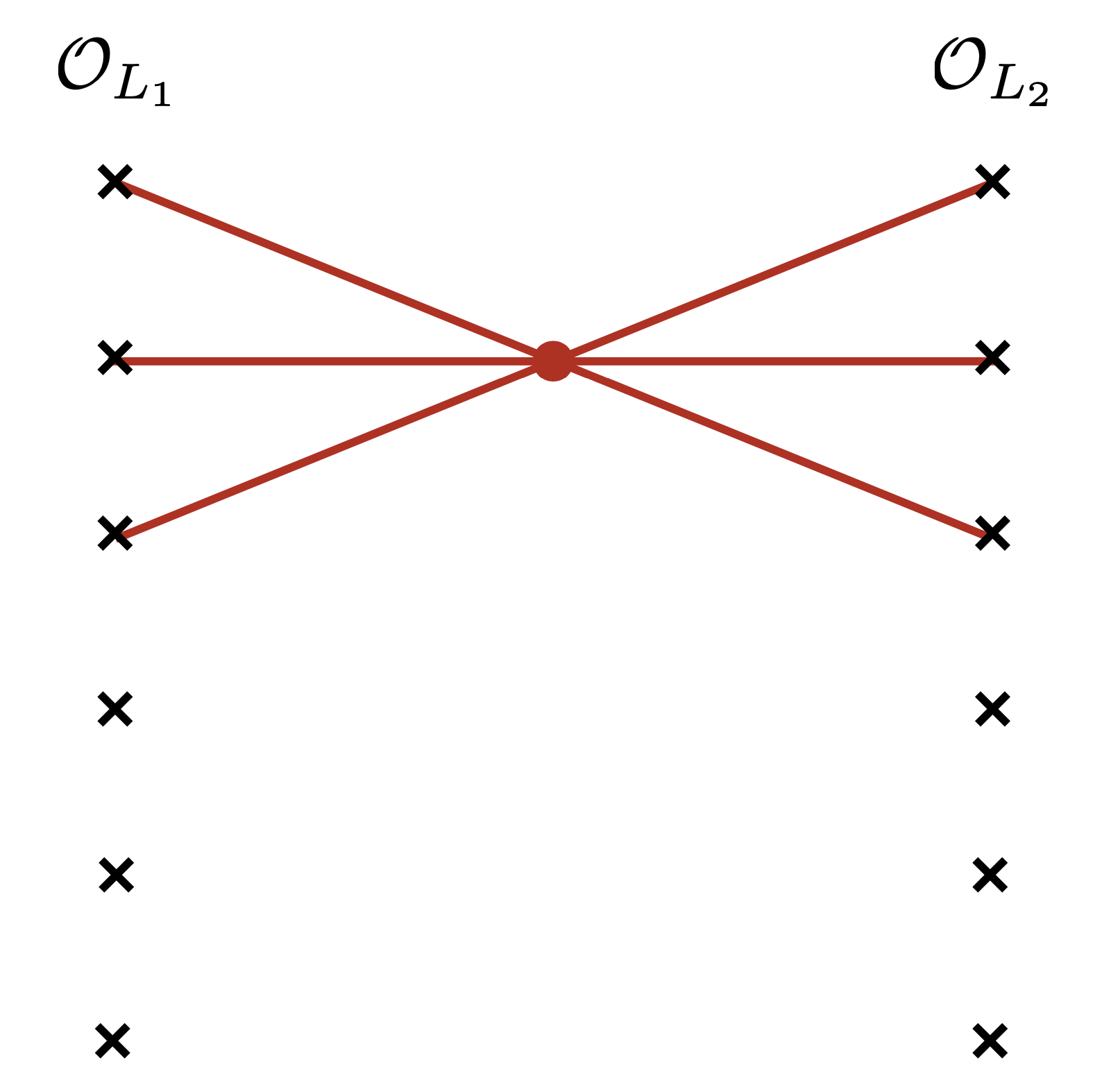}
        \caption{$L_1=L_2$}
        \label{fig:OO_6pt_delta0}
    \end{subfigure}\quad\quad
    \begin{subfigure}[b]{0.20\textwidth}
        \includegraphics[width=\textwidth]{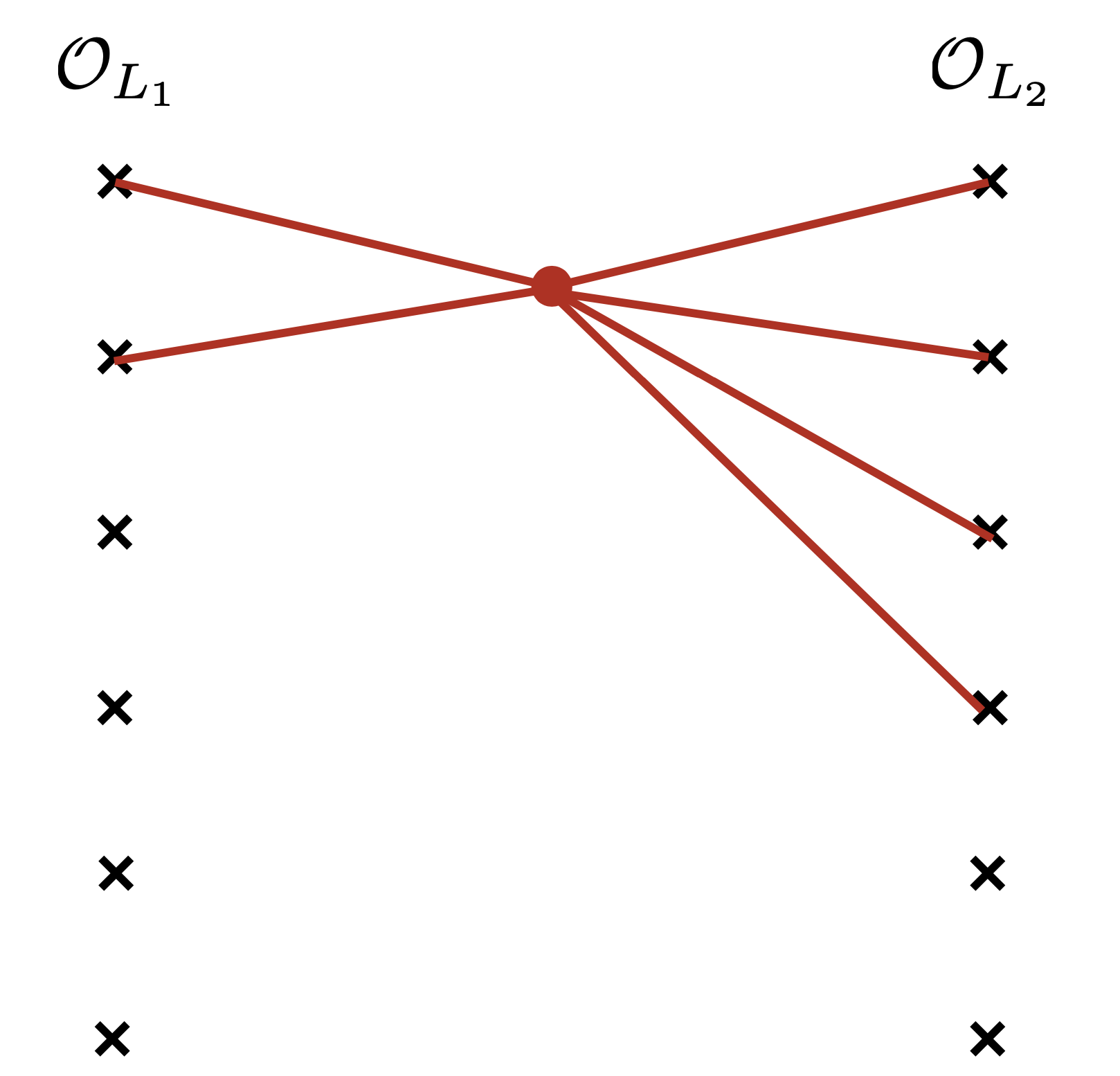}
         \caption{$L_1=L_2\pm 2$}
        \label{fig:OO_6pt_delta2}
    \end{subfigure}\quad\quad
    \begin{subfigure}[b]{0.20\textwidth}
        \includegraphics[width=\textwidth]{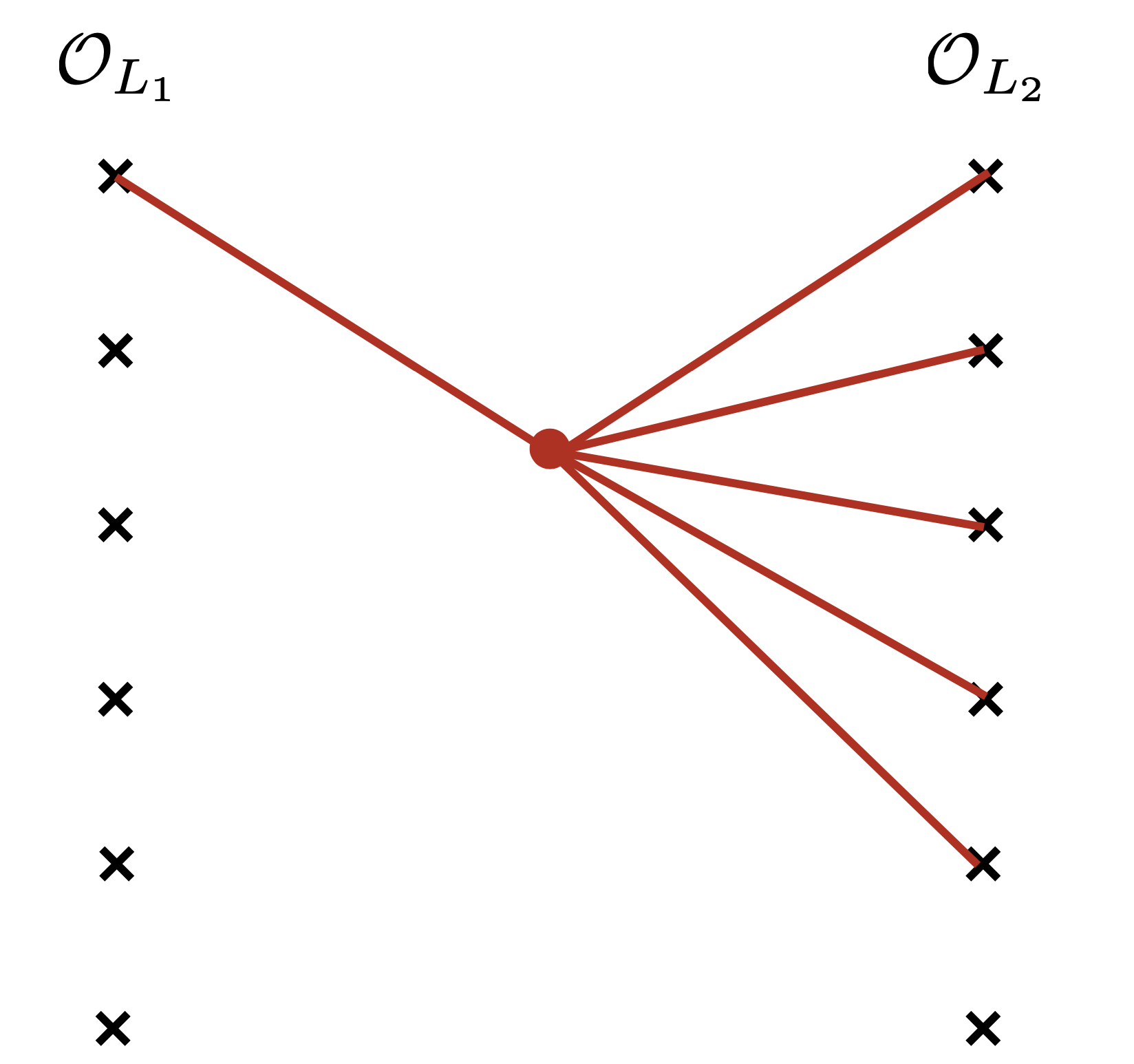}
        \caption{$L_1=L_2\pm 4$}
        \label{fig:OO_6pt_delta4}
    \end{subfigure}
\caption{Contributions to $\langle\mathcal{O}_{L_1}\mathcal{O}_{L_2}\rangle^{(2)}$ built using a six-point vertex.}
\label{fig:OO_6pt}
\end{figure}

\begin{figure}[hbt!]
    \centering
    \begin{subfigure}[b]{0.20\textwidth}
        \includegraphics[width=\textwidth]{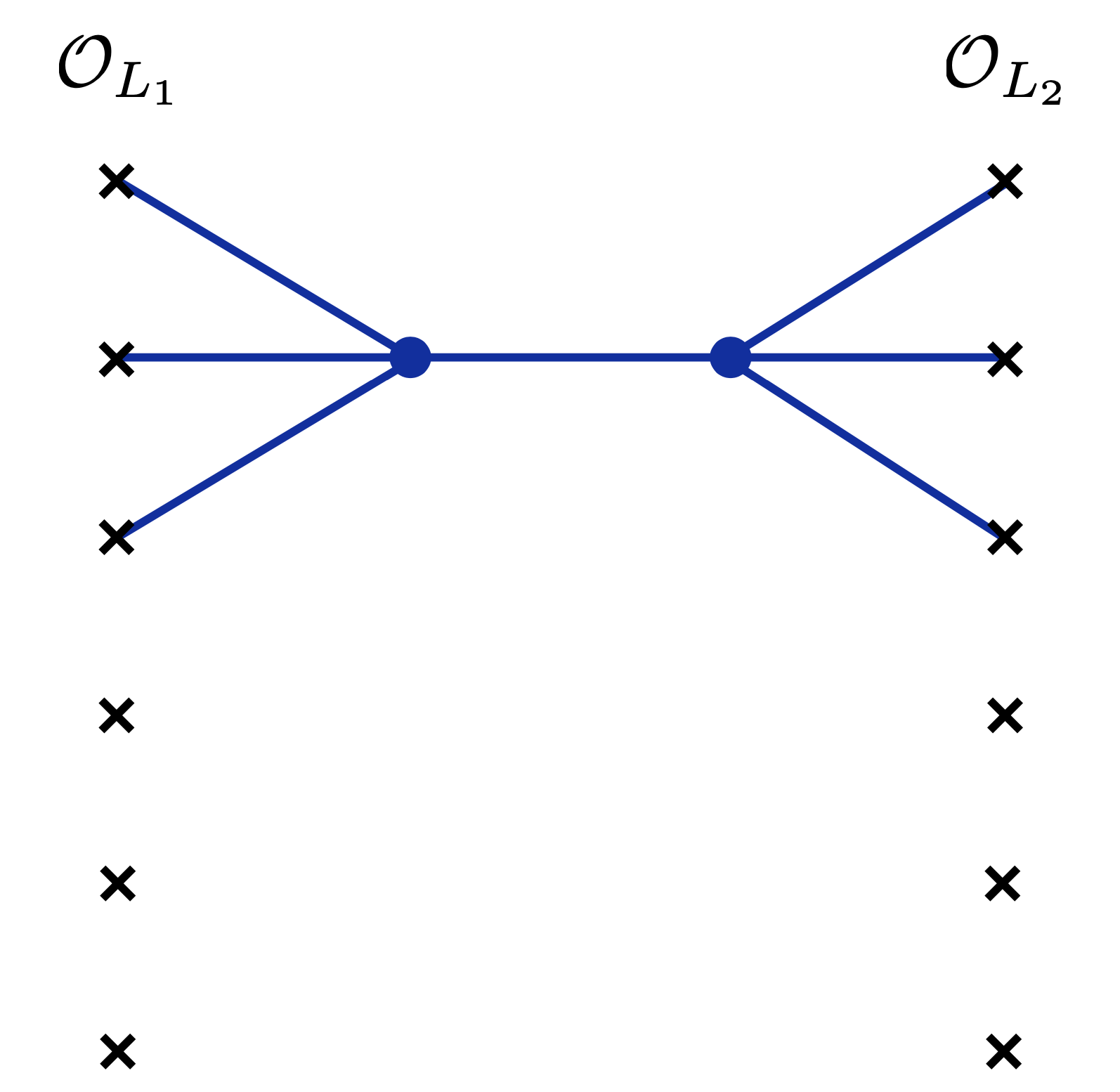}
        \caption{$L_1=L_2$}
        \label{fig:OO_4ptex_delta0}
    \end{subfigure}\quad\quad
    \begin{subfigure}[b]{0.20\textwidth}
        \includegraphics[width=\textwidth]{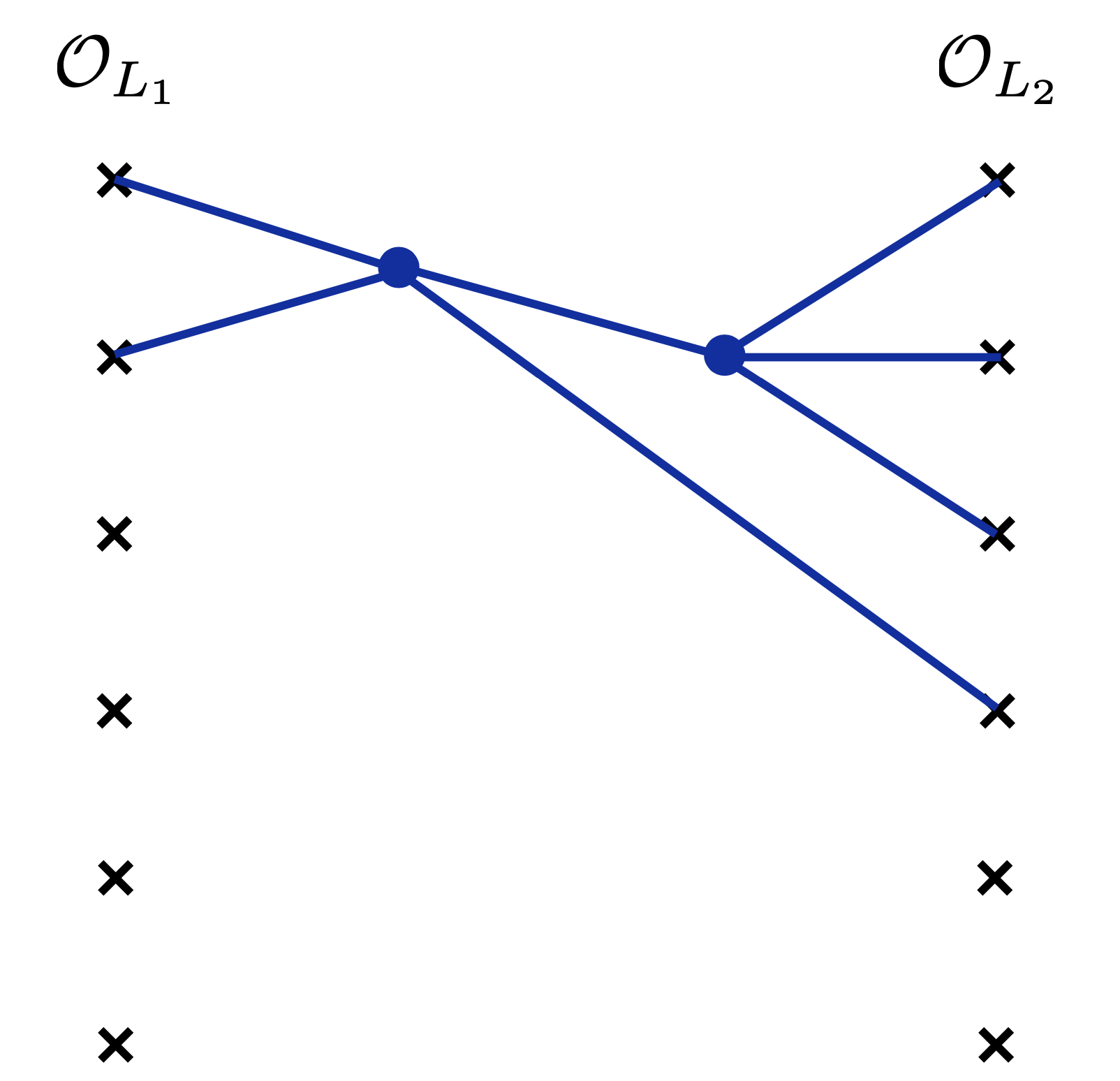}
         \caption{$L_1=L_2\pm 2$}
        \label{fig:OO_4ptex_delta2}
    \end{subfigure}\quad\quad
    \begin{subfigure}[b]{0.20\textwidth}
        \includegraphics[width=\textwidth]{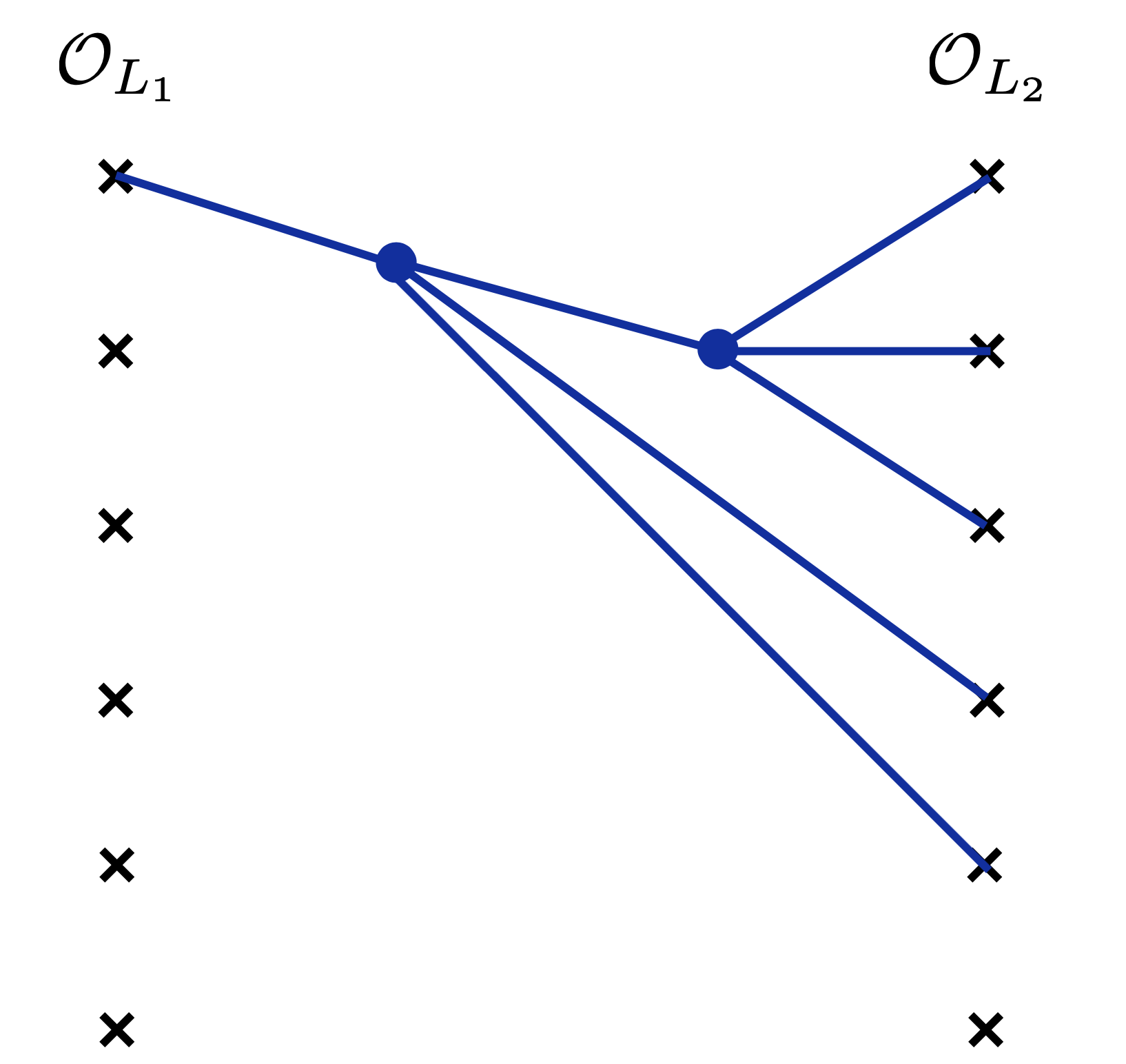}
        \caption{$L_1=L_2\pm 4$}
        \label{fig:OO_4ptex_delta4}
    \end{subfigure}
\caption{Contributions to $\langle\mathcal{O}_{L_1}\mathcal{O}_{L_2}\rangle^{(2)}$ built using the two four-point vertex twice to form an exchange diagram.}
\label{fig:OO_4ptex}
\end{figure}

\begin{figure}[hbt!]
    \centering
    \begin{subfigure}[b]{0.20\textwidth}
        \includegraphics[width=\textwidth]{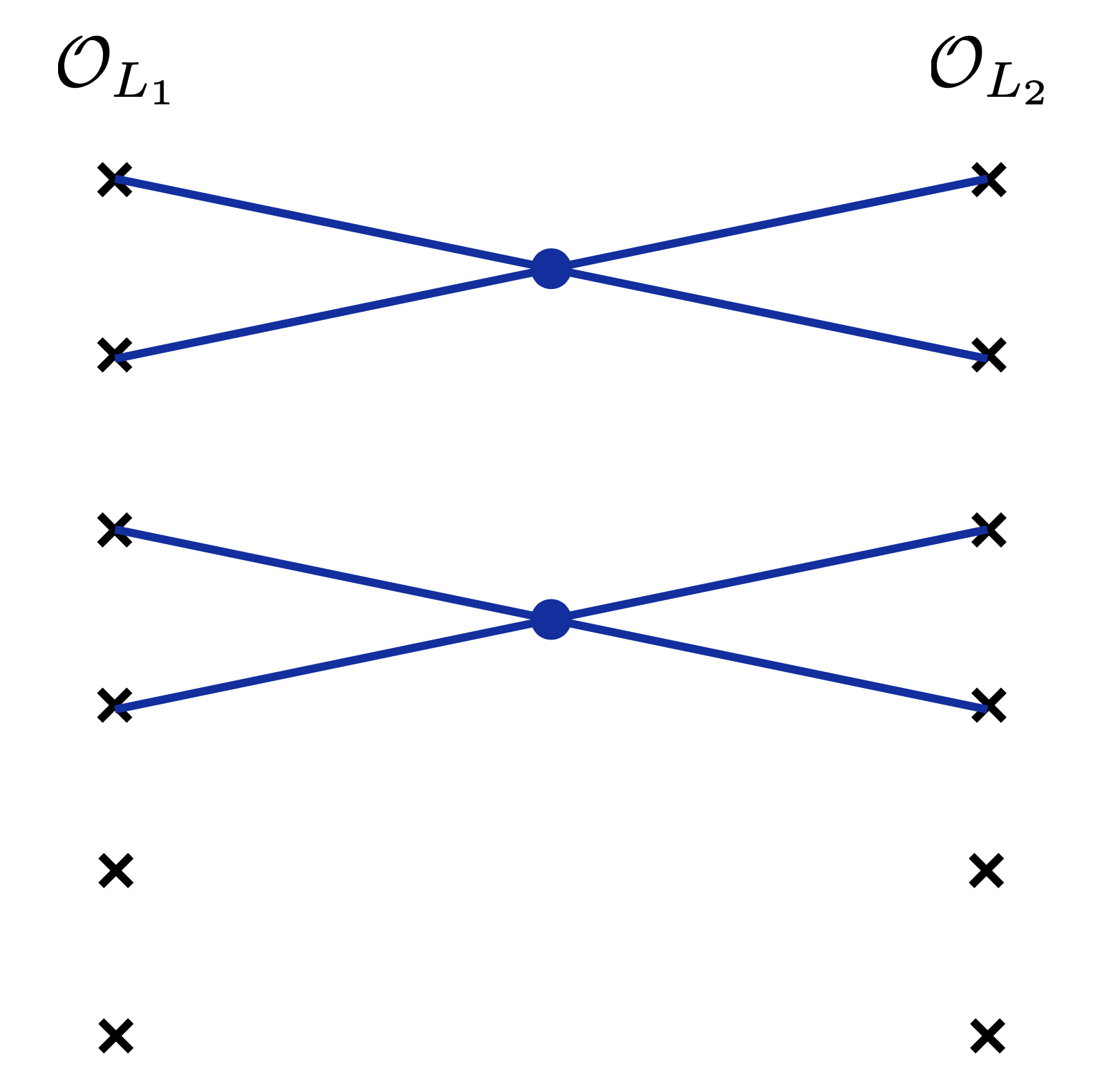}
        \caption{$L_1=L_2$}
        \label{fig:OO_4pt_2_delta0}
    \end{subfigure}\quad\quad
    \begin{subfigure}[b]{0.20\textwidth}
        \includegraphics[width=\textwidth]{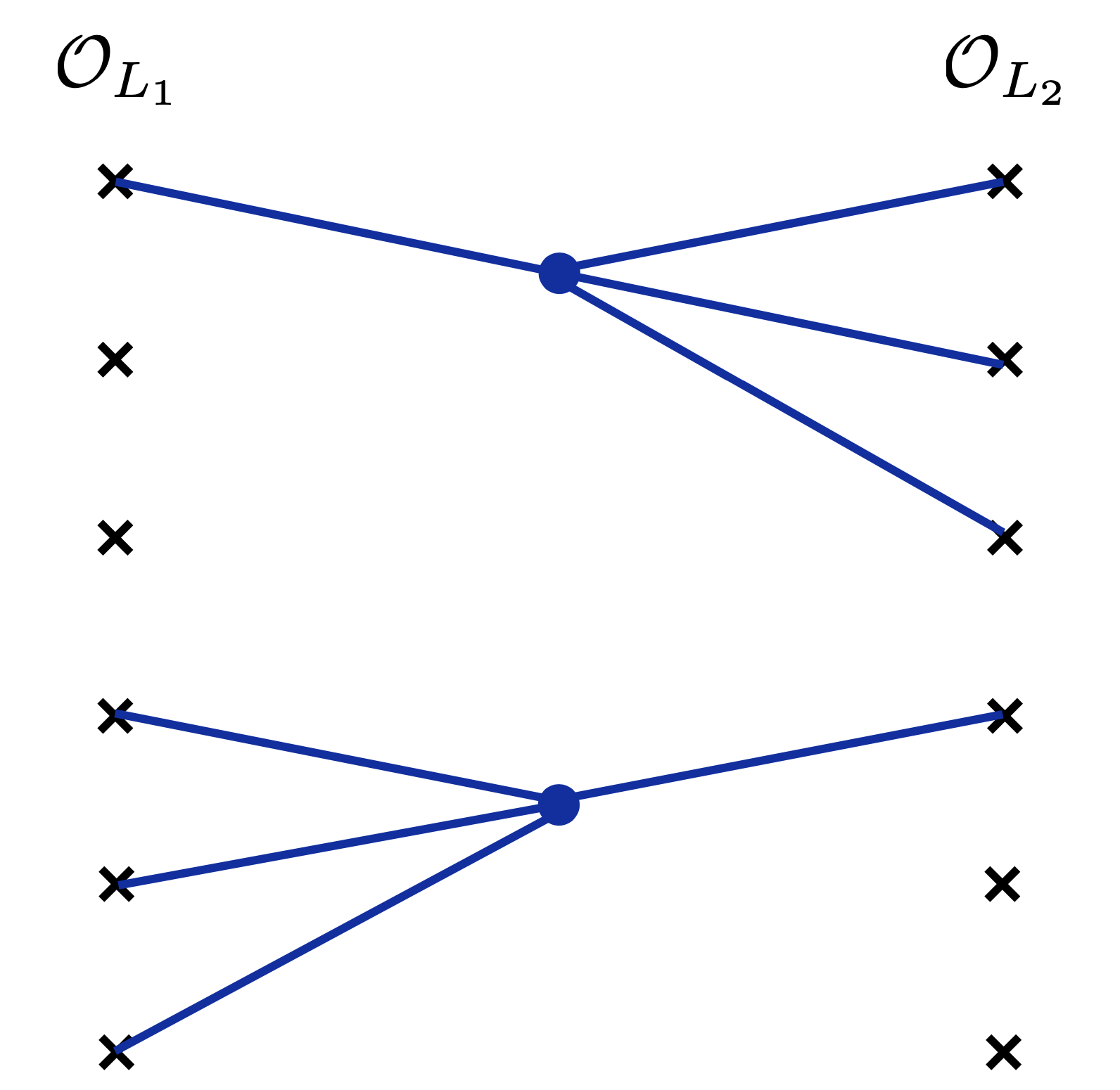}
         \caption{$L_1=L_2$}
        \label{fig:OO_4pt_2_delta2-2}
    \end{subfigure}\quad\quad
    \begin{subfigure}[b]{0.20\textwidth}
        \includegraphics[width=\textwidth]{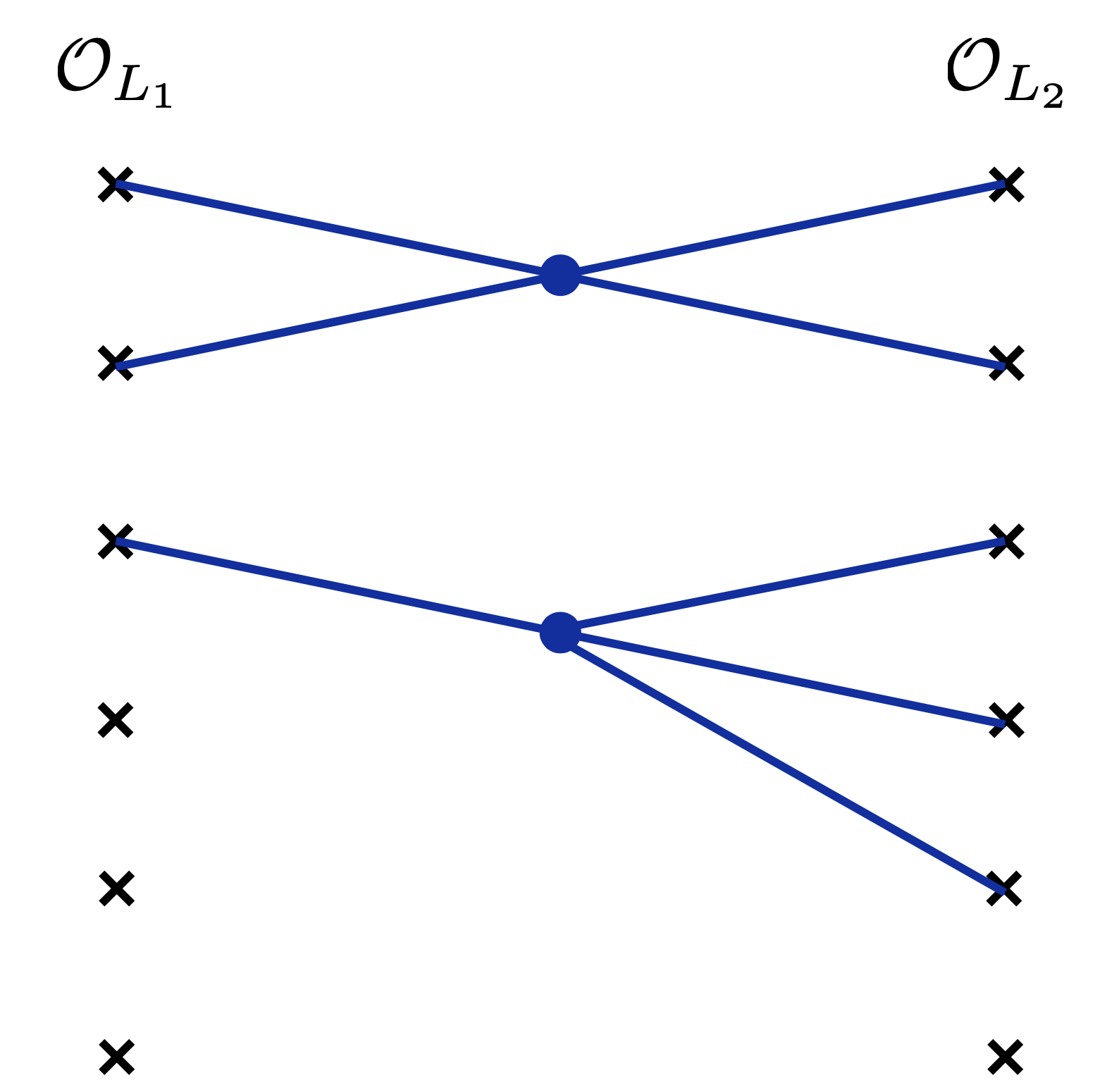}
         \caption{$L_1=L_2\pm 2$}
        \label{fig:OO_4pt_2_delta2}
    \end{subfigure}\quad\quad
    \begin{subfigure}[b]{0.20\textwidth}
        \includegraphics[width=\textwidth]{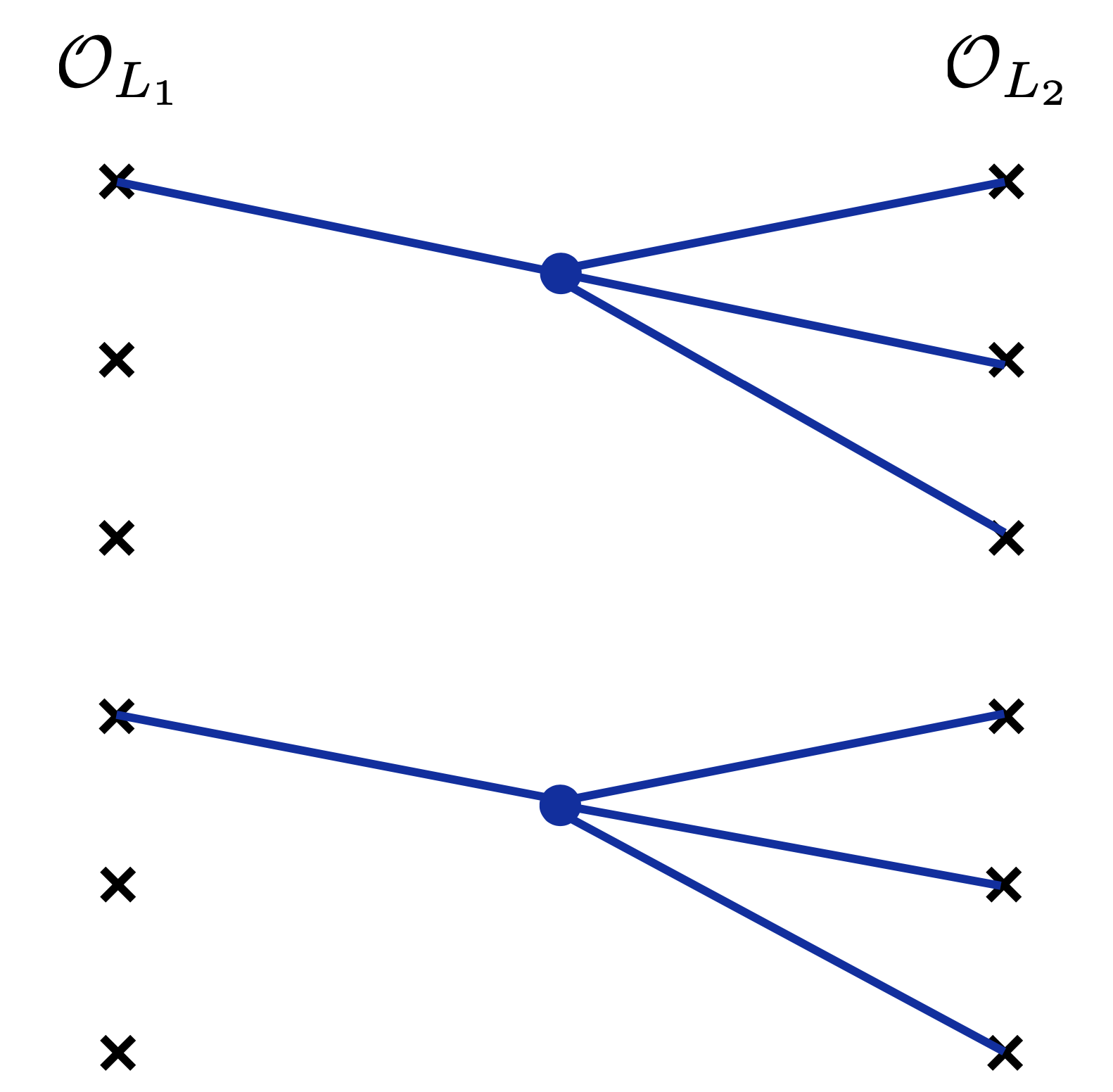}
        \caption{$L_1=L_2\pm 4$}
        \label{fig:OO_4pt_2_delta4}
    \end{subfigure}
\caption{Contributions to $\langle\mathcal{O}_{L_1}\mathcal{O}_{L_2}\rangle^{(2)}$ built using the two four-point vertex twice in a disconnected way.}
\label{fig:OO_4pt_2}
\end{figure}

\begin{figure}[hbt!]
    \centering
    \begin{subfigure}[b]{0.20\textwidth}
        \includegraphics[width=\textwidth]{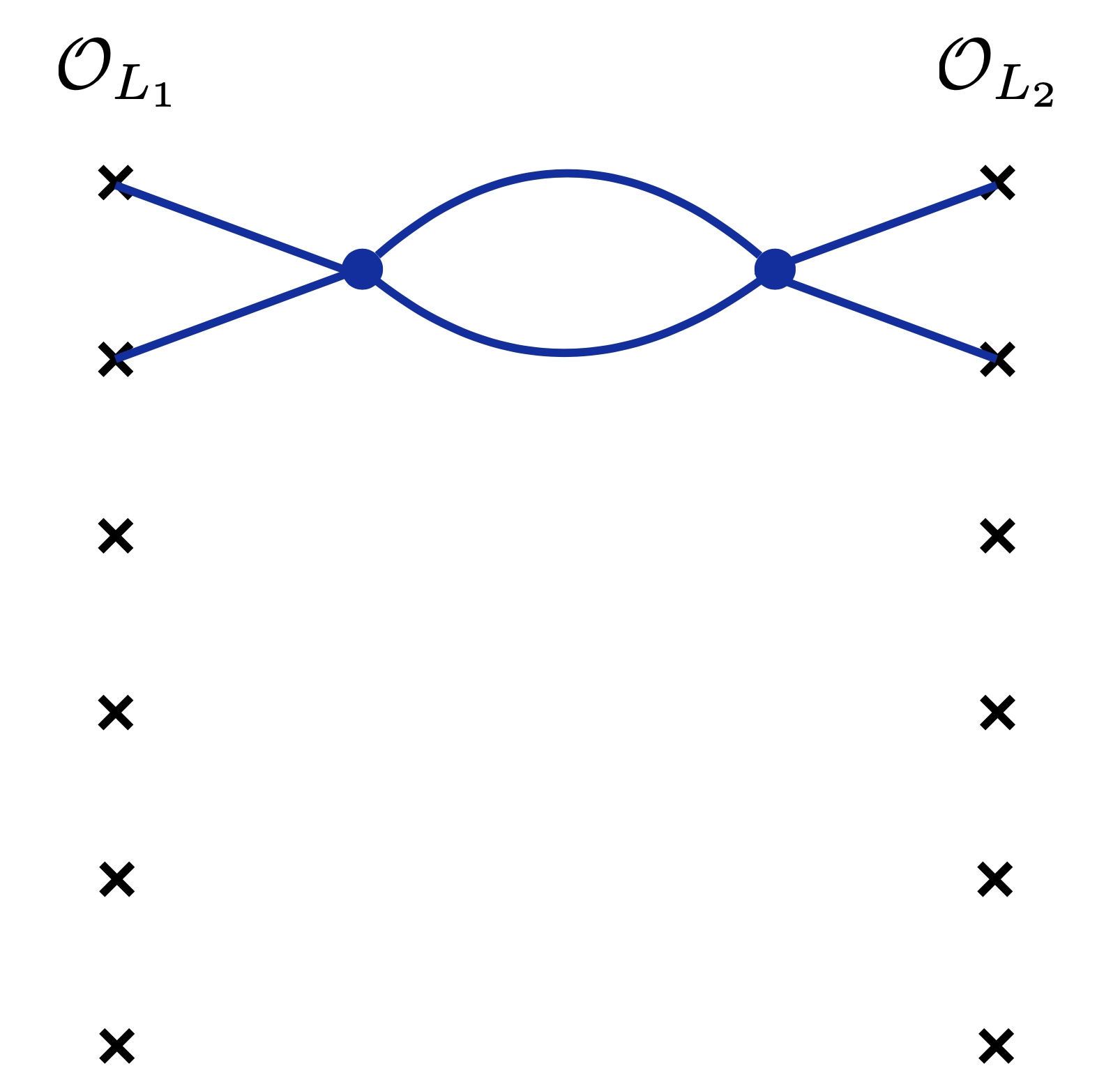}
        \caption{$L_1=L_2$}
        \label{fig:OO_loop_delta0}
    \end{subfigure}\quad\quad
    \begin{subfigure}[b]{0.20\textwidth}
        \includegraphics[width=\textwidth]{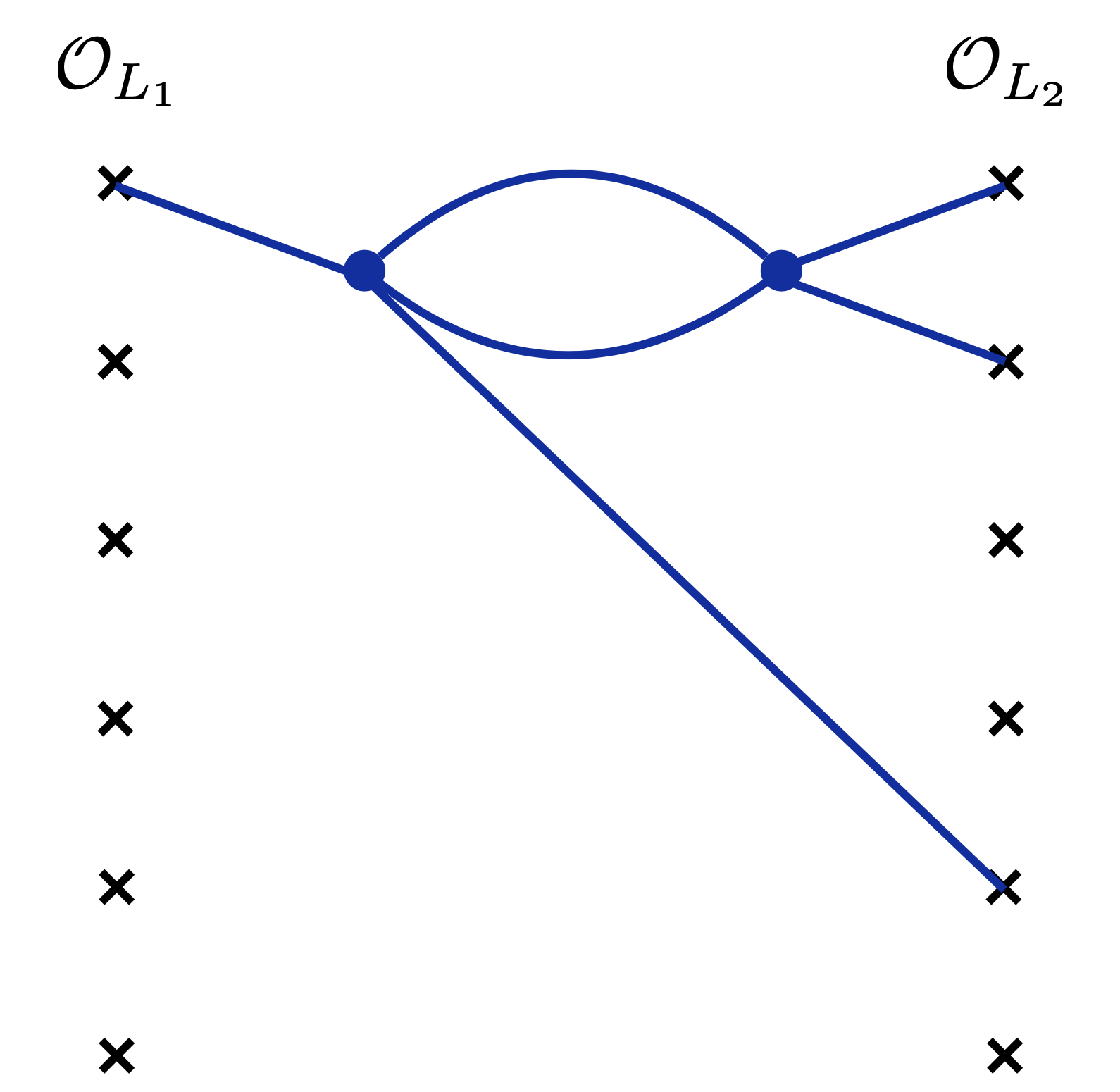}
         \caption{$L_1=L_2\pm 2$}
        \label{fig:OO_loop_delta2}
    \end{subfigure}\quad\quad
\caption{Contributions to $\langle\mathcal{O}_{L_1}\mathcal{O}_{L_2}\rangle^{(2)}$ built using the two four-point vertex twice to form a loop.}
\label{fig:OO_loop}
\end{figure}

We will dig further into the structure of the dilatation operator at first order in the second part of this section, while for the moment we move to second order. Clearly, now the type and number of diagrams will be more complicated, but we will still be able to draw similar conclusions. Let us discuss the type of diagrams that contribute at this order, with the help of Figures \ref{fig:OO_6pt}-\ref{fig:OO_loop}, where the crosses on the side represent the fundamental letters (elements of $\mathcal{D}_1$) used to build the operators, dots represent interactions and we have used red lines for the sextic vertex and blue lines for the quartic one. All crosses not connected by any line should be thought of as connected to a cross present in the other operator with a free propagator. The diagrams \ref{fig:OO_6pt_delta0}, \ref{fig:OO_4ptex_delta0}, \ref{fig:OO_4pt_2_delta0} and \ref{fig:OO_loop_delta0} give a non-vanishing contribution and change the length by zero units. The other relevant contribution comes from the diagrams \ref{fig:OO_6pt_delta2} and \ref{fig:OO_4ptex_delta2}, which change the length by two units. The two diagrams \ref{fig:OO_6pt_delta4} and \ref{fig:OO_4ptex_delta4} form the second-order contribution to a two-point function $\langle \mathcal{D}_1 X_{L=5}\rangle$, so they do not contribute to the anomalous dimensions as discussed for $\langle \mathcal{D}_1 X_{L=3}\rangle$ at first order. Finally, \ref{fig:OO_4pt_2_delta2-2}, \ref{fig:OO_4pt_2_delta2} and \ref{fig:OO_4pt_2_delta4} do not contribute since they are factorized with a factor being $\langle \mathcal{D}_1 X_{L=3}\rangle^{(1)}$, and similarly the contribution from \ref{fig:OO_loop_delta2} since it is the second order contribution to $\langle \mathcal{D}_1 X_{L=3}\rangle$.

All in all, we can draw the following conclusions regarding second-order contribution to \eqref{2pointLL'}: the two point functions can connect operators with either the same length ($L_1=L_2$) or length differing by two units ($L_1=L_2\pm 2$), and moreover the only diagrams that contribute to the maximal change in length are fully connected, tree-level diagrams, dressed with the proper amount of free theory propagators.  

So far, we have been sloppy about an important fact, namely whether the two operators in \eqref{2pointLL'} are superconformal primaries or descendants. In generic weakly coupled Lagrangian theories, certain short (super)conformal multiplets recombine into longer ones when the coupling is tuned from zero to a finite value. In this case, the (super)conformal primaries of certain multiplets can mix with the (super)conformal descendants of other multiplets.\footnote{A simple example is given by the operators $\Box\varphi$ and $\varphi^3$ in $\varphi^4$ theory in $d=4-\epsilon$ dimensions.} However, a special feature of the theory that we are investigating in this paper, and which is common in holographic settings, is that the spectrum at $\lambda=\infty$ is such that no recombination can happen. Indeed, as discussed at length in \cite{Ferrero:2023znz}, at $\lambda=\infty$ the only multiplets are either protected or long multiplets with dimension well above the unitarity bound. Given the absence of recombination, we then claim that the only two-point functions \eqref{2pointLL'} one has to consider is that between superconformal primaries, and those between (super)conformal descendants are fixed by symmetries accordingly. This also justifies expressions such as \eqref{defGammaDelta}, where we have written the anomalous dimensions matrix as a map between degeneracy spaces, which we have defined in \eqref{degeneracyspaces} as vector spaces of objects in a given superconformal multiplet of $\mathfrak{osp}(4^*|4)$, canonically represented by their superconformal primary.

\begin{figure}[h!]
\begin{center}
        \includegraphics[width=0.5\textwidth]{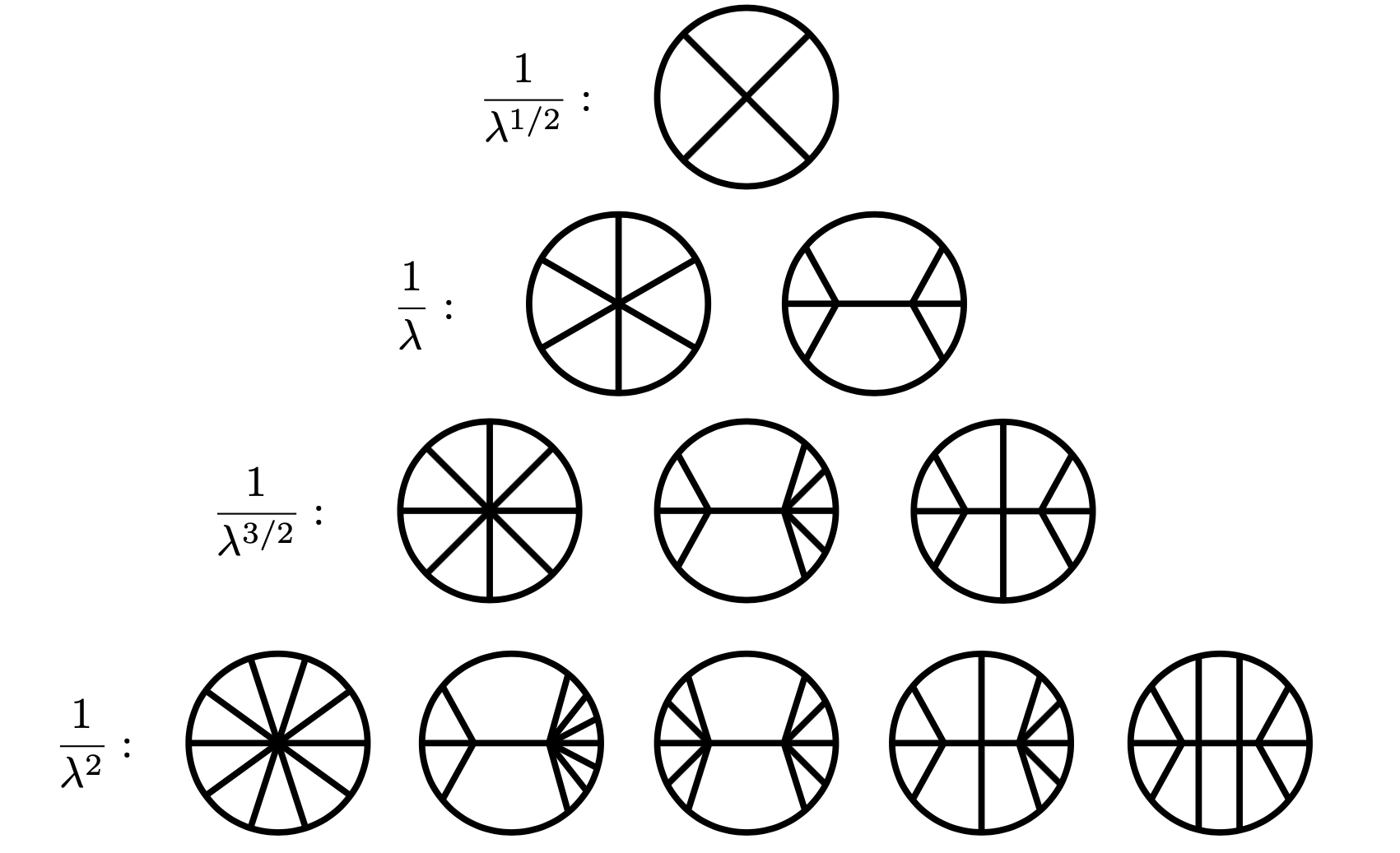}
\caption{Diagrams contributing to \eqref{npointPhi_connected} at the first few perturbative orders.}
\label{fig:npointconnected}
\end{center}
\end{figure}

Using analogous diagrammatic arguments to the ones above, it is actually possible to generalize similar observations to all perturbative orders, thus elucidating an interesting structure in the dilatation operator of this theory, which we will refer to as $\mathbb{D}$. The anomalous dimensions matrix $\Gamma_{\D}$ is then an explicit representation of $\mathbb{D}$ on a given degeneracy space $\mathtt{d}(\D)$. Given that $\mathbb{D}:\,\,\mathcal{H}\to \mathcal{H}$ and \eqref{Hilbertspacestrong}, we can decompose it as
\begin{align}
\mathbb{D}=\sum_{L_1,L_2}\mathbb{D}_{L_1\to L_2}\,,\qquad  \mathbb{D}_{L_1\to L_2}:\,\, \mathcal{H}_{L_1}\to \mathcal{H}_{L_2}\,.
\end{align}
The issues discussed above correspond to the question of which $\mathbb{D}_{L_1\to L_2}$ are trivial and which are not at a given order in perturbation theory. Expanding $\mathbb{D}$ perturbatively at large $\lambda$,
\begin{align}
\mathbb{D}=\sum_{\ell=0}^{\infty}\frac{1}{\lambda^{\ell/2}}\mathbb{D}^{(\ell)}\,,
\end{align}
generalizing the diagrammatic arguments of this section to higher orders one can prove that
\begin{align}
\mathbb{D}^{(\ell)}=\sum_L\sum_{\delta L=1-\ell}^{\ell-1}\mathbb{D}^{(\ell)}_{L\to L+2\delta L}\,,
\end{align}
or
\begin{align}
\mathbb{D}^{(\ell)}\,\,\,\,=\sum_{\substack{L_1,L_2\\ |L_1-L_2|\in\{0,2,\dots,2(\ell-1)\}}}
\mathbb{D}^{(\ell)}_{L_1\to L_2}\,,
\end{align}
or in words 
\begin{center}
{\it the action of the dilatation operator at $\ell$-th order ($\ell>0$) can only change the length of states by $\Delta L$ units, with $\Delta L=0,2,\ldots,2\ell-2$.}
\end{center}
We emphasize again that the special feature of the theory at hand is that the case $\Delta L=2\ell$ is disallowed by supersymmetry. Moreover, we would like to stress that the maximal change of length at given order\footnote{Interesting features of terms that change the length maximally at a given perturbative order in the case of planar $\mathcal{N}=4$ SYM at week coupling have been  observed in \cite{Zwiebel:2011bx}. } can be ascribed by a simple type of diagrams. Due to the structure of \eqref{AdSlagrangian}, at order $\mathcal{O}(\lambda^{-\ell/2})$ they can be extracted from the fully connected $(2\ell+2)$-point function of $\Phi$ (the superfield associated with $\mathcal{D}_1$, see \cite{Ferrero:2023znz}):
\begin{align}\label{npointPhi_connected}
\langle\Phi(1)\ldots\Phi(2\ell+2)\rangle_{\text{connected}}\,,
\end{align}
which only receives contributions from tree-type diagrams, like those represented in Figure \ref{fig:npointconnected} for the first few orders. The importance of the $(2\ell+2)$-point function of $\Phi$ at order $\mathcal{O}(\lambda^{-\ell/2})$ was already emphasized around \eqref{npointPhi} and in this case we are extracting its simplest part, which does not involve loop diagrams.

One might wonder if it is possible to obtain explicit expressions for the dilatation operator in perturbation theory, in the spirit of \cite{Beisert:2004ry}. It turns out that it is possible to obtain a particularly compact result at first order, while at second order we only explored certain components in Section \ref{sec:mixing}. Here we focus on the story at first order, where we have learned that there is only one relevant contribution to $\mathbb{D}^{(1)}$:
\begin{align}
\mathbb{D}^{(1)}=\sum_L \mathbb{D}^{(1)}_{L\to L}\,.
\end{align}
Given that at each order we are using a basis of operators that are defined in the free theory, where states are built using \eqref{Hilbertspacestrong}, $\mathbb{D}^{(1)}$ has a well-defined action on the letters that build each operator, and the fact that the only diagram contributing to it is \ref{fig:OO_4pt_delta0} implies that it must have the structure
\begin{align}
\label{D1assumofD22}
\mathbb{D}^{(1)}_{L\to L}=\sum_{1\le i<j\le L}(\mathbb{D}^{(1)}_{2\to 2})_{ij}\,,
\end{align}
where the indices $i,j$ indicate on which factors in the tensor product defining $\mathcal{H}_L$ the operator is acting (see \eqref{Hilbertspacestrong}). Note that $\mathbb{D}^{(1)}_{L\to L}$ is permutation invariant so its action on $\mathcal{H}_L$ is well defined. The full dilatation operator at first order can then be determined from $\mathbb{D}^{(1)}_{2\to 2}$, which we extract as follows. We start by observing that in the OPE $\mathcal{D}_1\times \mathcal{D}_1$ the only long multiplets that appear are the (non-degenerate) singlets of length $L=2$ (let $\D$ be their dimension): their first order anomalous dimensions can be extracted from $\langle\mathcal{D}_1\mathcal{D}_1\mathcal{D}_1\mathcal{D}_1\rangle^{(1)}$ and they read $\gamma^{(1)}_{\D}=-\tfrac{1}{2}\D(\D+3)$, which is proportional to the superconformal Casimir associated with the representation $\{\D,0,[0,0]\}$ (see Section \ref{sec:1111}). Using the fact that $\mathbb{D}^{(1)}$ commutes with the action of $\mathfrak{osp}(4^*|4)$ in the free theory, which can be argued from analogous observations to those of \cite{Beisert:2004ry}, one can use this information on the spectrum of $\mathbb{D}^{(1)}_{2\to 2}$ to conclude that the operator itself must be proportional to the quadratic Casimir of $\mathfrak{osp}(4^*|4)$:
\begin{align}
\label{D22ascasimir}
\mathbb{D}^{(1)}_{2\to 2}=-\frac{1}{2}\widehat{\mathfrak{C}}_2(J^{(12)})=:-J^{(1)}\cdot J^{(2)}\,, \qquad 
\widehat{\mathfrak{C}}_2(J^{(12)})=k(J,J)\,,
\end{align}
where $k$ is the Killing form of $\mathfrak{osp}(4^*|4)$, $J^{(1\ldots n)}=J^{(1)}+\ldots+ J^{(n)}$, each  $J^{(i)}$ corresponding to the representation of the single letter $\mathbb{V}_{\Phi}=\mathcal{D}_1$, and we introduced the notation $J^{(i)}\cdot J^{(j)}$. We have also used that the single letter representation $\mathbb{V}_{\Phi}$ has vanishing Casimir (see \cite{Ferrero:2023znz} for our conventions), namely $\widehat{\mathfrak{C}}_2(J^{(i)})=0$.
Inserting \eqref{D22ascasimir} in \eqref{D1assumofD22} and using again that the quadratic Casimir of $\mathcal{D}_1$ is zero, we obtain
\begin{align}
\mathbb{D}_{L\to L}^{(1)}=-\sum_{1 \leq i<j \leq L} J^{(i)} \cdot J^{(j)}=-\frac{1}{2} \sum_{1 \leq i, j \leq L} J^{(i)} \cdot J^{(j)}=-\frac{1}{2} \widehat{\mathfrak{C}}_2\left(J^{(12 \ldots L)}\right)\,,
\end{align}
that is we have just shown that
\begin{center}
{\it the first-order anomalous dimension of any operator $\mathcal{O}$ in a representation $\mathcal{R}_{\mathcal{O}}$ of $\mathfrak{osp}(4^*|4)$ is proportional to the quadratic superconformal Casimir eigenvalue of that representation, $\mathfrak{c}_2(\mathcal{R}_{\mathcal{O}})$.}
\end{center}
More precisely, we have
\begin{align}\label{gamma1casimir}
\gamma^{(1)}_{\mathcal{O}}=-\frac{1}{2}\mathfrak{c}_2(\mathcal{R}_{\mathcal{O}})=-\frac{1}{2}\left[\Delta\,(\Delta+3)+\frac{1}{4}\,s\,(s+2)-\frac{1}{2}\,a^2-a\,(b+2)-b\,(b+3)\right]\,,
\end{align}
where $\D$ is the conformal dimension of $\mathcal{O}$ for $\lambda=\infty$, $s$ is its transverse spin and $[a,b]$ the associated $\mathfrak{sp}(4)$ Dynkin labels, and we are following the conventions of \cite{Ferrero:2023znz}. Note that \eqref{gamma1casimir} implies in particular that operators that are degenerate at $\lambda=\infty$ (thus sharing the same $\D$, $s$, $[a,b]$) remain degenerate at first order, or in other words
\begin{center}
{\it the degeneracy of operators in the free theory at $\lambda=\infty$ is not broken at first order.}
\end{center}
As we shall discuss in Section \ref{sec:1111}, this fact is crucial because it allows to bootstrap correlators up to two loops without addressing the mixing problem, which only comes into the game at three loops from the bootstrap point of view. Note that we have reached these conclusions only from abstract algebraic considerations and the OPE analysis of $\langle\mathcal{D}_1\mathcal{D}_1\mathcal{D}_1\mathcal{D}_1\rangle^{(1)}$ given in Section \ref{sec:1111}, but we have {\it not} used it as an input for the bootstrap of $\langle\mathcal{D}_p\mathcal{D}_p\mathcal{D}_q\mathcal{D}_q\rangle^{(1)}$ in Section \ref{sec:ppqq}. The results of that section agree perfectly with the prediction \eqref{gamma1casimir}, thus providing strong evidence that our analysis is correct.

\section{Bootstrapping 1d CFTs}\label{sec:bootstrap}

The aim of this section is to present the strategy that we adopt to bootstrap correlation functions of local operators on the Maldacena Wilson line at strong coupling. The method, which was introduced in \cite{FernandoNotes,  Liendo:2018ukf,Ferrero:2019luz}, is based on a certain ansatz in terms of multiple polylogarithms and rational functions and has already been applied to various 1d CFTs \cite{Gimenez-Grau:2019hez,  Bianchi:2020hsz,Ferrero:2021bsb,FernandoNotes,Abl:2021mxo}. In this section we shall present such bootstrap algorithm in detail, highlighting some new features that have not been previously observed and formulating precisely the assumptions that the method is based on. We find it useful to first make some general remarks on 1d CFTs, so we start with a generic theory with $\mathfrak{sl}(2)$ invariance and spell out the conditions under which our method is applicable. Then, in the following subsections, we will apply the method to various types of four-point functions in the $\mathfrak{osp}(4^*|4)$-invariant theory we are interested in, involving both external short and long multiplets.

Although we shall make some general statements about CFTs in one dimension, our focus will be on theories with a perturbative parameter $\g$ (for the Wilson line defect at strong coupling, $g=1/\sqrt{\lambda}$). We refer to the theory at $\g=0$ as ``free theory'', although in a given Lagrangian description it might correspond to a strongly coupled point. We will adopt an analytic bootstrap approach and compute four-point functions order by order in the perturbation theory for small $\g$. 

\subsection{Generalities on 1d CFTs}

Let us start by considering a four-point function of identical operators $\varphi$ with dimension $\h_{\varphi}$ in a 1d CFT with $\mathfrak{sl}(2)$ invariance\footnote{See appendix A of \cite{Qiao:2017xif} for the set of axioms defining a 1d CFT.}. Is is easy to show that with $n$ points on a line one can form $n-3$ cross ratios, so a four-point correlator can be written in terms of a function of a single variable as
\begin{align}\label{1dCFT_fourpointsidentical}
\langle \varphi(t_1)\,\varphi(t_2)\,\varphi(t_3)\,\varphi(t_4)\rangle=\frac{1}{t_{12}^{2\h_{\varphi}}\,t_{34}^{2\h_{\varphi}}}\,G(\chi)\,,\qquad
\chi=\frac{t_{12}\,t_{34}}{t_{13}\,t_{14}}\,,
\end{align}
where $t_{ij}=t_i-t_j$ and $t_i\in \mathbb{R}$ are coordinates on a line. The presence of a unique cross ratios, $\chi$, can be compared with four-point functions in higher-dimensional CFTs, which depend on two cross ratios $u$ and $v$ (or, equivalently, $z$ and $\zb$) defined as
\begin{align}
u=z\,\zb=\frac{x^2_{12}\,x^2_{34}}{x^2_{13}\,x^2_{24}}\,,\qquad
v=(1-z)(1-\zb)=\frac{x^2_{14}\,x^2_{23}}{x^2_{13}\,x^2_{24}}\,,
\end{align}
where $x^2_{ij}=(x_i-x_j)^2$ are distances in $\mathbb{R}^d$ for $d>1$. It is straightforward to see that restricting all four operators to lie on the same line in $\mathbb{R}^d$, parametrized by the coordinate $t$ ({\it i.e.} $x_i\to t_i\in\mathbb{R}$) one obtains
\begin{align}
u=\chi^2\,, \qquad v=(1-\chi)^2\,.
\end{align}
We shall refer to this as the {\it diagonal limit} \cite{Hogervorst:2013kva,Sen:2019lec} of higher-dimensional CFTs, which can be obtained by taking $z=\zb=\chi$. A peculiar and well-known fact about 1d CFTs is that, due to the absence of a continuous group of rotations, the ordering of operators is important and in the definition of \eqref{1dCFT_fourpointsidentical} we should really specify which order we have chosen. In particular, for 
\begin{align}\label{order1234}
t_1<\dots<t_4 \quad \leftrightarrow \quad 0<\chi<1\,,
\end{align}
$G(\chi)$ reduces to a function $G_{(0)}(\chi)$ that is analytic for $\chi\in \mathbb{C}$, except for branch cuts at $\chi\in(-\infty,0]$ and $\chi \in [1,+\infty)$, as it can be argued from the OPE \cite{Hogervorst:2013sma,Qiao:2017lkv,Mazac:2018qmi}. Once the configuration \eqref{order1234} is fixed, the only transformations that leave it unchanged are cyclic permutations of the points on the (compactified) line
\begin{align}
t_i\to t_{i+1}\quad (t_5\equiv t_1)\quad \Rightarrow\quad  \chi \to 1-\chi\,,
\end{align} 
which in terms of $\chi$ have the property of mapping the interval $\chi\in(0,1)$ to itself, without crossing the branch cuts of $G_{(0)}(\chi)$. One thus obtains the relation
\begin{align}\label{cyclic1d}
(1-\chi)^{2\h_{\varphi}}\,G_{(0)}(\chi)=\chi^{2\h_{\varphi}}\,G_{(0)}(1-\chi)\,,
\end{align}
which is in general the only ``crossing symmetry'' satisfied by $G_{(0)}(\chi)$. To account for other configurations of points, following \cite{Mazac:2018qmi} one has to introduce three functions on three disjoint intervals,
\begin{align}
G(\chi)\,=\,
\begin{cases}
\,\,G_{(-)}(\chi)\quad &\text{for}\quad \chi\in(-\infty,0)\,,\\
\,\,G_{(0)}(\chi)\quad &\text{for}\quad \chi\in(0,1)\,,\\
\,\,G_{(+)}(\chi)\quad &\text{for}\quad \chi\in(1,+\infty)\,,\\
\end{cases}
\end{align}
which can be uniquely continued to complex values of $\chi$ (for example using the block expansion) but are {\it not} analytic continuations of each other, in general. Rather, they can be related by enforcing Bose symmetry of $G(\chi)$, defining
\begin{equation}\label{G0+-}
\begin{aligned}
G_{(-)}(\chi)&=G_{(0)}\big(\tfrac{\chi}{\chi-1}\big)\quad &&\text{for}\quad \chi\in(-\infty,0)\,,\\
G_{(+)}(\chi)&=\chi^{2\h_{\varphi}}\,G_{(0)}\left(\tfrac{1}{\chi}\right)\quad &&\text{for}\quad \chi\in(1,+\infty)\,.
\end{aligned}
\end{equation}
The outcome of this discussion is that in one dimension Bose symmetry is less powerful than in higher dimensional CFTs, in the following sense. Say that one wants to bootstrap $G_{(0)}(\chi)$ using Bose symmetry as a constraint: then the only transformation that acts as a symmetry of $G_{(0)}(\chi)$ is \eqref{cyclic1d}, while the other crossing symmetry transformations define independent functions. In other words, the crossing symmetry group $S_3$ of four points is reduced to a $\mathbb{Z}_2$ subgroup, so far as $G_{(0)}(\chi)$ is concerned.\footnote{Keeping $t_4$ to be the right-most operator on the line,  the six elements of $S_3$ permute $t_{1,2,3}$,  resulting in transformations for $\chi$ as
\begin{equation}\label{crossinggroupS3}
\begin{tabular}{|c|c|c|c|c|c|}
\hline
(1,2,3) & (2,1,3) & (3,2,1) & (1,3,2) & (2,3,1) & (3,1,2)\\
\hline
$\chi$ & $\tfrac{\chi}{\chi-1}$ & $1-\chi$ & $\tfrac{1}{\chi}$ & $\tfrac{\chi-1}{\chi}$ & $\frac{1}{1-\chi}$\\
\hline
$(0,1)$ & $(-\infty,0)$ & $(0,1)$ & $(1,+\infty)$ & $(-\infty,0)$ & $(1,+\infty)$\\
\hline
\end{tabular}
\end{equation}
where for completeness we have also added the images of the interval $(0,1)$ under the transformations.
} In particular, we will find it useful think of $S_3$ as generated by the identity and the two transformations $\chi\to \tfrac{\chi}{\chi-1}$ (obtained exchanging $t_1\leftrightarrow t_2$) and $\chi \to 1-\chi$ (obtained by cyclic transformations from our perspective),  which we shall refer to as braiding and cyclic transformations respectively. Then, we will say that $G_{(0)}(\chi)$ is generally not invariant under braiding.

To understand how this can happen, it is useful to consider an example, which can be provided by the same Maldacena Wilson line defect theory that we are interested in, but at weak coupling ($\lambda=0$) (see also appendix C of \cite{Liendo:2018ukf} for related discussions). Let us focus on the correlator between four insertions of one of the fundamental scalars of the theory that are not coupled to the line, which we call $\Phi_{\perp}^1$. At finite $N$, such four-point function receives contributions from three Feynman diagrams with the three independent trace structures. However, for the $SU(N)$ theory in the large $N$ limit
\begin{align}
\text{tr}(T^aT^aT^bT^b)\sim N^2\,\text{tr}(T^aT^bT^aT^b)\,,
\end{align}
so that one of the diagrams is suppressed in the limit and this spoils the symmetry under braiding transformations. Indeed, the result (in the configuration \eqref{order1234}) reads\footnote{The formula can also be obtained by setting $\alpha=\bar{\alpha}^{-1}=e^{i\,\pi/3}$ in eq. (56) of \cite{Kiryu:2018phb}.}
\begin{align}\label{1111weak}
\langle \Phi_{\perp}^1(t_1)\, \Phi_{\perp}^1(t_2)\, \Phi_{\perp}^1(t_3)\, \Phi_{\perp}^1(t_4)\rangle^{\text{weak}}=\frac{1}{t_{12}^2\,t_{34}^2}\,G_{(0)}^{\text{weak}}(\chi)=\frac{1}{t_{12}^2\,t_{34}^2}\left(1+\frac{\chi^2}{(1-\chi)^2}\right)\,,
\end{align}
which clearly satisfies \eqref{cyclic1d} but is not symmetric under $\chi \to \frac{\chi}{\chi-1}$. On the other hand, the result at strong coupling ($\lambda=\infty$) is computed by adding three disconnected Witten diagrams in AdS$_2$. Here all diagrams contribute with the same weight and the result reads\footnote{Similarly, this can be obtained by setting $\zeta_1=\zeta_2^{-1}=e^{i\,\pi/3}$ in eq. (6.9) of \cite{Liendo:2018ukf}.}
\begin{align}\label{1111strong}
\langle \Phi_{\perp}^1(t_1)\, \Phi_{\perp}^1(t_2)\, \Phi_{\perp}^1(t_3)\, \Phi_{\perp}^1(t_4)\rangle^{\text{strong}}=\frac{1}{t_{12}^2\,t_{34}^2}\,G_{(0)}^{\text{strong}}(\chi)=\frac{1}{t_{12}^2\,t_{34}^2}\left(1+\chi^2+\frac{\chi^2}{(1-\chi)^2}\right)\,,
\end{align}
which in addition to \eqref{cyclic1d} is also invariant under braiding transformations:
\begin{align}
G_{(0)}^{\text{strong}}(\chi)=G_{(0)}^{\text{strong}}\braid\,.
\end{align}
In particular, in this case $G_{(-)}(\chi)=G_{(+)}(\chi)=G_{(0)}(\chi)$ and one can easily argue that the difference between the two cases is related to the structure of the diagrammatic expansion.

The reason why this is relevant for us is related to the fact that, when one considers a 1d CFT with a dual weakly coupled description in AdS$_2$, order by order in perturbation theory correlation functions are defined by Witten diagrams in AdS$_2$, which are Bose-symmetrized by construction. Each Witten diagram is associated with a certain integral, which can be computed for instance in dimensional regularization setting the dimension to one only at the end. When doing so, it becomes manifest that there is nothing special in the one dimensional case, at least from the perspective of crossing symmetry. Hence we claim that for holographic correlators, order by order in the coupling constant, the three functions in $G_{(\pm)}(\chi)$ and $G_{(0)}(\chi)$ are not distinct but rather they can be obtained from a single function $G(\chi)$ which is invariant under the full $S_3$ Bose symmetry group which, although not holomorphic, is a single-valued function $G:\mathbb{C}\setminus\{0,1\}\to \mathbb{R}$, with singularities at $\chi=0,1,\infty$ dictated by the OPE. This property is going to be extremely useful since, as we shall discuss in detail, it puts additional constraints on correlation functions that will be crucial for our bootstrap strategy. Let us give two examples: one is a generalized free theory (GFT) correlator for operators of equal dimension $\h_{\varphi}$ (not necessarily integer). In this case
\begin{align}\label{GFT}
G(\chi)=1+|\chi|^{2\h_{\varphi}}+\left|\tfrac{\chi}{1-\chi}\right|^{2\h_{\varphi}},
\end{align}
the other is the one-dimensional box function 
\begin{align}\label{D1111_1d}
\bar{D}_{1111}=\left(\frac{\log|\chi|}{1-\chi}+\frac{\log|1-\chi|}{\chi}\right)\,,
\end{align}
both of which are invariant under all Bose symmetry transformations once absolute values are inserted in appropriate places. We will be always interested in the case when the external dimensions are (half-)integers in the free theory limit: in this case the absolute values in \eqref{GFT} are not needed, while we claim that absolute values can always be added inside the arguments of the logarithms in a suitable way as in \eqref{D1111_1d} so as to make the expressions manifestly crossing symmetric. In the case of the box function, this can be seen directly from the diagonal limit of the higher-dimensional box-function, which contains for instance 
\begin{align}
\log(z\,\zb)\to 2\log|\chi|\,.
\end{align}
Our tree level result will originate from a sum of contact diagrams, so at that order this follows from the properties of $D$-functions, but we anticipate here that at each perturbative order that we studied four-point functions on the Wilson line can be written purely in terms of functions arising from the diagonal limit of single-valued functions   of $z$ and $\zb$, hinting at a higher-dimensional origin.

So far we have only discussed the case of four identical operators, arguing that braiding transformations can be used as a symmetry to put constraints on correlation functions for theories with a weakly-coupled holographic description. In the case of non-identical operators, the same observations translate in the fact that correlators with operator insertions in different positions are related to each other by simple transformations of the argument, rather than being independent. In the following, we will only consider theories with this property.

\subsection{Logarithmic singularities, braiding and OPE coefficients}

The braiding symmetry of the free theory \eqref{1111strong} has an impact on the operators that can be exchanged in the OPE. To see this, consider the one-dimensional conformal blocks for a four-point function of identical operators
\begin{align}\label{1dblocks}
g_\h(\chi)=\chi^\h\,{}_2F_1(\h,\h,2\h;\chi)\,,
\end{align}
which are easily shown to satisfy
\begin{align}\label{braidingblocks}
g_\h(\chi)=(-1)^\h\,g_\h\braid\,,
\end{align}
which is only a symmetry for even $\h$. Hence, to reflect the braiding symmetry of \eqref{1111strong}, only operators with even dimension can be exchanged in the free theory. This parallels the fact that, in higher dimensions, in a four-point function of identical scalar operators only even spins can be exchanged, due to the fact that under exchange of the first two operators the blocks transform with
\begin{align}
g_{\Delta,l}(z,\zb)=(-1)^l\,g_{\Delta,l}\big(\tfrac{z}{z-1},\tfrac{\zb}{\zb-1}\big)\,.
\end{align}
While this is easily understood at the level of free theory, when perturbations are turned on the conformal dimensions of the exchanged operators need not be integers. Invariance under braiding, however, still bears implications on certain OPE coefficients also at the interacting level, as we shall discuss below.

In a perturbative theory with a small parameter $\g$ we consider a four-point function of identical local operators as in \eqref{1dCFT_fourpointsidentical} in the small $\g$ expansion
\begin{align}\label{G_pertexp}
G(\chi)=\sum_{\ell=0}^{\infty}G^{(\ell)}(\chi)\,,
\end{align}
where we neglect non-perturbative corrections and $\ell$ denotes the perturbative order. Imagine for simplicity that the dimensions of the external operators do not acquire perturbative corrections, although this case could be treated in a similar way as we discuss in Section \ref{sec:mixing}. Now consider the conformal blocks expansion of $G(\chi)$\footnote{Here we are going to focus on the direct channel OPE ($\chi \to 0$), focusing in particular on terms that arise in the perturbative expansion of the correlator which are not analytic at $\chi=0$. This is important to address the behavior of the four-point function under the transformation $\chi \to \tfrac{\chi}{\chi-1}$, which we refer to as braiding transformation. On the other hand, a similar analysis can be performed for all crossing transformations that do not map the interval $(0,1)$ to itself: for instance, to study the behavior of the correlator under the map $\chi \to \chi^{-1}$ one should repeat the same analysis considering the crossed channel OPE limit $\chi \to 1$.}
\begin{align}
G(\chi)=\sum_{\h}a_\h\,g_\h(\chi)\,,
\end{align}
where $a_h$ are squared OPE coefficients and the sum runs over the operators $\mathcal{O}_\h$ exchanged in the $\varphi\times \varphi$ OPE, with dimension $\h$. In a perturbative theory both the dimensions of the exchanged operators and the squared OPE coefficients can be expanded in powers of $\g$
\begin{align}
\h=\D+\sum_{\ell=1}^{\infty}\g^\ell\,\gamma^{(\ell)}_{\D}\,,
\qquad
a_h=\sum_{\ell=0}^{\infty}\g^\ell\,a^{(\ell)}_{\D}\,,
\end{align}
where we will use the convention that $\D$ denotes the free-theory value of the conformal dimensions of the operators exchanged by a certain four-point function, and therefore works as a label for the exchanged operators. We will be interested in the case when the set of allowed values for $\D$ are all integers (or half-integers for fermionic operators). While the hypergeometric function appearing in the conformal blocks \eqref{1dblocks} is analytic at $\chi=0$, the overall power of $\chi$ is not and produces logarithmic singularities via
\begin{align}\label{expansionpowerchi}
\chi^\h=\chi^{\D}\left(1+\g\,\gamma^{(1)}_{\D}\,\log\chi+\g^2(\gamma^{(2)}_{\D}\,\log\chi+\frac{1}{2}(\gamma^{(1)}_{\D})^2\,\log^2\chi)+\dots\right)\,,
\end{align}
where due to the assumption that $\D\in \mathbb{Z}$ the factor of $\chi^{\D}$ is regular at $\chi=0$ and the only non-analytic terms are explicit powers of $\log\chi$, which are higher at higher perturbative order. This leads to a natural decomposition of four-point functions at each order as
\begin{align}\label{Gpowerslog}
G^{(\ell)}(\chi)=\sum_{k=0}^{\ell}G^{(\ell)}_{\log^k}(\chi)\,\log^k\chi\,,
\end{align}
where each $G^{(\ell)}_{\log^k}(\chi)$ is analytic at $\chi=0$ and admits an expansion in conformal blocks and their derivatives\footnote{This decomposition was introduced in \cite{Liendo:2018ukf} in the context of 1d CFTs, see also \cite{Bissi:2020wtv,Bissi:2020woe} for a related analysis of all-order structures associated with logarithmic singularities in AdS$_5\times S^5$. }. For the leading logarithmic singularities, such expansions read\footnote{Here we neglect issues due to mixing, which we will come back to later in the paper.}
\begin{align}\label{logpowers_expansions}
\begin{split}
G^{(\ell)}_{\log^\ell}(\chi)=&\frac{1}{\ell!}\sum_{\D}a^{(0)}_{\D}\,\big(\gamma^{(1)}_{\D}\big)^{\ell}\,g_{\D}(\chi)\,,\\
G^{(\ell)}_{\log^{\ell-1}}(\chi)=&\frac{1}{(\ell-1)!}\sum_{\D}\left[\left(a^{(1)}_{\D}\big(\gamma^{(1)}_{\D}\big)^{\ell-1}+(\ell-1)\,a^{(0)}_{\D}\,\gamma^{(2)}_{\D}\,\big(\gamma^{(1)}_{\D}\big)^{\ell-2}\right)\,g_{\D}(\chi)+a^{(0)}_{\D}\,\big(\gamma^{(1)}_{\D}\big)^{\ell}\,g^{(1)}_{\D}(\chi)\right]\,,\\
G^{(\ell)}_{\log^{\ell-2}}(\chi)=&\frac{1}{(\ell-2)!}\sum_{\D}\left[\Big(a^{(2)}_{\D}\big(\gamma^{(1)}_{\D}\big)^{\ell-2}+(\ell-2)\,\big(a^{(0)}_{\D}\,\gamma^{(3)}_{\D}+a^{(1)}_{\D}\,\gamma^{(2)}_{\D}\big)\,\big(\gamma^{(1)}_{\D}\big)^{\ell-3}\right.\\
&\left.+\frac{1}{2}(\ell-2)(\ell-3)a^{(0)}_{\D}\,\big(\gamma^{(2)}_{\D}\big)^{2}\,\big(\gamma^{(1)}_{\D}\big)^{\ell-4}\Big)\,g_{\D}(\chi)\right.\\
&\left. +\left(a^{(1)}_{\D}\big(\gamma^{(1)}_{\D}\big)^{\ell-1}+(\ell-1)\,a^{(0)}_{\D}\,\gamma^{(2)}_{\D}\,\big(\gamma^{(1)}_{\D}\big)^{\ell-2}\right)\,g^{(1)}_{\D}(\chi)+\frac{1}{2}a^{(0)}_{\D}\,\big(\gamma^{(1)}_{\D}\big)^{\ell}\,g^{(2)}_{\D}(\chi)\right]\,,
\end{split}
\end{align}
where we have introduced the notation
\begin{align}
g^{(n)}_{\D}(\chi)=\chi^{\D}\,\big(\partial_{\D}\big)^{n}\,\chi^{-\D}\,g_{\D}(\chi)\,.
\end{align}
Closed form expressions for the expansion of $G^{(\ell)}_{\log^k}(\chi)$ for general $k$ and $\ell$ can be found in Appendix \ref{app:perturbativeOPE}.

We will come back to how precisely the CFT data appear in the expansions \eqref{logpowers_expansions} and the implication for our bootstrap algorithm as well as for the mixing problem. For the moment we would like to highlight two aspects of the decomposition \eqref{Gpowerslog} and its relation to braiding symmetry. First, as we stressed in the previous section, when we consider results that arise from a sum over AdS Witten diagrams it {\it must} be possible to write the results in such a way that the whole Bose symmetry group is manifest, simply due to the fact that we are summing over all permutations of external points by construction. One might argue that an expression like \eqref{Gpowerslog} is problematic from the point of view of braiding, since the map $\chi \to \tfrac{\chi}{\chi-1}$ implies that the image $\log\chi$ develops an imaginary part for $0<\chi<1$ and $G^{(\ell)}(\chi)$ cannot be symmetric under braiding at face value. Our remedy to this is to extend the definition of $G^{(\ell)}(\chi)$ for $\chi\in \mathbb{R}\setminus\{0,1\}$ by replacing $\log\chi \to \log|\chi|$ and similarly $\log(1-\chi)\to \log|1-\chi|$ for the logarithms arising in the crossed-channel OPE. In all the examples that we shall study this will be enough to define an extension $\bar{G}^{(\ell)}(\chi)$ of the correlator such that 
\begin{align}\label{braidingGbar}
\bar{G}^{(\ell)}\braid=\bar{G}^{(\ell)}(\chi)\,,
\end{align}
thus making Bose symmetry manifest. The small $\chi$ expansion of $\bar{G}^{(\ell)}(\chi)$ is obtained from \eqref{Gpowerslog} simply replacing $\log\chi$ with $\log|\chi|$ as described above,
\begin{align}
\bar{G}^{(\ell)}(\chi)=\sum_{k=0}^{\ell}G^{(\ell)}_{\log^k}(\chi)\,\log^k|\chi|\,,
\end{align}
which combined with \eqref{braidingGbar} gives\footnote{Later we will make an ansatz for $G^{(\ell)}(\chi)$ in terms of HPLs of transcendentality up to $\ell$, so that $G^{(\ell)}_{\log^k}(\chi)$ has transcendentality at most $\ell-k$. Note that both sides of \eqref{braiding1d} then have the same maximal transcendentality.}
\begin{align}\label{braiding1d}
G^{(\ell)}_{\log^k}(\chi)=\sum_{m=0}^{\ell-k}\binom{k+m}{m}(-1)^m\log^m(1-\chi)\,G^{(\ell)}_{\log^{k+m}}\braid\,,
\end{align}
where we remind the reader that this is an identity that make sense from the point of view of the small $\chi$ expansion and $G^{(\ell)}_{\log^k}(\chi)$ is analytic there. This can be also derived by looking at the transformation properties of conformal blocks and their derivatives when $\D$ is even, an approach that we consider in Appendix \ref{app:perturbativeOPE}. 

The other comment that we wish to make here is related to the OPE coefficients of exchanged operators with odd dimensions. In particular, at the start of this subsection we argued that due to the Bose symmetry of the free theory
\begin{align}
a^{(0)}_{\DOdd}=0\,,
\end{align}
and one can wonder whether this remains true at higher perturbative orders. Regardless of their anomalous dimension, operators with odd dimension at $\g=0$ with an OPE coefficient that is non-vanishing at a certain order $\bar{\ell}$ would appear in the OPE at that order with a term $a^{(\bar{\ell})}_{\DOdd}\,g_{\DOdd}(\chi)$, with derivatives of blocks and anomalous dimensions only appearing at higher orders. However, such term is manifestly antisymmetric under braiding, due to \eqref{braidingblocks} with $h\to \DOdd$. Thus, terms of this kind would contradict \eqref{braidingGbar} explicitly and cannot be generated in the OPE of perturbative correlators arising from Witten diagrams.  In the recent work \cite{Cavaglia:2023mmu} the authors introduced a parity operator that justifies the vanishing of certain OPE coefficients in the Wilson line defect CFT, based on its weak coupling description.

\subsection{The ansatz}

The bootstrap method that we use is based on making an ansatz for the spacetime expression of correlation functions order by order in perturbation theory. This should be seen as complementing the discussion of \cite{Ferrero:2021bsb} with more details. Let us start by laying out our assumptions, which can be summarized as follows:
\begin{enumerate}
\item[a.] We consider 1d CFTs with a small parameter $\g$, which could for instance arise as duals to perturbative QFTs in AdS$_2$ as in the case of interest here. 
\item[b.] We focus on correlation functions between (not necessarily identical) operators that, in the free theory limit $g=0$, have integer conformal dimension (or half-integer, if fermions). The external operators need not be protected: the method still applies when the external operators acquire anomalous dimensions perturbatively, with minor modifications and some caveats, as we discuss in Section \ref{sec:mixing}.
\item[c.] The free theory has an exact invariance under the full $S_3$ group of Bose symmetry transformations, in the sense described above. We also assume that this remains true at all perturbative orders, once a suitable prescription is given on how to define the correlators for $\chi<0$ or $\chi>1$ (see the discussion of the previous section).
\end{enumerate}

Given that we work order by order in perturbation theory, we shall make an ansatz for each $G^{(\ell)}(\chi)$ appearing in \eqref{G_pertexp}, where we refer to $\ell=0$ as the free theory, $\ell=1$ as tree level and $\ell>1$ as the result at $\ell-1$ loops, since we have in mind a corresponding expansion over Witten diagrams. For each fixed $\ell$, we make the ansatz that
\begin{align}\label{genericansatz}
G^{(\ell)}(\chi)=\sum_{i=1}^{N(\ell)}r_i(\chi)\, \mathcal{T}_i(\chi)\,,
\end{align}
where $r_i(\chi)$ are rational functions, while $\mathcal{T}_i(\chi)$ are transcendental functions from a chosen basis of dimension $N(\ell)$,  which depend on the perturbative order $\ell$. Furthermore, we claim\footnote{The authors are thankful to Luis Fernando Alday for sharing with us some unpublished notes suggesting this idea \cite{FernandoNotes} and for very useful discussions on this topic.} that the correct basis of transcendental functions for the problem at hand is given by multiple polylogarithms (MPLs) of argument $\chi$, whose only singularities for $\chi\in \mathbb{C}$ are located at $\chi\in\{0,1\}$. Following \cite{Dixon:2012yy,Drummond:2013nda} we refer to such functions as harmonic polylogarithms (HPLs) and they can be defined through iterated integrals using 
\begin{align}\label{defHPLs}
H(a_1,\dots,a_n;\chi)=\int_{0}^{\chi}\text{d}t\,f_{a_1}(t)\,H(a_2,\dots,a_n;t)\,, \qquad a_i\in\{0,1\}\,,
\end{align}
with 
\begin{align}
f_0(t)=\frac{1}{t}\,, \qquad f_1(t)=\frac{1}{1-t}\,.
\end{align}
This definition is complemented with $H(\chi)=1$ when all $a_i$ are zero and 
\begin{align}
H(\vec{0}_n;\chi)=\frac{1}{n!}\log^n\chi\,.
\end{align}
The number $n$ of indices of the {\it vector of singularities} $\vec{a}$ in \eqref{defHPLs} is called the {\it weight},  or {\it transcendentality},  of a certain HPL,  and will be denoted with the letter $\mathtt{t}$ (so in the definition above $\mathtt{t}=n$). One should then read the sum \eqref{genericansatz} as the instruction to sum over all possible HPLs up to a certain maximal transcendentality $\mathtt{t_{max}}(\ell)$, which for each $\ell$ depends on the specific problem that one is considering and should be seen as an external input:
\begin{align}
\mathcal{T}_i(\chi)\in \{\text{HPLs of transcendentality } \mathtt{t}\le  \mathtt{t_{max}}(\ell)\},
\end{align}
where we have stressed that $\mathtt{t_{max}}$ depend on the perturbative order $\ell$ that one is considering. We will discuss later how one can formulate a guess for $\mathtt{t_{max}}(\ell)$ case by case, for the moment let us just mention a couple of examples where the answer is known. In the case of $\varphi^4$-type derivative interactions in AdS$_2$ considered in \cite{Ferrero:2019luz} one has $\mathtt{t_{max}}(1)=1$,  $\mathtt{t_{max}}(2)=4$,  while for the half-BPS Wilson line defect theory that we consider here we shall argue that we expect $\mathtt{t_{max}}(\ell)=\ell$ for all $\ell$,  a claim which is supported by our explicit results for $\ell\le 4$. Given that for each $\mathtt{t}$ there are $2^{\mathtt{t}}$ independent HPLs (corresponding to the choices of $a_i\in\{0,1\}$),  the number $N(\ell)$ of functions in the basis appearing in \eqref{genericansatz} is given by
\begin{align}
N(\ell)=\sum_{\mathtt{t}=0}^{\mathtt{t_{max}}(\ell)} 2^{\mathtt{t}}=2^{1+\mathtt{t_{max}}(\ell)}-1\,.
\end{align}
More details on HPLs and their properties are presented in Appendix \ref{app:polylogs}.

Let us now spend some words of motivation for the ansatz \eqref{genericansatz}. First of all, the appearance of multiple polylogarithms in perturbative computations is ubiquitous in physics\footnote{A complete list of references on the appearance and use of polylogarithms in perturbative computations is beyond the scope of this work.  More detailed and complete lists of references can be found in the papers that we cite in this paragraph.} and therefore they provide a natural candidate for the transcendental functions $\mathcal{T}_i(\chi)$ in \eqref{genericansatz}. MPLs are well known to arise as the result of Feynman integrals in generic quantum field theories \cite{Duhr:2011zq},  in particular in the computation of scattering amplitudes.  This has been observed for a variety of models,  ranging from scattering in the Standard Model \cite{Gehrmann:2000zt,Vogt:2004mw,Moch:2004pa,Vermaseren:2005qc,Duhr:2012fh} to superstring theory \cite{Broedel:2013tta,Broedel:2014vla},  and especially for gluon amplitudes in $d=4$,   $\mathcal{N}=4$ SYM \cite{Bern:2005iz,Bern:2006ew,DelDuca:2009au,DelDuca:2010zg,Goncharov:2010jf,Dixon:2011pw,Golden:2013xva,Dixon:2013eka,Dixon:2014voa,Dixon:2014iba,Drummond:2014ffa,Dixon:2015iva,Dixon:2016nkn,Caron-Huot:2020bkp}.  Moreover,  multiple polylogarithms appear in perturbative correlators in various CFTs,  including the $\epsilon$-expansion \cite{Alday:2017zzv,Henriksson:2018myn,Guha:2019ipe} and again $d=4$,   $\mathcal{N}=4$ SYM,  both at weak \cite{Eden:2000mv,Eden:2011we,Drummond:2013nda,Chicherin:2015edu} and at strong \cite{Aprile:2017bgs,Bissi:2020woe,Huang:2021xws} coupling (in the planar limit).  This structure has been explored in particular for holographic CFTs,  where the AdS/CFT correspondence gives a recipe to compute correlation functions using Witten diagrams.  While the latter give rise to more complicated integrals than ordinary Feynman diagrams,  it appears that the results can still be expressed in terms of polylogarithms (or more general transcendental functions) in many cases; see, {\it e.g.}, \cite{Aharony:2016dwx,  Carmi:2019ocp, Ferrero:2019luz, Carmi:2021dsn,Ferrero:2021bsb}. In some cases, it is even possible to explicitly write Witten diagrams in terms of ordinary Feynman diagrams, making the analogy between the two even closer \cite{Heckelbacher:2022fbx}. For the Wilson line defect CFT, HPLs were also observed to appear in the large charge expansion of four-point functions between half-BPS operators \cite{Giombi:2022anm}, as well as at weak coupling \cite{Cavaglia:2022qpg} for the displacement four-point function.

Once convinced that the result should be expressed in terms of MPLs, the restriction to HPLs arises naturally from the properties inherited by $G^{(\ell)}(\chi)$ from the OPE structure of 1d CFTs. In particular, studying the convergence of the OPE it can be argued that $G^{(\ell)}(\chi)$ should be analytic for $\chi\in \mathbb{C}$ except for branch cuts at $\chi\in (-\infty,0]$ and $\chi\in [1,+\infty)$: for this reason one should restrict to MPLs with branch points located at $\chi\in\{0,1,\infty\}$, which leads precisely to HPLs. Moreover, as we review in Appendix \ref{app:polylogs}, HPLs form a closed set under Bose symmetry transformations $S_3$ defined in \eqref{crossinggroupS3}, {\it i.e.} once a basis for HPLs of a given transcendentality is chosen the action of $S_3$ is such that the functions of transformed argument can be always written in terms of the set of functions chosen as basis. 

A further motivation follows from the structure of the OPE in perturbative theories and in particular from the decomposition \eqref{Gpowerslog}. As we explained in the previous section, each $G^{(\ell)}_{\log^k}(\chi)$ has an expansion in terms of conformal blocks and their derivatives, with coefficients given by various combinations of OPE data as in \eqref{logpowers_expansions}. In some cases, as we shall explain in more detail in the next section, one can compute the sums \eqref{logpowers_expansions} at a given order from CFT data at previous orders, thus extracting a certain part of the correlator. In all examples that appeared in the literature the functions appearing as results in these sums are HPLs\footnote{Note that since we are making the powers of $\log\chi$ explicit in \eqref{Gpowerslog}, each $G^{(\ell)}_{\log^k}(\chi)$ can written in terms of HPLs that are regular at $\chi=0$.}, which as explained map to other HPLs after crossing symmetry transformations: case by case and order by order this provides a strong consistency check of our assumptions. Moreover, looking at the $G^{(\ell)}_{\log^k}(\chi)$ that can be computed from CFT data at previous orders also provides strong evidence of what the maximal transcendentality $\mathtt{t_{max}}(\ell)$ should be at each order, assuming that each $G^{(\ell)}_{\log^k}(\chi)\log^k\chi$ has the same maximal transcendentality $\mathtt{t_{max}}(\ell)$.

Finally let us mention that, for most correlators that we bootstrap, one can perform non-trivial self-consistency checks by looking at CFT data that can be extracted from multiple correlators. As we explain, in some cases we shall use data extracted from one correlator as an input to bootstrap a new one, but it is often the case that by implementing this procedure we encounter over-constrained systems of equations: the existence of a solution is thus a strong check on our procedure and results. Another powerful consistency check is the agreement between some of our CFT data and the strong coupling regime of analogous data that can be computed numerically using integrability \cite{Grabner:2020nis} and its combination with the numerical bootstrap \cite{Cavaglia:2021bnz,Cavaglia:2022qpg}.
 
Going back to the details of the ansatz \eqref{genericansatz}, let us comment on the other ingredient appearing in that formula: the rational functions $r_i(\chi)$.  Because of the analytic structure of $G^{(\ell)}(\chi)$ discussed above,  each $r_i(\chi)$ must be the ratio between a polynomial in $\chi$ and certain powers of $\chi$ and $1-\chi$,
\begin{align}\label{r_to_p}
r_i(\chi)=\frac{p_i(\chi)}{\chi^{a_i}\,(1-\chi)^{b_i}}\,.
\end{align}
This way,  the only poles are located at $\chi=0,1$ and the problem of finding the $r_i(\chi)$ is translated into that of finding $p_i(\chi)$,  for given values of $a_i$ and $b_i$. Once these and the degree of $p_i(\chi)$ are fixed (we will discuss later how to do so), it should be clear that the only unknowns in \eqref{genericansatz} are the coefficients of individual powers of $\chi$ in the polynomials $p_i(\chi)$. This drastically simplifies the problem: all constraints turn into linear systems of equations for the coefficients in the $p_i(\chi)$, as opposed to the non-linear problem arising if unknowns are present in the denominator as well. We would also like to emphasize that when applying Bose symmetry transformations to the arguments of HPLs,  and then relating the transformed functions to those that were chosen as basis,  one generates multiple zeta-values (MZVs),  {\it i.e.} the result of evaluating HPLs at unit argument (when they are regular at that point), see Appendix \ref{app:polylogs}.  One should then allow the coefficients in $p_i(\chi)$ to take values in a field given by the extension of $\mathbb{Q}$ by all the possible MZVs allowed by the maximal transcendentality of the correlator,  $\mathtt{t_{max}}(\ell)$\footnote{Since we are going to deal with HPLs of weight at most four,  the only MZVs that are relevant for this paper are ordinary zeta values $\zeta(n)$ with $n=2,3,4$.  For example, if for a certain value of $\ell$ one has $\mathtt{t_{max}}(\ell)=4$ and $\mathcal{T}_i(\chi)=\log\chi$ for a given value of $i$,  then we take
\begin{align}
r_i(\chi)=r_{i,1}(\chi)+\zeta(2)\,r^{(2)}_{i,2}(\chi)+\zeta(3)\,r^{(3)}_{i,3}(\chi)\,,
\end{align}
where $r_{i,j}(\chi)$ are rational functions whose numerators are polynomials in $\chi$ with rational coefficients.}. 

\subsection{The constraints}

Now that we have clarified all the ingredients that enter in the ansatz \eqref{genericansatz},  we are ready to list the constraints that we use to fix the rational functions $r_i(\chi)$.  We shall distinguish such constraints in two classes. The first class of constraints can be shown to be sufficient to compute, at each order, all terms in $G^{(\ell)}(\chi)$ that have transcendentality $\mathtt{t}\ge 2$. This fixes the result up to terms of transcendentality 0 or 1, namely combinations of $\{1,\log\chi,\log(1-\chi)\}$: in holographic theories they can be seen as due contact terms, as the diagonal limit of $\bar{D}$ functions has precisely this type of structure. This leads to the second class of constraints,  which allows to fix the remaining functions by inputting some additional information as we discuss below (in particular, an important role will be played by the Regge limit).  This procedure can be understood as the analogue of reconstructing loop level results from their discontinuities (given here by logarithmic singularities), which of course can be done only up to terms with no discontinuity, {\it i.e.} tree level contact terms in this case.

Let us start with the first type of constraints, which we enforce on an ansatz of the type \eqref{genericansatz} where all rational functions $r_i(\chi)$ are unknown. We proceed in two steps:
\begin{itemize}
\item {\it AdS unitarity method.} We note that in the decomposition \eqref{Gpowerslog} all functions $G^{(\ell)}_{\log^k}(\chi)$ with $k\ge 2$ can be obtained from CFT data that can be extracted from previous perturbative orders\footnote{This is true up to problems with operators mixing. If there is no mixing, considering a single correlator order by order is enough. If mixing is present, on the other hand, one should consider a system of correlators in order to extract the CFT data. This will be described in detail in Section \ref{sec:mixing}.}, by explicitly performing sums like those in \eqref{logpowers_expansions}. We will refer to these terms as the ``highest logarithmic singularities'' of the correlator. For all the sums of this type that we are able to perform, we find that
\begin{enumerate}
\item[1)] the $G^{(\ell)}_{\log^k}(\chi)$ can be expressed in terms of HPLs that are regular at $\chi=0$,
\item[2)] $G^{(\ell)}_{\log^k}(\chi)$ has $\mathtt{t}=\ell-k$, supporting our claim that $\mathtt{t_{max}}(\ell)=\ell$ in this case. 
\end{enumerate}
The idea that (at least part of) loop level results can be understood from results computed at previous orders in perturbation theory dates back to the work of Cutkosky \cite{Cutkosky:1960sp},  and was first introduced for the computation of holographic correlators in \cite{Aharony:2016dwx},  where it was termed ``AdS unitarity method''\footnote{See \cite{Meltzer:2019nbs, Meltzer:2020qbr} for recent works that put the AdS unitarity method on a firmer footing,  clarifying its precise connections to S-matrix theory.}.  This method has since been applied,  in various incarnations,  for the computation of holographic correlators at loop level in a variety of theories \cite{Aprile:2017bgs,Alday:2017xua,Alday:2019nin,Alday:2020tgi,Alday:2021ymb,Alday:2021ajh,Behan:2022uqr}, often in combination with the Lorentzian inversion formula \cite{Caron-Huot:2017vep}\footnote{An inversion formula that applies to one-dimensional CFTs has been developed in \cite{Mazac:2018qmi},  and it would be interesting to explore more precisely the connections with the statements made here.}. 

\item {\it Bose symmetry.} Once the highest logarithmic singularities are known, one can require that $G^{(\ell)}(\chi)$ is Bose-symmetric. As we discussed, cyclic transformations $\chi\to 1-\chi$ lead to the simple crossing equation \eqref{cyclic1d}, while for braiding transformations $\chi\to\tfrac{\chi}{\chi-1}$ all one needs to do is replace $\log\chi\to \log|\chi|$ in \eqref{Gpowerslog} and require invariance, which leads to equations like \eqref{braiding1d}. It turns out that, given the ansatz \eqref{genericansatz}, through the identities between HPLs at various arguments listed in appendix  \ref{app:polylogs}, Bose symmetry propagates the information  obtained from the highest logarithmic singularities to {\it all} rational functions multiplying HPLs (or MZVs) of transcendentality $\mathtt{t}\ge 2$, which can therefore be computed exactly without the need for an explicit ansatz for them. The only rational functions that cannot be fixed by this procedure are those multiplying $\{1,\log\chi,\log(1-\chi)\}$, which are only found to satisfy certain crossing equations that we shall discuss soon.
\end{itemize}

Once the procedure described above is used, one knows the correlator up to terms of the type 
\begin{align}\label{treeambiguity}
Q(\chi)=a(\chi)+b(\chi)\,\log(1-\chi)+c(\chi)\,\log\chi\,,
\end{align}
which can be interpreted as contact terms ambiguities,  given that this is precisely the structure of $D$-functions in $d=1$, where the rational functions defining $Q(\chi)$ satisfy the homogeneous crossing equations
\begin{gather}\nonumber
b(\chi)=\left(\tfrac{\chi}{1-\chi}\right)^{2\h_{\varphi}}c(1-\chi)\,, \quad
c(\chi)=c\big(\tfrac{\chi}{\chi-1}\big)\,, \quad 
a(\chi)=a\big(\tfrac{\chi}{\chi-1}\big)\,, \\ 
c(\chi)+\left(\tfrac{\chi}{1-\chi}\right)^{2\h_{\varphi}}\,c(1-\chi)+\chi^{2\h_{\varphi}}\,c\left(\tfrac{1}{1-\chi}\right)=0\,.
\end{gather}
As described in \cite{Liendo:2018ukf,Gimenez-Grau:2019hez,Ferrero:2019luz},  to solve this problem one makes an ansatz for the functions $a$ and $c$ (since $b$ is completely fixed by crossing in terms of $c$) as
\begin{align}\label{rationalansatz}
a(\chi)=\frac{p_a(\chi)}{\chi^{\mathtt{k}}\,(1-\chi)^{\mathtt{k}}}\,, \quad
c(\chi)=\frac{p_c(\chi)}{\chi^{\mathtt{k}}\,(1-\chi)^{\mathtt{k}}}\,,
\end{align}
where $p_a$ and $p_c$ are polynomials and ${\mathtt{k}}\ge 0$.  The denominators are chosen in such a way that the only poles are located at $\chi=0,1$,  in according with the structure of the OPE. A convenient strategy is then that of studying solutions for increasing values of ${\mathtt{k}}$ and, for each ${\mathtt{k}}$, vary the degree of the polynomials $p_a$ and $p_c$. As discussed in detail in \cite{Ferrero:2019luz}, there is a one-to-one correspondence between the values of ${\mathtt{k}}$ in solutions that are also compatible with the OPE at $\chi=0$ and the independent quartic contact terms with derivatives that one can write in AdS$_2$. To fix the correlator completely one should then input some external assumption to fix the coefficients of such derivatives contact terms. We do this by employing the second class of constraints, which can be summarized by the following two points:
\begin{itemize}
\item {\it Compatibility with the OPE.} The expansion of $Q(\chi)$ for small $\chi$ must be compatible with the structure of the OPE at the perturbative order $\ell$ that one is considering. In particular,  we have constraints of two kinds.  The first is related to the fact that we always know which is the lowest-dimensional operator exchanged in the OPE.  This sets a cutoff on the lowest power of $\chi$ that can appear in the small $\chi$ expansion of $G^{(\ell)}(\chi)$.  Due to this fact,  for each fixed ${\mathtt{k}}$ we can only find a finite number of solutions, regardless of how high the degree of $p_a$ and $p_c$ is. The other kind of constraint is related to the knowledge of certain OPE coefficients,  or anomalous dimensions,  of operators that appear in the OPE.  We shall encounter an example of this when discussing $\langle \mathcal{D}_p\mathcal{D}_p\mathcal{D}_q\mathcal{D}_q \rangle$ correlators in Section \ref{sec:ppqq}: in that case one exchanges short operators $\mathcal{D}_k$,  and all the OPE coefficients $\mathsf{C}_{pqr}\sim \langle \mathcal{D}_p\mathcal{D}_q\mathcal{D}_r\rangle$ are known from localization.  This fact also provides a definition of the coupling constant,  which is otherwise arbitrary in a bootstrap setup.  Another example will be discussed in Section \ref{sec:mixing},  where to bootstrap $\langle \mathcal{D}_1\mathcal{D}_1\mathcal{D}_2 \mathcal{L}\rangle$ correlators we will use as an input certain averaged anomalous dimensions that are known from other correlators.
\item {\it Regge limit.} As anticipated, the solutions for fixed ${\mathtt{k}}$ are related to the possibility of having contact terms of the type $(\nabla^n\varphi)^4$ in the AdS$_2$ effective action. For increasing value of ${\mathtt{k}}$ (or $n$) these have a more and more singular behavior in the Regge limit (as defined in \cite{Mazac:2018ycv}), which corresponds to a higher degree of divergence of the averaged anomalous dimensions $\langle\gamma^{(\ell)}\rangle_{\Delta}$ at large $\Delta$. We refer the reader to \cite{Ferrero:2019luz} for a more detailed discussion of these aspects, while here we limit to state our prescription, which consists in {\it assuming} that the correlators we are after have the mildest possible behavior in the Regge limit, corresponding to the mildest growth of $\langle\gamma^{(\ell)}\rangle_{\Delta}$ at large $\Delta$. In practice we find that, at least for $\ell\le 4$, this is given by
\begin{align}\label{gammasReggeGeneral}
\langle\gamma^{(\ell)}_{\Delta}\rangle\sim \Delta^{\ell+1}\,,\qquad \Delta\to \infty\,.
\end{align}
Roughly speaking, our prescription can be understood as demanding that one does not include unnecessary ({\it i.e.} with too high number of derivatives) counterterms to loop level results.
\end{itemize}

In the following sections we apply this method to various types of four-point functions between operators inserted along the Maldacena Wilson line, in a strong coupling expansion. Each time we will clarify the necessary modifications to the simple setup considered in this section, but the spirit of the procedure will remain unaltered.

\subsection{The mixing problem - a bootstrap perspective}\label{sec:mixing_bootstrap}

The applicability of the strategy discussed above rests on the fact that, when bootstrapping a certain correlator at fixed perturbative order, all CFT data at previous orders are known, so that the coefficients of the highest logarithmic singularities can be computed computed through sums of the type \eqref{logpowers_expansions}. However, in concrete examples of perturbative CFTs this is often much easier said than done, due to the presence of a degeneracy between inequivalent operators with the same quantum numbers in the free theory. A simple example of this is given by a free theory with a single scalar operator $\varphi$ of dimension $\h_{\varphi}=1$: in this case one can built two distinct operators of dimension four, which are schematically of the form $\partial^2\varphi^2$ and $\varphi^4$. As we now explain, one cannot distinguish the two from the OPE of a unique correlation function and is therefore led to analyze a multi-correlator analysis.

To understand why operators degeneracy is problematic from the point of view of the analytic bootstrap, imagine to be in a situation like the one just described, where in the interacting theory there are two operators $\mathcal{O}_{1,2}$ with the same quantum numbers under global and R-symmetries, whose dimensions $\h_{1,2}(g)$ are functions of the coupling $g$ such that, in the expansion for small $g$:
\begin{align}
h_i(g)=\Delta+g\,\gamma^{(1)}_{\mathcal{O}_i}+g^2\,\gamma^{(2)}_{\mathcal{O}_i}+\ldots\,,\qquad (i=1,2)\,,
\end{align}
where the free-theory dimension $\Delta$ is the same for both. Then, in the OPE of a given four-point function where these appear as intermediate operators, one has (at the non-perturbative level)
\begin{align}
G(\chi)\supset G_{\Delta}(\chi)\equiv a_1\,g_{h_1}(\chi)+a_2\,g_{h_2}(\chi)\,,
\end{align}
where $a_{1,2}\equiv \mu_{1,2}^2$ are squared OPE coefficients and for brevity we are denoting $a_1\equiv a_{\mathcal{O}_1}$ and so on. Moreover, we stress here that we are going to denote with $\mu$ OPE coefficients that involve at least one long multiplet, which are those relevant for the mixing problem discussed here. On the other hand, when dealing with three-point functions between half-BPS multiplets $\mathcal{D}_k$ we are going to use the symbol $\mathsf{C}$: these have the property that they are non-degenerate and moreover can be computed exactly from localization, as we shall discuss. Then, expanding for small coupling $g$ one finds
\begin{align}
\begin{split}
G_{\Delta}(\chi)&=[(\mu_1^{(0)})^2+(\mu_2^{(0)})^2]\,g_{\Delta}(\chi)\\
&+g\,\left\{[(\mu_1^{(0)})^2\,\gamma^{(1)}_1+(\mu_2^{(0)})^2\,\gamma^{(1)}_2]\,g_{\Delta}(\chi)\,\log\chi+\ldots\right\}\\
&+g^2\,\left\{\frac{1}{2}[(\mu_1^{(0)})^2\,(\gamma^{(1)}_1)^2+(\mu_2^{(0)})^2\,(\gamma^{(1)}_2)^2]\,g_{\Delta}(\chi)\,\log^2\chi+\ldots\right\}+\mathcal{O}(g^3)\,,
\end{split}
\end{align}
where we have only emphasized certain terms that we are interested in, so that one is led to the introduction of the quantities
\begin{align}\label{averagesCFTdata}
\begin{split}
\langle a^{(0)}\rangle_{\Delta}&\equiv (\mu_1^{(0)})^2+(\mu_2^{(0)})^2\,,\\
\langle a^{(0)}\,\gamma^{(1)}\rangle_{\Delta}&\equiv (\mu_1^{(0)})^2\,\gamma^{(1)}_1+(\mu_2^{(0)})^2\,\gamma^{(1)}_2\,,\\
\langle a^{(0)}\,(\gamma^{(1)})^2\rangle_{\Delta}&\equiv(\mu_1^{(0)})^2\,(\gamma^{(1)}_1)^2+(\mu_2^{(0)})^2\,(\gamma^{(1)}_2)^2\,,
\end{split}
\end{align}
where we are essentially computing weighted averages with weight given by squared OPE coefficients. The latter depend on the external operators (as opposed to the anomalous dimensions), so the result of the average depends on the four-point function that one is considering.

Now, say that we know $G^{(0)}(\chi)$ and $G^{(1)}(\chi)$ and we would like to bootstrap $G^{(2)}(\chi)$ using the method described in this section. From $G^{(0)}(\chi)$ one can extract $\langle a^{(0)}\rangle_{\Delta}$ and from $G^{(1)}(\chi)$ the OPE gives access to the value of $\langle a^{(0)}\,\gamma^{(1)}\rangle_{\Delta}$, but these are not sufficient to compute the quantity $\langle a^{(0)}\,(\gamma^{(1)})^2\rangle_{\Delta}$ which is necessary to compute the $\log^2\chi$ singularity of $G^{(2)}(\chi)$. This explains the well-known fact that, when degeneracy is present in the free theory, studying a unique correlator in perturbation theory is not enough to access all CFT data, but rather only some weighted averages thereof. The approach that is usually taken to bypass this issue is to consider, at a given order, not just one correlator but a large enough family in such a way as to obtain many data points for the averages \eqref{averagesCFTdata}. If one obtains enough independent equations, it becomes possible to extract the individual CFT data, thus ``resolving'' the mixing problem. Note also that the OPE coefficients $\mu^{(0)}_{1,2}$ appear in a rotationally-invariant combination in $\langle a^{(0)}\rangle_{\Delta}$, so that by only looking at free-theory correlators one could only compute such OPE coefficients up to a possible rotation in the space $(\mu^{(0)}_{1},\mu^{(0)}_{2})$. Such ambiguity can only be solved by looking at the averages $\langle a^{(0)}\,\gamma^{(1)}\rangle_{\Delta}$, which manifestly break such rotational invariance (when $\gamma^{(1)}_1\neq \gamma^{(1)}_2$): the computation of free-theory OPE coefficients is then closely related to the mixing problem at first order. Similarly,  with this approach at higher orders one can only fully resolve the OPE coefficients $\mu^{(\ell)}_i$ together with the anomalous dimensions $\gamma^{(\ell+1)}_i$. 

However, note that as anticipated our case is special, in the sense that the first-order correction to the dilatation operator does not break the free-theory degeneracy. In the example, this corresponds to the case where $\gamma^{(1)}_1= \gamma^{(1)}_2$, so that it is impossible to completely disentangle $\mu^{(0)}_{1,2}$ just by looking at the averages $\langle a^{(0)}\rangle_{\Delta}$ and $\langle a^{(0)}\,\gamma^{(1)}\rangle_{\Delta}$, even if an infinite amount of correlators is considered. Rather, one should move to the following order and consider $\langle a^{(0)}\,\gamma^{(2)}\rangle_{\Delta}$: this time we will find $\gamma^{(2)}_1\neq \gamma^{(2)}_2$ and studying enough averaged data allows one to compute $\mu^{(0)}_{1,2}$ and $\gamma^{(2)}_{1,2}$. Similarly, in our setup at higher orders the computation of $\mu^{(\ell)}_{1,2}$ is closely tied to the resolution of mixing for $\gamma^{(\ell+2)}_{1,2}$.

It is sometimes useful to choose a basis of operators which is {\it not} necessarily the basis of eigenstates of the dilatation operator, and we will exploit this freedom in Section \ref{sec:mixing}. This is particularly true when one has a Lagrangian description at the free theory point $g=0$, which allows for an explicit construction of operators in terms of elementary fields: for the case at hand this is discussed in detail in \cite{Ferrero:2023znz}. The operators constructed in this way are not necessarily eigenstates of the dilatation operator, but this does not constitute a problem and one simply ends up with a non-diagonal dilatation operator in each degeneracy space. In terms of our simple $2\times 2$ example, we can define a vector of OPE coefficients and an anomalous dimension matrix, which in the eigenstates basis read
\begin{align}
\mu^{(0)}_{\alpha}=(\mu^{(0)}_{1},\,\mu^{(0)}_{2})_{\alpha}\,,\qquad
(\Gamma^{(1)})^{\alpha\beta}=
\begin{pmatrix}
\gamma^{(1)}_1 & 0\\
0 & \gamma^{(1)}_2
\end{pmatrix}_{\alpha\beta}\,.
\end{align}
In terms of these objects, the averages can be rewritten in matrix notation as
\begin{align}\label{averagesCFTdata_tensors}
\begin{split}
\langle a^{(0)}\rangle_{\Delta}&=\mu^{(0)}_{\alpha}\,(\mathtt{g}^{-1})^{\alpha\beta}\mu^{(0)}_{\beta}\,,\\
\langle a^{(0)}\,\gamma^{(1)}\rangle_{\Delta}&=\mu^{(0)}_{\alpha}\,(\Gamma^{(1)})^{\alpha\beta}\mu^{(0)}_{\beta}\,,\\
\langle a^{(0)}\,(\gamma^{(1)})^2\rangle_{\Delta}&=\mu^{(0)}_{\alpha}\,(\Gamma^{(1)})^{\alpha\beta}\,\mathtt{g}_{\beta\gamma}\,(\Gamma^{(1)})^{\gamma\delta}\mu^{(0)}_{\delta}\,,
\end{split}
\end{align}
where we have introduced the matrix of norms,
\begin{align}
\mathtt{g}_{\alpha\beta}=
\begin{pmatrix}
\langle\mathcal{O}_1|\mathcal{O}_1\rangle & \langle\mathcal{O}_1|\mathcal{O}_2\rangle\\
\langle\mathcal{O}_2|\mathcal{O}_1\rangle & \langle\mathcal{O}_2|\mathcal{O}_2\rangle
\end{pmatrix}_{\alpha\beta}=\delta_{\alpha\beta}\,.
\end{align}
The latter is a metric tensor in the degeneracy space and in the basis of normalized eigenstates of the dilatation operator that we have adopted is simply the identity. It is now clear that one can perform an arbitrary change of basis in the degeneracy space by means of a matrix $M\in GL(2)$, acting as
\begin{align}
\widehat{\mathcal{O}}_{\alpha}=M_{\alpha}^{\,\,\,\beta}\,{\mathcal{O}}_{\beta}\,,
\end{align}
where $\widehat{\mathcal{O}}_{\alpha}$ is now a completely arbitrary basis of (not necessarily normalized) operators, where the metric tensor is no longer the identity, the anomalous dimensions matrix is not diagonal and the usual rules of linear algebra give\footnote{Note that there is no contradiction with \eqref{defGammaDelta}, since $\widehat{\Gamma}^{\alpha\beta}$ in this section is always used with upper indices, which is related to $\widehat{\Gamma}_{\alpha\beta}$ in \eqref{defGammaDelta} by the action of the metric $\widehat{\mathtt{g}}$.}
\begin{align}
\begin{split}
\widehat{\mu}_{\alpha}=M_{\alpha}^{\,\,\,\beta}\,\mu_{\beta}\,, \quad
(\widehat{\Gamma}^{(1)})^{\alpha\beta}=(M^{-1}\Gamma^{(1)}(M^{-1})^T)^{\alpha\beta}\,,\quad
\widehat{\mathtt{g}}_{\alpha\beta}=(M\mathtt{g} M^T)_{\alpha\beta}\,,
\end{split}
\end{align}
which give us a way to interpret the averages \eqref{averagesCFTdata_tensors} in an arbitrary basis. This is the approach that we will adopt in Section \ref{sec:mixing}, where we will make a convenient choice of operator basis in the free theory, that allows us to compute their OPE coefficients from Wick contractions. The dilatation operator in such basis then takes the form of a non-diagonal matrix, whose eigenvalues represent the anomalous dimensions of the physical eigenstates, and we explore the entries of such matrix by studying a system of multiple correlators. We anticipate here that the choice of basis that we found convenient is a basis where operators are words of fixed length in terms of the fundamental letters contained in $\mathcal{D}_1$ (see \cite{Ferrero:2023znz} for more details). From now on, we will keep the distinction between operators $\mathcal{O}$ that diagonalize the dilatation operator and operators $\widehat{\mathcal{O}}$, which have well-defined length but are not necessarily eigenstates.

\section{Bootstrapping $\langle \mathcal{D}_1\mathcal{D}_1\mathcal{D}_1\mathcal{D}_1\rangle$ at three loops}\label{sec:1111}

In the rest of the paper we are going to present various bootstrap results for four-point functions on the Wilson line. The kinematics, superconformal blocks and Ward identities for the correlators that we will discuss are presented in the companion paper \cite{Ferrero:2023znz}, so we will not repeat the derivations in detail but only quote some results. We start in this section by presenting our main result: the bootstrap for the four-point function of the super-displacement operator, $\langle \mathcal{D}_1\mathcal{D}_1\mathcal{D}_1\mathcal{D}_1\rangle$, up to fourth order in perturbation theory, corresponding to three-loop Witten diagrams. In doing so, we shall use some results that are derived in the following sections, where more general correlators between half-BPS operators, as well as four-point functions including long multiplets, are studied. The results up to one loop will be mostly a review of those already presented in \cite{Liendo:2018ukf},  with some additional justifications of why they are correct despite the fact that the mixing problem was not taken into account in that paper, while at two and three loops we provide a detailed derivation of the results presented in \cite{Ferrero:2021bsb}.

\subsection{Kinematics and free theory}

Let us start by briefly reviewing some kinematics from \cite{Liendo:2018ukf,Ferrero:2023znz}. While all four-point functions between half-BPS operators are constrained by superconformal Ward identities (SCWI) -- see \eqref{SCWI} -- the four-point function $\langle \mathcal{D}_1\mathcal{D}_1\mathcal{D}_1\mathcal{D}_1\rangle$ is special in that the SCWI can be solved in terms of a number $\mathsf{f}$ and a single function $f(\chi)$ of the cross-ratio $\chi$
\begin{align}\label{solWI_1111}
\frac{\langle \mathcal{D}_1\mathcal{D}_1\mathcal{D}_1\mathcal{D}_1\rangle}{\langle \mathcal{D}_1\mathcal{D}_1\rangle\langle\mathcal{D}_1\mathcal{D}_1\rangle}=\mathsf{f}\,\frac{\chi^2}{\zeta_1\,\zeta_2}+\mathbb{D}\,f(\chi)\,,
\end{align}
where $\zeta_{1,2}$ are R-symmetry cross ratios and $\mathbb{D}$ is a differential operator which can be written as
\begin{align}\label{D_1111}
\mathbb{D}=v_1+v_2-v_1\,v_2\,\chi^2\partial_{\chi}\,,\qquad v_i=\chi^{-1}-\zeta_i^{-1}\,,
\end{align}
see \cite{Ferrero:2023znz} for more details. We remind the reader that the number $\mathsf{f}$ is a datum of the topological algebra associated with 1d CFTs with $\mathfrak{osp}(4^*|4)$ symmetry. In the specific case of the half-BPS Wilson line in planar $\mathcal{N}=4$ SYM it can be computed using localization and its non-perturbative expression as a function of the 't~Hooft coupling $\lambda$ reads \cite{Drukker:2009sf,Giombi:2009ds,Giombi:2018qox,Liendo:2018ukf}
\begin{align}\label{f_fromW}
\mathsf{f}=(\mathsf{C}_{112})^2=3\,\frac{\mathcal{W}(\lambda)\,\mathcal{W}''(\lambda)}{(\mathcal{W}'(\lambda))^2}\,,
\end{align}
where $\mathcal{W}(\lambda)$ is the expectation value of the circular one-half BPS Wilson loop in planar $\mathcal{N}=4$ SYM,  given by
\begin{align}
\mathcal{W}(\lambda)=\frac{2}{\sqrt{\lambda}}\,I_1(\sqrt{\lambda})\,,
\end{align}
with $I_1$ a modified Bessel function of the first kind. $\mathsf{C}_{112}$ in \eqref{f_fromW} is the OPE coefficient $\langle \mathcal{D}_1\mathcal{D}_1\mathcal{D}_2\rangle$. Neglecting non-perturbative terms proportional to $e^{-\sqrt{\lambda}}$ in the expansion of $I_1(\sqrt{\lambda})$,  one finds at large $\lambda$ 
\begin{align}\label{f1111_3loops}
\mathsf{f}=\sum_{\ell=0}^{\infty}\frac{\mathsf{f}^{(\ell)}}{\lambda^{\ell/2}}=3-\frac{3}{\lambda^{1/2}}+\frac{45}{8\,\lambda^{3/2}}+\frac{45}{4\,\lambda^2}+\frac{1215}{128\,\lambda^{5/2}}-\frac{135}{8\,\lambda^3}+\mathcal{O}(\lambda^{-7/2})+\mathcal{O}(e^{-\sqrt{\lambda}})\,,
\end{align}
Note that from the point of view of the conformal bootstrap,  the knowledge of $\mathsf{f}$ provides a definition of the coupling constant $\lambda$ at all orders,  which would otherwise be arbitrary from a purely bootstrap perspective. The other crucial ingredient in \eqref{solWI_1111} is the function $f(\chi)$, which we shall sometimes refer to as {\it reduced correlator}, and it is the object to which we will devote our attention in this section.

Given the OPE
\begin{align}\label{D1D1OPE}
\mathcal{D}_1\times \mathcal{D}_1 =\mathcal{I}+ \mathcal{D}_2+\sum_{\h}\mathcal{L}^{\h}_{0,[0,0]}\,,
\end{align}
we remind the expression for the reduced conformal blocks associated to the three types of exchanged multiplets, which read \cite{Liendo:2018ukf,Ferrero:2023znz}
\begin{align}\label{blocks1111}
\begin{split}
\mathfrak{f}_{\mathcal{I}}(\chi)&=\chi\,,\\
\mathfrak{f}_{\mathcal{D}_2}(\chi)&=\chi-\chi\, {}_2F_1(1,2,4;\chi)\,,\\
\mathfrak{f}_{\h}(\chi)&=\frac{\chi^{1+\h}}{1-\h}\,_2F_1(1+\h,2+\h,4+2\h;\chi)\,.
\end{split}
\end{align}
Note that the reduced blocks for long operators are eigenfunctions of the reduced quadratic Casimir 
\begin{align}\label{reducedcasimir}
\widehat{\mathcal{C}}=(1-\chi)\partial_{\chi}\left(\chi^2\,\partial_{\chi}\right)-2\,,
\end{align}
namely they satisfy
\begin{align}\label{casimironreducedblocks}
\widehat{\mathcal{C}}\,\mathfrak{f}_{\h}(\chi)=\JJh\,\mathfrak{f}_{h}(\chi)\,,\qquad
\JJh=\h(\h+3)\,,
\end{align}
where $\JJh$ is the quadratic Casimir eigenvalue for singlets of R symmetry and transverse spin. A property of the reduced conformal blocks that will play a crucial role in the following is their behavior under braiding transformations
\begin{align}\label{braidingreducedblocks}
\begin{split}
\mathfrak{f}_{\mathcal{I}}(\chi)+\mathfrak{f}_{\mathcal{I}}\big(\tfrac{\chi}{\chi-1}\big)&=\frac{\chi^2}{\chi-1}\,,\\
\mathfrak{f}_{\mathcal{D}_2}(\chi)+\mathfrak{f}_{\mathcal{D}_2}\big(\tfrac{\chi}{\chi-1}\big)&=\frac{\chi^2}{\chi-1}\,,\\
\mathfrak{f}_{\h}(\chi)+(-1)^{\h}\,\mathfrak{f}_{\h}\big(\tfrac{\chi}{\chi-1}\big)&=0\,.
\end{split}
\end{align}

Let us now discuss the free theory result for $\langle \mathcal{D}_1\mathcal{D}_1\mathcal{D}_1\mathcal{D}_1\rangle$, which can be computed using Wick contractions and reads
\begin{align}\label{free1111}
\mathcal{G}^{(0)}_{\{1,1,1,1\}}=1+\frac{\chi^2}{\zeta_1\,\zeta_2}+\frac{\chi^2\,(1-\zeta_1)\,(1-\zeta_2)}{(1-\chi)^2\,\zeta_1\,\zeta_2}\,,
\end{align}
which correctly gives $\mathsf{f}^{(0)}=3$, as well as
\begin{align}\label{free1111}
f^{(0)}(\chi)=\chi+\frac{\chi^2}{\chi-1}\,.
\end{align}
Under cyclic transformations this satisfies
\begin{align}\label{crossing1111free}
(1-\chi)^2\,f^{(0)}(\chi)=\chi^2\,f^{(0)}(1-\chi)\,,
\end{align}
while under braiding we have
\begin{align}\label{braiding1111free}
f^{(0)}(\chi)+f^{(0)}\braid=\mathsf{f}^{(0)}\,\frac{\chi^2}{\chi-1}\,.
\end{align}
Given \eqref{braidingreducedblocks}, this implies that only long multiplets with even $\h$ can contribute to the $\mathcal{D}_1\times \mathcal{D}_1$ OPE in the free theory. We can confirm this result by expanding in blocks
\begin{align}
f^{(0)}(\chi)=\mathfrak{f}_{\mathcal{I}}(\chi)+(\mathsf{C}^{(0)}_{112})^2\,\mathfrak{f}_{\mathcal{D}_2}(\chi)+\sum_{\D}\langle a^{(0)}\rangle_{\D}\,\mathfrak{f}_{\D}(\chi)\,,
\end{align}
where $\langle a^{(0)}\rangle_{\D}$ are averages of squared OPE coefficients, and we have used the notation $\D$ introduced in Section \ref{sec:bootstrap} to denote the set of conformal dimensions exchanged in the free theory. In this case \eqref{braiding1111free} implies that $\D \in 2\mathbb{N}_{0}$ and we find
\begin{align}\label{a0free1111}
\langle a^{(0)}\rangle_{\D}=\frac{\Gamma[3+\D]\Gamma[1+\D](\D-1)}{\Gamma[2+2\D]}\,.
\end{align}

In the following sections we will turn on perturbations and the exchanged operators acquire anomalous dimensions order by order in perturbation theory for large $\lambda$,
\begin{align}\label{hTOdelta}
h=\D+\sum_{\ell=0}^{\infty}\frac{\gamma^{(\ell)}_{\D}}{\lambda^{\ell/2}}\,,
\end{align}
with an analogous expansion for the averaged OPE coefficients. While the symmetry under cyclic transformations remains exact at each order and can be written more generally as
\begin{align}\label{crossing1111}
(1-\chi)^2\,f^{(\ell)}(\chi)+\chi^2\,f^{(\ell)}(1-\chi)=0\,,
\end{align}
for braiding one should expand as in \eqref{Gpowerslog}
\begin{align}\label{logsexpansion1111}
f^{(\ell)}(\chi)=\sum_{k=0}^{\ell}f^{(\ell)}_{\log^k}(\chi)\,\log^k\chi\,,
\end{align}
and for the functions $f^{(\ell)}_{\log^k}(\chi)$ we have
\begin{align}\label{braiding1111}
f^{(\ell)}_{\log^k}(\chi)+\sum_{m=0}^{\ell-k}\binom{k+m}{m}(-1)^m\,\log^m(1-\chi)\,f^{(\ell)}_{\log^{k+m}}\big(\tfrac{\chi}{\chi-1}\big)=\delta_{k,0}\,\frac{\chi^2}{\chi-1}\,\mathsf{f}^{(\ell)}\,,
\end{align}
which is just \eqref{braiding1d} adapted to this case. We will also highlight that, at each order, one can pick a basis of functions such that, just replacing $\log\chi \to \log|\chi|$ and $\log (1-\chi)\to \log|1-\chi|$ one obtains a function $\bar{f}^{(\ell)}(\chi)$ such that braiding becomes completely manifest, as in \eqref{braidingGbar}:
\begin{align}\label{braidingfbar}
\bar{f}^{(\ell)}\big(\tfrac{\chi}{\chi-1}\big)+\bar{f}^{(\ell)}(\chi)=\frac{\chi^2}{\chi-1}\,\mathsf{f}^{(\ell)}\,.
\end{align}

\subsection{Warm up: solving functional relations}

Before attacking the bootstrap problem for $\langle \mathcal{D}_1\mathcal{D}_1\mathcal{D}_1\mathcal{D}_1\rangle$, let us briefly consider two sets of functional relations that will prove useful in the discussion of the bootstrap problem at all orders. As we anticipated in Section \ref{sec:bootstrap}, our strategy at each order consists in computing the coefficients of the leading logarithmic singularities at each order, which after requiring cyclic and braiding invariance fixes the whole correlator up to a function satisfying the ``tree-level'' crossing equations. It is therefore useful to study here the most general solution to these equations, assuming that the result has transcendentality one, as it should be for contact terms in AdS$_2$, see \cite{Ferrero:2019luz} (note that the Lagrangian \eqref{AdSlagrangian} does not contain cubic vertices). At each order, we will then refer to the solutions found here to fix the contact term ambiguity.

Consider the function
\begin{align}
Q(\chi)=q_1(\chi)+q_2(\chi)\,\log|1-\chi|+q_3(\chi)\,\log|\chi|\,,
\end{align}
subject to the crossing equations
\begin{align}\label{toycrossing}
(1-\chi)^2\,Q(\chi)+\chi^2\,Q(1-\chi)=0\,, \qquad
Q(\chi)+Q\big(\tfrac{\chi}{\chi-1}\big)=0\,,
\end{align}
with $\chi\in \mathbb{R}\setminus \{0,1\}$, where $q_{1,2,3}$ are rational functions.  Requiring the coefficient of each HPL in \eqref{toycrossing} to vanish,  one obtains that $q_3$ can be eliminated using 
\begin{align}
q_3(\chi)=-\frac{\chi^2}{(1-\chi)^2}\,q_2(1-\chi)\,,
\end{align}
with the remaining two functions constrained by
\begin{align}\label{systemh1}
(1-\chi)^2\,q_1(\chi)+\chi^2\,q_1(1-\chi)=0\,, \qquad
q_1(\chi)+q_1\big(\tfrac{\chi}{\chi-1}\big)=0\,,
\end{align}
and
\begin{align}\label{systemh2}
(1-\chi)^2\,\left(q_2(\chi)-q_2\big(\tfrac{\chi}{\chi-1}\big)\right)=\chi^2\,q_2(1-\chi)\,, \qquad
q_2(\chi)+\chi^2\,q_2\big(\tfrac{1}{\chi}\big)=0\,.
\end{align}
For the reasons explained in Section \ref{sec:bootstrap}, we are interested in rational solutions to \eqref{systemh1} and \eqref{systemh2} that are holomorphic functions of $\chi$ except for,  at most,  poles at $\chi=0,1$. One can then make an ansatz of the type \eqref{rationalansatz},  namely
\begin{align}
q_i(\chi)=\frac{p_i(\chi)}{\chi^{\mathtt{k}}\,(1-\chi)^{\mathtt{k}}}\,,
\end{align}
and study the solutions varying the degree of the polynomial $p_i$ in the numerators and the parameter $\mathtt{k}$. We find that there are no solutions for $\mathtt{k}\le 0$, while for each fixed $\mathtt{k}>0$ there is only a finite number of solutions, regardless of how high the degree of $p_i$ is taken to be in the ansatz. It is then simple to explore all solutions, and we find that both for $q_1$ and for $q_2$ they can be described by one infinite family, labeled by an integer $n$. For $q_1$ we find
\begin{align}\label{h1star}
q^{(n)}_1(\chi)=\frac{(2-\chi)(1+\chi)(1-2\chi)(1-\chi+\chi^2)^{3n-1}}{\chi^{2n-1}\,(1-\chi)^{2n+1}}\,, \quad n\ge 1\,,
\end{align}
while for $q_2$ we have
\begin{align}\label{h2star}
q^{(n)}_2(\chi)=\frac{1-(-1)^n\,\chi^{3n-2}}{\chi^{n-2}\,(1-\chi)^n}\,, \quad n\ge 1\,.
\end{align}
Note that we have made a particularly simple choice of basis which allows to write all solutions in closed form,  but since we are dealing with a linear and homogeneous problem all linear combinations of the functions chosen as basis still provide a valid solution. An important aspect that we would like to stress is that the appearance of a new solution (increasing $\mathtt{k}$) is closely related to a more singular behavior at $\chi=0,1,\infty$, so that restrictions on this behavior provide a powerful constraint. In the concrete problems that we shall study in the next subsections, such constraints will come from the OPE, which sets the lowest power of $\chi$ that can appear in the expansion around $\chi=0$, and from the Regge limit, which is an expansion around $\chi=\infty$ and is related to the behavior of the anomalous dimensions associated to a certain solution at large $\D$,  see \cite{Ferrero:2019luz}.

\subsection{Tree level}

Let us now turn on the perturbation and examine results at tree level.  According to the discussion of Section \ref{sec:bootstrap} we make an ansatz for $f^{(1)}(\chi)$ with transcendentality one
\begin{align}\label{tree1111ansatz}
\begin{split}
f^{(1)}(\chi)=\,&f^{(1)}_{\log^0}(\chi)+f^{(1)}_{\log^1}(\chi)\,\log\chi\,,\\
f^{(1)}_{\log^0}(\chi)=\,&r_1(\chi)+r_2(\chi)\,\log(1-\chi)\,,\\
f^{(1)}_{\log^1}(\chi)=\,&r_3(\chi)\,,
\end{split}
\end{align}
where $r_i$ ($i=1,2,3$) are rational functions and from the results of Appendix \ref{app:perturbativeOPE} one has the expansions
\begin{align}\label{treeOPE1111}
\begin{split}
f^{(1)}_{\log^1}(\chi)&=\sum_{\D}\langle a^{(0)}\,\gamma^{(1)} \rangle_{\D}\,\mathfrak{f}_{\D}(\chi)\,\\
f^{(1)}_{\log^0}(\chi)&=\mathsf{f}^{(1)}\,\mathfrak{f}_{\mathcal{D}_2}(\chi)+\sum_{\D}\left[\langle a^{(1)} \rangle_{\D}\,\mathfrak{f}_{\D}(\chi)+\langle a^{(0)}\,\gamma^{(1)}\rangle_{\D}\,\mathfrak{f}^{(1)}_{\D}(\chi)\right]\,,
\end{split}
\end{align}
where we have introduced the notation 
\begin{align}
\mathfrak{f}^{(n)}_{\D}(\chi)=\chi^{\D}\,\left(\partial_{\D}\right)^n\,\chi^{-\D}\,\mathfrak{f}_{\D}(\chi)\,.
\end{align}
Given that there are no terms proportional to $\log^k\chi$ with $k\ge 2$ in \eqref{tree1111ansatz},  there are no constraints coming from the highest logarithmic singularities,  as one should expect for a tree-level result. This is the equivalent of contact terms in higher-dimensional AdS spaces having vanishing double discontinuity, in the language of the Lorentzian inversion formula \cite{Caron-Huot:2017vep}. On the other hand,  demanding that $f^{(1)}$ satisfies the crossing equations \eqref{crossing1111} and \eqref{braiding1111} leads to the functional relations
\begin{gather}
\nonumber
(1-\chi)^2\,r_1(\chi)=\chi^2\,r_1(1-\chi)\,,\quad 
(1-\chi)^2\,r_2(\chi)=\chi^2\,r_3(1-\chi)\,,\quad r_1(\chi)+r_1\big(\tfrac{\chi}{\chi-1}\big)=\mathsf{f}^{(1)}\,\frac{\chi^2}{\chi-1}\\ \label{functionalconstr1111tree}
r_2(\chi)+\chi^2\,r_2\big(\tfrac{1}{\chi}\big)=0\,, \quad
(1-\chi)^2\,\left(r_2(\chi)-r_2\big(\tfrac{\chi}{\chi-1}\big)\right)=\chi^2\,r_2(1-\chi)\,,
\end{gather}
which allow us to eliminate $r_2$ in terms or $r_3$,  leading to a non-homogeneous version of the equations \eqref{systemh1} (with $r_1 \leftrightarrow q_1$) and \eqref{systemh2} (with $r_2 \leftrightarrow q_2$), with the non-homogeneous term provided by the topological OPE data $\mathsf{f}^{(1)}$. The solution to \eqref{functionalconstr1111tree} is then given by any solution of the non-homogeneous equation plus an arbitrary linear combination of all solutions to the homogeneous equation. Hence, we can express the solution to \eqref{functionalconstr1111tree} as
\begin{align}\label{allsolr123}
\begin{split}
r_1(\chi)&=\frac{\mathsf{f}^{(1)}}{3}\frac{\chi(1-2\chi)}{1-\chi}+\sum_{n=1}^{\infty}c_1^{(n)}\,q^{(n)}_1(\chi)\,,\\
r_2(\chi)&=\sum_{n=1}^{\infty}c_2^{(n)}\,q^{(n)}_2(\chi)\,, \quad
r_3(\chi)=-\frac{\chi^2}{(1-\chi)^2}\,r_2(1-\chi)\,.
\end{split}
\end{align}
This is obviously far from a unique solution, as we are yet to impose appropriate boundary conditions. First, the OPE \eqref{treeOPE1111} dictates that expanding around $\chi=0$
\begin{align}\label{OPEconstr1111}
f^{(1)}(\chi)=\frac{3}{2}\chi^2+\mathcal{O}(\chi^3)\,,
\end{align}
which can be shown to fix all the coefficients $c^{(n)}_2$ in terms of the $c^{(n)}_1$.  While finding a closed-form expression for all solutions to this constraint seems non-trivial,  we can give some examples:
\begin{itemize}
\item If $c^{(n)}_1=0$, then $c^{(n)}_2=\big\{-\tfrac{2}{3}\,\mathsf{f}^{(1)},\tfrac{1}{3}\,\mathsf{f}^{(1)},0,0,...\big\}$.
\item If $c^{(n)}_1=\delta_{n,1}$, then $c^{(n)}_2=\big\{-\tfrac{2}{3},\tfrac{31}{3},-10,2,0,0,...\big\}$.
\item If $c^{(n)}_1=\delta_{n,2}$, then $c^{(n)}_2=\big\{-\tfrac{89}{45},\tfrac{3299}{90},-\tfrac{191}{3},\tfrac{139}{3},-16,2,0,0,...\big\}$.
\end{itemize}
It is important to stress that only a finite number of $c^{(n)}_2$ is non-zero for each $c^{(n)}_1$ that is taken to be non-zero. Because of crossing symmetry, no new constraint arises expanding around $\chi=1$. On the other hand,  a crucial constraint comes from the expansion around $\chi=\infty$. As discussed in \cite{Ferrero:2019luz}, this is equivalent to studying the behavior of the anomalous dimensions $\langle\gamma^{(1)}\rangle_{\D}$ associated to a certain solution via \eqref{treeOPE1111} at large $\D$.  In particular, we find
\begin{align}\label{gammastreeall}
\langle \gamma^{(1)}\rangle_{\D} :=\langle a^{(0)} \rangle_{\D}^{-1} \langle a^{(0)}\gamma^{(1)} \rangle_{\D}=\frac{\mathsf{f}^{(1)}}{6}\,\JJ+\sum_{n=1}^{\infty}c_1^{(n)}\,\langle\gamma^{(1)}_n\rangle_{\D}\,,
\end{align}
where $\JJ=\D(\D+3)$ was introduced in \eqref{casimironreducedblocks}, and
\begin{align}\label{largeDeltatree}
\langle\gamma^{(1)}_n\rangle_{\D}\sim \D^{4n+2}\,, \quad (\D\to \infty)\,.
\end{align}
Given the correspondence between the growth of $\langle \gamma^{(1)}\rangle_{\D}$ at large $\Delta$ and the number of derivatives in tree-level contact interactions \cite{Heemskerk:2009pn} (see also \cite{Ferrero:2019luz} for the $d=1$ case), we conclude that the term proportional to $\mathsf{f}^{(1)}$ in \eqref{gammastreeall} is generated by quartic contact terms with at most four derivatives (precisely as in \eqref{AdSlagrangian}), while each $\langle\gamma^{(1)}_n\rangle_{\D}$ corresponds to contact terms with $4n+4$ derivatives. These cannot be excluded simply by bootstrap arguments as they all provide solutions to the functional equations \eqref{functionalconstr1111tree} with the boundary condition \eqref{OPEconstr1111}, but they can be set to zero by requiring that our solution describes quartic interactions with at most four derivatives, which should be the case if we want to reproduce the theory described by the Lagrangian \eqref{AdSlagrangian}. Note that a suitable adaptation of this prescription is usually employed in Mellin space when computing holographic correlators in higher dimensions, where one has a constraint on the number of derivatives appearing in the supergravity Lagrangian, translating to polynomials Mellin amplitudes of a given degree -- see, {\it e.g.}, \cite{Alday:2020dtb,Alday:2021odx} or \cite{Bissi:2022mrs} for a review. Here we are using very similar ideas, just implementing them directly in spacetime rather than in Mellin space.

Having clarified the structure of the general solution at tree level,  it should now be clear how to pinpoint the one corresponding to the AdS$_2$ Lagrangian \eqref{AdSlagrangian} is selected. We simply have to set to zero all coefficients corresponding to contact terms with more than four derivatives, which means choosing 
\begin{align}
c^{(n)}_1=0\quad \forall n \ge 1\,, \quad c^{(1)}_2=-\tfrac{2}{3}\,\mathsf{f}^{(1)}\,, \quad c^{(2)}_2=\tfrac{1}{3}\,\mathsf{f}^{(1)}\,, \quad
c^{(n)}_2=0\quad \forall n\ge 3\,,
\end{align}
in \eqref{allsolr123}. The solution can be written as
\begin{align}\label{tree1111reduced}
f^{(1)}(\chi)=-\frac{\chi(1-2\chi)}{1-\chi}-(1-\chi^2)\,\log(1-\chi)+\frac{\chi^3(2-\chi)}{(1-\chi)^2}\,\log\chi\,,
\end{align}
with associated CFT data
\begin{align}\label{tree1111CFTdata}
\langle\gamma^{(1)}\rangle_{\D}=-\frac{1}{2}\JJ\,, \qquad
\langle a^{(1)}\rangle=\partial_{\D}\,\langle a^{(0)}\,\gamma^{(1)}\rangle_{\D}\,,
\end{align}
where we have also used $\mathsf{f}^{(1)}=-3$ from \eqref{f1111_3loops}. These are precisely the results already presented in \cite{Liendo:2018ukf}.  Note that the first order correction to the squared OPE coefficients satisfies the ``derivative rule'' first observed in \cite{Heemskerk:2009pn}.  However,  this is just an accident of the fact that the result for $\langle \mathcal{D}_1\mathcal{D}_1\mathcal{D}_1\mathcal{D}_1 \rangle^{(0)}$ agrees with that of a generalized free theory,  for which the derivative rule is known to apply,  while general $\langle \mathcal{D}_p\mathcal{D}_p\mathcal{D}_q\mathcal{D}_q \rangle^{(0)}$ arise from Wick contractions in a free theory,  so there will be corrections to the derivative rule.

Finally, let us present the final result for the tree-level correlator. As anticipated, we can write define a function $\bar{f}^{(1)}(\chi)$ that is Bose-symmetric and reduces to $f^{(1)}(\chi)$ for $\chi\in(0,1)$ simply by replacing $\log a \to \log|a|$. Given our choice of basis, it is natural to express the result in a form that makes cyclic invariance manifest, namely
\begin{align}\label{final1111tree}
\bar{f}^{(1)}(\chi)=\big[f^{(1)}_{\log}(\chi)\,\log|\chi|\big]-\tfrac{\chi^2}{(1-\chi)^2}\big[\chi \to 1-\chi\big]+\frac{\mathsf{f}^{(1)}}{3}\,f^{(0)}(\chi)\,,
\end{align}
where $f^{(0)}(\chi)$ is the free theory reduced correlator introduced in \eqref{free1111}, while 
\begin{align}
f^{(1)}_{\log}(\chi)=\frac{\chi^3(2-\chi)}{(1-\chi)^2}=\widehat{\mathcal{C}}\left[f^{(0)}(\chi)+\frac{3\chi^2}{2(1-\chi)}\right]\,,
\end{align}
and $\widehat{\mathcal{C}}$ is the Casimir operator introduced in \eqref{reducedcasimir}. Note that although only cyclic invariance is manifest in \eqref{final1111tree}, the correlator is invariant under the full Bose symmetry group $S_3$ and defined for all $\chi\in \mathbb{R}\setminus\{0,1\}$. 

\subsection{One loop}

Let us now move to the second perturbative correction,  which corresponds to one loop diagrams in AdS$_2$. Starting from this order, it is {\it a priori} not clear what the maximal transcendentality $\mathtt{t_{max}}(\ell)$ of the ansatz \eqref{genericansatz} should be.  Let us then take a little detour, that will give us an educated guess for what $\mathtt{t_{max}}(\ell)$ should be.  One-loop four-point functions in generic scalar $\varphi^4$ theories in AdS$_2$ lead to functions of transcendentality four, as it can be found in \cite{Mazac:2018ycv,Ferrero:2019luz}. On the other hand, the results found in \cite{Liendo:2018ukf} suggest that one loop correlators on the Wilson line should have transcendentality two. This discrepancy can be understood as follows. Let us assume that all terms in \eqref{logsexpansion1111} contain HPLs of the maximal transcendentality,  then in particular $f^{(\ell)}_{\log^{\ell}}(\chi)$ should have transcendentality $\mathtt{t_{max}}(\ell)-\ell$,  and it can be evaluated using (see Appendix \ref{app:perturbativeOPE})
\begin{align}
f^{(\ell)}_{\log^{\ell}}(\chi)=\frac{1}{\ell!}\sum_{\D}\langle a^{(0)}\,\big(\gamma^{(1)}\big)^{\ell}\rangle_{\D}\,\mathfrak{f}_{\D}(\chi)\,.
\end{align}
Now, computing this sum at all orders is non-trivial due to the presence of operators degeneracy. In \cite{Mazac:2018ycv,Ferrero:2019luz} the absence of mixing was assumed but not justified. On the Wilson line, on the other hand, we have already anticipated in Section \ref{sec:setup} that the first-order perturbation does not lift the free theory degeneracy. Hence, in both cases one is using that
\begin{align}\label{nodegeneracy1111}
\langle a^{(0)}\,\big(\gamma^{(1)}\big)^{\ell}\rangle_{\D}=\langle a^{(0)} \rangle_{\D}\langle \gamma^{(1)} \rangle^{\ell}_{\D}\,.
\end{align}
Now, for the Wilson line defect theory at tree level one has the anomalous dimensions \eqref{tree1111CFTdata}, which are proportional to the Casimir eigenvalue $\JJ$ and therefore
\begin{align}\label{fl_logl_1111}
f^{(\ell)}_{\log^{\ell}}(\chi)=\frac{1}{\ell!}\,\Big(-\tfrac{1}{2}\,\widehat{\mathcal{C}}\Big)^{\ell-1}\,f^{(1)}_{\log^1}(\chi)\,.
\end{align}
Since $f^{(1)}_{\log^1}$ is a rational function (found in \eqref{tree1111reduced}) and $\widehat{\mathcal{C}}$ is a differential operator, then all $f^{(\ell)}_{\log^{\ell}}$ are rational functions. Their degree of transcendentality is then zero, so from our initial statement we can read off that
\begin{align}\label{tmax(l)}
\mathtt{t}\big[f^{(\ell)}_{\log^{\ell}}\big]=\mathtt{t_{max}}(\ell)-\ell=0 \,, \qquad \Rightarrow \qquad
\mathtt{t_{max}}(\ell)=\ell\,,
\end{align}
at all orders. On the other hand, for the $\varphi^4$ theory with $\D_{\varphi}=1$ considered in \cite{Mazac:2018ycv} one has $\gamma^{(1)}_{\D}\sim j^{-2}_{\D}$, so that the leading logarithmic singularities at higher loops are obtained by acting with the {\it inverse} Casimir operator on the tree level result.  Since this is an integral operator, it raises the degree of transcendentality of the functions it acts on, so that even if the coefficient of $\log\chi$ in the tree-level result is a rational function, the coefficient of $\log^{\ell}\chi$ at $\ell-1$ loops is transcendental.  We can therefore immediately see how the defect theory defined on the Wilson line has a much simpler perturbative structure than a generic $\varphi^4$-type theories in AdS$_2$, and at the computational level this is due to the fact that the tree-level anomalous dimensions are polynomial, rather than rational, functions of the dimension $\D$ of the exchanged operators. Note that this has been observed to happen for other 1d defect theories as well, such as the Wilson line in ABJM \cite{Bianchi:2020hsz} and AdS$_2\times S^2$ holographic correlators \cite{Abl:2021mxo}.

Now that we have established our expectation for $\mathtt{t_{max}}(\ell)$, we can make an ansatz for the one loop reduced correlator $f^{(2)}$ as a combination of HPLs of weight up to two\footnote{Note that we keep using the same symbol $r_i(\chi)$ for the rational functions in the ansatz,  as we did already for the tree level ansatz \eqref{tree1111ansatz},  in order to avoid a cumbersome notation.  We will keep doing so in every ansatz that is spelled out explicitly and there should be no confusion as each of them appears in different subsections, without cross-referencing.}
\begin{align}
\begin{split}
f^{(2)}(\chi)=& f^{(2)}_{\log^0}(\chi)+f^{(2)}_{\log^1}(\chi)\,\log\chi+f^{(2)}_{\log^2}(\chi)\,\log^2\chi\\
 f^{(2)}_{\log^0}(\chi)=&r_1(\chi)+r_2(\chi)\,\log(1-\chi)+r_3(\chi)\,\log^2(1-\chi)+r_4(\chi)\,\Li_2(\chi)\,,\\
f^{(2)}_{\log^1}(\chi)=&r_5(\chi)+r_6(\chi)\,\log(1-\chi)\,,\\
f^{(2)}_{\log^2}(\chi)=&r_7(\chi)\,,
\end{split}
\end{align}
where all the numerators in the rational functions are polynomials with coefficients in $\mathbb{Q}$,  except for\footnote{Note that the terms $\Li_2$ and $\zeta(2)$ were excluded {\it a priori} from the ansatz used in \cite{Liendo:2018ukf}. Here we allow for their presence, but we will show that their coefficients actually vanish in the final result.}
\begin{align}
r_1(\chi)=r_{1,1}(\chi)+r_{1,2}(\chi)\,\zeta(2)\,,
\end{align}
where we have allowed explicitly the possibility of terms proportional to $\zeta(2)$, the only MZV of weight two.  Following Appendix \ref{app:perturbativeOPE}, we also have the following expansions in blocks and derivatives
\begin{align}\label{1loopOPE1111}
\begin{split}
f^{(2)}_{\log^2}(\chi)&=\sum_{\D}\frac{1}{2!}\langle a^{(0)}\,(\gamma^{(1)})^2 \rangle_{\D}\,\mathfrak{f}_{\D}(\chi)\,,\\
f^{(2)}_{\log^1}(\chi)&=\sum_{\D}\Big[\langle a^{(0)}\gamma^{(2)}+a^{(1)}\gamma^{(1)}\rangle_{\D}\,\mathfrak{f}_{\D}(\chi)+\langle a^{(0)}\,(\gamma^{(1)})^2 \rangle_{\D}\,\mathfrak{f}^{(1)}_{\D}(\chi)\Big]\,,\\
f^{(2)}_{\log^0}(\chi)&=\mathsf{f}^{(2)}\,\mathfrak{f}_{\mathcal{D}_2}(\chi)+
\sum_{\D}\Big[\langle a^{(2)} \rangle_{\D}\mathfrak{f}_{\D}(\chi)+\langle a^{(0)}\gamma^{(2)}+a^{(1)}\gamma^{(1)} \rangle_{\D}\,\mathfrak{f}^{(1)}_{\D}(\chi)+\frac{1}{2}\langle a^{(0)}\,(\gamma^{(1)})^2 \rangle_{\D}\,\mathfrak{f}^{(2)}_{\D}(\chi)\Big].
\end{split}
\end{align}
We are then ready to apply the algorithm described in Section \ref{sec:bootstrap}.  First, reiterating what we said at the start of this section, we can use the fact that the operator degeneracy is not lifted at tree level: $\langle \gamma^{(1)}_{\D} \rangle$ given in \eqref{tree1111CFTdata} is the exact anomalous dimension of {\it any} operator $\mathcal{L}^{\D}_{0,[0,0]}$ at first order, and degenerate operators in the free theory are still degenerate at first order.  This allows us to write \eqref{nodegeneracy1111} and therefore setting $\ell=2$ in \eqref{fl_logl_1111} we find
\begin{align}\label{r71loop}
r_7(\chi)=f^{(2)}_{\log^{2}}(\chi)=-\frac{\chi^3\,(2-\chi)\,(5-5\chi+3\chi^2)}{2(1-\chi)^3}\,.
\end{align}
The next step is to use Bose symmetry, which we do by considering the crossing equations \eqref{crossing1111} and \eqref{braiding1111} and setting the coefficient of each HPL in the chosen basis to zero. Note that we consider $\zeta(2)$ to be a basis element on its own, being a MZV of transcendentality two.

Let us focus first on the rational functions multiplying HPLs of $\mathtt{t}\ge 2$, for which we find
\begin{gather}\label{1loopt>1}
r_3(\chi)=-\frac{\chi^2}{(1-\chi)^2}\,r_7(1-\chi)\,, \quad
r_4(\chi)=0\,,\quad
r_6(\chi)=-r_7(\chi)+\frac{\chi^2}{(1-\chi)^2}\,r_7(1-\chi)+\chi^2\,r_7\big(\tfrac{1}{1-\chi}\big)\,,
\end{gather}
implying that these terms are completely fixed in terms of \eqref{r71loop}. The remaining functions are constrained by
\begin{gather}\label{1loopzeta2}
r_{1,2}(\chi)+r_{1,2}\big(\tfrac{\chi}{\chi-1}\big)=0, \quad
(1-\chi)^2\,r_{1,2}(\chi)+\chi^2\,r_{1,2}(1-\chi)=0\,,
\end{gather}
as well as
\begin{align}\label{1looptree}
\begin{split}
(1-\chi)^2\,r_2(\chi)+\chi^2\,r_5(1-\chi)=0\,,\\
(1-\chi)^2\,r_{1,1}(\chi)+\chi^2\,r_{1,1}(1-\chi)=0\,,\quad r_{1,1}(\chi)+r_{1,1}\big(\tfrac{\chi}{\chi-1}\big)=\mathsf{f}^{(2)}\,\frac{\chi^2}{\chi-1}\,,\\
(1-\chi)^2\,\left(r_{2}(\chi)-r_{2}\big(\tfrac{\chi}{\chi-1}\big)\right)=\chi^2\,r_{2}(1-\chi)\,, \quad
r_{2}(\chi)+\chi^2\,r_{2}\big(\tfrac{1}{\chi}\big)=0\,,
\end{split}
\end{align}
which are just the crossing equations for a tree-level problem.  Their solution can be expressed in terms of the functions $q_{1,2}^{(n)}$ introduced in (\ref{h1star}-\ref{h2star}) and reads 
\begin{align}
\begin{split}
r_{1,1}(\chi)&=\sum_{n=1}^{\infty}c_{1,1}^{(n)}\,q_1^{(n)}(\chi)\,,\quad
r_{1,2}(\chi)=\sum_{n=1}^{\infty}c_{1,2}^{(n)}\,q_1^{(n)}(\chi)\,,\\
r_2(\chi)&=\sum_{n=1}^{\infty}c_{2}^{(n)}\,q_2^{(n)}(\chi)\,,\quad r_5(\chi)=-\frac{\chi^2}{(1-\chi)^2}r_2(1-\chi)\,,
\end{split}
\end{align}
where we have used that $\mathsf{f}^{(2)}=0$ from \eqref{f1111_3loops}.

We can now fix the coefficients by requiring the correct behavior in the direct OPE channel ($\chi \to 0$) and in the Regge limit. In particular, we demand that
\begin{align}
f(\chi)=\mathcal{O}(\chi^3)\,,
\end{align}
as well as requiring the mildest possible growth for $\langle \gamma^{(2)}\rangle_{\D}$ at large $\D$. While not obvious {\it a priori}, this turns out to be
\begin{align}
\langle \gamma^{(2)}\rangle_{\D}\sim \D^3\,, \quad (\D \to \infty)\,.
\end{align}
These two conditions, together, fix 
\begin{align}
c_{1,1}^{(n)}=0\quad \forall n\ge 1\,,\quad
c_{1,2}^{(n)}=0\quad \forall n\ge 1\,,\quad
c_2^{(1)}=-\frac{11}{4}\,, \quad
c_2^{(2)}=\frac{3}{2}\,, \quad
c_2^{(n)}=0\quad \forall n\ge 3\,,
\end{align}
or in other words
\begin{align}
r_{1,1}(\chi)=0=r_{1,2}(\chi)\,, \quad
r_2(\chi)=\frac{(1+\chi)(6-11\chi+6\chi^2)}{4(1-\chi)}\,,
\end{align}
which fixes completely the solution at one loop to be
\begin{align}\label{1loop1111reduced}
\begin{split}
f^{(2)}(\chi)=&\frac{(1+\chi)(6-11\chi+6\chi^2)}{4(1-\chi)}\log(1-\chi)+\frac{(1-\chi^2)(3-\chi+3\chi^2)}{2\chi}\log^2(1-\chi)\\
&-\left[\frac{\chi(2-\chi)(1-\chi+6\chi^2)}{4(1-\chi)^2}+\frac{(1-2\chi)(1+3\chi^2-6\chi^3+3\chi^4)}{2(1-\chi)^2}\log(1-\chi)\right]\,\log\chi\\
&-\frac{\chi^3(2-\chi)(5-5\chi+3\chi^2)}{2(1-\chi)^3}\,\log^2\chi\,.
\end{split}
\end{align}
The associated CFT data,  which can be extracted from \eqref{1loopOPE1111},  read\footnote{Note that it is the combination $\langle \gamma^{(2)}-\gamma^{(1)}\,\partial_{\D}\gamma^{(1)} \rangle_{\D}$,  and not simply $\langle\gamma^{(2)}\rangle_{\D}$,  that according to the reciprocity principle \eqref{1dreciprocityappendix} should have an expansion in powers of $\JJ$.  This is indeed the case,  since it is true for all terms in \eqref{1loopCFTdata1111},  including $H_{\D+1}$.}
\begin{align}\label{1loopCFTdata1111}
\begin{split}
\langle\gamma^{(2)}\rangle_{\D}=&\gamma^{(1)}_{\D}\,\partial_{\D}\gamma^{(1)}_{\D}+\frac{\JJ}{8}\left(-11-\frac{6}{\JJ+2}+4\,H_{1+\D}\right)\,,\\
\langle a^{(2)} \rangle_{\D}=&\partial_{\D}\langle a^{(0)}\,\gamma^{(2)}+a^{(1)}\,\gamma^{(1)}\rangle_{\D}-\frac{1}{2}\partial_{\D}^2\langle a^{(0)}\,(\gamma^{(1)})^2 \rangle_{\D}\\
&+\langle a^{(0)} \rangle_{\D}\,\left(\frac{\JJ(\JJ-2)}{2}(S_{-2}(\D)+\frac{1}{2}\zeta(2))+\frac{24+48\D-17\D^2-44\D^4-11\D^5-\D^6}{4(\JJ+2)}\right)\,,
\end{split}
\end{align}
where we have introduced the (generalized) harmonic numbers
\begin{align}
H^{(m)}_n=\sum_{k=1}^n\frac{1}{k^m}\,,  \quad
H_n\equiv H^{(1)}_n\,,
\end{align}
and the harmonic sum\footnote{Note that the same notation is conventionally used for harmonic sums and for Nielsen polylogarithms,  which are discussed in Appendix \ref{app:polylogs}.  We hope that in this work the distinction will be clear from the context,  as Nielsen polylogarithms will be used to describe correlators as functions of $\chi$,  while harmonic sums appear in the expression of CFT data as functions of $\D$.}
\begin{align}\label{Sminus2_harmonic}
S_{-2}(n)=\sum_{k=1}^n\frac{(-1)^k}{k^2}=\frac{(-1)^n}{4}\left(H^{(2)}_{n/2}-H^{(2)}_{(n-1)/2}\right)-\frac{1}{2}\zeta(2)\,.
\end{align}

Once again, we can extend the definition of $f^{(2)}(\chi)$ to a function $\bar{f}^{(2)}(\chi)$ simply inserting absolute values in the arguments of the logarithms. The resulting function is defined on the whole real axis, except for the OPE limits $\chi=0,1$, and is Bose symmetric. It can be expressed as
\begin{align}\label{final1111_1loop}
\begin{split}
\bar{f}^{(2)}(\chi)=&\left[\tfrac{1}{2!}(-\tfrac{1}{2}\widehat{\mathcal{C}})[f^{(1)}_{\log}(\chi)]\,\log^2|\chi|+\tfrac{\chi(-2+3\chi-13\chi^2+6\chi^3)}{4(1-\chi)^2}\,\log|\chi|\right]-\tfrac{\chi^2}{(1-\chi)^2}\Big[\chi \to 1-\chi\Big]\\
&+\left[-\left(\tfrac{1}{2\chi(1-\chi)}+\tfrac{3\chi(1-\chi)}{2}\right)\,\log|1-\chi|\,\log|\chi|+\tfrac{\mathsf{f}^{(2)}}{3}\right]\,f^{(0)}(\chi)\\
&=-\tfrac{1}{4}\left[\left(\widehat{\mathcal{C}}[f^{(1)}(\chi)]\,-6\chi^2\right)\,\log|\chi|\right]-\tfrac{\chi^2}{(1-\chi)^2}\Big[\chi \to 1-\chi\Big]\,,
\end{split}
\end{align}
with the same notation and similar properties to \eqref{final1111tree}. Note that not all allowed HPLs at this order actually appear in the answer: we started from a basis of dimension seven but it is actually possible to express the result in terms of six basis functions, due to the fact that the coefficient of $\Li_2(\chi)$ turns out to vanish. We will see more examples of this phenomenon at higher orders, where the simplification in the final answer is less trivial and potentially suggestive of a general structure.

\subsection{Two loops}

According to our guess \eqref{tmax(l)}, moving to two loops we should make an ansatz based on HPLs of maximal transcendentality three. Making use of the explicit basis discussed in Appendix \ref{app:polylogs}, we then write
\begin{align}\label{2loopansatz1111}
\begin{split}
f^{(3)}(\chi)=&f^{(3)}_{\log^0}(\chi)+f^{(3)}_{\log^1}(\chi)\,\log\chi+f^{(3)}_{\log^2}(\chi)\,\log^2\chi+f^{(3)}_{\log^3}(\chi)\,\log^3\chi\\
f^{(3)}_{\log^0}(\chi)=&r_1(\chi)+r_2(\chi)\,\log(1-\chi)+r_3(\chi)\,\log^2(1-\chi)+r_4(\chi)\,\Li_2(\chi)\\
& + r_5(\chi)\,\log^3(1-\chi)+r_6(\chi)\,\Li_2(\chi)\,\log(1-\chi)+r_7(\chi)\,\Li_3(\chi)+r_8(\chi)\,S_{1,2}(\chi)\,,\\
f^{(3)}_{\log^1}(\chi)=&r_9(\chi)+r_{10}(\chi)\,\log(1-\chi)+r_{11}(\chi)\,\log^2(1-\chi)+r_{12}(\chi)\,\Li_2(\chi)\,,\\
f^{(3)}_{\log^2}(\chi)=&r_{13}(\chi)+r_{14}(\chi)\,\log(1-\chi)\,,\\
f^{(3)}_{\log^2}(\chi)=&r_{15}(\chi)\,,
\end{split}
\end{align}
where $S_{1,2}(\chi)$ is a Nielsen polylogarithm and we can introduce explicit MZVs by expanding
\begin{align}
\begin{split}
r_1(\chi)&=r_{1,1}(\chi)+r_{1,2}(\chi)\,\zeta(2)+r_{1,3}(\chi)\,\zeta(3)\,,\\
r_2(\chi)&=r_{2,1}(\chi)+r_{2,2}(\chi)\,\zeta(2)\,,\\
r_9(\chi)&=r_{9,1}(\chi)+r_{9,2}(\chi)\,\zeta(2)\,.
\end{split}
\end{align}
Following Appendix \ref{app:perturbativeOPE} we can also write the blocks expansion for each $f^{(\ell)}_{\log^k}$ as
\begin{align}\label{2loopOPE1111}
\begin{split}
f^{(3)}_{\log^3}(\chi)=&\sum_{\D}\frac{1}{3!}\langle a^{(0)}\,(\gamma^{(1)})^3\rangle_{\D}\,\mathfrak{f}_{\D}(\chi)\,,\\
f^{(3)}_{\log^2}(\chi)=&\sum_{\D}\left[\langle \frac{1}{2}a^{(1)}\,(\gamma^{(1)})^2+a^{(0)}\,\gamma^{(1)}\,\gamma^{(2)} \rangle_{\D}\,\mathfrak{f}_{\D}(\chi)+\frac{1}{2}\langle a^{(0)}\,(\gamma^{(1)})^3\rangle_{\D}\,\mathfrak{f}^{(1)}_{\D}(\chi)\right]\,,\\
f^{(3)}_{\log^1}(\chi)=&\sum_{\D}\left[
\langle a^{(2)}\,\gamma^{(1)}
+a^{(1)}\,\gamma^{(2)} 
+a^{(0)}\,\gamma^{(3)} \rangle_{\D} \mathfrak{f}_{\D}(\chi)
+\langle a^{(1)}\,(\gamma^{(1)})^2
+2\,a^{(0)}\,\gamma^{(1)}\,\gamma^{(2)} \rangle_{\D} \mathfrak{f}^{(1)}_{\D}(\chi)
\right.\\
&\left.  +\frac{1}{2} \langle a^{(0)}\,(\gamma^{(1)})^3 \rangle_{\D} \mathfrak{f}^{(2)}_{\D}(\chi)
\right]\,,\\
f^{(3)}_{\log^0}(\chi)=&\,\mathsf{f}^{(3)}\,\mathfrak{f}_{\mathcal{D}_2}(\chi)+\sum_{\D}\left[
\langle a^{(3)}\rangle_{\D}  \mathfrak{f}_{\D}(\chi)+
\langle a^{(2)}\,\gamma^{(1)}
+a^{(1)}\,\gamma^{(2)}
+a^{(0)}\,\gamma^{(3)} \rangle_{\D} \mathfrak{f}^{(1)}_{\D}(\chi)\right.\\
&\left.  
+\frac{1}{2}\langle a^{(1)}\,(\gamma^{(1)})^2
+2\,a^{(0)}\,\gamma^{(1)}\,\gamma^{(2)} \rangle_{\D} \mathfrak{f}^{(2)}_{\D}(\chi)
+\frac{1}{3!} \langle a^{(0)}\,(\gamma^{(1)})^3 \rangle_{\D} \mathfrak{f}^{(3)}_{\D}(\chi)
\right]\,.
\end{split}
\end{align}
Again, the first step of the algorithm is to determine the functions that multiply the highest logarithmic singularities, {\it i.e.} $f^{(3)}_{\log^3}(\chi)$ and $f^{(3)}_{\log^2}(\chi)$. The former is easily obtained from equation \eqref{fl_logl_1111},  which is valid to all loops and exploits the fact that the operators degeneracy present in the free theory is not lifted by tree level corrections.  We thus obtain
\begin{align}\label{f3_log3}
r_{15}(\chi)=f^{(3)}_{\log^3}(\chi)=\frac{\chi^3}{6\,(1-\chi)^4}\left(50-125\chi+214\chi^2-196\chi^3+93\chi^4-18\chi^5\right)\,.
\end{align} 
At two loops there is another term that can be computed exactly from previous order data, provided that we can correctly account for the mixing problem. Luckily, this is done simply by observing that we can rewrite the second line in \eqref{2loopOPE1111} as
\begin{align}
f^{(3)}_{\log^2}(\chi)=&\sum_{\D}\left[\left( \frac{1}{2}\langle a^{(1)}\rangle_{\D} \,(\gamma^{(1)}_{\D})^2+\langle a^{(0)}\,\gamma^{(2)}\rangle_{\D} \gamma^{(1)}_{\D} \right)\,\mathfrak{f}_{\D}(\chi)+\frac{1}{2} \langle a^{(0)}\rangle_{\D}(\gamma^{(1)}_{\D})^3\,\mathfrak{f}^{(1)}_{\D}(\chi)\right]\,,
\end{align}
which is exactly computable given the averages of CFT data at previous orders.  We obtain
\begin{align}\label{f3_log2}
\begin{split}
f^{(3)}_{\log^2}(\chi)=&r_{13}(\chi)+r_{14}(\chi)\,\log(1-\chi)\\
=&\frac{\chi^3}{4\,(1-\chi)^3}\left[(2-\chi)(35-35\chi+23\chi^2)-2(25-78\chi+110\chi^2-72\chi^3+18\chi^3)\,\log(1-\chi)\right]\,.
\end{split}
\end{align}
From these results one can already see that, as opposed to the one-loop result where the coefficient of $\Li_2(\chi)$ is set to zero by crossing and braiding, here ordinary logarithms are not enough and one needs the functions $\Li_2(\chi)$, $\Li_3(\chi)$ and $S_{1,2}(\chi)$ for crossing and braiding to be symmetries of $f^{(3)}(\chi)$, given the results \eqref{f3_log3} and \eqref{f3_log2}.

Let us be more explicit and consider the functional constraints that arise when one separately considers the coefficient of each HPL in the basis, as well as the coefficients of the MZVs $\zeta(2)$ and $\zeta(3)$, in the crossing and braiding relations. As for the one-loop case, we can see that the rational functions multiplying HPLs of $\mathtt{t}\ge 2$ are completely fixed in terms of the highest logarithmic singularities, in this case \eqref{f3_log3} and \eqref{f3_log2}. In particular, we find
\begin{align}
\begin{split}
r_3(\chi)=-\frac{\chi^2}{(1-\chi)^2}r_{13}(\chi)\,, \quad
r_4(\chi)=0=r_6(\chi)\,, \quad
r_5(\chi)=\frac{\chi^2}{3(1-\chi)^2}\left(r_{14}(1-\chi)-r_{14}\big(\tfrac{\chi-1}{\chi}\big)\right)\,,\\
r_7(\chi)=-r_{12}(\chi)=\frac{\chi^2}{(1-\chi)^2}r_{8}(1-\chi)=-r_{14}(\chi)-r_{14}\big(\tfrac{\chi}{\chi-1}\big)+\chi^2\,r_{14}\big(\tfrac{1}{1-\chi}\big)+\frac{\chi^2}{(1-\chi)^2}r_{14}(1-\chi)\,,\\
r_{10}(\chi)=-r_{13}(\chi)-\chi^2\,r_{13}\big(\tfrac{1}{\chi}\big)+\frac{\chi^2}{(1-\chi)^2}r_{13}(1-\chi)\,, \quad
r_{11}(\chi)=-\frac{1}{2}r_8(\chi)-\frac{\chi^2}{(1-\chi)^2}r_{14}(1-\chi)\,.
\end{split}
\end{align}
Next, we turn to the rational functions multiplying MZVs and here we notice a new phenomenon: the presence of $\Li_3(\chi)$ and $S_{1,2}(\chi)$ generates terms proportional to $\zeta(3)$ in the crossing relations, which source the functional equation for $r_{1,3}(\chi)$. For this function, we can write
\begin{align}
r_{1,3}(\chi)=\tilde{r}_{1,3}(\chi)+\frac{\chi^2}{(1-\chi)^2}\left(r_{14}\big(\tfrac{\chi-1}{\chi}\big)-(1-\chi)^2r_{14}\big(\tfrac{1}{\chi}\big)\right)\,,
\end{align}
where we have solved explicitly for $r_{1,3}$ in terms of the source $r_{14}$, which can be done only up to a function $\tilde{r}_{1,3}$ satisfying homogeneous functional equations. Indeed, this and the remaining rational functions satisfy
\begin{align}
\begin{split}
r_{9,1}(\chi)=-\frac{\chi^2}{(1-\chi)^2}r_{2,1}(1-\chi)\,, \quad
r_{9,2}(\chi)=-\frac{\chi^2}{(1-\chi)^2}r_{2,2}(1-\chi)\,,\\ 
(1-\chi)^2r_{1,1}(\chi)+\chi^2r_{1,1}(1-\chi)=0\,, \quad r_{1,1}(\chi)+r_{1,1}\big(\tfrac{\chi}{\chi-1}\big)=\mathsf{f}^{(3)}\frac{\chi^2}{\chi-1}\,, \\
(1-\chi)^2r_{1,2}(\chi)+\chi^2r_{1,2}(1-\chi)=0\,,\quad
r_{1,2}(\chi)+r_{1,2}\big(\tfrac{\chi}{\chi-1}\big)=0\,,\\
(1-\chi)^2\tilde{r}_{1,3}(\chi)+\chi^2\tilde{r}_{1,3}(1-\chi)=0\,,
\quad
\tilde{r}_{1,3}(\chi)+\tilde{r}_{1,3}\big(\tfrac{\chi}{\chi-1}\big)=0\,,\\
(1-\chi)^2\,\left(r_{2,1}(\chi)-r_{2,1}\big(\tfrac{\chi}{\chi-1}\big)\right)=\chi^2\,r_{2,1}(1-\chi)\,, \quad
r_{2,1}(\chi)+\chi^2\,r_{2,1}\big(\tfrac{1}{\chi}\big)=0\,,\\
(1-\chi)^2\,\left(r_{2,2}(\chi)-r_{2,2}\big(\tfrac{\chi}{\chi-1}\big)\right)=\chi^2\,r_{2,2}(1-\chi)\,, \quad
r_{2,2}(\chi)+\chi^2\,r_{2,2}\big(\tfrac{1}{\chi}\big)=0\,,
\end{split}
\end{align}
which is just a set of tree-level crossing equations for the function
\begin{align}
\begin{split}
Q^{(3)}(\chi)=&r_{1,1}(\chi)+r_{1,2}(\chi)\,\zeta(2)+\tilde{r}_{1,3}(\chi)\,\zeta(3)+\left(r_{2,1}(\chi)+r_{2,2}(\chi)\,\zeta(2)\right)\,\log(1-\chi)\\
&+\left(r_{9,1}(\chi)+r_{9,2}(\chi)\,\zeta(2)\right)\,\log(\chi)\,,
\end{split}
\end{align}
taking into account that $\mathsf{f}^{(3)}$ is a rational number, so it does not source the crossing equations for rational functions multiplying MZVs.

The bootstrap problem is therefore solved, up to an infinite number of contact term ambiguities, as for one loop. Using again the functions \eqref{h1star} and \eqref{h2star}, we write
\begin{align}
\begin{split}
r_{1,1}(\chi)&=\frac{\mathsf{f}^{(3)}}{3}\,\frac{\chi  (1-2 \chi )}{1-\chi }+\sum_{n=1}^{\infty}c_{1,1}^{(n)}q_1^{(n)}(\chi)\,, \quad 
r_{1,2}(\chi)=\sum_{n=1}^{\infty}c_{1,2}^{(n)}q_1^{(n)}(\chi)\,, \quad
\tilde{r}_{1,3}(\chi)=\sum_{n=1}^{\infty}c_{1,3}^{(n)}q_1^{(n)}(\chi)\,,  \\
r_{2,1}(\chi)&=\sum_{n=1}^{\infty}c_{2,1}^{(n)}q_2^{(n)}(\chi)\,,  \quad
r_{2,2}(\chi)=\sum_{n=1}^{\infty}c_{2,2}^{(n)}q_2^{(n)}(\chi)\,,  \\
r_{9,1}(\chi)&=-\frac{\chi^2}{(1-\chi)^2}r_{2,1}(1-\chi)\,, \quad
r_{9,2}(\chi)=-\frac{\chi^2}{(1-\chi)^2}r_{2,2}(1-\chi)\,,
\end{split}
\end{align}
and all is left to do is to fix the coefficients $c^{(n)}_{i,j}$. To do so, we demand the correct behavior in the direct OPE channel ($\chi\to 0$),
\begin{align}
f^{(3)}(\chi)=-\frac{45}{16}\chi^2+\mathcal{O}(\chi^3)\,,
\end{align}
as well as the correct behavior in the Regge limit. We now find that the mildest possible Regge behavior is given by
\begin{align}
\langle\gamma^{(3)}\rangle_{\D}\sim \D^4\,, \quad (\D\to\infty)\,,
\end{align}
which we then impose as a constraint on our results. Taking into account that in our conventions $c^{(n)}_{i,j}$ are rational numbers, and so the coefficient of each MZV should vanish independently, we find a unique solution to these two constraints, given by
\begin{align}
\begin{split}
c_{1,1}^{(n)}=0\quad \forall n\ge 1\,, \quad
c_{1,2}^{(n)}=0\quad \forall n\ge 1\,, \quad
c_{1,3}^{(n)}=0\quad \forall n\ge 1\,, \\
c_{2,1}^{(1)}=\frac{5}{2}\,, \quad
c_{2,1}^{(2)}=-1\,, \quad c_{2,1}^{(n)}=0\quad \forall n\ge 3\,, \quad
c_{2,2}^{(n)}=0\quad \forall n\ge 1\,, 
\end{split}
\end{align}
or in other words
\begin{align}
\begin{split}
r_{1,1}(\chi)&=\frac{15}{8}\frac{\chi(1-2\chi)}{1-\chi}\,,\quad
r_{1,2}(\chi)=0\,,\quad
\tilde{r}_{1,3}(\chi)=0\,,\\
r_{2,1}(\chi)&=-\frac{(2-\chi)(1+\chi)(1-2\chi)}{2(1-\chi)}\,,\quad
r_{2,2}(\chi)=0\,.
\end{split}
\end{align}
This fixes completely the solution, whose expression might not look particularly illuminating at face value, but still contains some interesting features. In particular, one would once again like to define a Bose-symmetric function $\bar{f}^{(3)}(\chi)$, but as opposed to the previous orders we now have polylogarithms into the game, in addition to ordinary logarithms, and it is a priory not clear how to do this. However, we observe that the functions appearing in $f^{(3)}(\chi)$ are such that a special combination of HPLs appears naturally:
\begin{align}
\Li_3(\chi)-\Li_2(\chi)\,\log\chi-\frac{1}{3}\log(1-\chi)\,\log^2\chi\,.
\end{align}
This is nothing but the restriction to $\chi\in(0,1)$ of a weight three Lewin polylogarithm \cite{Lewin1985TheOO}, defined for $\chi\in\mathbb{R}\setminus\{0,1\}$ as\footnote{Note that of course $\Li_1(\chi)=-\log(1-\chi)$, but we chose to emphasize that this originates from the general expression \eqref{generalLewinLn}, and moreover that this specific logarithm should be taken {\it not} to have an absolute value in its argument.}
\begin{align}\label{L3_firsttime}
L_3(\chi)=\Li_3(\chi)-\Li_2(\chi)\,\log|\chi|+\frac{1}{2}\Li_1(\chi)\,\log^2|\chi|+\frac{1}{6}\log|1-\chi|\,\log^2|\chi|\,.
\end{align}
We review some properties of this function, together with its higher-weight cousins, in Appendix \ref{app:polylogs}. For the moment let us just remark that its appearance is quite natural from the point of view of Bose symmetry, since $L_3$ satisfies (by construction) ``clean'' functional equations
\begin{align}\label{crossingL3}
L_3(\chi)=L_3\big(\tfrac{1}{\chi}\big)\,,\quad
L_3(\chi)+L_3(1-\chi)+L_3\big(\tfrac{\chi}{\chi-1}\big)=\zeta(3)\,,
\end{align}
that do not involve polylogarithms of lower degree. Using such interesting new function, and in particular trading in our basis $\Li_3(\chi)$ for $L_3(\chi)$ and $S_{1,2}(\chi)$ for $L_3(1-\chi)$, we can express $\bar{f}^{(3)}(\chi)$ as
\begin{align}\label{final1111_2loop}
\begin{split}
\bar{f}^{(3)}(\chi)=&\left[\tfrac{1}{3!}\left(-\tfrac{1}{2}\widehat{\mathcal{C}}\right)^2[f^{(1)}_{\log}(\chi)]\,\log^3|\chi|+\tfrac{-1+3\chi+\chi^2-7\chi^3+15\chi^4-11\chi^5+3\chi^6}{2(1-\chi)^3}\,L_3(\chi)\right.\\
&\left. +\tfrac{-1+3\chi+\chi^2-82\chi^3+249\chi^4-341\chi^5+219\chi^6-54\chi^7}{6(1-\chi)^3}\,\log^2|\chi|\,\log|1-\chi|\right.\\
&\left. -\tfrac{\chi^3(-70+105\chi-81\chi^2+23\chi^3)}{4(1-\chi)^3}\,\log^2|\chi|-\tfrac{\chi(2-3\chi-3\chi^2+2\chi^3)}{2(1-\chi)^2}\,\log|\chi|\right]-\tfrac{\chi^2}{(1-\chi)^2}\Big[\chi \to 1-\chi]\\
&+\left[\left(\tfrac{1}{4\chi(1-\chi)}+\tfrac{23\chi(1-\chi)}{4}\right)\,\log|1-\chi|\,\log|\chi|+\left(\tfrac{3}{2\chi(1-\chi)}-2\right)\tfrac{\zeta(3)}{\chi(1-\chi)}+\tfrac{\mathsf{f}^{(3)}}{3}\right]\,f^{(0)}(\chi)\,.
\end{split}
\end{align}
Note in particular that we were able to express the result only using the building blocks $\log$ and $L_3$ (as well as their products), evaluated at arguments $\chi$ and $1-\chi$. Once again, it was possible to express the result in such a way that $\Li_2$ never appears. We will comment more on this in Section \ref{sec:1111/remarks}.

To conclude the discussion of the result at two loops,  let us present the associated CFT data.  In principle one would expect to be able to extract,  from $f^{(3)}(\chi)$,  the average anomalous dimensions $\langle \gamma^{(3)}\rangle_{\D}$.  However,  as it should be clear from \eqref{2loopOPE1111},  from $f^{(3)}_{\log^1}(\chi)$ one can only obtain the combination $\langle a^{(1)}\, \gamma^{(2)}+a^{(0)}\, \gamma^{(3)}\rangle_{\D}$,  as we do not have an independent determination of $\langle a^{(1)}\, \gamma^{(2)}\rangle_{\D}$ at this stage.  We find
\begin{align}
\begin{split}
\langle a^{(1)}&\, \gamma^{(2)}+a^{(0)}\, \gamma^{(3)}\rangle_{\D}-\langle a^{(1)}\rangle_{\D}\langle \gamma^{(2)}\rangle_{\D}=\langle a^{(0)}\rangle_{\D}\,\left[\partial_{\D}(\gamma^{(1)}_{\D}\,\langle\gamma^{(2)}\rangle_{\D})-\frac{1}{6}\partial^2_{\D}(\gamma^{(1)}_{\D})^3+\tilde{\gamma}^{(3)}_{\D}\right]\,,\\
\tilde{\gamma}^{(3)}_{\D}=&-\frac{(1+\JJ)(3+2\JJ)(16+5\JJ)}{4(2+\JJ)^2}+\frac{12+13\JJ+5j^4}{2(2+\JJ)}\,H_{1+\D}\\&
+\frac{1}{4}\JJ(\JJ-2)\,S_{-2}(1+\D)-\frac{\JJ}{2}\,H_{1+\D}^2+\frac{j^4}{4}\,\zeta(2)\,,
\end{split}
\end{align}
where the harmonic sum $S_{-2}(n)$ was introduced in \eqref{Sminus2_harmonic}.  On the other hand,  we were not able to find a closed-form expression for the third-order correction to the average of the squared OPE coefficients, so we only give a list of the first few values:
\begin{align}
\langle a^{(3)}\rangle_{\D} \,:\quad \left\{ \frac{11195}{1728}+4 \zeta(3),\,-\frac{33246449}{5103000}+13\zeta(3),\, -\frac{83060873856120557}{3304614047454720}+\frac{90}{11}\zeta(3),\,\dots \right\}\,,
\end{align}
where the $n$-th element of the list corresponds to $\D=2n$.  Note that since for $\D=2$ there is no degeneracy,  we have $\langle a^{(3)}\rangle_{\D=2}=a^{(3)}_{\D=2}$.

\subsection{Three loops}

Finally,  let us turn to the derivation of the main result of this paper: the $\langle \mathcal{D}_1\mathcal{D}_1\mathcal{D}_1\mathcal{D}_1\rangle$ correlator at three loops,  or $\mathcal{O}(\lambda^{-2})$.  Following \eqref{tmax(l)} we make an ansatz of transcendentality four, which reads
\begin{align}\label{3loopansatz1111}
\begin{split}
f^{(4)}(\chi)=&f^{(4)}_{\log^0}(\chi)+f^{(4)}_{\log^1}(\chi)\,\log\chi+f^{(4)}_{\log^2}(\chi)\,\log^2\chi+f^{(4)}_{\log^3}(\chi)\,\log^3\chi+f^{(4)}_{\log^4}(\chi)\,\log^4\chi\,,\\
f^{(4)}_{\log^0}(\chi)=&r_1(\chi)+r_2(\chi)\,\log(1-\chi)+r_3(\chi)\,\log^2(1-\chi)+r_4(\chi)\,\Li_2(\chi)+ r_5(\chi)\,\log^3(1-\chi)\\
&  +r_6(\chi)\,\Li_2(\chi)\,\log(1-\chi)+r_7(\chi)\,\Li_3(\chi)+r_8(\chi)\,S_{1,2}(\chi)+r_9(\chi)\,\log^4(1-\chi)\\
& +r_{10}(\chi)\,\Li_2(\chi)\,\log^2(1-\chi)+r_{11}(\chi)\,\Li_2(\chi)^2+r_{12}(\chi)\,\Li_3(\chi)\,\log(1-\chi)\\
& +r_{13}(\chi)\,S_{1,2}(\chi)\,\log(1-\chi)+r_{14}(\chi)\,\Li_4(\chi)+r_{15}(\chi)\,S_{2,2}(\chi)+r_{16}(\chi)\,S_{1,3}(\chi)\,,\\
f^{(4)}_{\log^1}(\chi)=&r_{17}(\chi)+r_{18}(\chi)\,\log(1-\chi)+r_{19}(\chi)\,\log^2(1-\chi)+r_{20}(\chi)\,\Li_2(\chi)\\
&  + r_{21}(\chi)\,\log^3(1-\chi)+r_{22}(\chi)\,\Li_2(\chi)\,\log(1-\chi)+r_{23}(\chi)\,\Li_3(\chi)+r_{24}(\chi)\,S_{1,2}(\chi)\,,\\
f^{(4)}_{\log^2}(\chi)=&r_{25}(\chi)+r_{26}(\chi)\,\log(1-\chi)+r_{27}(\chi)\,\log^2(1-\chi)+r_{28}(\chi)\,\Li_2(\chi)\,,\\
f^{(4)}_{\log^3}(\chi)=&r_{29}(\chi)+r_{30}(\chi)\,\log(1-\chi)\,,\\
f^{(4)}_{\log^4}(\chi)=&r_{31}(\chi)\,,
\end{split}
\end{align}
where we take the rational functions to have rational coefficients in their numerators, except for the following MZVs that we introduce explicitly
\begin{align}
\begin{split}
r_1(\chi)&=r_{1,1}(\chi)+r_{1,2}(\chi)\,\zeta(2)+r_{1,3}(\chi)\,\zeta(3)+r_{1,4}(\chi)\,\zeta(4)\,,\\
r_2(\chi)&=r_{2,1}(\chi)+r_{2,2}(\chi)\,\zeta(2)+r_{2,3}(\chi)\,\zeta(3)\,,\\
r_3(\chi)&=r_{3,1}(\chi)+r_{3,2}(\chi)\,\zeta(2)\,,\\
r_4(\chi)&=r_{4,1}(\chi)+r_{4,2}(\chi)\,\zeta(2)\,,\\
r_{17}(\chi)&=r_{17,1}(\chi)+r_{17,2}(\chi)\,\zeta(2)+r_{17,3}(\chi)\,\zeta(3)\,.
\end{split}
\end{align}
Following Appendix \ref{app:perturbativeOPE} we can also give the expansion of each $f^{(4)}_{\log^k}$ in terms of blocks and derivatives, which reads
\begin{align}\label{3loopOPE1111}
\begin{split}
f^{(4)}_{\log^4}(\chi)=&\sum_{\D}\frac{1}{4!}\langle a^{(0)}\,(\gamma^{(1)})^4\rangle_{\D}\,\mathfrak{f}_{\D}(\chi)\,,\\
f^{(4)}_{\log^3}(\chi)=&\sum_{\D}\left[\langle \frac{1}{6}a^{(1)}\,(\gamma^{(1)})^3+\frac{1}{2}a^{(0)}\,(\gamma^{(1)})^2\,\gamma^{(2)}\rangle_{\D}\,\mathfrak{f}_{\D}(\chi)+\frac{1}{6}\langle a^{(0)}\,(\gamma^{(1)})^4\rangle_{\D}\,\mathfrak{f}^{(1)}_{\D}(\chi)\right]\,,\\
f^{(4)}_{\log^2}(\chi)=&\sum_{\D}\left[
\langle \frac{1}{2}a^{(2)}\,(\gamma^{(1)})^2
+a^{(1)}\,\gamma^{(1)}\,\gamma^{(2)} 
+a^{(0)}\,\gamma^{(1)}\,\gamma^{(3)} + \frac{1}{2}a^{(0)}\,(\gamma^{(2)})^2\rangle_{\D} \mathfrak{f}_{\D}(\chi)\right.\\
&\left.
+\langle \frac{1}{2}a^{(1)}\,(\gamma^{(1)})^3
+\frac{3}{2}a^{(0)}\,(\gamma^{(1)})^2\,\gamma^{(2)} \rangle_{\D} \mathfrak{f}^{(1)}_{\D}(\chi)
 +\frac{1}{4} \langle a^{(0)}\,(\gamma^{(1)})^4 \rangle_{\D} \,\mathfrak{f}^{(2)}_{\D}(\chi)
\right]\,,\\
f^{(4)}_{\log^1}(\chi)=&\,\sum_{\D}\left[
\langle a^{(3)}\,\gamma^{(1)}+a^{(2)}\,\gamma^{(2)}+a^{(1)}\,\gamma^{(3)}+a^{(0)}\,\gamma^{(4)}\rangle_{\D}\,  \mathfrak{f}_{\D}(\chi)\right.\\
&\left.  
+\langle a^{(2)}\,(\gamma^{(1)})^2
+2\,a^{(1)}\,\gamma^{(1)}\,\gamma^{(2)} 
+2\,a^{(0)}\,\gamma^{(1)}\,\gamma^{(3)} 
+a^{(0)}\,(\gamma^{(2)})^2\rangle_{\D}\, \mathfrak{f}^{(1)}_{\D}(\chi)\right.\\
&\left.  
+\langle \frac{1}{2}\,a^{(1)}\,(\gamma^{(1)})^3
+\frac{3}{2}\,a^{(0)}\,(\gamma^{(1)})^2\,\gamma^{(2)} \rangle_{\D}\, \mathfrak{f}^{(2)}_{\D}(\chi)
+\langle\frac{1}{6}a^{(0)}\,(\gamma^{(1)})^4  \rangle_{\D}\, \mathfrak{f}^{(3)}_{\D}(\chi)
\right]\,,\\
f^{(4)}_{\log^0}(\chi)=&\,\mathsf{f}^{(4)}\,\mathfrak{f}_{\mathcal{D}_2}(\chi)+\sum_{\D}\left[
\langle a^{(4)}\rangle_{\D} \, \mathfrak{f}_{\D}(\chi)+
\langle a^{(3)}\,\gamma^{(1)}
+a^{(2)}\,\gamma^{(2)}
+a^{(1)}\,\gamma^{(3)}
+a^{(0)}\,\gamma^{(4)}\rangle_{\D}\, \mathfrak{f}^{(1)}_{\D}(\chi)\right.\\
&\left.  
+\langle \frac{1}{2}\,a^{(2)}\,(\gamma^{(1)})^2
+a^{(1)}\,\gamma^{(1)}\,\gamma^{(2)} +a^{(0)}\,\gamma^{(1)}\,\gamma^{(3)}+\frac{1}{2} a^{(0)}\,(\gamma^{(2)})^2 \rangle_{\D} \,\mathfrak{f}^{(2)}_{\D}(\chi)\right.\\
&\left.+\langle \frac{1}{6} a^{(1)}\,(\gamma^{(1)})^3+\frac{1}{2}a^{(0)}\,(\gamma^{(1)})^2\,\gamma^{(2)}\rangle_{\D}\,\mathfrak{f}^{(3)}_{\D}(\chi)+\frac{1}{4!} \langle a^{(0)}\,(\gamma^{(1)})^4 \rangle_{\D} \mathfrak{f}^{(4)}_{\D}(\chi)
\right]\,.
\end{split}
\end{align}
As for the previous orders, we start by determining the highest logarithmic singularities, here given by the three functions $f^{(4)}_{\log^4}(\chi)$,  $f^{(4)}_{\log^3}(\chi)$ and $f^{(4)}_{\log^2}(\chi)$. The first is again the simplest, as thanks to \eqref{fl_logl_1111} we know $f^{(\ell)}_{\log^{\ell}}(\chi)$ at each order. At three loops, we have 
\begin{align}\label{f4log4}
\begin{split}
r_{31}(\chi)&=f^{(4)}_{\log^4}(\chi)\\
&=-\frac{\chi^3}{(1-\chi)^5}\left(250-875\chi+2871\chi^2-4990\chi^3+5273\chi^4
-3357\chi^5+1188\chi^6-180\chi^7\right)\,.
\end{split}
\end{align}
We can compute also $f^{(4)}_{\log^3}(\chi)$ without worrying about the mixing problem, since we can rewrite the second line in \eqref{3loopOPE1111} as
\begin{align}
f^{(4)}_{\log^3}(\chi)=\sum_{\D}\left[\left( \frac{1}{6}\langle a^{(1)}\rangle_{\D}\,(\gamma^{(1)}_{\D})^3+\frac{1}{2}\langle a^{(0)}\,\gamma^{(2)}\rangle_{\D}\,( \gamma^{(1)}_{\D})^2 \right)\,\mathfrak{f}_{\D}(\chi)+\frac{1}{6}\langle a^{(0)}\rangle_{\D}\,(\gamma^{(1)}_{\D})^4\,\mathfrak{f}^{(1)}_{\D}(\chi)\right]\,,
\end{align}
which can in fact be computed from averages of CFT data at previous orders,  with the result
\begin{align}\label{f4log3}
\begin{split}
f^{(4)}_{\log^3}(\chi)=&r_{29}(\chi)+r_{30}(\chi)\,\log(1-\chi)\\
=&\frac{1}{24(1-\chi)^4}\left[(2-\chi)\chi(1-3\chi-585\chi^2+1175\chi^3-2214\chi^4
+1626\chi^5-549\chi^6)\right.\\
&\left. +2(1-4\chi+14\chi^2+222\chi^3-1226\chi^4+3490\chi^5
-5260\chi^6+4419\chi^7\right.\\
&\left. -1962\chi^8+360\chi^9)\,\log(1-\chi) \right]\,.
\end{split}
\end{align}
Finally,  for $f^{(4)}_{\log^2}(\chi)$ one can rewrite the relevant OPE in \eqref{3loopOPE1111} as
\begin{align}\label{f4log2}
\begin{split}
f^{(4)}_{\log^2}(\chi)=&\sum_{\D}\left[
 \left(\frac{1}{2}\langle a^{(2)}\rangle_{\D}\,(\gamma^{(1)}_{\D})^2
+ \gamma^{(1)}_{\D} \langle a^{(1)}\,\gamma^{(2)} 
+a^{(0)}\,\gamma^{(3)}\rangle_{\D} +\langle a^{(0)}\,(\gamma^{(2)})^2\rangle_{\D}\right) \mathfrak{f}_{\D}(\chi)\right.\\
&\left.
+\left( \frac{1}{2}\langle a^{(1)}\rangle_{\D}\,( \gamma^{(1)}_{\D})^3
+\frac{3}{2}\langle a^{(0)}\,\gamma^{(2)}\rangle_{\D}\,(\gamma^{(1)}_{\D})^2 \right) \mathfrak{f}^{(1)}_{\D}(\chi)
 +\frac{1}{4} \langle a^{(0)}\rangle_{\D}\,(\gamma^{(1)}_{\D})^4  \,\mathfrak{f}^{(2)}_{\D}(\chi)
\right]\,,
\end{split}
\end{align}
where all terms can be obtained from averages of CFT data in $\langle \mathcal{D}_1 \mathcal{D}_1 \mathcal{D}_1 \mathcal{D}_1\rangle$ at previous orders,  except for 
\begin{align}
\langle a^{(0)}\,(\gamma^{(2)})^2\rangle_{\D} \neq 
\langle a^{(0)}\rangle_{\D}\,\langle\gamma^{(2)}\rangle_{\D}^2\,.
\end{align}
This is the main obstacle that we face in our bootstrap problem, since it requires a thorough study of the operators degeneracy. This is a familiar issue in perturbative bootstrap computations, discussed for instance for higher-dimensional holographic correlators in \cite{Alday:2017xua,Aprile:2017bgs,Aprile:2017xsp,Alday:2020tgi,Abl:2021mxo,Alday:2021ymb,Alday:2021ajh,Behan:2022uqr} or in \cite{Alday:2017zzv} for the $\epsilon$-expansion.  Note however that typically the degeneracy of operators present in the free theory is already lifted at first order, while here this only happens at second order, which makes it possible to obtain a three-loop result with relative simplicity. We shall discuss the details of how we tackle the mixing problem in Section \ref{sec:mixing}, where we will show that (see \eqref{deltaGamma2sq})
\begin{align}\label{gamma2squared}
\begin{split}
\frac{\langle a^{(0)}\,(\gamma^{(2)})^2\rangle_{\D}}{\langle a^{(0)}\rangle_{\D}}=&\langle \gamma^{(2)}\rangle_{\D}^2+\frac{1}{2}\JJ(\JJ-2)S_{-2}(1+\D)+\frac{1}{8}\JJ(3\JJ-4)H_{1+\D}^2\\
&+\left(-j^4_{\D}+\frac{3}{4}\left(5+\frac{2}{\JJ+2}\right)\right)H_{1+\D}+\frac{1}{32}\left(-156+50 \JJ+29 j^4_{\D}+\frac{24}{\JJ+2}\right)\,.
\end{split}
\end{align}
We postpone the proof of this to Section \ref{sec:mixing} and for the moment limit to use \eqref{gamma2squared} to compute
\begin{align}\label{f4log2}
\begin{split}
f^{(4)}_{\log^2}(\chi)=&r_{25}(\chi)+r_{26}(\chi)\,\log(1-\chi)+r_{27}(\chi)\,\log^2(1-\chi)+r_{28}(\chi)\,\Li_2(\chi)\\
=&\frac{1}{32\,\chi\,(1-\chi)^4}\left[(2-\chi)(1-\chi)\chi^2(5-10\chi-486\chi^2+491\chi^3-408\chi^4)\right.\\
&\left.+2(1-\chi)\chi(20-62\chi-90\chi^2+1462\chi^3-4619\chi^4
+6541\chi^5-4332\chi^6\right.\\
&\left.+1098\chi^7)\,\log(1-\chi)+2(1-\chi)(9-32\chi+42\chi^2-28\chi^3-366\chi^4+2016\chi^5\right.\\
&\left. -4728\chi^6+5508\chi^7-3168\chi^8+720\chi^9)\,\log^2(1-\chi)+12\chi(1-4\chi+14\chi^2-28\chi^3\right.\\
&\left. +158\chi^4-274\chi^5+256\chi^6-123\chi^7+24\chi^8)\,\Li_2(\chi)\right]\,.
\end{split}
\end{align}
Now that we have computed the coefficients of the highest logarithmic singularities, we can use the crossing equations to fix the result up to ambiguities arising from contact terms, which we shall fix at the end. As usual, after applying cyclic or braiding transformations to the HPLs we map them back to the ones chosen as basis in Appendix \ref{app:polylogs}, and we consider each MZV as an independent function. Once again, this fixes the coefficients of all HPLs of transcendentality $\mathtt{t}\ge 2$ in terms of the highest logarithmic singularities, encoded in the functions $r_i(\chi)$ with $i=25,\dots,31$. The explicit relations are rather long and read
\begingroup\makeatletter\def\f@size{8}\check@mathfonts
\begin{align}\label{crossing3loopT>2}
\begin{split}
r_{3,1}(\chi)=&-\frac{\chi^2}{(1-\chi)^2}r_{25}(\chi)\,, \quad r_{3,2}(\chi)=r_{4,1}(\chi)=r_6(\chi)=r_{10}(\chi)=r_{11}(\chi)=0\,,\\
r_{4,2}(\chi)=&-4\left(r_{27}(\chi)+r_{27}\big(\tfrac{\chi}{\chi-1}\big)+\frac{\chi^2}{(1-\chi)^2}r_{27}(1-\chi)+\frac{\chi^2}{(1-\chi)^2}r_{27}\big(\tfrac{\chi-1}{\chi}\big)+\chi^2 r_{27}\big(\tfrac{1}{\chi}\big)+\chi^2 r_{27}\big(\tfrac{1}{1-\chi}\big)\right)\\
&-3\left(r_{30}(\chi)+r_{30}\big(\tfrac{\chi}{\chi-1}\big)+\frac{\chi^2}{(1-\chi)^2}r_{30}(1-\chi)+\frac{\chi^2}{(1-\chi)^2}r_{30}\big(\tfrac{\chi-1}{\chi}\big)+\chi^2 r_{30}\big(\tfrac{1}{\chi}\big)+\chi^2 r_{30}\big(\tfrac{1}{1-\chi}\big)\right)\,,\\
r_5(\chi)=&\frac{\chi^2}{3(1-\chi)^2}\left(r_{26}(1-\chi)-r_{26}\big(\tfrac{\chi-1}{\chi}\big)\right)\,,\quad 
r_9(\chi)=\frac{\chi^2}{3(1-\chi)^2}\left(r_{30}(1-\chi)-r_{30}\big(\tfrac{\chi-1}{\chi}\big)\right)\,,\\
r_7(\chi)=&\frac{\chi^2}{(1-\chi)^2} r_8(1-\chi)=-r_{20}(\chi)=-r_{26}(\chi)-r_{26}\big(\tfrac{\chi}{\chi-1}\big)+\chi^2 r_{26}\big(\tfrac{1}{1-\chi}\big)+\frac{\chi^2}{(1-\chi)^2}r_{26}(1-\chi)\,,\\
r_{12}(\chi)=&-r_{22}(\chi)=-4r_{27}(\chi)+2\chi^2 r_{27}\big(\tfrac{1}{\chi}\big)-\frac{2\chi^2}{(1-\chi)^2}\left(r_{27}(1-\chi)+r_{27}\big(\tfrac{\chi-1}{\chi}\big)\right)\,,\\
r_{13}(\chi)=&2 r_{27}(\chi)+2\chi^2r_{27}\big(\tfrac{1}{\chi}\big)+\frac{\chi^2}{(1-\chi)^2}\left(4 r_{27}(1-\chi)+4 r_{27}\big(\tfrac{\chi-1}{\chi}\big)+3 r_{30}(1-\chi)+3 r_{30}\big(\tfrac{\chi-1}{\chi}\big)\right)\,,\\
r_{14}(\chi)=&\frac{\chi^2}{(1-\chi)^2}r_{16}(1-\chi)=4\chi^2\left(r_{27}\big(\tfrac{1}{1-\chi}\big)+\frac{r_{27}(1-\chi)}{(1-\chi)^2}\right)+12\left(r_{27}(\chi)+r_{27}\big(\tfrac{\chi}{\chi-1}\big)+r_{30}(\chi)+r_{30}\big(\tfrac{\chi}{\chi-1}\big)\right)\,,\\
r_{15}(\chi)=&-4\left(2 r_{27}(\chi)+r_{27}\big(\tfrac{\chi}{\chi-1}\big)+2\frac{\chi^2}{(1-\chi)^2}r_{27}(1-\chi)+\frac{\chi^2}{(1-\chi)^2}r_{27}\big(\tfrac{\chi-1}{\chi}\big)-\chi^2 r_{27}\big(\tfrac{1}{\chi}\big)-\chi^2 r_{27}\big(\tfrac{1}{1-\chi}\big)\right)\\
&-6\left(r_{30}(\chi)+r_{30}\big(\tfrac{\chi}{\chi-1}\big)+\frac{\chi^2}{(1-\chi)^2}r_{30}(1-\chi)+\frac{\chi^2}{(1-\chi)^2}r_{30}\big(\tfrac{\chi-1}{\chi}\big)-\chi^2 r_{30}\big(\tfrac{1}{\chi}\big)-\chi^2 r_{30}\big(\tfrac{1}{1-\chi}\big)\right)\,,\\
r_{18}(\chi)=&-r_{25}(\chi)-\chi^2r_{25}\big(\tfrac{1}{\chi}\big)+\frac{\chi^2}{(1-\chi)^2}r_{25}(1-\chi)\,,\quad
r_{19}(\chi)=-\frac{\chi^2}{2(1-\chi)^2}(r_7(1-\chi)+2 r_{26}(1-\chi))\,,\\
r_{21}(\chi)=&-\frac{1}{3}\left(r_{27}(\chi)+\chi^2 r_{27}\big(\tfrac{1}{\chi}\big)\right)-\frac{\chi^2}{2(1-\chi)^2}\left(r_{30}(1-\chi)-r_{30}\big(\tfrac{\chi-1}{\chi}\big)\right)\,,\\
r_{23}(\chi)=&-2\chi^2 r_{27}\big(\tfrac{1}{1-\chi}\big)-\frac{2\chi^2}{(1-\chi)^2}r_{27}(1-\chi)-8 r_{27}(\chi)-8 r_{27}\big(\tfrac{\chi-1}{\chi}\big)-9 r_{30}(\chi)-9 r_{30}\big(\tfrac{\chi-1}{\chi}\big)\,,\\
r_{24}(\chi)=&6r_{27}(\chi)+2r_{27}\big(\tfrac{\chi}{\chi-1}\big)+\frac{4\chi^2}{(1-\chi)^2}r_{27}(1-\chi)+\frac{4\chi^2}{(1-\chi)^2}r_{27}\big(\tfrac{\chi-1}{\chi}\big)-4\chi^2 r_{27}\big(\tfrac{1}{\chi}\big)-2\chi^2 r_{27}\big(\tfrac{1}{1-\chi}\big)\\
&+6r_{30}(\chi)+3r_{30}\big(\tfrac{\chi}{\chi-1}\big)+\frac{3\chi^2}{(1-\chi)^2}r_{30}(1-\chi)+\frac{6\chi^2}{(1-\chi)^2}r_{30}\big(\tfrac{\chi-1}{\chi}\big)-6\chi^2 r_{30}\big(\tfrac{1}{\chi}\big)-3\chi^2 r_{30}\big(\tfrac{1}{1-\chi}\big)\,.
\end{split}
\end{align}
\endgroup
Besides the explicit expressions, we would like to make two remarks. The first is that, as anticipated, all terms of transcendentality $\mathtt{t}\ge 2$ are fixed in terms of the highest logarithmic singularities of $f^{(4)}(\chi)$, specified by the rational functions $r_i(\chi)$ with $i=25,...,31$. The second comment is that these functions are found to satisfy ``unexpected'' functional relations that significantly simplify the final results. In particular we note that, in addition to the functions that were already found to vanish in \eqref{crossing3loopT>2}, the results \eqref{f4log4}, \eqref{f4log3} and \eqref{f4log2} are such that we also have
\begin{align}\label{simpler3loop}
r_{14}(\chi)=r_{15}(\chi)=r_{16}(\chi)=0\,,\quad
(1-\chi)^2r_{24}(\chi)=\chi^2r_{12}(\chi)\,,\quad
(1-\chi)^2r_{23}(\chi)=\chi^2r_{13}(\chi)
\end{align}
which in particular implies that the new HPLs that we introduced at transcendentality four ($\Li_4(\chi)$, $S_{2,2}(\chi)$ and $S_{1,3}(\chi)$) all drop out of the final result. It is important to stress that the simplifications \eqref{simpler3loop} only arise once mixing is properly taken into account in computing \eqref{gamma2squared}: if one were to neglect mixing and take $\langle(\gamma^{(2)})^2\rangle_{\D}=\langle\gamma^{(2)}\rangle_{\D}^2$, the bootstrap computation would still make sense but it would produce a result where none of the identities \eqref{simpler3loop} is satisfied. In particular, functions of transcendentality four like $\Li_4$ would be present in the final result.

We can now turn to analyzing the remaining functional equations. As for the two loops case, the coefficients of certain MZVs satisfy non-homogeneous equations that we can solve in terms of the highest logarithmic singularities only up to new rational functions, satisfying homogeneous equations. In particular, we have
\begin{align}
\begin{split}
r_{1,3}(\chi)=&\tilde{r}_{1,3}(\chi)+\frac{10}{7}\left(r_{26}\big(\tfrac{\chi}{\chi-1}\big)-r_{26}(\chi)-\chi^2 r_{26}\big(\tfrac{1}{1-\chi}\big)+\frac{\chi^2}{(1-\chi)^2}r_{26}(1-\chi)\right)\\
&+\frac{3}{7}\left(\chi^2 r_{26}\big(\tfrac{1}{\chi}\big)-\frac{\chi^2}{(1-\chi)^2}r_{26}\big(\tfrac{\chi-1}{\chi}\big)\right)\,,\\
r_{1,4}(\chi)=&\tilde{r}_{1,4}(\chi)-r_{27}(\chi)+r_{27}\big(\tfrac{\chi}{\chi-1}\big)+3\chi^2 r_{27}\big(\tfrac{1}{1-\chi}\big)+2\chi^2 r_{27}\big(\tfrac{1}{\chi}\big)-3 \frac{\chi^2}{(1-\chi)^2}r_{27}(1-\chi)\\
&-2 \frac{\chi^2}{(1-\chi)^2}r_{27}\big(\tfrac{\chi-1}{\chi}\big)+3\chi^2 \left(r_{30}\big(\tfrac{1}{\chi}\big)+r_{30}\big(\tfrac{1}{1-\chi}\big)\right)-3\frac{\chi^2}{(1-\chi)^2}\left(r_{30}(1-\chi)+r_{30}\big(\tfrac{\chi-1}{\chi}\big)\right)\,,\\
r_{2,3}(\chi)=&\tilde{r}_{2,3}(\chi)-\frac{1}{5}r_{27}(\chi)-\frac{11\chi^2}{5(1-\chi)^2}r_{27}(1-\chi)-\frac{9}{5}\left(\chi^2 r_{27}\big(\tfrac{1}{\chi}\big)+\frac{\chi^2}{(1-\chi)^2} r_{27}\big(\tfrac{\chi-1}{\chi}\big)\right)\\
&+\frac{9}{5}\left(\chi^2 r_{30}\big(\tfrac{1}{\chi}\big)+\frac{\chi^2}{(1-\chi)^2} r_{30}\big(\tfrac{\chi-1}{\chi}\big)-r_{30}(\chi)\right)-\frac{24\chi^2}{5(1-\chi)^2}r_{30}(1-\chi)\,,\\
r_{17,3}(\chi)&=-\frac{\chi^2}{(1-\chi)^2}r_{2,3}(1-\chi)\,,
\end{split}
\end{align}
where $\tilde{r}_{1,3}$, $\tilde{r}_{1,4}$ and $\tilde{r}_{2,3}$ satisfy the usual tree-level homogeneous crossing relations, as we shall discuss below. Note that, in addition to the simplifications \eqref{simpler3loop}, the functions $r_i(\chi)$ with $i=25,...,31$, corresponding with the highest logarithmic singularities of $f^{(4)}(\chi)$, are such that
\begin{align}
r_{1,4}(\chi)=\tilde{r}_{1,4}(\chi)\,,
\end{align}
which as we will discuss below will turn out to be zero. Thus, in correspondence with the fact that there are no terms with $\Li_4$, all terms with $\zeta(4)$ also drop out of the final result. This is the same that happens at one loop with $\Li_2$ and $\zeta(2)$. We can then define a function containing all terms which satisfy the same homogeneous crossing equations as the tree level result, which here reads
\begin{align}\label{treeambiguity3loops}
\begin{split}
Q^{(4)}(\chi)=&r_{1,1}(\chi)+r_{1,2}(\chi)\,\zeta(2)+\tilde{r}_{1,3}(\chi)\,\zeta(3)+\tilde{r}_{1,4}(\chi)\,\zeta(4)\\
&+(r_{2,1}(\chi)+r_{2,2}(\chi)\,\zeta(2)+\tilde{r}_{2,3}(\chi))\,\log(1-\chi)\\
&+(r_{17,1}(\chi)+r_{17,2}(\chi)\,\zeta(2)+\tilde{r}_{17,3}(\chi))\,\log\chi\,.
\end{split}
\end{align}
Here it should be understood that the function $Q^{(4)}(\chi)$ has the same Bose symmetry properties as a tree-level correlator, so we can as usual express the solution for the functions appearing in \eqref{treeambiguity3loops} in terms of \eqref{h1star} and \eqref{h2star} as
\begin{align}
\begin{split}
r_{1,1}(\chi)&=\frac{\mathsf{f}^{(4)}}{3}\frac{\chi(1-2\chi)}{1-\chi}+\sum_{n=1}^{\infty}c_{1,1}^{(n)}\,q_1^{(n)}(\chi)\,, \quad
r_{1,2}(\chi)=\sum_{n=1}^{\infty}c_{1,2}^{(n)}\,q_1^{(n)}(\chi)\,, \\
\tilde{r}_{1,3}(\chi)&=\sum_{n=1}^{\infty}c_{1,3}^{(n)}\,q_1^{(n)}(\chi)\,, \quad 
\tilde{r}_{1,4}(\chi)=\sum_{n=1}^{\infty}c_{1,3}^{(n)}\,q_1^{(n)}(\chi)\,, \\
r_{2,1}(\chi)&=\sum_{n=1}^{\infty}c_{2,1}^{(n)}\,q_2^{(n)}(\chi)\,, \quad 
r_{2,2}(\chi)=\sum_{n=1}^{\infty}c_{2,2}^{(n)}\,q_2^{(n)}(\chi)\,, \quad
\tilde{r}_{2,3}(\chi)=\sum_{n=1}^{\infty}c_{2,3}^{(n)}\,q_2^{(n)}(\chi)\,, \\
r_{17,1}(\chi)&=-\frac{\chi^2}{(1-\chi)^2}r_{2,1}(1-\chi)\,, \quad
r_{17,2}(\chi)=-\frac{\chi^2}{(1-\chi)^2}r_{2,2}(1-\chi)\,, \\
\tilde{r}_{17,3}(\chi)&=-\frac{\chi^2}{(1-\chi)^2}\tilde{r}_{2,3}(1-\chi)\,.
\end{split}
\end{align}
All that is left to do is then fix all the free parameters appearing in the expressions above.As usual, we do so by requiring that in the OPE limit $\chi\to 0$ the correlator satisfies
\begin{align}
f^{(4)}(\chi)=-\frac{45}{8}\chi^2+\mathcal{O}(\chi^3)\,,
\end{align}
as well as requiring the mildest possible growth from the anomalous dimension for large $\D$, which at this order reads
\begin{align}
\langle\gamma^{(4)}\rangle_{\D}\sim \D^5\,, \quad (\D\to \infty)\,.
\end{align}
These conditions are enough to fix a unique solution,which is specified by
\begin{align}
\begin{split}
c_{1,1}^{(n)}&=0\quad\forall n\ge 1\,, \quad
c_{1,2}^{(n)}=0\quad\forall n\ge 1\,, \quad
c_{1,3}^{(n)}=0\quad\forall n\ge 1\,, \quad
c_{1,4}^{(n)}=0\quad\forall n\ge 1\,, \\
c_{2,1}^{(1)}&=\frac{37}{64}\,, \quad 
c_{2,1}^{(2)}=-\frac{21}{32}\,, \quad
c_{2,1}^{(n)}=0\quad\forall n\ge 3\,, \quad
c_{2,2}^{(n)}=0\quad\forall n\ge 1\,,\\
c_{2,3}^{(1)}&=-\frac{3}{5}\,, \quad 
c_{2,3}^{(2)}=\frac{3}{10}\,, \quad
c_{2,3}^{(n)}=0\quad\forall n\ge 3\,, 
\end{split}
\end{align}
or equivalently by
\begin{align}
\begin{split}
r_{1,1}(\chi)&=\frac{15\chi(1-2\chi)}{4(1-\chi)}\,, \quad
r_{1,2}(\chi)=\tilde{r}_{1,3}(\chi)=\tilde{r}_{1,4}(\chi)=0\,,\\
r_{2,1}(\chi)&=-\frac{(1+\chi)(42-37\chi+42\chi^2)}{64(1-\chi)}\,, \quad r_{2,2}(\chi)=0\,, \quad
\tilde{r}_{2,3}(\chi)=\frac{3}{10}\chi(2-\chi)\,.
\end{split}
\end{align}

The three-loops reduced correlator $f^{(4)}(\chi)$ is now completely fixed. As one might expect, its expression is even more complicated than that for $f^{(3)}(\chi)$, but very interestingly we observe here a completely analogous structure in terms of HPLs. As we stressed, despite the complicated crossing equations \eqref{crossing3loopT>2}, the highest logarithmic singularities dictate a simpler structure in the rational functions, in particular implying the simplifications \eqref{simpler3loop}. This, in turn, implies that the coefficients of $\Li_4$, $S_{2,2}$ and $S_{1,3}$ are all vanishing. As for the result at two loops, we can then trade terms of the type $\log a$ with $\log|a|$, $\Li_3(\chi)$ with $L_3(\chi)$ and $S_{1,2}(\chi)$ with $L_3(1-\chi)$. The result is a function, $\bar{f}^{(4)}(\chi)$, which is defined for all $\chi\in \mathbb{R}\setminus\{0,1\}$ and that restricts to $f^{(4)}(\chi)$ for $\chi\in (0,1)$, which is fully Bose symmetric. As done at previous orders, we express the result in a way that makes cyclic invariance manifest, although the result is also braiding invariant:\begin{align}\label{final1111_3loop}
\begin{split}
\bar{f}^{(4)}(\chi)=&\left[\tfrac{1}{4!}\left(-\tfrac{1}{2}\widehat{\mathcal{C}}\right)^3[f^{(1)}_{\log}(\chi)]\,\log^4|\chi|-\tfrac{3(28-84\chi+94\chi^2-48\chi^3+157\chi^4-147\chi^5+46\chi^6)}{16(1-\chi)^3}\,L_3(\chi)\right.\\
&\left. 
-\tfrac{3(9-32\chi+42\chi^2-28\chi^3-2\chi^4-48\chi^5+104\chi^6 -84\chi^7+24\chi^8)}{8\chi(1-\chi)^3}\,L_3(\chi)\,\log|1-\chi|\right.\\
&\left. 
+\tfrac{3(1-4\chi+14\chi^2-28\chi^3+158\chi^4-274\chi^5+256\chi^6 -123\chi^7+24\chi^8)}{8(1-\chi)^4}\,L_3(\chi)\,\log|\chi|\right.\\
&\left.
+\tfrac{-1+4\chi-14\chi^2+528\chi^3-2926\chi^4 +7802\chi^5-11288\chi^6+9207\chi^7-3996\chi^8 +720\chi^9}{24(1-\chi)^4}\,\log^3|\chi|\,\log|1-\chi|
\right.\\
&\left.
+\tfrac{\chi(\chi-2)(-1+3\chi+585\chi^2-1175\chi^3+2214\chi^4-1626\chi^5
+549\chi^6}{24(1-\chi)^4}\,\log^3|\chi|
\right.\\
&\left.
+\tfrac{-4+11\chi-92\chi^2+755\chi^3-2388\chi^4 +3344\chi^5-2189\chi^6+549\chi^7}{8(1-\chi)^3}\,\log^2|\chi|\,\log|1-\chi|\right.\\
&\left.
+\tfrac{\chi(2-\chi)(-5+10\chi+486\chi^2-491\chi^3+408\chi^4)}{32(1-\chi)^3}\,\log^2|\chi|
-\tfrac{\chi(-94+141\chi-131\chi^2+42\chi^3)}{64(1-\chi)^2}\,\log|\chi|\right.\\
&\left.
-\tfrac{-45+201\chi-354\chi^2+294\chi^3-202\chi^4 +58\chi^5-32\chi^6+8\chi^7}{8\chi(1-\chi)^4}\,\zeta(3)\,\log|\chi|
\right]-\tfrac{\chi^2}{(1-\chi)^2}[\chi \to 1-\chi]\\
&+\left[\left(\tfrac{3}{8\chi^2(1-\chi)^2}-\tfrac{1}{2\chi(1-\chi)}+\tfrac{3\chi(1-\chi)}{4}-\tfrac{45\chi^2(1-\chi)^2}{2}\right)\,\log^2|1-\chi|\,\log^2|\chi|\right.\\
&\left.+\tfrac{1}{16}\left(-5+\tfrac{121}{\chi(1-\chi)}-204\chi(1-\chi)\right)\,\log|1-\chi|\,\log|\chi| \right.\\
&\left.  +\left(\tfrac{249}{16}-\tfrac{69}{8\chi(1-\chi)}+\tfrac{\chi(1-\chi)}{2}\right)\tfrac{\zeta(3)}{\chi(1-\chi)}+\tfrac{\mathsf{f}^{(4)}}{3}\right]\,f^{(0)}(\chi)\,.
\end{split}
\end{align}
All polylogarithms of even weight are absent and the correlator can be written purely in terms of sums and products of the two basic building blocks $L_3$ and $\log$, evaluated at $\chi$ and $1-\chi$.

In principle,  one could perform the OPE and find an expression for the CFT data $\langle\gamma^{(4)}\rangle_{\D}$ and $\langle a^{(4)}\rangle_{\D}$ at three loops.  However,  due to the presence of mixing,  rather than the former from $f^{(4)}_{\log^1}(\chi)$ one can only extract the combination $\langle a^{(3)}\,\gamma^{(1)}+a^{(2)}\,\gamma^{(2)}+a^{(1)}\,\gamma^{(3)}+a^{(0)}\,\gamma^{(4)}\rangle_{\D}$ (see \eqref{3loopOPE1111}),  for which we were not able to find a closed-form expression.  On the other hand,  to extract $\langle a^{(4)}\rangle_{\D}$ from the OPE one should study the mixing problem to compute quantities such as $\langle a^{(2)}\,\gamma^{(2)}+a^{(1)}\,\gamma^{(3)}\rangle_{\D}$,  which is a problem that we have not attempted to solve.  However,  since for $\D=2$ there is no degeneracy,  we can drop the average symbol around CFT data and use the first terms in the small $\chi$ expansion of $f^{(4)}(\chi)$ and \eqref{3loopOPE1111} to obtain
\begin{align}
\begin{split}
\gamma^{(4)}_{\D=2}=\frac{351845}{13824 }- \frac{75}{2}\,\zeta(3) \,,\qquad
a^{(4)}_{\D=2}=-\frac{1705}{96}-\frac{1613}{24} \,\zeta (3)\,.
\end{split}
\end{align}

We can now to summarize the CFT data that we have computed analytically for the superconformal primary of the multiplet $\mathcal{L}^{\D=2}_{0,[0,0]}$,  which we refer to schematically as $\varphi^2$.  Its dimension at strong coupling is
\begin{align}
\h_{\varphi^2}=2-\frac{5}{\lambda^{1/2}}+\frac{295}{24}\frac{1}{\lambda}-\frac{305}{16}\frac{1}{\lambda^{3/2}}+\left(\frac{351845}{13824 }- \frac{75}{2}\,\zeta(3) \right)\frac{1}{\lambda^2}+\mathcal{O}(\lambda^{-5/2})\,,
\end{align}
while the OPE coefficient $\mu_{11\varphi^2}$,  such that $\mu_{11\varphi^2}^2=a_{\D=2}$,  has the expansion
\begin{align}
\begin{split}
\mu_{11\varphi^2}=\sqrt{\frac{2}{5}}&\left[1-\frac{43}{24}\frac{1}{\lambda^{1/2}}-\frac{649}{1152}\frac{1}{\lambda}+\left(\frac{7259}{1024}+5\,\zeta(3)\right)\frac{1}{\lambda^{3/2}}\right.\\
&\left.-\left(\frac{25635205}{2654208}+\frac{7205}{96}\,\zeta(3)\right)\frac{1}{\lambda^{2}}\right]+\mathcal{O}(\lambda^{-5/2})\,.
\end{split}
\end{align}

\subsection{Concluding remarks}\label{sec:1111/remarks}

To end this section, let us make some comments on the results that we have obtained for the four-point function of the super-displacement operator in perturbation theory at strong coupling. First, as we have pointed out in several places along the way, although our results might naively appear complicated, they are actually much simpler than they could potentially be given our initial ansatz. In particular, while at order $\ell$ we are allowing for the presence of $2^{\ell+1}-1$ independent HPLs (including the identity), we found that with a convenient choice of basis one can actually express the results in terms of fewer functions. A summary is contained in table \ref{tab:independentfunctions1111}. 
\begin{table}[h!]
\centering
\begin{tabular}{c || c | c | c | c} 
$\ell$ & 1 & 2 & 3 & 4\\
\hline
$2^{\ell+1}-1$ & 3 & 7 & 15 & 31 \\  
\hline
$\#$ functions & 3 & 6 & 12 & 21
\end{tabular}
\caption{Number of independent functions used to express our results for $f^{(\ell)}(\chi)$ in \eqref{final1111tree}, \eqref{final1111_1loop}, \eqref{final1111_2loop} and \eqref{final1111_3loop}, compared to the number of independent HPLs of transcendentality $\mathtt{t}\le \ell$.}
\label{tab:independentfunctions1111}
\end{table}
Note that this feature of our results only becomes apparent once a particular basis is adopted, which in our case up to three loops only involves the functions $\log$ and $L_3$ (introduced in \eqref{L3_firsttime}), evaluated at arguments $\chi$ and $1-\chi$. It is then straightforward to identify the reason for the mismatch between the last two lines in table \ref{tab:independentfunctions1111}: it is due to the fact that in the chosen basis the functions $\Li_2$ and $\Li_4$ never appear in the final result, although they are {\it a priori} allowed by our ansatz following the observations of Section \ref{sec:bootstrap}.

Let us then further investigate the properties of HPLs that could justify this result. First, we note that since Bose symmetry plays of course a crucial role in our derivation, it is natural to expect functions with simple transformation properties to make the result simpler. Once a basis of HPLs is chosen, the basis functions evaluated at arguments that are mapped into each other by $S^3$ transformations \eqref{crossinggroupS3} can be always written as a linear combination of the functions chosen as basis, evaluated at the original argument, but such relations are generally complicated and involve products of HPLs of lower weight than the original one, see {\it e.g.} \eqref{S3NielsenPolylogs}. As shown in \eqref{crossingL3}, the function $L_3$ on the other hand has the property that it satisfies ``clean'' functional relations, so that it is a more natural object to use than $\Li_3$ to emphasize the transformation properties under crossing symmetry. However, the same could be said of a modification of $\Li_2$, namely the so-called Rogers dilogarithm, which appears for instance in weak coupling computations on the Wilson line \cite{Kiryu:2018phb,Barrat:2021tpn,Barrat:2022eim}
\begin{align}\label{Rogers}
L_R(\chi)=\Li_2(\chi)+\tfrac{1}{2}\,\log \chi \,\log(1-\chi)\,,
\end{align}
which for $\chi\in(0,1)$ satisfies for instance
\begin{align}\label{RogersCrossing}
L_R(\chi)+L_R(1-\chi)=\zeta(2)\,,\qquad L_R(\chi)+L_R\big(\tfrac{\chi}{\chi-1}\big)=-\tfrac{i\,\pi}{2}\log(1-\chi)\,.
\end{align}
It turns out that $L_R$ and $L_3$ are actually related, as one can introduce a whole family of functions $L_n$, known as Lewin polylogarithms \cite{Lewin1985TheOO} (for $n>1$)
\begin{align}\label{generalLewinLn_maintext}
L_n(\chi)=\sum_{k=0}^{n-1}\frac{(-1)^k}{k!}\Li_{n-k}(\chi)\,\log^k|\chi|-\frac{(-1)^n}{n!}\log|1-\chi|\,\log^{n-1}|\chi|\,,
\end{align} 
and note that $L_2$ reduces to $L_R$ for $\chi\in(0,1)$ but has the correct prescription for the absolute values such that the imaginary part drops from the second equation in \eqref{RogersCrossing}. However, while the functional relations \eqref{crossingL3} hold for all real values of $\chi$, those for $L_2$ receive corrections when one tries to extend them beyond $\chi\in(0,1)$. This is a general difference between Lewin polylogarithms of even and odd weight, which satisfy
\begin{align}
n\,\,\,\text{odd}:\,\,\,L_n(\chi)-L_n(1/\chi)=0\,, \qquad
n\,\,\,\text{even}:\,\,\,L_n(\chi)+L_n(1/\chi)=
\begin{cases}
-2(1-2^{1-n})\zeta(n)\,\quad \chi<0\,,\\
2\zeta(n)\,,\hspace{2.1cm} \chi>0\,.
\end{cases}
\end{align}
Besides functional relations (that we have discussed for real $\chi$), another interesting question is related to the analytic structure of $L_n$ for complex $\chi$. In particular, one can show that for all $n>1$ Lewin's polylogarithms have no discontinuities for $\chi=0,1$:
\begin{align}\label{LewinDisc}
L_n(\chi+i\,0^+)=L_n(\chi+i\,0^-)\,,
\end{align}
for all $\chi\in \mathbb{R}$. Note that not only the precise coefficients and functions in \eqref{generalLewinLn_maintext}, but also the choice of absolute values, are crucial for \eqref{LewinDisc} to hold. While from this perspective there is no difference between even and odd weight $L_n$, an interesting difference arises when one thinks of 1d correlators as the diagonal limit of higher-dimensional four-point functions, which in terms of $d>1$ cross ratios $z$, $\bar{z}$ corresponds to taking $z=\bar{z}=\chi$. In particular, higher-dimensional four-point functions are single-valued in the Euclidean configuration $\bar{z}=z^*$, so one can wonder whether some of the $L_n$ (if any) have such ``higher-dimensional'' origin. We address this question in some detail in Appendix \ref{app:polylogs}, where we conclude that such origin only exists for the odd-weight functions. In particular, there is an interesting connection between the ladder integrals \cite{USSYUKINA1993363,Usyukina:1993ch,Isaev:2003tk,Aprile:2017bgs} 
\begin{align}\label{defLadder_maintext}
\Phi^{(L)}(z,\bar{z})=-\frac{(1-z)(1-\bar{z})}{z-\bar{z}}\sum_{k=0}^L(-1)^r\frac{(2L-k)!}{k!L!(L-k)!}\,\log^{k}(z\bar{z})\,\left(\Li_{2L-k}(z)-\Li_{2L-k}(\bar{z})\right)\,,
\end{align}
which often appear in perturbative computations of scattering amplitudes and correlation functions, and the Lewin polylogarithms. The ladder integrals are single-valued in the Euclidean configuration in the sense that 
\begin{align}
\text{disc}_{z=0}\Phi^{(L)}=\text{disc}_{\bar{z}=0}\Phi^{(L)}\,, \qquad 
\text{disc}_{z=1}\Phi^{(L)}=\text{disc}_{\bar{z}=1}\Phi^{(L)}\,,
\end{align}
and here we make an observation that, to the best of our knowledge, has not appeared in the literature before. Namely, the diagonal limit of ladder integrals can be written as a linear combination of Lewin polylogarithms of {\it odd} weight:
\begin{align}
\lim_{z\to\bar{z}\equiv\chi}\Phi^{(L)}(z,\bar{z})=\frac{(1-\chi)^2}{\chi}\,\sum_{k=1}^{L-1}\frac{(-1)^{L-k}\,(2k+2)!}{(k+1)!L!(L-k-1)!}\,\log^{2L-2k-2}|\chi|\,L_{2k+1}(\chi)\,.
\end{align}

We have thus shown that all our results can be expressed in terms of functions that arise in the diagonal limit of single-valued functions of two complex variables $z,\bar{z}$. Note that this is {\it not} true for general 1d correlators: it does not apply for instance to correlators on the Wilson line at weak coupling \cite{Kiryu:2018phb,Barrat:2021tpn,Barrat:2022eim} or for 1d CFTs that arise as dual of simple AdS$_2$ Lagrangians with $\varphi^4$ interactions \cite{Mazac:2018ycv,Ferrero:2019luz}. It would be interesting to investigate the meaning of this fact in greater detail and to understand whether it can be used as a constraint on results at higher orders, in particular restricting the space of allowed HPLs in the ansatz. A more restrictive ansatz could allow to bypass, at least partially, the necessity to resolve the mixing problem at higher orders. A similar strategy was recently employed in \cite{Huang:2021xws,Drummond:2022dxw} to compute the two-loop four-graviton scattering amplitude in AdS$_5\times S^5$ from $\mathcal{N}=4$ SYM, where a good intuition of the correct ansatz allowed to avoid resolving the mixing problem for triple-trace operators. At this stage it is not clear what functions should appear at the next order: should we only add $L_5(x)$ or more general HPLs as well? Our observations also seem to agree with the comments of \cite{Giombi:2022anm} (see section 5.2), where the Wilson line defect CFT correlators are studied in a large charge expansion. Those results seem to indicate that only products of ordinary logarithms and polylogarithms appear at higher orders, which would indeed suggest a special role played by the functions $L_n(x)$ for odd $n$. It would be interesting to investigate this more in detail in the future.

Let us now move to another point, still related to the idea of understanding the general structure of $f^{(\ell)}(\chi)$ at each order. A crucial role in our bootstrap algorithm is played by the idea of computing the coefficients of the highest logarithmic singularities ($\log^k\chi$ with $k\ge 2$), which as we argued (and proved in Appendix \ref{app:perturbativeOPE}) are controlled at each order by CFT data at previous perturbative orders. Computing such data at low orders, one can then obtain certain all-loop structures in $\langle \mathcal{D}_1 \mathcal{D}_1 \mathcal{D}_1 \mathcal{D}_1\rangle$, very much in the spirit of \cite{Bissi:2020wtv,Bissi:2020woe}. The simplest of such structures is the leading logarithmic singularity ($k=\ell$)
\begin{align}
f^{(\ell)}_{\log^{\ell}}(\chi)=\frac{1}{\ell!}\sum_{\D}\langle a^{(0)}(\gamma^{(1)})^{\ell}\rangle_{\D}\,\mathfrak{f}_{\D}(\chi)\,,
\end{align}
and was already observed in \eqref{fl_logl_1111}. To reiterate, one has that the free theory degeneracy is not lifted at tree level and moreover the anomalous dimensions are proportional to the eigenvalue of the Casimir operator acting on long superconformal blocks, so that, all in all,
\begin{align}
\langle a^{(0)}(\gamma^{(1)})^{\ell}\rangle_{\D}\,\mathfrak{f}_{\D}(\chi)=\langle a^{(0)}\rangle_{\D}\left(-\frac{1}{2}\widehat{\mathcal{C}}\right)^{\ell}\,\mathfrak{f}_{\D}(\chi)\,,
\end{align}
thus leading to 
\begin{align}
f^{(\ell)}_{\log^{\ell}}(\chi)=&-\frac{1}{2\ell}\,\widehat{\mathcal{C}}f^{(\ell-1)}_{\log^{\ell-1}}(\chi)
=\frac{1}{\ell!}\left(-\frac{1}{2}\right)^{\ell}\widehat{\mathcal{C}}^{\ell}f^{(0)}_L(\chi)\,,
\end{align}
where we have introduced $f^{(0)}_L(\chi)=f^{(0)}(\chi)-3\chi$, satisfying
\begin{align}
\widehat{\mathcal{C}}f^{(0)}_L(\chi)=\widehat{\mathcal{C}}\sum_{\D}\langle a^{(0)}(\gamma^{(1)})^{\ell}\rangle_{\D}\,\mathfrak{f}_{\D}(\chi)\,.
\end{align}
All-order structures for the subleading logarithmic singularities ($k<\ell$) are harder to obtain, given the presence of derivatives of superconformal blocks. To deal with those, it is useful to take derivatives of the Casimir equation 
\begin{align}\label{CasimirODE1111}
\widehat{\mathcal{C}}\,\mathfrak{f}_{\D}(\chi)=\D(\D+3)\,\mathfrak{f}_{\D}(\chi)\,,
\end{align}
with respect to $\D$, which leads to
\begin{align}\label{CasimirODE1111derivatives}
\widehat{\mathcal{C}}\mathfrak{f}^{(n)}_{\D}(\chi)=\D(\D+3)\,\mathfrak{f}^{(n)}_{\D}(\chi)-n (2+2\D+\chi)\,\mathfrak{f}^{(n-1)}_{\D}(\chi)+2n\chi(1-\chi)\partial_{\chi}\mathfrak{f}^{(n-1)}_{\D}(\chi)-n(n-1)\chi \mathfrak{f}^{(n-2)}_{\D}(\chi)\,.
\end{align}
In particular, this allows to derive a general result for the first subleading logarithmic singularities ($k=\ell-1$) at all orders, which have an OPE (for $\ell\ge 1$)
\begin{align}\label{subleadinglogOPE}
f^{(\ell)}_{\log^{\ell-1}}(\chi)=\sum_{\D}\frac{(\gamma^{(1)}_{\D})^{\ell-2}}{(\ell-1)!}\left(\langle a^{(1)}\gamma^{(1)}+(\ell-1)a^{(0)}\gamma^{(2)}\rangle_{\D} \mathfrak{f}_{\D}(\chi)+\langle a^{(0)}(\gamma^{(1)})^2\rangle_{\D} \mathfrak{f}^{(1)}_{\D}(\chi)\right)\,.
\end{align}
Massaging this expression, with a suitable use of \eqref{CasimirODE1111derivatives} (for $n=1$) as well as $\gamma^{(1)}_{\D}=-\JJ/2$, one can derive
\begin{align}\label{subleadinglog}
\begin{split}
f^{(\ell)}_{\log^{\ell-1}}(\chi)=-\frac{1}{2(\ell-1)}\widehat{\mathcal{C}}\,f^{(\ell-1)}_{\log^{\ell-2}}(\chi)+\frac{1}{(\ell-1)!}\left(-\frac{1}{2}\right)^{\ell}\big[4\,\widehat{\mathcal{C}}^{\ell-2} f^{(2)}_{\tilde{\gamma},\log^1}(\chi)+(1-\chi)(2\chi\partial_{\chi}+1)\widehat{\mathcal{C}}^{\ell-1}f^{(0)}_L(\chi)\big]\,,
\end{split}
\end{align}
where
\begin{align}
f^{(2)}_{\tilde{\gamma},\log^1}(\chi)=\sum_{\D}\langle a^{(0)}(\gamma^{(2)}-\gamma^{(1)}\partial_{\D}\gamma^{(1)})\rangle_{\D}\,\mathfrak{f}_{\D}(\chi)\,,
\end{align}
and $\gamma^{(2)}-\gamma^{(1)}\partial_{\D}\gamma^{(1)}$ is the combination of anomalous dimensions that is a function of $\JJ$ at one loop, see Appendix \ref{app:perturbativeOPE}.  Remarkably,  it can be summed exactly,  and we find
\begin{align}
f^{(2)}_{\tilde{\gamma},\log^1}(\chi)=\frac{(2-\chi)\chi(1-\chi+6\chi^2)}{4(1-\chi)^2}+\frac{(1-\chi+\chi^2)^2}{2(1-\chi)^2}\log(1-\chi)\,.
\end{align}
Note that \eqref{subleadinglog} should be seen as a recursion relation that holds for $\ell\ge 2$, the starting point of the relation being 
\begin{align}
f^{(1)}_{\log^0}(\chi)=\frac{(2-\chi)(-9+9\chi+\chi^2)}{(1-\chi)\chi}-\frac{(1-\chi)(3+\chi)(6-2\chi+\chi^2)}{\chi^2}\log(1-\chi)\,,
\end{align}
which is the case $\ell=1$ in \eqref{subleadinglogOPE}.

Finally, we would like to emphasize once again that the braiding symmetry \eqref{braiding1111} (or \eqref{braidingfbar}) has a strong impact on the OPE coefficients of the exchanged operators at all orders in perturbation theory. In particular, we have already observed how in the conformal blocks decomposition of the free theory correlator only operators with even $\D$ appear, which is a consequence of braiding symmetry. While this relation is obvious at infinite $\lambda$ due to the transformation property of conformal blocks, it might be less obvious at higher perturbative orders where derivatives of blocks appear. However, a basic observation is that if a three-point function $\langle \mathcal{D}_1\mathcal{D}_1\mathcal{O}_{\DOdd}\rangle$ was non-vanishing at a certain order $\bar{\ell}$, its first contribution to the OPE would be in the expansion of $f^{(\bar{\ell})}(\chi)$ through a term that is simply $a^{(\bar{\ell})}_{\DOdd}\,\mathfrak{f}_{\DOdd}(\chi)$, with no anomalous dimensions or derivatives. But since 
\begin{align}
\mathfrak{f}_{\DOdd}(\chi)-\mathfrak{f}_{\DOdd}\big(\tfrac{\chi}{\chi-1}\big)=0\,,
\end{align}
this would spoil the braiding symmetry of $f^{(\bar{\ell})}(\chi)$, which we have assumed to hold at all orders. This implies that if the OPE coefficients $\langle \mathcal{D}_1\mathcal{D}_1\mathcal{O}_{\DOdd}\rangle$ are non-vanishing, they must necessarily be non-perturbative from the point of view of the large $\lambda$ expansion. In fact, as shown in \cite{Cavaglia:2023mmu} they actually vanish non-perturbatively (in the planar theory).

\section{$\langle \mathcal{D}_p\mathcal{D}_p\mathcal{D}_q\mathcal{D}_q \rangle$ correlators to one loop}\label{sec:ppqq}

Let us now switch to four-point functions between more general half-BPS operators $\mathcal{D}_k$. For simplicity, rather than considering the most general case with four arbitrary weights, we take them to be pairwise equal. This simplifies the bootstrap analysis and still allows us to draw certain crucial conclusions on the structure of the dilatation operator in perturbation theory. However, let us stress that with the same methods used here one could in principle bootstrap $\langle \mathcal{D}_{k_1}\mathcal{D}_{k_2}\mathcal{D}_{k_3}\mathcal{D}_{k_4} \rangle$ correlators as well. One initial reason to study this family of four-point functions is that it naively seems to be the natural analogue of holographic correlators between half-BPS operators in higher dimensions, which in that context are used to attack the mixing problem at first order \cite{Alday:2017xua,Aprile:2017bgs,Aprile:2017xsp,Alday:2020tgi,Abl:2021mxo,Alday:2021ymb,Alday:2021ajh,Behan:2022uqr}. However, here we have a crucial difference: while in the most common AdS/CFT setups one has an internal space and the half-BPS operators of the CFT as Kaluza-Klein (KK) modes of some fundamental supergravity multiplet on the internal space, this is not the case in our model. In particular, while KK modes are independent fields, the half-BPS operators on the Wilson line are simply defined as suitable normal-ordered products of the displacement operator and are therefore not independent. This implies that their correlation functions can be obtained from those of $\mathcal{D}_1$, thus providing no additional information. This is what leads us to consider even more general four-point functions in Section \ref{sec:mixing}, where having external long multiplets turns out to be key for the solution of the mixing problem.

In this section we take a bootstrap approach to the study of $\langle \mathcal{D}_p\mathcal{D}_p\mathcal{D}_q\mathcal{D}_q \rangle$ four-point functions, up to one loop, finding perfect agreement with the considerations above. We start from the free theory, discussing some aspects of the OPE and its relation with the braiding symmetry already observed for $\langle \mathcal{D}_1\mathcal{D}_1\mathcal{D}_1\mathcal{D}_1 \rangle$, as well as reviewing the structure of the protected topological sector present in the correlators on the Wilson line. We then proceed to the actual bootstrap, where our results clearly show that at tree level the correlators are fixed by the four-point function of $\mathcal{D}_1$, while at one loop one also needs the six-point function. This is related to the fact that in \eqref{AdSlagrangian} only quartic vertices contribute at $\mathcal{O}(\lambda^{-1/2})$, while sextic vertices come into the game at $\mathcal{O}(\lambda^{-1})$. 

Besides the explicit form of the four-point functions, one of the main results of our analysis is that it clearly shows how the tree-level anomalous dimensions of all operators are proportional to the superconformal Casimir eigenvalue of the associated $\mathfrak{osp}(4^*|4)$ representation, thus proving that the free theory degeneracy is not lifted at tree level -- a fact that was used repeatedly in the previous sections. On the other hand, at one loop these correlators do not provide enough information on the dilatation operator due to the reasons mentioned above, hence we will consider long multiplets in Section \ref{sec:mixing}.

\subsection{Free theory and topological sector}

At infinitely strong coupling the SCFT that we are describing becomes free, and its local operators are built by taking normal ordered products of the fundamental fields in the displacement multiplet $\mathcal{D}_1$,  as described in \cite{Ferrero:2023znz}.  Correlation functions are then defined and computed by Wick contractions using the two-point function of the $\mathcal{D}_1$. Here we present some explicit results in the free theory, highlighting some structures that are relevant for the mixing problem.

\begin{figure}[hbt!]
\begin{center}
        \includegraphics[width=0.5\textwidth]{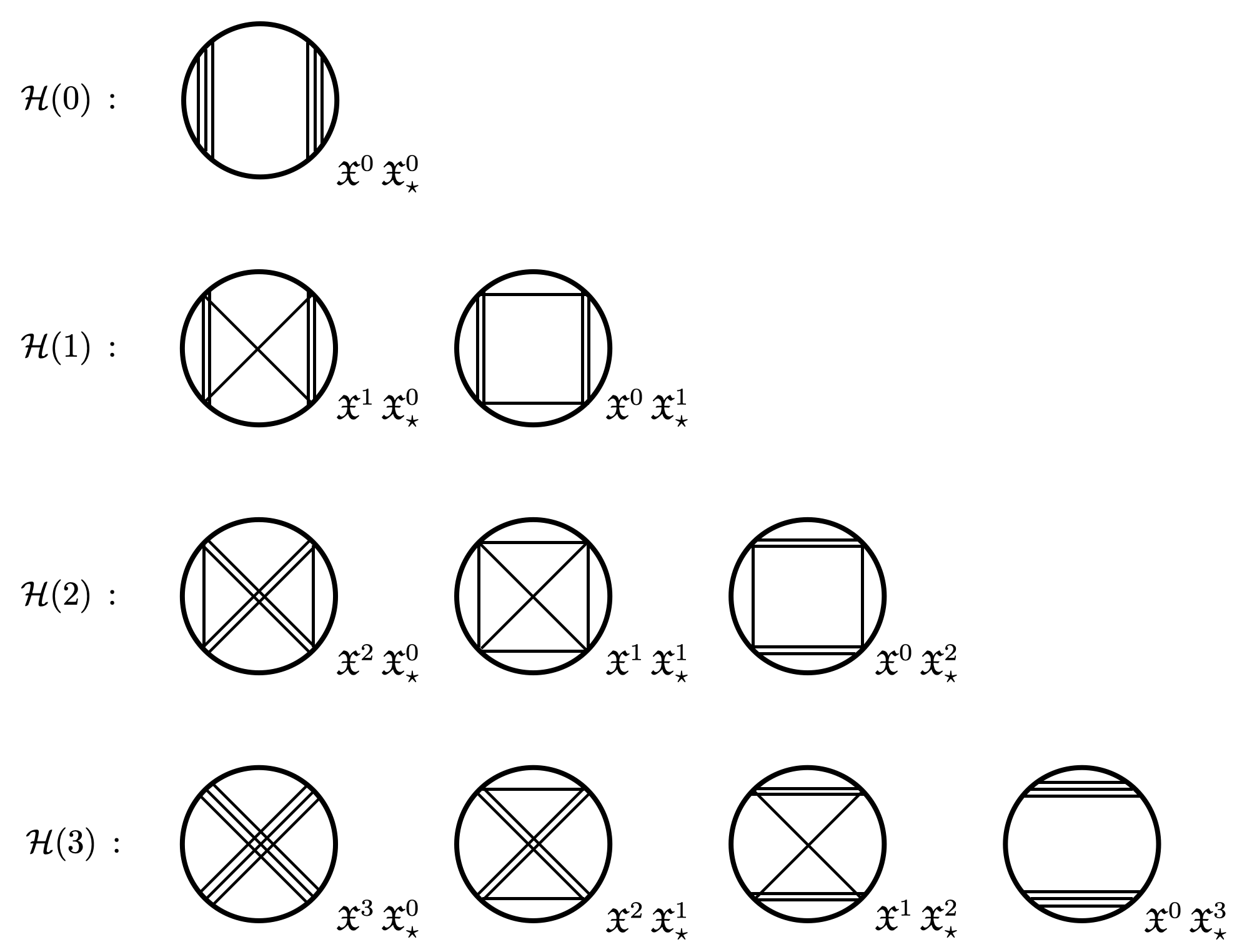}
\caption{Diagrams contributing to the free $\langle \mathcal{D}_3\mathcal{D}_3\mathcal{D}_3\mathcal{D}_3\rangle$ correlator, grouped by their contribution to each higher-spin conformal block $\mathcal{H}(n)$ and labelled with the powers of  $\sX$ and $\sXs$ that they contribute with to the expression of $\mathcal{G}^{(0)}_{\{3,3,3,3\}}$. It is understood that lines that cross each other do not correspond to vertices, since we are dealing with a free theory. Although the labels are omitted, our conventions (to match with the discussion in the text) is that the first point is located at the top left and the order increases anti-clockwise.}
\label{fig:3333diagrams}
\end{center}
\end{figure}

The kinematics and superconformal blocks for four-point functions of arbitrary half-BPS operators were studied in \cite{Liendo:2018ukf,Ferrero:2023znz}, where it is shown that
\begin{align}
\langle \mathcal{D}_{k_1}\mathcal{D}_{k_2}\mathcal{D}_{k_3}\mathcal{D}_{k_4} \rangle=K_{\{k_1,k_2,k_3,k_4\}}\,\times \,\mathcal{G}_{\{k_1,k_2,k_3,k_4\}}(\chi,\zeta_1,\zeta_2)\,,
\end{align}
where the prefactor $K$ accounts for the superconformal weights and $\mathcal{G}$ is a function of the 1d cross ratio $\chi$ and the two R symmetry cross ratios $\zeta_{1,2}$, which is constrained by the SCWI
\begin{align}\label{SCWI}
\left.(\partial_{\zeta_1}+\tfrac{1}{2}\partial_{\chi})\,\mathcal{G}\,\right|_{\zeta_1=\chi}=0=\left.(\partial_{\zeta_2}+\tfrac{1}{2}\partial_{\chi})\,\mathcal{G}\,\right|_{\zeta_2=\chi}\,.
\end{align}
To describe the free theory, it is useful to introduce the superconformal cross ratios \cite{Liendo:2018ukf}
\begin{align}
\sX=\frac{\chi^2}{\zeta_1\,\zeta_2}\,, \qquad
\sXt=\frac{(1-\chi)^2}{(1-\zeta_1)\,(1-\zeta_2)}\,, \qquad
\sXs=\frac{\sX}{\sXt}\,,
\end{align}
each of them satisfying the SCWI independently.  Moreover,  note that they are mapped into each other by braiding and cyclic transformations:
\begin{align}
\begin{split}
\text{braiding (1 $\leftrightarrow$ 2)}&: \quad (\chi,\zeta_1,\zeta_2)\rightarrow \left(\frac{\chi}{\chi-1},\frac{\zeta_1}{\zeta_1-1},\frac{\zeta_2}{\zeta_2-1}\right)\quad \Rightarrow \quad \sX \rightarrow \sXs\,,\\
\text{cyclic ($i \rightarrow i+1$)}&: \quad (\chi,\zeta_1,\zeta_2)\rightarrow \left(1-\chi,1-\zeta_1,1-\zeta_2\right)\quad \Rightarrow \quad \sX \rightarrow \sXt\,,
\end{split}
\end{align}
$\langle \mathcal{D}_p\mathcal{D}_p\mathcal{D}_q\mathcal{D}_q \rangle$ correlators in the free theory are symmetric polynomials in $\sX$ and $\sXs$,  and are conveniently expressed in terms of so-called higher-spin conformal blocks \cite{Alday:2016njk}
\begin{align}\label{higherspinblocks}
\mathcal{H}_n(x,y)=x^n\,  _2F_1(-n,-n,1;y/x)=\sum_{k=0}^n \binom{n}{k}^2\,x^{n-k}\,y^k\,,
\end{align}
which are clearly symmetric in $x$ and $y$.  In terms of these,  we have
\begin{align}\label{allfreeppqq}
\begin{split}
\mathcal{G}_{\{p,p,q,q\}}^{(0)}(\chi,\zeta_1,\zeta_2)&=\sum_{n=0}^{\text{min}(p,q)}\binom{p}{n}\binom{q}{n}\,\mathcal{H}_n(\sX,\sXs)\\
&=F_4(-p,-q;1,1;\sX,\sXs)\,,
\end{split}
\end{align}
where $F_4$ is an Appell function, defined by
\begin{align}
F_4(a,b;c_1,c_2;x,y)=\sum_{m,n=0}^{\infty}\frac{(a)_{m+n}\,(b)_{m+n}}{(c_1)_m\,(c_2)_n\,m!\,n!}x^m\,y^n\,,
\end{align}
and note that for the case at hand the sum truncates at $m=n=\text{min}(p,q)$. For each fixed $p\le q$, free theory four-point functions are polynomials in $q$ of degree $p$. In the above each $\mathcal{H}(n)$ corresponds to the contribution of all graphs with $\text{min}(p,q)-n$ propagators connecting points 1 and 2, where each power of $\sX$ ($\sXs$) corresponds to a propagator between 1 and 3 (1 and 4). The $\mathcal{H}(n)$ are normalized in such a way that the terms $\sX^n$ and $\sXs^n$ appear with unit coefficient, and the relative combinatorics of the other terms can be easily worked out by considering $\langle \mathcal{D}_n\mathcal{D}_n\mathcal{D}_n\mathcal{D}_n\rangle$ correlators, where $\mathcal{H}(n)$ contributes with unit coefficient: there, a power of $\sXs^k$ has multiplicity $\binom{n}{k}$ corresponding to the number of way to connect points 1 and 4 with $k$ lines, with the remaining $n-k$ connecting 1 and 3 (since in the $\mathcal{H}(n)$ contribution to $\langle nnnn\rangle$ there are no lines between 1 and 2). Similarly, $\sX^{n-k}$ has a combinatorial factor of $\binom{n}{n-k}=\binom{n}{k}$, hence the squared binomial coefficients in \eqref{higherspinblocks}. To work out the coefficient in front of each $\mathcal{H}(n)$ in $\langle \mathcal{D}_p\mathcal{D}_p\mathcal{D}_q\mathcal{D}_q\rangle$, one can then take the graph representing the contribution of $\mathcal{H}(n)$ to $\langle  \mathcal{D}_n\mathcal{D}_n\mathcal{D}_n\mathcal{D}_n\rangle$ and dress it with the required number of lines, which gives a further factor of $\binom{p}{n}\binom{q}{n}$. As a simple example, in figure \ref{fig:3333diagrams} we have drawn the diagrams contributing to each $\mathcal{H}(n)$ for $0\le n\le 3$ in the correlator $\langle  \mathcal{D}_3\mathcal{D}_3\mathcal{D}_3\mathcal{D}_3\rangle$, as well as the monomial in $\sX$ and $\sXs$ correspond to each graph.

A crucial fact about the free theory that we are discussing is that, as anticipated, it enjoys an exact braiding symmetry
\begin{align}\label{freetheorybraiding}
\mathcal{G}^{(0)}_{\{p,p,q,q\}}(\chi,\zeta_1,\zeta_2)=
\mathcal{G}^{(0)}_{\{p,p,q,q\}}\big(\tfrac{\chi}{\chi-1},\tfrac{\zeta_1}{\zeta_1-1},\tfrac{\zeta_2}{\zeta_2-1}\big)\,,
\end{align}
which is manifest in \eqref{allfreeppqq} since the higher-spin blocks \eqref{higherspinblocks} satisfy
\begin{align}
\mathcal{H}_n(x,y)=\mathcal{H}_n(y,x)\,.
\end{align}
Moreover,  combining cyclic and braiding transformations we can define
\begin{align}
\begin{split}
\mathcal{G}^{(0)}_{\{p,q,q,p\}}(\chi,\zeta_1,\zeta_2)&=\sXs^{\frac{p+q}{2}}\,\sXt^{\frac{p-q}{2}}\,\mathcal{G}^{(0)}_{\{p,p,q,q\}}(1-\chi,1-\zeta_1,1-\zeta_2)\,,\\
\mathcal{G}^{(0)}_{\{p,q,p,q\}}(\chi,\zeta_1,\zeta_2)&=\sX^{\frac{p+q}{2}}\,\mathcal{G}^{(0)}_{\{p,p,q,q\}}(\chi^{-1},\zeta_1^{-1},\zeta_2^{-1})\,,
\end{split}
\end{align}
both of which become symmetries of \eqref{allfreeppqq} in the case $p=q$.

Let us now discuss the superconformal blocks expansion of the free-theory correlators \eqref{allfreeppqq}.  In the absence of semi-short representations,  which is the case at strong coupling (see \cite{Ferrero:2023znz} for a proof),  the OPE selection rules for half-BPS operators read \cite{Liendo:2018ukf}\footnote{It is also useful to recall the $\mathfrak{sp}(4)$ selection rule
\begin{align}
[0,p]\,\otimes [0,q]=\bigoplus_{i=0}^{\text{min}(p,q)}\bigoplus_{j=0}^{i}\,[2i-2j,\,2j+|p-q|]\,.
\end{align}}
\begin{align}\label{OPEpq}
\mathcal{D}_p \times \mathcal{D}_q=\mathcal{I}\,\delta_{p,q}+\sum_{k=|p-q|, \,\,\text{step 2}}^{p+q}\mathcal{D}_k+\sum_{i=0}^{\text{min}(p-1,q-1)}\sum_{j=0}^i\sum_{\Delta}\mathcal{L}^{\Delta}_{0,\left[2i-2j,2j+|p-q|\right]}\,,
\end{align}
where $\mathcal{I}$ is the identity operator,  so that the blocks expansion in the direct channel reads
\begin{align}\label{freeblockexpansion}
\begin{split}
\mathcal{G}_{\{p,p,q,q\}}^{(0)}(\chi,\zeta_1,\zeta_2)=
1+\sum_{\substack{k=2\\ \text{step}\,\,2}}^{2\,\text{min}(p,q)}&\mathsf{C}^{(0)}_{ppk}\mathsf{C}^{(0)}_{qqk}\,\mathfrak{G}_{\mathcal{D}_k}(\chi,\zeta_1,\zeta_2)
\\&+\sum_{\substack{a,b=0,\,\,\text{step}\,\,2\\
a+b\le 2\,\text{min}(p,q)-2}}^{2\,\text{min}(p,q)-2}\langle a^{(0)}\rangle^{(p,q)}_{\mathcal{L}^{\Delta}_{0,[a,b]}}\,\mathfrak{G}_{\Delta,[a,b]}(\chi,\zeta_1,\zeta_2)\,,
\end{split}
\end{align}
where $\mathfrak{G}$ are superconformal blocks,  whose explicit expression can be found \cite{Ferrero:2023znz}.  In \eqref{freeblockexpansion},  $\mathsf{C}^{(0)}_{pqr}$ are OPE coefficients between half-BPS operators,  whose expression we shall discuss below,  while $\langle a^{(0)}\rangle^{(p,q)}_{\mathcal{L}^{\Delta}_{0,[a,b]}}$ are averaged products of OPE coefficients,  defined by\footnote{Here we are expressing the result in terms of a basis of normalized eigenvalues of the dilatation operator, but recall that one can pick any basis of operators to compute the average.}
\begin{align}
\langle a^{(0)}\rangle^{(p,q)}_{\mathcal{L}^{\Delta}_{0,[a,b]}}=\sum_{\mathcal{O}\,\text{s.t.}\,\omega_{\mathcal{O}}=\{\Delta,0,[a,b]\}}\mu^{(0)}_{pp\mathcal{O}}\,\mu^{(0)}_{qq\mathcal{O}}\,.
\end{align}
The braiding symmetry \eqref{freetheorybraiding} of the free theory is closely related to the type of operators that are exchanged in the OPE. We have already seen this for $\langle \mathcal{D}_1\mathcal{D}_1\mathcal{D}_1\mathcal{D}_1\rangle$, where the long operators exchanged in the OPE have {\it even} conformal dimensions in the free theory, which makes all the superconformal blocks appearing in the expansion of the correlator symmetric under braiding. A similar phenomenon is responsible for the braiding symmetry of $\langle \mathcal{D}_p\mathcal{D}_p\mathcal{D}_q\mathcal{D}_q\rangle$, although with a slight difference. In particular, short superconformal blocks $\mathfrak{G}_{\mathcal{D}_k}$ appearing in \eqref{freeblockexpansion} are all symmetric under braiding for even $k$ (which is always the case in \eqref{freeblockexpansion}), while the long superblocks with even $a$ and $b$ (which again is the case relevant for \eqref{freeblockexpansion}) satisfy
\begin{align}\label{superblockbraiding}
\mathfrak{G}_{\Delta,[a,b]}(\chi,\zeta_1,\zeta_2)=(-1)^{\Delta+\frac{3a}{2}+b}\,\mathfrak{G}_{\Delta,[a,b]}\big(\tfrac{\chi}{\chi-1},\tfrac{\zeta_1}{\zeta_1-1},\tfrac{\zeta_2}{\zeta_2-1}\big)\,,
\end{align}
and the sum over $\Delta$ in \eqref{freeblockexpansion} runs only over
\begin{align}\label{Deltastrong[a,b]}
\Delta=\frac{3a}{2}+b+2n\,, \qquad (n\ge 1)\,,
\end{align}
in such a way that braiding is always a symmetry.  The fact that only these operators appear in the OPE \eqref{OPEpq} at strong coupling has been recently understood in \cite{Cavaglia:2023mmu} in terms of a parity that flips the orientation of the Wilson line, which can be defined in terms of the weak coupling description of the theory.

Finally,  let us note that an efficient way of computing the averaged products of OPE coefficients $\langle a^{(0)}_{\Delta}\rangle_{(p,q)}$ is to expand the individual higher-spin blocks \eqref{higherspinblocks} as a sum over superconformal blocks. Since as previously discussed (see figure \ref{fig:3333diagrams}) each $\mathcal{H}(n)$ corresponds to a diagram where points 1 and 2 are connected by exactly $p-n$ propagators, so that $2n$ propagators are connecting one of points 1 and 2 to one of points 3 and 4, then the contribution of each $\mathcal{H}(n)$ in the OPE corresponds to that of exchanged operators of length $L=2n$. This can be also argued from the fact that the OPE $\mathcal{D}_p\times\mathcal{D}_p$ contains operators with length $L\le 2p$ and in \eqref{allfreeppqq} only $\mathcal{H}(n)$ with $n\le p$ contribute. We can expand each $\mathcal{H}(n)$ using
\begin{align}
\begin{split}
\mathcal{H}_n(\sX,\sXs)&=\binom{2n}{n}\,\mathfrak{G}_{\mathfrak{D}_n}(\chi,\zeta_1,\zeta_2)+\sum_{\substack{a,b=0\,\,\text{step}\,\,2\\
a+b\le 2n-2}}^{2n-2} h^{[a,b]}_{\Delta}(n)\,\mathfrak{G}_{\Delta,[a,b]}(\chi,\zeta_1,\zeta_2)\,,
\end{split}
\end{align}
so that from the $h^{[a,b]}_{\Delta}(n)$ we obtain
\begin{align}\label{a0fromh[n]}
\langle a^{(0)}\rangle^{(p,q)}_{\mathcal{L}^{\Delta}_{0,[a,b]}}=\sum_{n=1}^p \binom{p}{n}\binom{q}{n}h^{[a,b]}_{\Delta}(n)\,.
\end{align}
This is convenient because the individual higher-spin blocks are simpler than the full free theory correlator,  and the coefficients $h^{[a,b]}_{\Delta}(n)$ admit a relatively simple closed-form expression for each $a$ and $b$ of interest. For example,  for the singlet representation $[a,b]=[0,0]$ we find
\begin{align}
\begin{split}
h_{\Delta}^{[0,0]}(n)=&
\frac{3\pi\,2^{4n-2(1+\Delta)}\,\Gamma\left[n+\tfrac{3+\Delta}{2}\right]\,\Gamma\left[\tfrac{\Delta}{2}\right]\,\Gamma[3+\Delta]}{(\Delta+4)\,\Gamma[3+2n]\,\Gamma\left[1-n+\tfrac{\Delta}{2}\right]\,\Gamma\left[\tfrac{3}{2}+\Delta\right]\,\Gamma\left[\tfrac{5+\Delta}{2}\right]}    \\
&\Big[4\,(n+1)\,\, _3\tilde{F}_2\left(-\tfrac{3+\Delta }{2},\tfrac{\Delta }{2},-n;\tfrac{1}{2}-n,n;1\right)\\
&-n\,\Delta(\Delta+3)\,\, _3\tilde{F}_2\left(-\tfrac{1+\Delta }{2},1+\tfrac{\Delta }{2},1-n;\tfrac{3}{2}-n,n+1;1\right)\Big]\,,
\end{split}
\end{align}
where $_3\tilde{F}_2$ is a regularized hypergeometric function.  Similar expressions in terms of $_3\tilde{F}_2$ are possible for all representations $[a,b]$,  but we have not been able to find a closed form in terms of $a$ and $b$. 

The expression \eqref{a0fromh[n]} is related to an important piece of information about the structure of the OPE in the free theory, which ties in with the discussion of the spectrum of the free theory of \cite{Ferrero:2023znz} as well as with the mixing problem, which we consider in Section \ref{sec:mixing}. Recall that it is useful to label the operators in the free theory by their length $L$, related by the index $n$ in \eqref{a0fromh[n]} by $L=2n$. As we discussed, long operators in the free theory fall into degeneracy spaces for fixed quantum numbers and we are free to pick a basis of operators with fixed length. Given a certain representation of $\mathfrak{osp}(4^*|4)$ and a length $L=2n$, one can pick an arbitrary basis of operators $\widehat{\mathcal{O}}_{L=2n,\alpha}$ to span the degeneracy space $\mathtt{d}_L$ (we are suppressing the $\mathfrak{osp}(4^*|4)$ quantum numbers since the present discussion is independent of those). Then, the set of OPE coefficients between two half-BPS operators $\mathcal{D}_p$ and one of the $\widehat{\mathcal{O}}_{L=2n,\alpha}$ can be thought of a vector in the space $\mathtt{d}_L$:
\begin{align}\label{ppO}
\langle \mathcal{D}_p\mathcal{D}_p\widehat{\mathcal{O}}_{L=2n,\alpha}\rangle^{(0)} \sim \mu^{(0)}_{pp\widehat{\mathcal{O}}_{L=2n,\alpha}}\,,\quad \alpha=1,\ldots,\text{dim}(\mathtt{d}_L)\,.
\end{align}
For any such operator, one can simply argue from Wick contractions that
\begin{align}\label{C_pplong_Wickproperty}
\binom{p}{n}\,\mu^{(0)}_{qq\widehat{\mathcal{O}}_{L=2n,\alpha}}
=\binom{q}{n}\,\mu^{(0)}_{pp\widehat{\mathcal{O}}_{L=2n,\alpha}}\,,
\end{align}
for all $\alpha$, that is the two are parallel vectors in $\mathtt{d}_L$. In particular, one has
\begin{align}
\mu^{(0)}_{pp\widehat{\mathcal{O}}_{L=2n,\alpha}}=\binom{p}{n}\,\mu^{(0)}_{nn\widehat{\mathcal{O}}_{L=2n,\alpha}}\,.
\end{align}
The upshot is that while one might expect that varying the first two operators in \eqref{ppO} allows to span independent directions in $\mathtt{d}_L$, this turns out not to be the case, due to the fact that the half-BPS operators $\mathcal{D}_p$ are obtained from normal-ordered products of $\mathcal{D}_1$, in contrast with more familiar holographic theories where half-BPS operators dual to KK modes are all independent of each other. This is particularly important for us, as one of the main results of this paper is the study of the dilatation operator at one loop for long operators in the singlet representation of $\mathfrak{sp}(4)\oplus \mathfrak{su}(2)$, which only appear in $\mathcal{D}_p\times \mathcal{D}_q$ when $q=p$. Hence, this is telling us something that we have already anticipated in the introduction: to explore the degeneracy spaces, and then probe all components of the dilatation operator, we need to consider more general correlators than $\langle \mathcal{D}_p\mathcal{D}_p\mathcal{D}_q\mathcal{D}_q\rangle$. Going back to eq. \eqref{a0fromh[n]}, dropping the labels for the $\mathfrak{osp}(4^*|4)$ representation of the exchanged operators and assuming as usual $p\le q$ we can write
\begin{align}\label{a0fromCppnCqqn}
\langle a^{(0)} \rangle^{(p,q)}=\sum_{n=1}^{p}\,\,\sum_{\alpha,\beta}\mu^{(0)}_{pp\widehat{\mathcal{O}}_{L=2n,\alpha}}\,(\mathtt{g}^{-1})^{\alpha\beta}\,\mu^{(0)}_{pp\widehat{\mathcal{O}}_{L=2n,\beta}}\,,
\end{align}
where we are highlighting the fact that the average $\langle a^{(0)} \rangle$ can be expressed as a sum of averages over degeneracy spaces of operators with fixed length $L=2n$. Using \eqref{C_pplong_Wickproperty}, we can rewrite the above as
\begin{align}
\langle a^{(0)} \rangle^{(p,q)}=\sum_{n=1}^{p}\,\binom{p}{n}\binom{q}{n}\,\sum_{\alpha,\beta}\mu^{(0)}_{nn\widehat{\mathcal{O}}_{L=2n,\alpha}}\,(\mathtt{g}^{-1})^{\alpha\beta}\,\mu^{(0)}_{nn\widehat{\mathcal{O}}_{L=2n,\beta}}\equiv \sum_{n=1}^{p}\,\binom{p}{n}\binom{q}{n}\,\langle a^{(0)}_{nn\widehat{\mathcal{O}}_{L=2n}}\rangle\,,
\end{align}
where we have introduced the average $\langle a^{(0)}_{nn\widehat{\mathcal{O}}_{L=2n}}\rangle$, which by definition only depends on $n$. Comparing with \eqref{a0fromh[n]} then immediately gives
\begin{align}\label{avg_nnOn}
\langle a^{(0)}_{nn\widehat{\mathcal{O}}_{L=2n}}\rangle=h^{[a,b]}_{\Delta}(n)\,,
\end{align}
where all the $\widehat{\mathcal{O}}_{L=2n,\alpha}$ are taken to have dimension $\Delta$ and $\mathfrak{sp}(4)$ representation $[a,b]$. Given that for all $p$ the coefficients $\mu^{(0)}_{pp\widehat{\mathcal{O}}_{L,\alpha}}$ are parallel vectors, for every fixed $L$ one could choose a basis in $\mathtt{d}_L$ where one of the basis elements is along that vector, while the others are orthogonal (thus have zero ope coefficient with $\mathcal{D}_p\times \mathcal{D}_p$). In that basis, one could even drop the average symbol from \eqref{avg_nnOn} and the only non-vanishing OPE coefficient is precisely $\left[h^{[a,b]}_{\Delta}(n)\right]^{1/2}$.

Let us now turn to the discussion of the topological sector of the theory,  whose existence follows directly from the SCWI \eqref{SCWI}.  As discussed in \cite{Liendo:2018ukf},  the four-point functions $\langle \mathcal{D}_p\mathcal{D}_p\mathcal{D}_q\mathcal{D}_q\rangle$ become topological when one sets $\chi=\zeta_1=\zeta_2$ ({\it i.e.} they no longer depend on the position of operators),  and the only contribution to the four-point function in this configuration comes from OPE coefficients of short operators,  as it can be argued from the fact that superconformal blocks satisfy 
\begin{align}
\mathfrak{G}_{\mathfrak{D}_k}(\chi,\chi,\chi)=1\,, \quad
\mathfrak{G}_{\Delta,[a,b]}(\chi,\chi,\chi)=0\,,
\end{align}
so that at all orders in perturbation theory
\begin{align}
\mathcal{G}_{\{p,p,q,q\}}(\chi,\chi,\chi)=1+\sum_{k=2,\,\,\text{step 2}}\mathsf{C}_{ppk}\mathsf{C}_{qqk}\,.
\end{align}
Remarkably,  the OPE coefficients $\mathsf{C}_{k_1k_2 k_3}$ can be computed exactly using supersymmetric localization \cite{Pestun:2007rz,  Giombi:2009ds,  Giombi:2018qox},  to all orders in $\lambda$ (including non-perturbative corrections).  We refer to section 3 of \cite{Liendo:2018ukf} for more details on how to extract the OPE coefficients from the localization result,  while here we shall limit to give the expression for the expansion of $\mathsf{C}_{k_1 k_2 k_3}$ at large $\lambda$ (neglecting non-analytic $e^{-\sqrt{\lambda}}$ terms).  The first few terms in the expansion read
\begin{align}\label{shortOPE}
\mathsf{C}_{k_1 k_2 k_3} = \mathsf{C}^{(0)}_{k_1 k_2 k_3}\left(1+\tfrac{1}{\lambda^{1/2}}\, \delta \mathsf{C}^{(1)}_{k_1 k_2 k_3}+\tfrac{1}{\lambda}\, \delta \mathsf{C}^{(2)}_{k_1 k_2 k_3}+ \mathcal{O}(\lambda^{-3/2}) \right)+\mathcal{O}(e^{-\sqrt{\lambda}})\,,
\end{align}
where $\mathsf{C}^{(0)}_{k_1 k_2 k_3}$ is the free theory result
\begin{align}
\mathsf{C}^{(0)}_{k_1 k_2 k_3}=\frac{\alpha_1!\,\alpha_2!\,\alpha_3!}{\sqrt{k_1!\,k_2!\,k_3!}}\,,
\end{align}
and 
\begin{align}
\alpha_1=\frac{-k_1+k_2+k_3}{2}\,.	\quad
\alpha_2=\frac{k_1-k_2+k_3}{2}\,,\quad
\alpha_3=\frac{k_1+k_2-k_3}{2}\,,
\end{align}
while the first two perturbative corrections are
\begin{align}
\begin{split}
\delta \mathsf{C}^{(1)}_{k_1 k_2 k_3}=&-\frac{3}{4}\,(\alpha_1\,\alpha_2+\alpha_1\,\alpha_3+\alpha_2\,\alpha_3)\,,\\
\delta \mathsf{C}^{(2)}_{k_1 k_2 k_3}=&\frac{1}{4}\,\delta \mathsf{C}^{(1)}_{k_1 k_2 k_3}+\frac{1}{2}\,\left(\delta \mathsf{C}^{(1)}_{k_1 k_2 k_3}\right)^2\\
&-\frac{3}{16}\left[3\,\alpha_1\,\alpha_2\,\alpha_3+\alpha_1\,\alpha_2\,\alpha_3\,(\alpha_1+\alpha_2+\alpha_3)\right]\,.
\end{split}
\end{align}
For the special case of $\mathsf{C}_{112}$,  \eqref{f_fromW} gives a result valid to all orders, where $\mathsf{f}=1+(\mathsf{C}_{112})^2$ and its perturbative expansion at large $\lambda$ is given in \eqref{f1111_3loops}.

\subsection{Bootstrapping $\langle \mathcal{D}_p\mathcal{D}_p\mathcal{D}_q\mathcal{D}_q\rangle$}

In this section we explain how the strategy outlined in Section \ref{sec:bootstrap} can be adapted to the study of $\langle \mathcal{D}_p\mathcal{D}_p\mathcal{D}_q\mathcal{D}_q\rangle$, which we then bootstrap at tree level and at one loop. The main difference with the correlators that we have discussed so far is that, for generic $p$ and $q$, we have to include the dependency on the R-symmetry cross ratios $\zeta_{1,2}$. We find it convenient to work in the case $q\ge p$ (from which $q<p$ can easily be obtained) and proceed by increasing the value of $p$ by one unit at the time. As we shall see, at each step the information of all correlators with $p$ smaller than the one under inspection will provide valuable information to fix all the coefficients in our ansatz. Taking into account of the R-symmetry dependence, at each order the ansatz reads
\begin{align}\label{ansatzppqq}
\mathcal{G}^{(\ell)}_{\{p,p,q,q\}}(\chi,\zeta_1,\zeta_2)=\sum_{i=1}^{2^{1+\ell}-1}\sum_{m,n=0}^{p}\frac{r_i^{(m,n)}(\chi)}{\zeta_1^{m}\,\zeta_2^{n}}\,\mathcal{T}_i(\chi)\,,
\end{align}
which is just a generalization of \eqref{genericansatz}. Here $r_i^{(m,n)}(\chi)$ are rational functions of $\chi$ and since $\langle \mathcal{D}_p\mathcal{D}_p\mathcal{D}_q\mathcal{D}_q\rangle$ is symmetric in $\zeta_1$ and $\zeta_2$, we take $r_i^{(m,n)}(\chi)=r_i^{(n,m)}(\chi)$. $\mathcal{T}_i(\chi)$ are the familiar HPLs and we have already specified the choice of a maximal transcendentality at each order, which in agreement with the observations of Section \ref{sec:1111} is $\mathtt{t_{max}}(\ell)=\ell$, so that for fixed $\ell$ the basis of HPLs has dimension $2^{1+\ell}-1$.  An important point to stress is that the spacetime and R-symmetry dependence of these correlators are not at all arbitrary, as the SCWI \eqref{SCWI} must be satisfied. This provides a fundamental constraint that allows to fix all free parameters in the ansatz, which was somehow implicit in the discussion of the $\langle \mathcal{D}_1\mathcal{D}_1\mathcal{D}_1\mathcal{D}_1\rangle$ correlator since in that case we were working directly in terms of the function that solves the SCWI automatically. 

Given the ansatz \eqref{ansatzppqq}, let us briefly discuss how the various constraints of Section \ref{sec:bootstrap} are adapted to the current setting. We observe here that the main modification arises from the fact that, as for the $\langle 112 \mathcal{L}\rangle$ correlators of Section \ref{sec:mixing}, we are no longer dealing with a four-point functions of identical operators and therefore we should consider constraints coming from the OPE both in the direct ($\chi\to 0$) and from the crossed ($\chi\to 1$) channel.
\begin{itemize}
\item {\it AdS unitarity method.} The set of constraints coming from AdS unitarity, namely the fact that one can compute the highest logarithmic singularities from CFT data at previous orders (up to resolving degeneracies), is essentially the same in this context but with a small caveat: the direct ($d$) and crossed ($c$) channel OPE are no longer related to each other by crossing symmetry, so one should consider a decomposition of the correlators analogous to \eqref{Gpowerslog}, both as an expansion in powers of $\log\chi$ and one in powers of $\log(1-\chi)$ (independently):
\begin{align}\label{logexpppqq}
\begin{split}
\mathcal{G}^{(\ell)}_{\{p,p,q,q\}}(\chi,\zeta_1,\zeta_2)&=\sum_{k=0}^{\ell}\mathcal{G}^{d,(k,\ell)}_{\{p,p,q,q\}}(\chi,\zeta_1,\zeta_2)\,\log^k\chi\\
&=\sum_{k=0}^{\ell}\mathcal{G}^{c,(k,\ell)}_{\{p,p,q,q\}}(\chi,\zeta_1,\zeta_2)\,\log^k(1-\chi)\,,
\end{split}
\end{align}
where $\mathcal{G}^{d,(k,\ell)}_{\{p,p,q,q\}}(\chi,\zeta_1,\zeta_2)$ are holomorphic at $\chi=0$ and $\mathcal{G}^{c,(k,\ell)}_{\{p,p,q,q\}}(\chi,\zeta_1,\zeta_2)$ are holomorphic at $\chi=1$. The statement of AdS unitarity is then that we can obtain all $\mathcal{G}^{d,(k,\ell)}_{\{p,p,q,q\}}(\chi,\zeta_1,\zeta_2)$ and $\chi=0$ and $\mathcal{G}^{c,(k,\ell)}_{\{p,p,q,q\}}(\chi,\zeta_1,\zeta_2)$ for $k\ge 2$ from CFT data at previous orders. We will not repeat the expansion for each such function over superconformal blocks, as it can be easily argued from \eqref{OPEpq} using the results of Appendix \ref{app:perturbativeOPE}. 
\item {\it Bose symmetry.} Again it is useful to focus on the two transformations that generate the Bose symmetry group, namely braiding and cyclic ones. Given the exact braiding symmetry \eqref{freetheorybraiding} of the free theory (due to the fact that we are considering pairwise equal operators), we can use braiding as a constraint at each order in perturbation theory precisely in the spirit of \eqref{braidinglogpieces}, since the presence of anomalous dimensions (related to powers of $\log\chi$ in the direct channel OPE) once again one breaks the symmetry in a ``weak'' sense. Hence, in full analogy with \eqref{braidinglogpieces}, we have (note that this is a statement about the OPE in the direct channel)
\begin{align}\label{braidinglogpieces}
\mathcal{G}^{d,(k,\ell)}_{\{p,p,q,q\}}(\chi,\zeta_1,\zeta_2)=\sum_{m=0}^{\ell-k}\binom{k+m}{m}(-1)^m\,\log^m(1-\chi)\,\mathcal{G}^{d,(k+m,\ell)}_{\{p,p,q,q\}}\big(\tfrac{\chi}{\chi-1},\tfrac{\zeta_1}{\zeta_1-1},\tfrac{\zeta_2}{\zeta_2-1}\big)\,.
\end{align}
On the other hand, since the four operators in $\langle \mathcal{D}_p\mathcal{D}_p\mathcal{D}_q\mathcal{D}_q\rangle$ are not all identical for $q\neq p$, cyclic transformations are no longer a symmetry, as one could expect. Instead, we can use these transformations to define the configuration $\langle \mathcal{D}_p\mathcal{D}_q\mathcal{D}_q\mathcal{D}_p\rangle$ via
\begin{align}\label{crossingppqq}
\mathcal{G}^{(\ell)}_{\{p,q,q,p\}}(\chi,\zeta_1,\zeta_2)=\sXs^{\frac{p+q}{2}}\,\sXt^{\frac{p-q}{2}}\,\mathcal{G}^{(\ell)}_{\{p,p,q,q\}}(1-\chi,1-\zeta_1,1-\zeta_2)\,,
\end{align}
whose direct channel OPE is completely equivalent to the crossed channel OPE for $\langle \mathcal{D}_p\mathcal{D}_p\mathcal{D}_q\mathcal{D}_q\rangle$. The superconformal blocks of \cite{Ferrero:2023znz} are in fact the correct ones for the expansion of $\langle \mathcal{D}_p\mathcal{D}_q\mathcal{D}_q\mathcal{D}_p\rangle$ for small $\chi$.
\item {\it Compatibility with the OPE.} The requirement of compatibility of our ansatz \eqref{ansatzppqq} with the OPE in {\it both} channels (direct and crossed) contains a richer structure than one could naively expect. Indeed, we have three types of constraints. The first two are analogous to what we had for $\langle \mathcal{D}_1\mathcal{D}_1\mathcal{D}_1\mathcal{D}_1\rangle$: expanding the ansatz \eqref{ansatzppqq} around $\chi=0$ ($\chi=1$), the OPE gives information about the smallest power of $\chi$ ($1-\chi$) that can appear in the expansion, since this is related to the dimension of the lightest exchanged operator. Moreover, we can also exploit the information coming from the knowledge of all OPE coefficients between short operators \eqref{shortOPE}, which again can be used as a constraint both in the direct and in the crossed channel. The third constraint, on the other hand, is new and deserves a more detailed explanation.

The spectrum of the free theory at strong coupling is such that certain operators are not degenerate, {\it i.e.} for certain given representations of $\mathfrak{osp}(4^*|4)$ there is only one operator, even in the free theory where all operators have integer (if bosonic) or half-integer (if fermionic) dimension. As it can be argued looking at the tables in \cite{Ferrero:2023znz},  this is the case for long superconformal primary operators with weights\footnote{Note that for $b=0$ there is actually no operator with $\Delta=\tfrac{3a}{2}+3$.}
\begin{align}
\omega=\{\Delta^*,0,[a,b]\}\,, \quad
\text{and}\quad \omega=\{\Delta^*+1,0,[a,b]\}\,, \qquad
\Delta^*=\tfrac{3a}{2}+b+2\,,
\end{align}
which in particular are the lightest long operators with spin zero and $\mathfrak{sp}(4)$ representation $[a,b]$ appearing in the spectrum of the theory at strong coupling. The importance of the existence of non-degenerate operators rests on the fact that their anomalous dimension can be computed from any arbitrary correlator where they appear in the OPE. On the contrary,  where degeneracy is present,  one should consider a family of correlators and diagonalize an anomalous dimension matrix to obtain the exact conformal dimension of each operator in the degenerate family.  We are going to exploit this fact working for one fixed value of $p$ at the time and increasing $p$ by one unit at each step, as follows.  Start from $p=1$: the OPE of $\langle \mathcal{D}_1\mathcal{D}_1\mathcal{D}_q\mathcal{D}_q \rangle$ in the crossed channel contains operators in the $\mathfrak{sp}(4)$ representation $[0,q-1]$ (see \eqref{OPEpq}),  and the two with lowest dimension ($\Delta=q+1$ and $\Delta=q+2$) are non-degenerate.  We can then extract their exact anomalous dimension from $\langle \mathcal{D}_1\mathcal{D}_1\mathcal{D}_q\mathcal{D}_q \rangle$ alone and use this information as a constraint on $\langle \mathcal{D}_2\mathcal{D}_2\mathcal{D}_q\mathcal{D}_q \rangle$, where the operators in representations of the type $[0,q]$ are exchanged and we now know the anomalous dimension of some of them. This, together with the previous constraints and the observations on the Regge limit that we will mention below, seems to be enough to obtain $\langle \mathcal{D}_2\mathcal{D}_2\mathcal{D}_q\mathcal{D}_q \rangle$ at all orders. From this one can extract the anomalous dimensions of new non-degenerate operators, which in turn allow to derive $\langle \mathcal{D}_3\mathcal{D}_3\mathcal{D}_q\mathcal{D}_q \rangle$. The procedure continues increasing $p$ at each step and using all the information about non-degenerate operators coming from correlators with lower values of $p$. Analyzing the OPE for a few $\langle \mathcal{D}_p\mathcal{D}_p\mathcal{D}_q\mathcal{D}_q \rangle$ correlators one can also guess a closed-form expression for the one-loop anomalous dimension of non-degenerate operators:
\begin{align}\label{gamma2nondegenerate}
\begin{split}
\gamma^{(2)}_{\mathcal{L}^{\Delta^*}_{0,[a,b]}}&=\frac{(a+2)}{192}(1180+723a+752b+107a^2+216ab+112b^2)\,,\\
\gamma^{(2)}_{\mathcal{L}^{\Delta^*+1}_{0,[a,b]}}&=\gamma^{(2)}_{\mathcal{L}^{\Delta^*}_{0,[a,b]}}+\frac{1}{96}(1088+692a+328b+105a^2+104ab-4b^2)\,.
\end{split}
\end{align}
\item {\it Regge limit.} To conclude, the observations on the Regge limit are completely analogous to those of Section \ref{sec:bootstrap}, in that we require that for each representation $[a,b]$ of $\mathfrak{sp}(4)$ the averaged anomalous dimensions at any fixed order have the mildest possible growth at large $\Delta$. Such growth turns out to be the same as that found for $\langle \mathcal{D}_1\mathcal{D}_1\mathcal{D}_1\mathcal{D}_1 \rangle$ in Section \ref{sec:1111} and in particular does not depend on $a$ and $b$:
\begin{align}
\langle \gamma^{(\ell)}\rangle^{(p,q)}_{\mathcal{L}^{\Delta}_{0,[a,b]}}\sim \Delta^{\ell+1}\,,\quad (\Delta\to\infty)\,.
\end{align}
\end{itemize}

In the next two sections we will briefly comment on how this method is applied to bootstrap tree-level and one-loop correlators, but we shall skip most of the technical details as they are analogous to what we already discussed in the previous sections. Instead, we will focus on the solution of the bootstrap problem, which has an interesting structure that can be described uniformly in terms of $p$ and $q$ and is related to the expression of the correlator as a sum of Witten diagrams arising from the Lagrangian \eqref{AdSlagrangian}.

\subsection{Tree level}

\begin{figure}[hbt!]
    \centering
    \begin{subfigure}[b]{0.20\textwidth}
        \includegraphics[width=\textwidth]{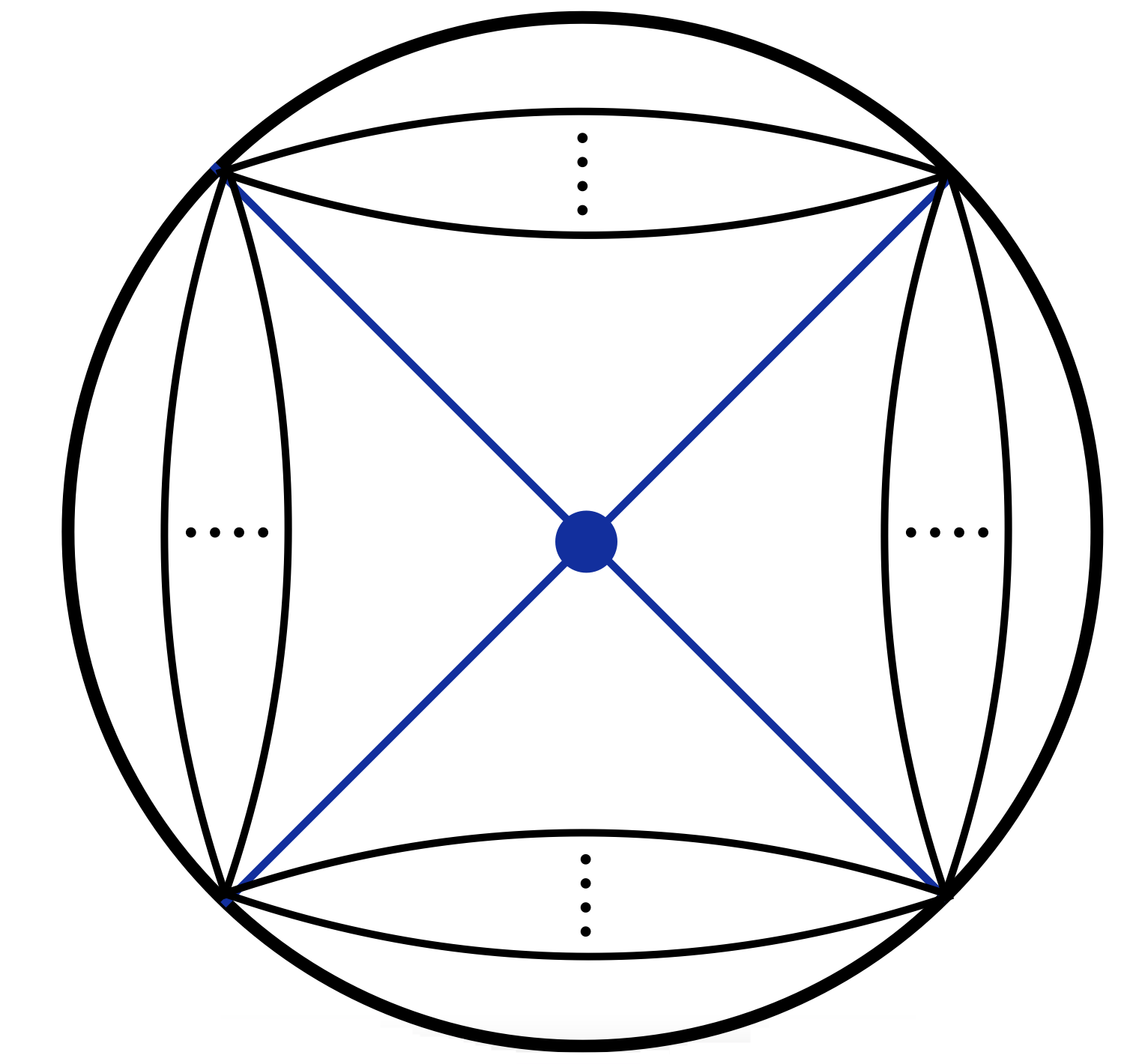}
        \caption{Diagrams contributing to the first term in \eqref{treeppqq}.}
        \label{fig:treediagramsppqqA}
    \end{subfigure}\qquad\qquad
    \begin{subfigure}[b]{0.20\textwidth}
        \includegraphics[width=\textwidth]{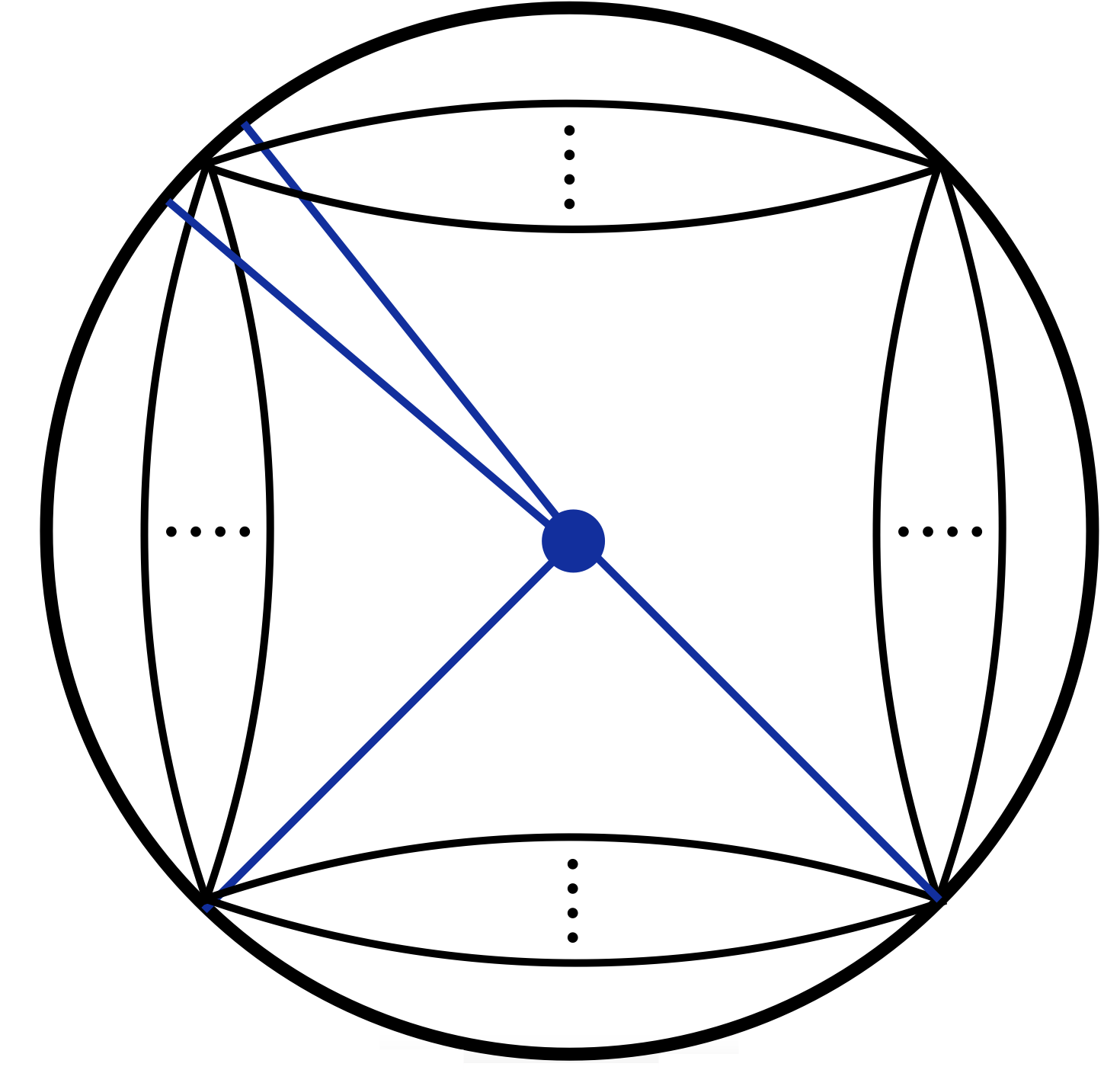}
        \caption{Diagrams contributing to the second term in \eqref{treeppqq}.}
        \label{fig:treediagramsppqqB}
    \end{subfigure}
\caption{Schematic representation of the Witten diagrams contributing to $\langle \mathcal{D}_p\mathcal{D}_p\mathcal{D}_q\mathcal{D}_q\rangle$ at tree level. Black lines represent free theory propagators, while blue lines crossing in a thick dot represent lines connected by the quartic vertex in \eqref{AdSlagrangian}.}
\label{fig:treediagramsppqq}
\end{figure}

Let us consider now the case of tree level correlators, for which according to \eqref{ansatzppqq} we should make an ansatz of transcendentality one
\begin{align}\label{treeansatzppqq}
\mathcal{G}^{(1)}_{\{p,p,q,q\}}(\chi,\zeta_1,\zeta_2)=\sum_{m,n=0}^{p}\frac{1}{\zeta_1^{m}\,\zeta_2^{n}}\,\left(r_1^{(m,n)}(\chi)+r_2^{(m,n)}(\chi)\,\log(1-\chi)+r_3^{(m,n)}(\chi)\,\log\chi\right)\,,
\end{align}
where we remind that all rational functions $r_i^{(m,n)}(\chi)$ are symmetric in $m$ and $n$ and, as in section \eqref{sec:bootstrap}, are only allowed to have powers of $\chi$ and $1-\chi$ in their denominators. For all values of $p$, using all constraints except the Regge behavior still leads to an infinite amount of solutions corresponding, as in the previous sections, to contact terms with higher and higher number of derivatives in the quartic vertex. However, assuming the mildest possible Regge behavior for each $p$ and using, step by step, all the available information on non-degenerate operators, it is possible to pinpoint a unique solution for each $p$, with a simple polynomial dependence in $q$ (of degree $p+1$, taking into account that we assume $p\le q$). Remarkably, it is possible to express the result in a closed form as a function of $p$ and $q$, which reads
\begin{align}\label{treeppqq}
\mathcal{G}_{\{p,p,q,q\}}^{(1)}=p\,q\,\mathcal{G}_{\{1,1,1,1\}}^{(1)}\,\mathcal{G}_{\{p-1,p-1,q-1,q-1\}}^{(0)}+\mathsf{P}^{(1)}(\sX,\,\sXs)\,,
\end{align}
where $\mathsf{P}^{(1)}$ are polynomials given by
\begin{align}\label{P1treelevel}
\begin{split}
\mathsf{P}^{(1)}(x,y)=\,3\,\mathcal{Q}^{(1)}\,\,\mathcal{G}^{(0)}_{\{p,p,q,q\}}(x,y)\,.
\end{split}
\end{align}
where we have introduced the differential operator
\begin{align}
\mathcal{Q}^{(1)}=\left(x\,\partial_x\right)^2+\left(y\,\partial_y\right)^2+\,x\,y\,\partial_x\,\partial_y-\frac{p+q}{2}\left(x\,\partial_x+y\,\partial_y\right)\,,
\end{align}
and in the argument of $\mathcal{G}^{(0)}_{\{p,p,q,q\}}$ we replace $\sX\to x$ and $\sXs\to y$. This result can be seen as arising naturally from the Witten diagrams generated by the Lagrangian \eqref{AdSlagrangian} and from the fact that the half-BPS operators $\mathcal{D}_k$ are simply given by normal-ordered products of $k$ copies of $\mathcal{D}_1$ (with an appropriate symmetrization in the R-symmetry indices). Given that at $\mathcal{O}(\lambda^{-1/2})$ there is only a quartic vertex including for $\varphi$'s (the superprimary of the $\mathcal{D}_1$ multiplet), there are schematically two types of diagram that can contribute to any $\langle \mathcal{D}_p\mathcal{D}_p\mathcal{D}_q\mathcal{D}_q\rangle$ at tree level, represented in figure \ref{fig:treediagramsppqq}.

\begin{figure}[hbt!]
\begin{center}
        \includegraphics[width=0.5\textwidth]{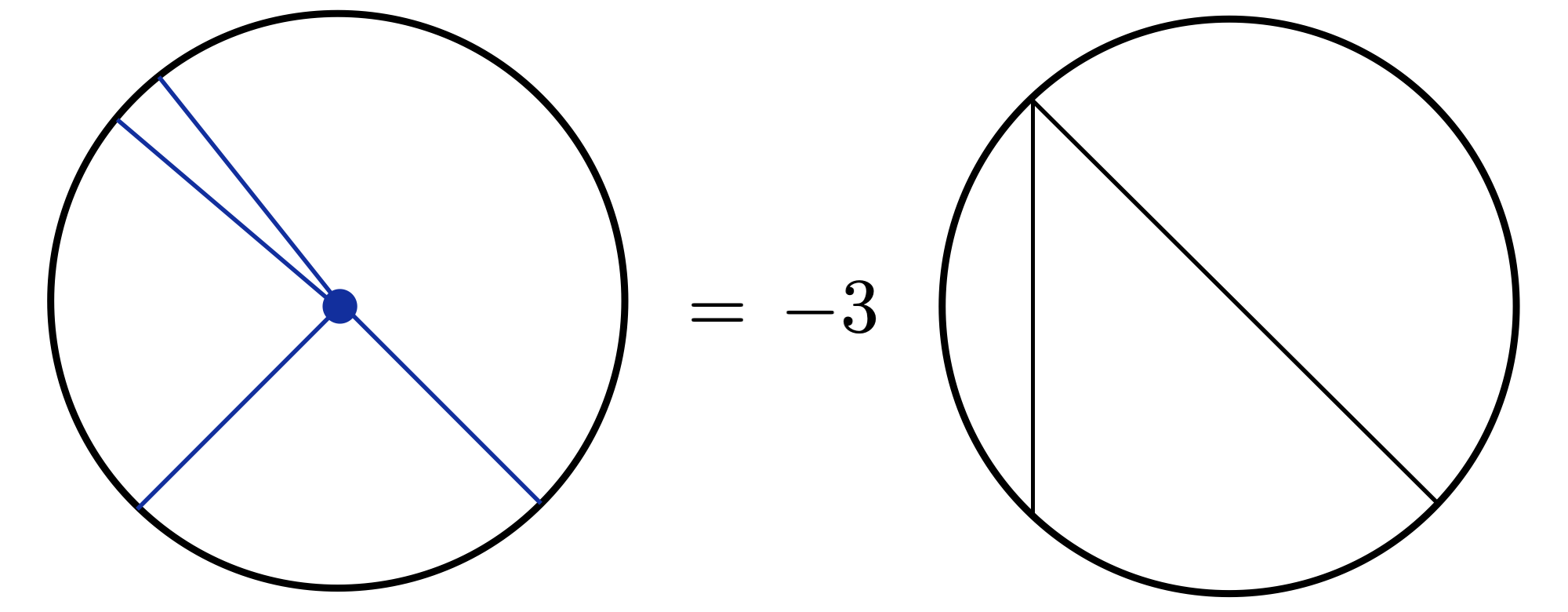}
\caption{Collapsing two points in a quartic vertex produces the same effect as a couple of free theory propagators.}
\label{fig:collapsequarticvertexonce}
\end{center}
\end{figure}

Diagrams of the type represented in figure \ref{fig:treediagramsppqqA} are such that each external operator is connected to a quartic vertex, which carries a factor $\lambda^{-1/2}$. The dependence on the coupling constant is therefore saturated and the remaining operators must be connected by free theory propagators. This kind of diagrams is therefore completely factorized, with the quartic vertex contributing as in $\langle \mathcal{D}_1\mathcal{D}_1\mathcal{D}_1\mathcal{D}_1\rangle$, multiplied by a free theory term corresponding to a $\langle\mathcal{D}_{p-1}\mathcal{D}_{p-1}\mathcal{D}_{q-1}\mathcal{D}_{q-1}\rangle$ four-point function, hence the first term in \eqref{treeppqq}. On the other hand, diagrams of the type represented in figure \ref{fig:treediagramsppqqB} have one operator connected to the quartic vertex with two lines, two with one line and one operator which is only connected to the others by free theory propagators. Understanding the contribution of these diagrams to \eqref{treeppqq} is simple if one focuses on the lines connected to the quartic vertex. With the position of the lines as in \ref{fig:treediagramsppqqB}, one has
\begin{align}\label{collapsequarticvertexonce}
\lim_{\substack{t_*\to t_1\\
y_*\to y_1}}\langle\mathcal{D}_1(t_1,y_1)\mathcal{D}_1(t_*,y_*)\mathcal{D}_1(t_2,y_2)\mathcal{D}_1(t_3,y_3)\rangle^{(1)}=-3\,(12)\,(13)\,,
\end{align}
where notice that the limit gives a finite result since we are taking $y_*\to y_1$, which corresponds to defining a protected operator $\mathcal{D}_2$ at position $(t_1,y_1)$. Diagrammatically, we can express the identity \eqref{collapsequarticvertexonce} as in figure \ref{fig:collapsequarticvertexonce}, which notice is precisely the kind of diagram that one should evaluate to compute the OPE coefficient $\mathsf{C}^{(1)}_{112}$, where the superscript denotes a tree level contribution.
In particular, we can see that the effect of collapsing two points in a quartic vertex produced the same result as the product of two free-theory propagators. Therefore, diagrams of the type \ref{fig:treediagramsppqqB} are actually equivalent to free theory diagrams, just with different combinatorial factors (and up to a $\lambda$-dependent overall factor). Therefore, they contribute to the second term in \eqref{treeppqq}, which can be seen as the action of an operator on $\langle \mathcal{D}_p\mathcal{D}_p\mathcal{D}_q\mathcal{D}_q\rangle^{(0)}$, as described by \eqref{P1treelevel}, where the operator has the role of suppressing certain diagrams and appropriately modifying combinatorial factors. Hence, it would have been relatively simple to argue the result \eqref{treeppqq} directly by looking at the structure of Witten diagrams (once the result for $\langle \mathcal{D}_1\mathcal{D}_1\mathcal{D}_1\mathcal{D}_1\rangle^{(1)}$ is known), and it is reassuring that the bootstrap result confirms the expectations.

The structure of \eqref{treeppqq} has an important implication for the first order correction to the spectrum of the free theory. As we have already stressed, the $\ell$-th order anomalous dimensions of the operators exchanged in a certain correlator at order $\lambda^{-\ell/2}$ can be extracted from the part of the correlator that is proportional to $\log\chi$. In \eqref{treeppqq}, the second term only depends on $\chi$ polynomially and therefore all the terms proportional to $\log\chi$ arise from $\langle  \mathcal{D}_1\mathcal{D}_1\mathcal{D}_1\mathcal{D}_1\rangle^{(1)}$. Similar diagrammatic arguments immediately show that a four-point function between {\it any} four operators only receives contributions proportional to $\log\chi$ through a product of free-theory correlators and $\langle \mathcal{D}_1\mathcal{D}_1\mathcal{D}_1\mathcal{D}_1\rangle^{(1)}$. One can then expect that the anomalous dimensions of all operators can be extracted equivalently from any correlator, with averages of anomalous dimensions over different four-point functions leading to identical results. This can only happen if operators that are degenerate in the free theory remain such when the first-order perturbation is turned on, thus justifying the claims of Section \ref{sec:setup} on the structure of the dilatation operator at tree level. Therefore, at tree level all the symbols of averages can be dropped around anomalous dimensions. The only way this can happen is if the first-order correction to the dilatation operator is insensitive to the detailed structure of the operators and only depends on certain simple properties such as their quantum numbers. We find that this is precisely the case and the dependence is particularly simple and tree-level anomalous dimensions of operators in a certain representation of $\mathfrak{osp}(4^*|4)$ are proportional to the quadratic superconformal Casimir eigenvalue of that representation. More precisely, if an operator $\mathcal{O}$ is, in the free theory, in a representation $\mathcal{R}$ with weights $\omega=\{\Delta,s,[a,b]\}$ (with $\Delta$, as usual, the unperturbed dimension), then its anomalous dimension at first order is
\begin{align}\label{treegammas}
\gamma^{(1)}_{\mathcal{O}}=-\frac{1}{2}\mathfrak{c}_2(\mathcal{R})=-\frac{1}{2}\left[\Delta\,(\Delta+3)+\frac{1}{4}\,s\,(s+2)-\frac{1}{2}\,a^2-a\,(b+2)-b\,(b+3)\right]\,,
\end{align}
using the conventions of \cite{Ferrero:2023znz}. This can be argued by inspecting the anomalous dimensions extracted from $\langle \mathcal{D}_p\mathcal{D}_p\mathcal{D}_q\mathcal{D}_q\rangle$ or by observing that the action of the quadratic Casimir differential operator $\widehat{\mathfrak{C}}_{\mathfrak{osp}(4^*|4)}$ is such that
\begin{align}\label{logtreefromCasimir}
\left.\langle \mathcal{D}_p\mathcal{D}_p\mathcal{D}_q\mathcal{D}_q\rangle^{(1)}\right|_{\log\chi}=-\frac{1}{2}\,\widehat{\mathfrak{C}}_{\mathfrak{osp}(4^*|4)}\,\langle \mathcal{D}_p\mathcal{D}_p\mathcal{D}_q\mathcal{D}_q\rangle^{(0)}\,,
\end{align}
since the superconformal blocks are eigenfunctions of $\widehat{\mathfrak{C}}_{\mathfrak{osp}(4^*|4)}$ with eigenvalue $\mathfrak{c}_2$.

This is a crucial fact about the spectrum of the theory at strong coupling, that we have heavily exploited in the previous sections. In particular, we have used the fact that when the expression of correlation functions as sums over conformal blocks is expanded perturbatively for large $\lambda$, certain products of anomalous dimensions and OPE coefficients arise, which should in principle interpreted as averages over a certain correlator. However, due to the fact that the free theory degeneracy is not lifted one has, for any representation,
\begin{align}\label{degnoliftedtree}
\langle (\gamma^{(1)})^{k}\rangle_{\mathcal{L}^{\Delta}_{s,[a,b]}}=\langle \gamma^{(1)}\rangle_{\mathcal{L}^{\Delta}_{s,[a,b]}}^k\,,
\end{align}
with the average taken on any correlator. 

\subsection{One loop}

Moving to one loop, one should make an ansatz of transcendentality two. Given the results found for $\langle \mathcal{D}_1\mathcal{D}_1\mathcal{D}_1\mathcal{D}_1\rangle$, for simplicity we can directly exclude terms with $\Li_2$ and $\zeta(2)$, but it is also possible to include them and then show that their coefficient should vanish once all the constraints are imposed. To start, one should compute the terms contributing to the highest logarithmic singularities in the two OPE channels ($\chi\to 0$ and $\chi \to 1$). At one loop, one only has a $\log^2\chi$ singularity, which in the direct channel ($\chi \to 0$) has an expansion over superconformal blocks given by
\begin{align}
\mathcal{G}^{d,(\ell,\ell)}_{\{p,p,q,q\}}=\frac{1}{\ell!}\sum_{\Delta,[a,b]}\langle a^{(0)}\,\big(\gamma^{(1)}\big)^{\ell}\rangle_{\mathcal{L}^{\Delta}_{0,[a,b]}} \mathfrak{G}_{\Delta,[a,b]}\,,
\end{align}
for $\ell=2$, using the notation of \eqref{logexpppqq}. We have written the equation for general $\ell$ as this it gives the $\log^{\ell}\chi$ contribution to an $\ell$-th order correlator in a uniform way. Now, thanks to \eqref{degnoliftedtree} we can drop the symbol of average from the anomalous dimensions any $\ell$,  and using \eqref{treegammas} as well as the superconformal Casimir equation (see \cite{Ferrero:2023znz}) we obtain
\begin{align}
\mathcal{G}^{d,(\ell,\ell)}_{\{p,p,q,q\}}=\frac{1}{\ell!}\sum_{\Delta,[a,b]}\langle a^{(0)}\rangle_{\mathcal{L}^{\Delta}_{0,[a,b]}}\,\left(\gamma^{(1)}_{\mathcal{L}^{\Delta}_{0,[a,b]}}\right)^{\ell} \mathfrak{G}_{\Delta,[a,b]}
=\left(-\frac{1}{2}\,\widehat{\mathfrak{C}}_{\mathfrak{osp}(4^*|4)}\right)^{\ell}\,\mathcal{G}_{\{p,p,q,q\}}^{(0)}\,,
\end{align}
meaning that as observed for $\langle \mathcal{D}_1\mathcal{D}_1\mathcal{D}_1\mathcal{D}_1\rangle$ in Section \ref{sec:1111} the highest logarithmic singularity at each order can be obtain simply by repeatedly acting with the quadratic Casimir operator on the free theory result. Analogous expressions can be obtained for terms proportional to $\log^2(1-\chi)$ in the crossed channel expansion, around $\chi=1$. 

Once the terms proportional to $\log^2\chi$ and $\log^2(1-\chi)$ are computed as above, one can use a braiding transformation to obtain the rational function multiplying $\log\chi\,\log(1-\chi)$, so that assuming no $\Li_2(\chi)$ is present we have computed all terms of transcendentality two. All is left is therefore the ambiguity given by tree-level contact terms, namely the infinite set of solutions to the tree-level bootstrap problem. As usual, a unique solution for each $p$ can be obtained by exploiting all information on the OPE and non-degenerate operators obtained from correlators with lower $p$, as well as assuming the mildest possible Regge behavior for our correlators, corresponding to
\begin{align}
\langle\gamma^{(2)}\rangle_{\mathcal{L}^{\Delta}_{0,[a,b]}}\sim \Delta^3\,,\quad (\Delta\to\infty)\,,
\end{align}
for an average over any correlator and for all values of $a$ and $b$.

\begin{figure}[hbt!]
    \centering
    \begin{subfigure}[b]{0.20\textwidth}
        \includegraphics[width=\textwidth]{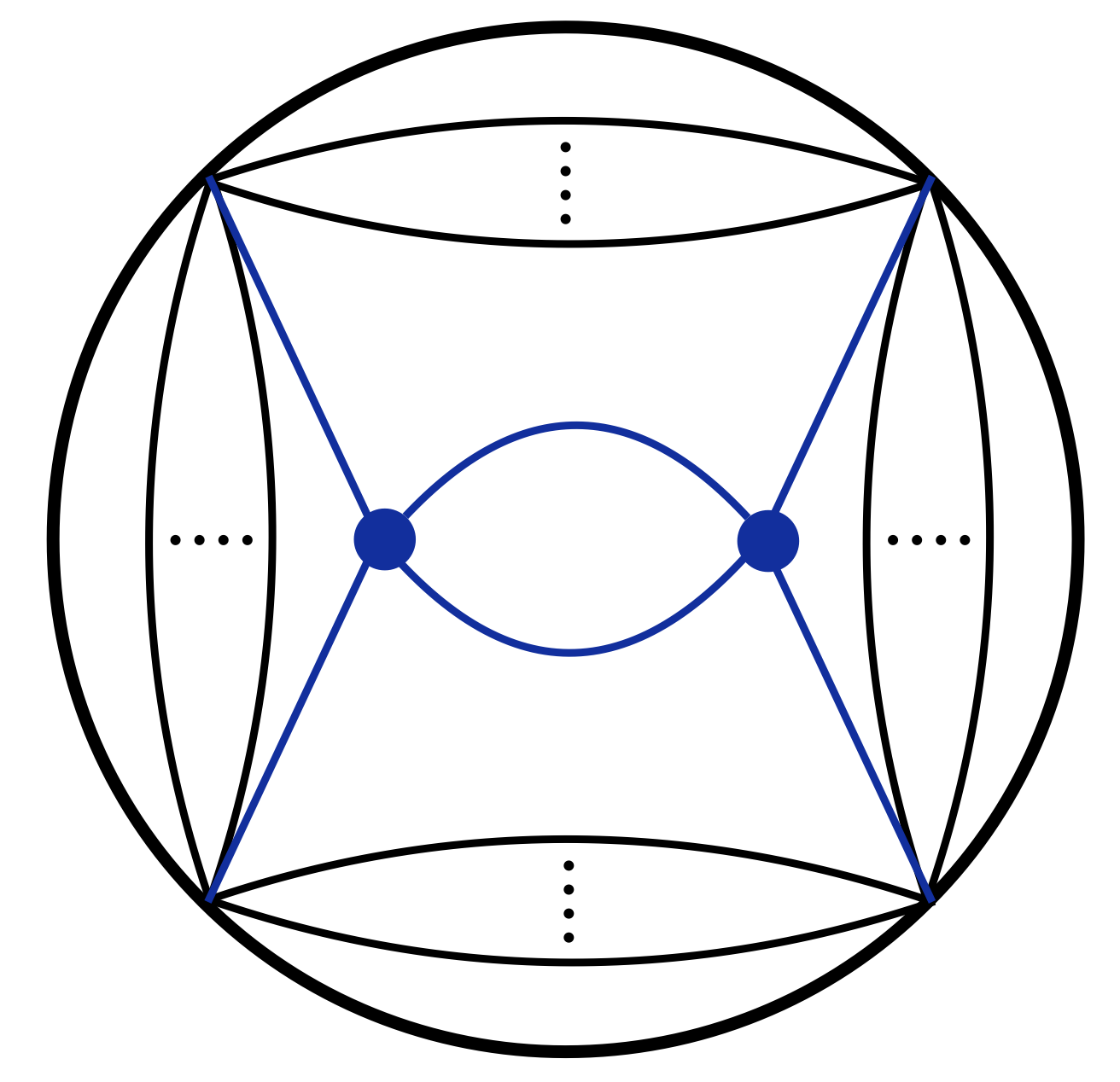}
        \caption{Contribution from one-loop diagrams with quartic vertices only.}
        \label{fig:1loop_from_loop}
    \end{subfigure}\quad\quad
    \begin{subfigure}[b]{0.20\textwidth}
        \includegraphics[width=\textwidth]{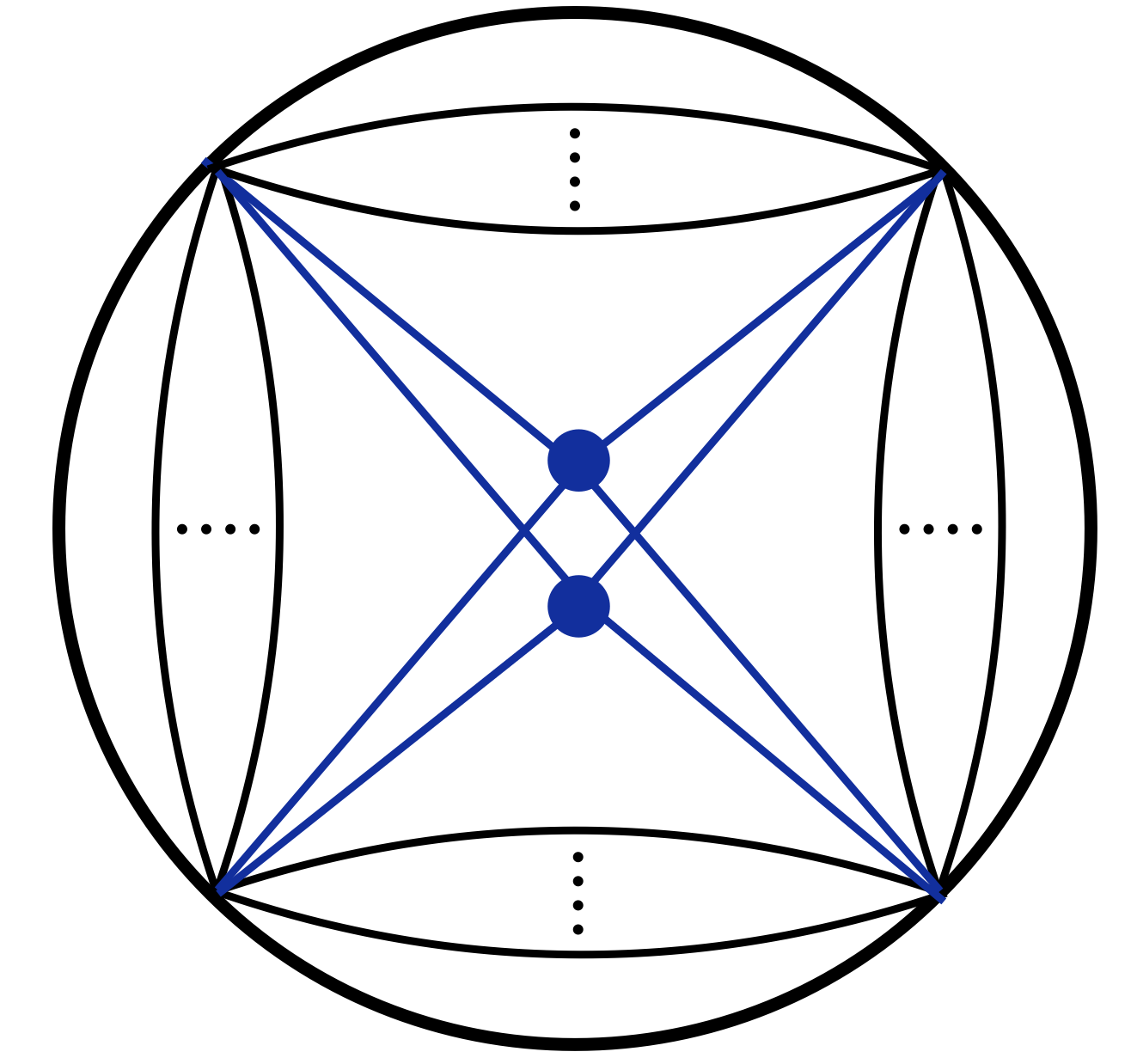}
        \caption{Contribution from the product of two tree-level quartic interactions.}
        \label{fig:1loop_from_tree_2}
    \end{subfigure}
\caption{Type of diagrams contributing to the first line of equation \eqref{1loop_ppqq}. We have highlighted four-point vertices in blue and six-point vertices in red.}
\label{fig:1loopppqqA}
\end{figure}

\begin{figure}[hbt!]
    \centering
    \begin{subfigure}[b]{0.20\textwidth}
        \includegraphics[width=\textwidth]{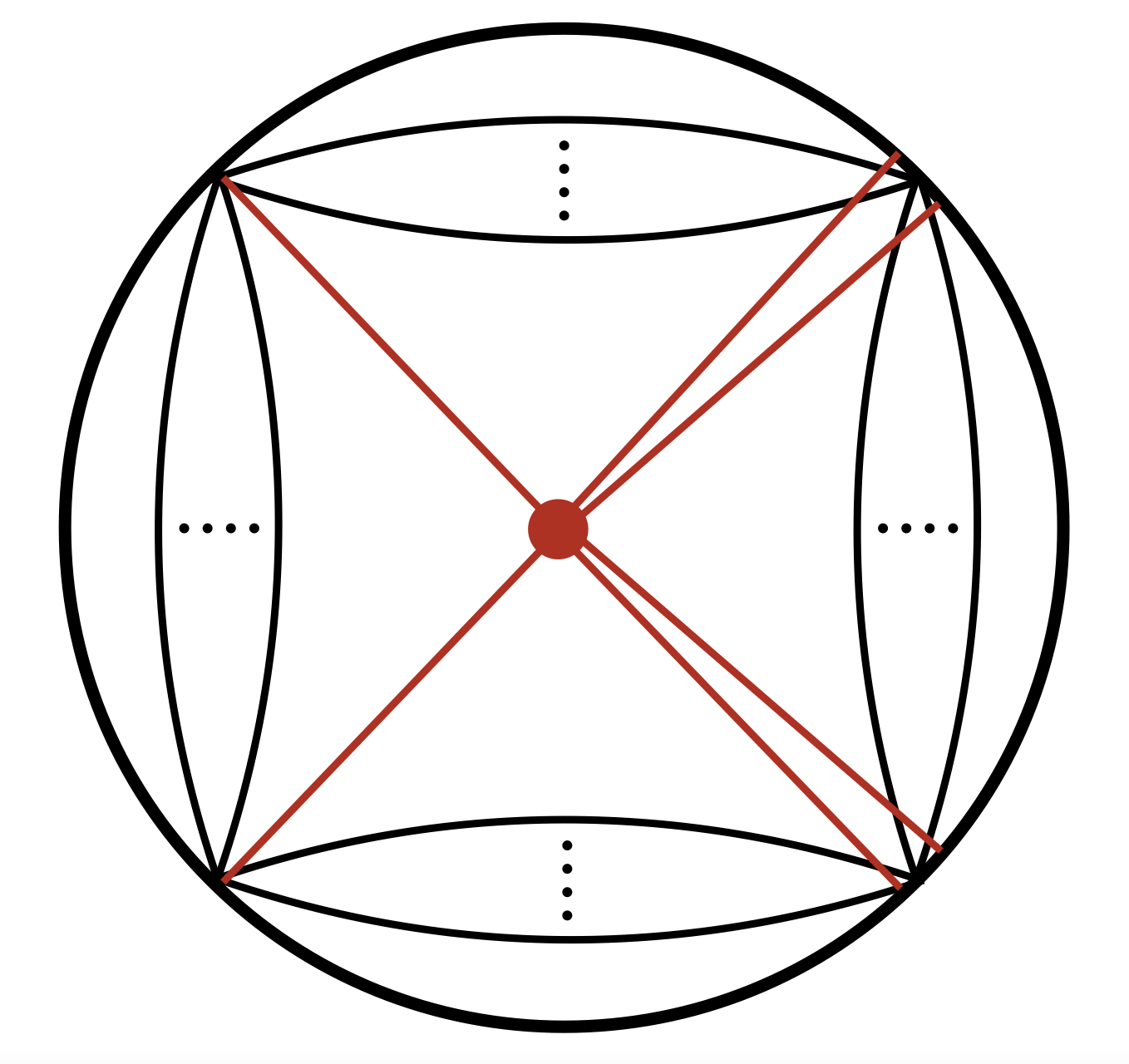}
        \caption{Contribution from tree diagrams with sextic interactions.}
        \label{fig:1loop_from_6pt}
    \end{subfigure}\quad\quad
    \begin{subfigure}[b]{0.20\textwidth}
        \includegraphics[width=\textwidth]{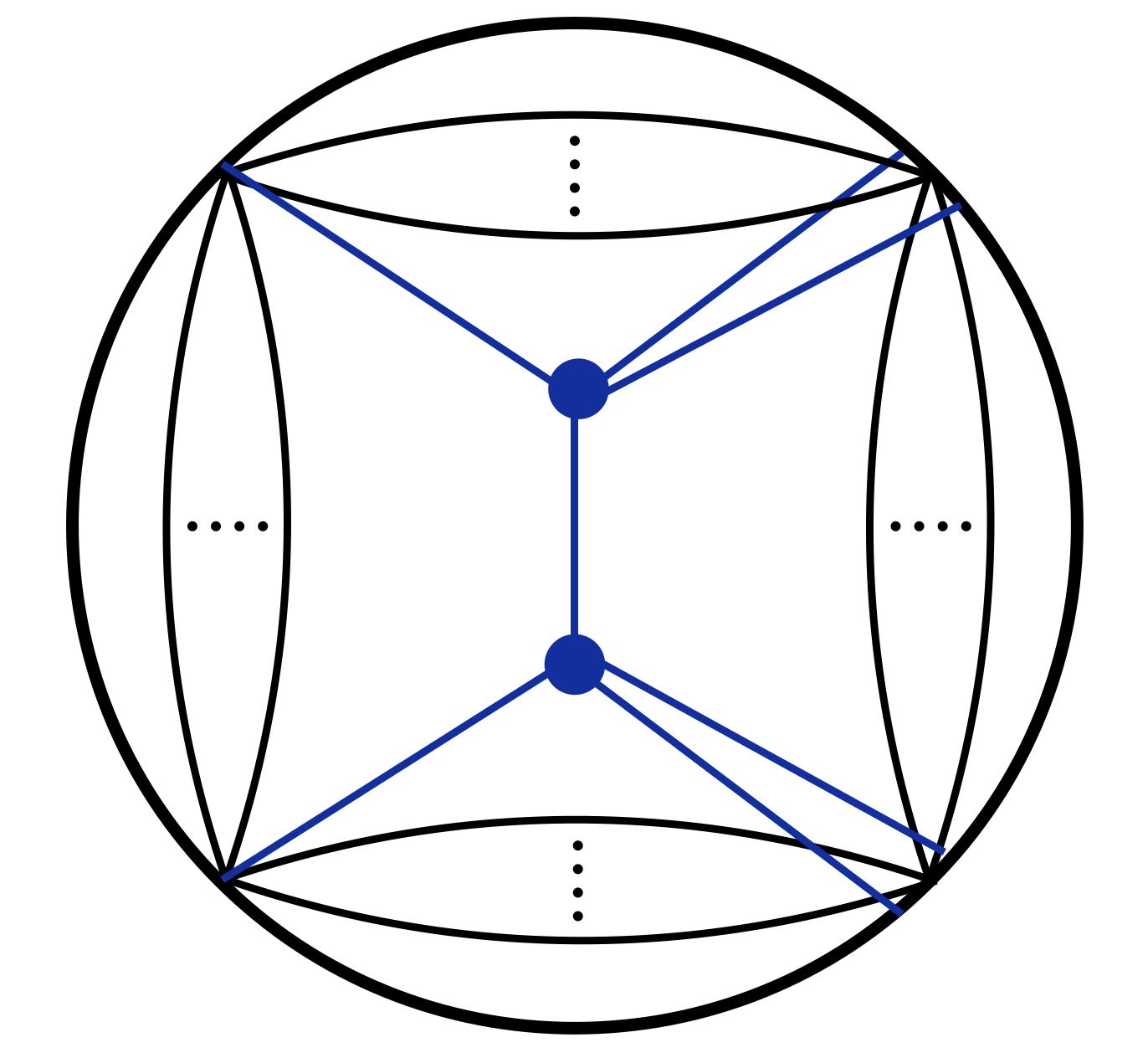}
        \caption{Contribution from collapsed six-point exchange diagrams.}
        \label{fig:1loop_from_exchange}
    \end{subfigure}
\caption{Type of diagrams arising from a six-point function and contributing to \eqref{1loop_ppqq} as in \eqref{Gfrom6pt}. We have highlighted four-point vertices in blue and six-point vertices in red.}
\label{fig:1loopppqqB}
\end{figure}

\begin{figure}[hbt!]
    \centering
    \begin{subfigure}[b]{0.20\textwidth}
        \includegraphics[width=\textwidth]{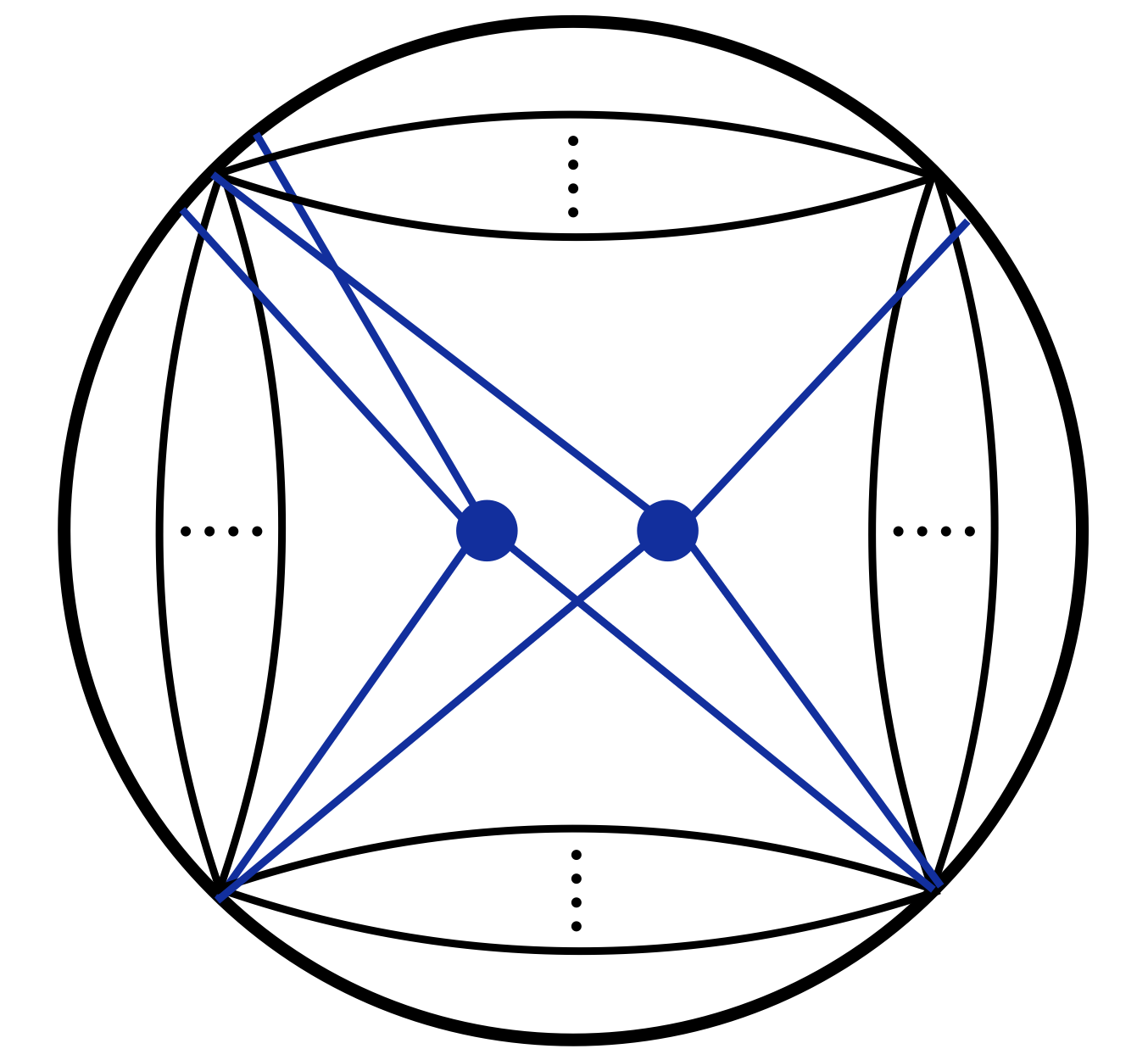}
        \caption{Contribution from a ``full'' quartic vertex and a collapsed one. }
        \label{fig:1loop_from_tree_1}
    \end{subfigure}\quad\quad
    \begin{subfigure}[b]{0.19\textwidth}
        \includegraphics[width=\textwidth]{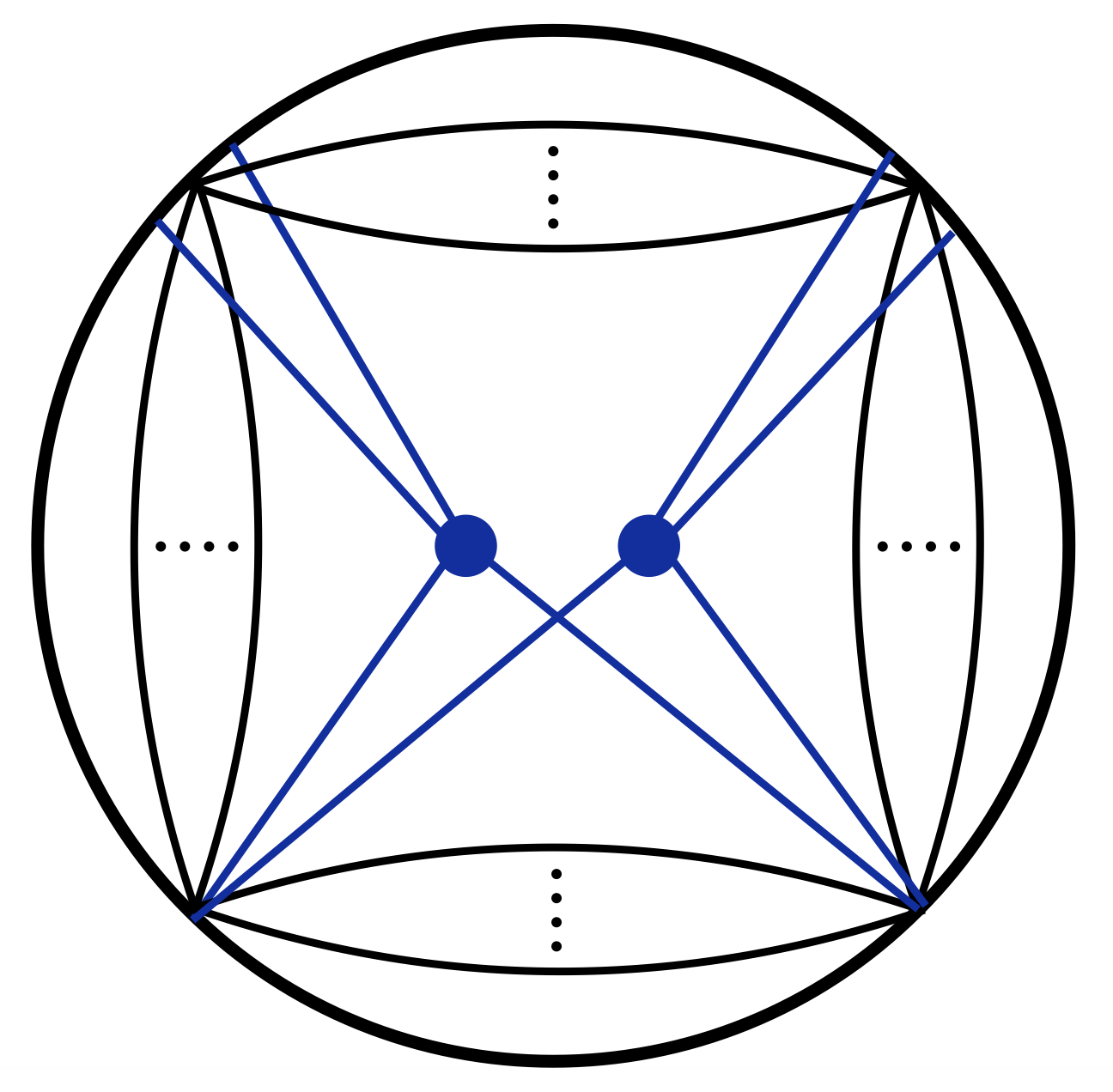}
        \caption{Contribution from two quartic vertices with two points collapsed.}
        \label{fig:1loop_from_treecollapsed}
    \end{subfigure}\quad\quad
    \begin{subfigure}[b]{0.20\textwidth}
        \includegraphics[width=\textwidth]{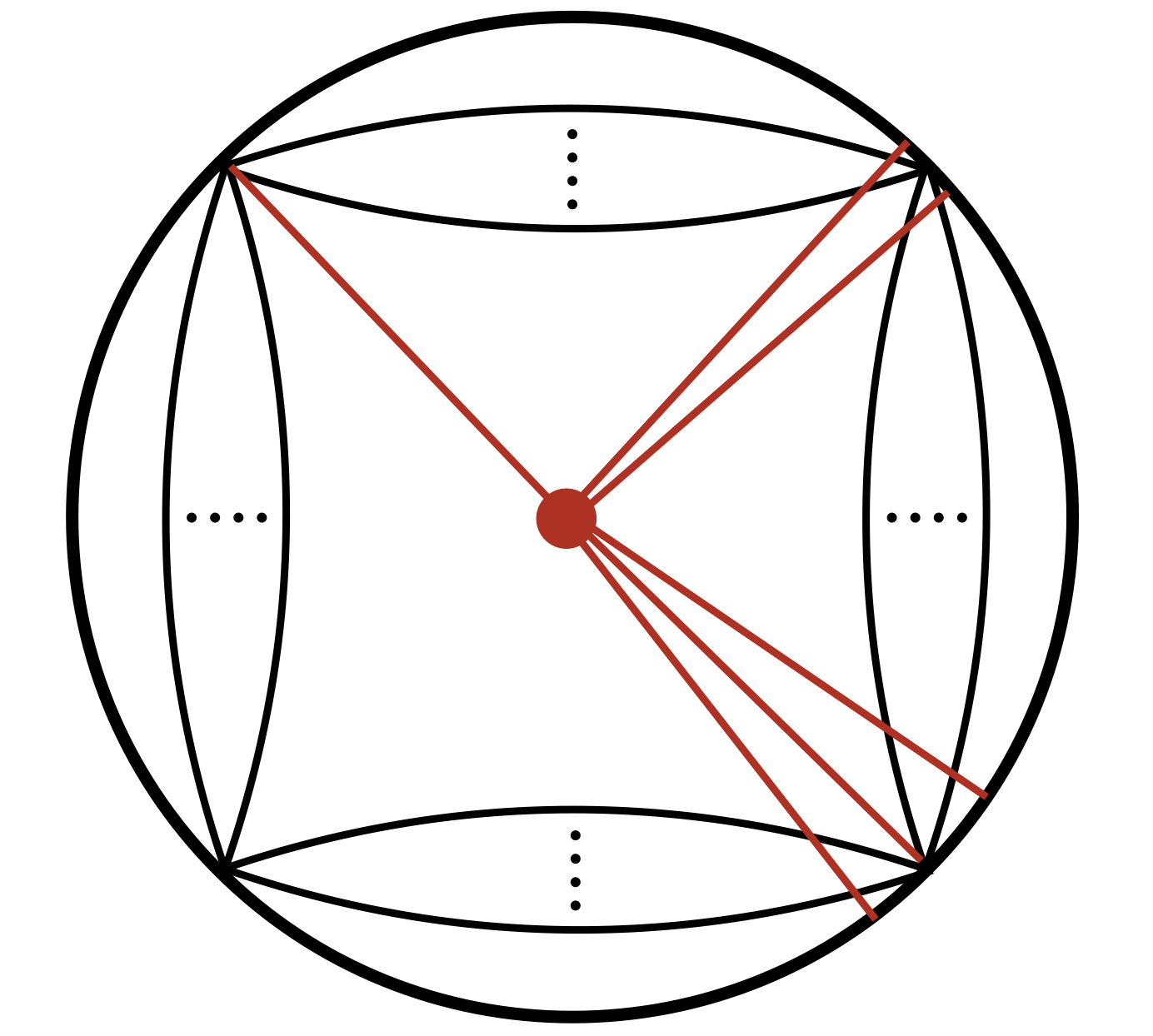}
        \caption{Contribution from a sextic vertex with three legs collapsed.}
        \label{fig:1loop_from_6ptcollapsed}
    \end{subfigure}
\caption{Type of diagrams contributing to the last two terms in the second line of \eqref{1loop_ppqq}. We have highlighted four-point vertices in blue and six-point vertices in red. Diagrams of the type \ref{fig:1loop_from_tree_1} contribute to $\mathsf{P}^{(2)}_1\,\mathcal{G}^{(1)}$, while diagrams \ref{fig:1loop_from_treecollapsed} and \ref{fig:1loop_from_6ptcollapsed} only contribute to to $\mathsf{P}^{(2)}_2$.}
\label{fig:1loopppqqC}
\end{figure}

As for tree-level correlators, the final result admits a closed-form expression in terms of $p$ and $q$, which is closely related to the structure of the associated Witten diagrams. We can express our results as 
\begin{align}\label{1loop_ppqq}
\begin{split}
\mathcal{G}^{(2)}_{\{p,p,q,q\}}=&\,p\,q\,\mathcal{G}^{(2)}_{\{1,1,1,1\}}\,\mathcal{G}^{(0)}_{\{p-1,p-1,q-1,q-1\}}+2\,\binom{p}{2}\binom{q}{2}\,\left(\mathcal{G}^{(1)}_{\{1,1,1,1\}}\right)^2\,\mathcal{G}^{(0)}_{\{p-2,p-2,q-2,q-2\}}\\
&+\mathcal{G}^{(\text{6pt})}_{\{p,p,q,q\}}+\mathsf{P}^{(2)}_1(\sX,\sXs)\,\mathcal{G}^{(1)}_{\{1,1,1,1\}}+\mathsf{P}^{(2)}_2(\sX,\sXs)\,,
\end{split}
\end{align}
where $\mathcal{G}^{(1)}_{\{1,1,1,1\}}$ and $\mathcal{G}^{(2)}_{\{1,1,1,1\}}$ are the tree-level and one-loop four-point functions of the displacement, $\mathcal{G}^{(\text{6pt})}_{\{p,p,q,q\}}$ can be computed from the first connected contribution to the six-point function of the displacement, while $\mathsf{P}^{(2)}_1$ and $\mathsf{P}^{(2)}_2$ are certain polynomials given below. More precisely\footnote{Here, when evaluating $\mathcal{G}_{\star}(\chi^{-1},\zeta_1^{-1},\zeta_2^{-1})$, we consider that all the arguments of $\log$ have an absolute value, in agreement with the observations about braiding symmetry of Section \ref{sec:1111}. This allows us to write, for example, $\log|1-\chi^{-1}|=\log|1-\chi|-\log|\chi|$.}
\begin{align}\label{Gfrom6pt}
\begin{split}
\mathcal{G}^{(\text{6pt})}_{\{p,p,q,q\}}=&
\,\frac{p}{2}\binom{q}{2}\,\mathcal{G}_{\star}(\chi,\zeta_1,\zeta_2)\,\mathcal{G}^{(0)}_{\{p-1,p-1,q-2,q-2\}}
+\frac{q}{2}\binom{p}{2}\,\mathcal{G}_{\star}(\chi,\zeta_1,\zeta_2)\,\mathcal{G}^{(0)}_{\{p-2,p-2,q-1,q-1\}}\\
&+2\binom{p}{2}\binom{q}{2}\,\sX\,\mathcal{G}_{\star}(\chi^{-1},\zeta_1^{-1},\zeta_2^{-1})\,\mathcal{G}^{(0)}_{\{p-2,p-1,q-2,q-1\}}\\
&+2\binom{p}{2}\binom{q}{2}\,\sXs\,\mathcal{G}_{\star}(1-\chi,1-\zeta_1,1-\zeta_2)\,\mathcal{G}^{(0)}_{\{p-1,p-2,q-2,q-1\}}\,,
\end{split}
\end{align}
with 
\begin{align}\label{Gstar}
\begin{split}
\mathcal{G}_{\star}(\chi,\zeta_1,\zeta_2)=&\mathbb{D}\,f_{\star}(\chi)\,,\\
f_{\star}(\chi)=&\left[-4\chi(2-\chi)+2(2-2\chi+\chi^2)\,\log(1-\chi)\right]\,\frac{\chi^2\,\log\chi}{(1-\chi)^2}\\
&+\left[\chi(5-5\chi-2\chi^2+4\chi^3)+2(1-\chi)^2(2+\chi+\chi^2)\,\log(1-\chi)\right]\,\frac{\log(1-\chi)}{\chi(1-\chi)}\,.
\end{split}
\end{align}
On the other hand,  $\mathsf{P}^{(2)}_1$ and $\mathsf{P}^{(2)}_2$ are polynomials,  given by
\begin{align}
\begin{split}
\mathsf{P}^{(2)}_1(x,y)=&
\,\mathcal{Q}^{(2)}_1\,\mathcal{G}^{(0)}_{\{p-1,p-1,q-1,q-1\}}(x,y)\,,
\end{split}
\end{align}
and 
\begin{align}
\begin{split}
\mathsf{P}^{(2)}_2(x,y)=&\,\mathcal{Q}^{(2)}_2\,\mathcal{G}^{(0)}_{\{p,p,q,q\}}(x,y)\,,
\end{split}
\end{align}
where the differential operators $\mathcal{Q}^{(2)}_1$ and $\mathcal{Q}^{(2)}_2$ are
\begin{align}
\begin{split}
\mathcal{Q}^{(2)}_1=&\,3\,p\,q\left[\mathcal{Q}^{(1)}+\left(x\,\partial_x+y\,\partial_y\right)\right]\,,\\
\mathcal{Q}^{(2)}_2=&\left[\frac{9}{2}\,\mathcal{Q}^{(1)}\,\left(\mathcal{Q}^{(1)}+\frac{p+q-2}{36}\right)+\frac{p\,q}{4}(p+q-2)-\frac{9}{16}(p-q)^2\left(x\,\partial_x+y\,\partial_y\right)\right.\\
&\left.+\frac{33}{4}\,x\,y\,\partial_x\,\partial_y\,\left(x\,\partial_x+y\,\partial_y-\frac{p+q}{2}\right)\right]\,.
\end{split}
\end{align}
As in the previous section, to understand this structure one should look at the Witten diagrams that contribute to $\langle \mathcal{D}_p\mathcal{D}_p\mathcal{D}_q\mathcal{D}_q\rangle$ at two loops. The first term in \eqref{1loop_ppqq} can be traced back to diagrams of the type \ref{fig:1loop_from_loop}, where two quartic vertices are used to form a loop and all the remaining $\mathcal{D}_1$ insertions are connected by free theory propagators, so that these diagrams are completely factorized as products of the one-loop contribution to $\langle \mathcal{D}_1\mathcal{D}_1\mathcal{D}_1\mathcal{D}_1\rangle$ and a free theory four-point function where all external operators have weights lowered by one unit.  The first line of \eqref{1loop_ppqq} is completed by a contribution coming from diagrams of the type \ref{fig:1loop_from_tree_2}, where eight of the external $\mathcal{D}_1$ insertions are connected by two pairs of four-point vertices, giving rise to the square of $\langle \mathcal{D}_1\mathcal{D}_1\mathcal{D}_1\mathcal{D}_1\rangle^{(1)}$, with the remaining insertions connected by propagators to form a free theory four-point function where all external operators have weights lowered by two units. Moving to the second line of \eqref{1loop_ppqq}, the first term originates from the same type of diagrams that would contribute to a $\langle \mathcal{D}_1\mathcal{D}_1\mathcal{D}_1\mathcal{D}_1\mathcal{D}_1\mathcal{D}_1\rangle$ six-point function at $\mathcal{O}(\lambda^{-1})$, with two pairs of $\mathcal{D}_1$ insertions collapsed to the same point: a six-point contact term, which appears in \eqref{AdSlagrangian} with a factor of $\lambda^{-1}$ and is represented in figure \ref{fig:1loop_from_6pt}, as well as an exchange diagram drawn using two times the quartic vertex, represented in \ref{fig:1loop_from_exchange}. The four terms in \eqref{Gfrom6pt} reflect the four inequivalent ways to collapse the six-point function to four points, each dressed by suitable free-theory four-point functions connecting the remaining $\mathcal{D}_1$ insertions. The second-last term in the second line of \eqref{1loop_ppqq} arises from diagrams of the type \ref{fig:1loop_from_tree_1} where one uses two times the quartic vertex, but one of them has two legs collapsed to one. Finally, the last term is a rational functions of $\chi$ and receives contributions from two types of diagrams: one where one uses two collapsed quartic vertices, as in figure \ref{fig:1loop_from_treecollapsed}, and one where a collapsed six-point vertex is used, as in \eqref{fig:1loop_from_6ptcollapsed}.

In this case, while as we have just observed the general structure can be argued looking at the Witten diagrams that contribute at this order, the expression of one-loop $\langle \mathcal{D}_p\mathcal{D}_p\mathcal{D}_q\mathcal{D}_q\rangle$ requires the knowledge of the six-point function $\langle \mathcal{D}_1\mathcal{D}_1\mathcal{D}_1\mathcal{D}_1\mathcal{D}_1\mathcal{D}_1\rangle^{(2)}$, in addition to $\langle \mathcal{D}_1\mathcal{D}_1\mathcal{D}_1\mathcal{D}_1\rangle^{(2)}$. Since we have not computed any six-point functions, which would be considerably harder than the four-point ones discussed here\footnote{See \cite{Barrat:2021tpn} for the computation of higher-point functions on the Wilson line at weak coupling and \cite{Giombi:2023zte} at strong coupling.}, our bootstrap approach should be seen as an even greater simplification of the traditional diagrammatic algorithm. Moreover, the contribution of a collapsed six-point function to $\langle \mathcal{D}_p\mathcal{D}_p\mathcal{D}_q\mathcal{D}_q\rangle^{(2)}$ finally lifts the degeneracy of operators in the free theory and the anomalous dimensions $\langle\gamma^{(2)}\rangle_{\mathcal{L}^{\Delta}_{0,[a,b]}}$ extracted from \eqref{1loop_ppqq} should really be seen as averages, which depend on the correlator from which they are computed. As an example, for long operators in the singlet representation of $\mathfrak{sp}(4)\oplus \mathfrak{su}(2)$, from $\langle \mathcal{D}_1\mathcal{D}_1\mathcal{D}_q\mathcal{D}_q\rangle$ correlators we find
\begin{align}
\langle\gamma^{(2)}\rangle_{\mathcal{L}^{\Delta}_{0,[0,0]}}-\gamma^{(1)}_{\Delta}\,\partial_{\Delta}\,\gamma^{(1)}_{\Delta}=\frac{\JJ}{2}\,\left[q\,H_{1+\Delta}-\frac{2(5+9q)+(4+7q)	\,\JJ}{4(2+\JJ)}\right]\,,
\end{align}
where the dependence of the anomalous dimensions of the exchanged operators on the external weight $q$ is a clear sign of operators mixing. 

We would like to conclude this section with some comments on the relation between the structure of $\langle \mathcal{D}_p\mathcal{D}_p\mathcal{D}_q\mathcal{D}_q\rangle$ and that of the dilatation operator. At tree level, we observed that the terms in the correlator that produce anomalous dimensions (those proportional to $\log\chi$) are all generated by the quartic vertex in \eqref{AdSlagrangian}, as depicted in figure \ref{fig:treediagramsppqqA}. Hence, for the purposes of understanding the dilatation operator at tree level, it is essentially enough to know the four-point function $\langle \mathcal{D}_1\mathcal{D}_1\mathcal{D}_1\mathcal{D}_1\rangle^{(1)}$, with all logarithmic terms in $\langle \mathcal{D}_p\mathcal{D}_p\mathcal{D}_q\mathcal{D}_q\rangle^{(1)}$ fixed in terms of that, as clearly expressed by \eqref{treeppqq}. In particular, as we discussed in Section \ref{sec:setup}, the fact that the dilatation operator $\mathbb{D}^{(1)}$ is fixed by a quartic vertex implies that $\mathbb{D}^{(1)}$ cannot change the length of operators and it can be decomposed as a sum of blocks acting on spaces $\mathcal{H}_L$ of operators with fixed length:
\begin{align}
\mathbb{D}^{(1)}=\bigoplus_L\mathbb{D}^{(1)}_L\,, \qquad
\mathbb{D}^{(1)}_L:\,\,\mathcal{H}_L\to \mathcal{H}_L\,.
\end{align}

On the other hand, at one loop we have seen that the knowledge of $\langle \mathcal{D}_1\mathcal{D}_1\mathcal{D}_1\mathcal{D}_1\rangle^{(2)}$ is no longer sufficient to determine all $\langle \mathcal{D}_p\mathcal{D}_p\mathcal{D}_q\mathcal{D}_q\rangle^{(2)}$. On the contrary, as clear from \eqref{1loop_ppqq}, one also requires the knowledge of the six-point function $\langle \mathcal{D}_1\mathcal{D}_1\mathcal{D}_1\mathcal{D}_1\mathcal{D}_1\mathcal{D}_1\rangle^{(2)}$, which is due to the presence of a six-point vertex in \eqref{AdSlagrangian}, that cannot be detected by $\langle \mathcal{D}_1\mathcal{D}_1\mathcal{D}_1\mathcal{D}_1\rangle^{(2)}$. The presence of a sextic vertex in the Lagrangian also has implications on the one-loop dilatation operator $\mathbb{D}^{(2)}$, as it is clear also from the presence of $\log\chi$ in \eqref{Gstar} (which therefore contribute to the anomalous dimensions). This is closely related to a crucial property of $\mathbb{D}^{(2)}$ that we have introduced in Section \ref{sec:setup}: the fact that this time it can in fact change the length of operators, but only by two units at most. We could write
\begin{align}\label{dilatation1loop}
\mathbb{D}^{(2)}=\bigoplus_L\,\,\left[\mathbb{D}^{(2)}_{L-2}\oplus \mathbb{D}^{(2)}_L\oplus \mathbb{D}^{(2)}_{L+2}\right]\,, \qquad \mathbb{D}^{(2)}_{L+\delta L}:\,\, \mathcal{H}_L\to \mathcal{H}_{L+\delta L}\,.
\end{align}
Similarly, at higher orders more vertices in \eqref{AdSlagrangian} would contribute, further complicating the structure of the dilatation operator (and that of $\langle \mathcal{D}_p\mathcal{D}_p\mathcal{D}_q\mathcal{D}_q\rangle$). In particular, at $\mathcal{O}(\lambda^{-\ell/2})$ in order to determine all $\langle \mathcal{D}_p\mathcal{D}_p\mathcal{D}_q\mathcal{D}_q\rangle^{(\ell)}$ one would need to know the $(2\ell+2)$-pt function of $\mathcal{D}_1$ at that order, with the associated structure of $\mathbb{D}^{(\ell)}$ given by
\begin{align}
\mathbb{D}^{(\ell)}=\bigoplus_L\bigoplus_{a=-(\ell-1)}^{\ell-1}\,\,\mathbb{D}^{(\ell)}_{L+2a}\,,
\end{align}
with the same notation as \eqref{dilatation1loop}. This is precisely the structure that we discussed in Section \ref{sec:setup}.

\subsection{Two loops}

The strategy applied so far could be in principle carried out for all $\langle \mathcal{D}_p\mathcal{D}_p\mathcal{D}_q\mathcal{D}_q\rangle$ at two loops as well, without addressing the mixing problem (which would not be true at three loops). However, the result is not really necessary for our purposes and the problem starts getting more complicated. We have so far limited to computing the two-loop answer for $\langle \mathcal{D}_1\mathcal{D}_1\mathcal{D}_q\mathcal{D}_q\rangle$ and $\langle \mathcal{D}_2\mathcal{D}_2\mathcal{D}_2\mathcal{D}_2\rangle$, which we present here. We will not go into the details of the derivation, since the results follow from the usual strategy and a transcendentality-three ansatz completely analogous to that used for $\langle \mathcal{D}_1\mathcal{D}_1\mathcal{D}_1\mathcal{D}_1\rangle$ at two loops.

For $\langle \mathcal{D}_1\mathcal{D}_1\mathcal{D}_q\mathcal{D}_q\rangle$ one can actually solve the Ward identities in terms of a single function like in the $q=1$ case. As in \eqref{solWI_1111} we have
\begin{align}
\frac{\langle \mathcal{D}_1\mathcal{D}_1\mathcal{D}_q\mathcal{D}_q\rangle}{\langle \mathcal{D}_1\mathcal{D}_1\rangle\langle\mathcal{D}_q\mathcal{D}_q\rangle}=\mathsf{f}_q\,\frac{\chi^2}{\zeta_1\,\zeta_2}+\mathbb{D}\,f_q(\chi)\,.
\end{align}
The topological data has an expansion up to two loops
\begin{align}
\mathsf{f}_q=1+2q-\frac{3q^2}{\lambda^{1/2}}+\frac{3q(q-1)(2q-1)}{4\lambda}+\frac{q(14-2q+28q^2+5q^3)}{8\lambda^{3/2}}+\mathcal{O}(1/\lambda^2)\,,
\end{align}
while the reduced correlator at two loops can be expressed as
\begin{align}
f_q^{(3)}(\chi)=\sum_{n=1}^4\binom{q}{n}\,f_{q,n}^{(3)}(\chi)\,,
\end{align}
where by definition $f_{q,1}^{(3)}(\chi)=f^{(3)}(\chi)$ given in \eqref{final1111_2loop}, while for the higher orders we have
\begin{align}
\begin{split}
f_{q,2}^{(3)}(\chi)&=-\tfrac{\chi ^2 \left(\chi ^2-2 \chi +2\right)}{(\chi -1)^2}\,L_3(\chi)-\tfrac{4 \chi ^5-8 \chi ^4+4 \chi ^2-5 \chi +2}{(\chi -1)^2 \chi }\,L_3(1-\chi)\\
&-\tfrac{(1-\chi) \left(9 \chi ^4+5 \chi ^3+5 \chi ^2+2 \chi +27\right)}{3 \chi ^2}\,\log^3(1-\chi)+\tfrac{\chi ^2 \left(9 \chi ^4-34 \chi ^3+48 \chi ^2-28 \chi +14\right)}{3 (\chi -1)^3}\,\log(1-\chi)\,\log^2\chi\\
&-\frac{18 \chi ^6-47 \chi ^5+40 \chi ^4-12 \chi ^3-8 \chi ^2+10 \chi -4}{3 (\chi -1)^2 \chi }\,\log^2(1-\chi)\,\log\chi\\
&+\tfrac{(\chi -2) \chi ^3 \left(8 \chi ^2-11 \chi +11\right)}{2 (\chi -1)^3}\,\log^2\chi+\tfrac{(\chi -1) \left(8 \chi ^3+18 \chi ^2+15 \chi +29\right)}{2 \chi }\,\log^2(1-\chi)\\
&-\tfrac{16 \chi ^5-25 \chi ^4-\chi ^3+22 \chi ^2+2 \chi -1}{2 (\chi -1)^2}\,\log(1-\chi)\,\log\chi+\tfrac{(\chi -2) \chi  \left(\chi ^2-\chi +1\right)}{4 (\chi -1)^2}\,\log\chi\\
&-\tfrac{(\chi -2) \left(\chi ^2-2 \chi +3\right)}{4 (\chi -1)}\,\log(1-\chi)+\tfrac{4 \chi ^2-5 \chi +2}{(\chi -1)^2 \chi }\,\zeta(3)+\tfrac{\chi  (131 \chi -28)}{8 (\chi -1)}\,,\\
f_{q,3}^{(3)}(\chi)&=-\tfrac{3 (\chi -2) \chi ^3}{(\chi -1)^2}\,L_3(1-\chi)+\tfrac{(\chi -1) \left(\chi ^3+\chi ^2+6\right)}{\chi ^2}\,\log^3(1-\chi)-\tfrac{(\chi -2) \chi ^3}{(\chi -1)^2}\,\log^2(1-\chi)\,\log\chi\\
&+\tfrac{9 (\chi -1) \left(\chi ^2+\chi +2\right)}{2 \chi }\,\log^2(1-\chi)-\tfrac{9 \chi ^2 \left(\chi ^2-2 \chi +2\right)}{2 (\chi -1)^2}\,\log(1-\chi)\,\log\chi\\
&+\tfrac{3 \left(\chi ^2-\chi +1\right)}{2 (\chi -1)}\,\log(1-\chi)+\tfrac{3 \chi  (15 \chi -1)}{2 (\chi -1)}\,,\\
f_{q,4}^{(3)}(\chi)&=\frac{15 \chi ^2}{2 (\chi -1)}\,.
\end{split}
\end{align}

To express the result for $\langle \mathcal{D}_2\mathcal{D}_2\mathcal{D}_2\mathcal{D}_2\rangle$ at two loops it is convenient to first solve the Ward identities, which in this case can be done in terms of one constant and three functions of $\chi$. Explicitly, we have
\begin{align}
\frac{\langle \mathcal{D}_2\mathcal{D}_2\mathcal{D}_2\mathcal{D}_2\rangle}{\langle \mathcal{D}_2\mathcal{D}_2\rangle\langle\mathcal{D}_2\mathcal{D}_2\rangle}=\mathsf{f}_2\,\frac{\chi^2}{\zeta_1\zeta_2}+\mathbb{D}_1\,f_1(\chi)+\mathbb{D}_2\,f_2(\chi)+\mathbb{D}_3\,f_3(\chi)\,,
\end{align}
where in terms of the OPE coefficients $\mathsf{C}_{pqr}\sim \langle\mathcal{D}_p\mathcal{D}_q\mathcal{D}_r\rangle$
\begin{align}
\mathsf{f}_2=1+\mathsf{C}_{222}+\mathsf{C}_{224}\,,
\end{align}
while the differential operators $\mathbb{D}_i$ are given by
\begin{align}
\begin{split}
\mathbb{D}_1=\chi^2\,\left(\bar{v}-v_1\,v_2\,\chi\,\partial_{\chi}\right)\,,\qquad
\mathbb{D}_2=\frac{\chi^2}{\zeta_1\,\zeta_2}\,\mathbb{D}_1\,,\qquad 
\mathbb{D}_3=\chi^5\,v_1\,v_2\,,
\end{split}
\end{align}
where $v_i=\chi^{-1}-\zeta_i^{-1}$ as in \eqref{D_1111}, while $\bar{v}=\chi^{-2}-\zeta_1^{-1}\,\zeta_2^{-1}$. At two loops we have
\begin{align}
\mathsf{f}^{(3)}_2=\frac{1341}{4}\,,
\end{align}
while
\begin{align}
\begin{split}
f_1^{(3)}(\chi)&=-\tfrac{2 \left(8 \chi ^7-41 \chi ^6+77 \chi ^5-49 \chi ^4-3 \chi ^3+29 \chi ^2-13 \chi +1\right) }{(\chi -1)^4 \chi }L_3(\chi)\\
&-\tfrac{2 \left(8 \chi ^7-15 \chi ^6-\chi ^5-\chi ^4+31 \chi ^3-49 \chi ^2+35 \chi -9\right)}{(\chi -1)^3 \chi ^2}\,L_3(1-\chi)\\
&-\tfrac{2\chi^2  \left(21 \chi ^6-137 \chi ^5+374 \chi ^4-556 \chi ^3+480 \chi ^2-337 \chi +50\right) }{3 (\chi -1)^5}\log ^3\chi \\
&+\tfrac{\left(2 \left(63 \chi ^8-305 \chi ^7+584 \chi ^6-557 \chi ^5+265 \chi ^4+6 \chi ^3-53 \chi ^2+13 \chi -1\right)\right) }{3 (\chi -1)^4 \chi }\log ^2\chi \log (1-\chi )\\
&-\tfrac{2 \left(63 \chi ^8-199 \chi ^7+213 \chi ^6-70 \chi ^5-25 \chi ^4-29 \chi ^3+104 \chi ^2-73 \chi +15\right) }{3 (\chi -1)^3 \chi ^2} \log \chi \log ^2(1-\chi )\\
&+\tfrac{2 \left(21 \chi ^6+11 \chi ^5+4 \chi ^4+10 \chi ^3+\chi ^2+108 \chi -105\right) }{3 \chi ^3}\log ^3(1-\chi )\\
&+\tfrac{\chi ^2 \left(78 \chi ^4-350 \chi ^3+604 \chi ^2-569 \chi +114\right)}{(\chi -1)^4} \log ^2\chi +\tfrac{\left(78 \chi ^4+38 \chi ^3+22 \chi ^2+99 \chi -123\right) }{\chi ^2}\log ^2(1-\chi )\\
&-\tfrac{\left(156 \chi ^6-468 \chi ^5+474 \chi ^4-168 \chi ^3+53 \chi ^2-47 \chi +3\right)}{(\chi -1)^3 \chi }\log\chi \log (1-\chi )\\
&+\tfrac{\left(68 \chi ^4-88 \chi ^3-89 \chi ^2+154 \chi -57\right) }{2 (\chi -1)^2 \chi }\log (1-\chi )-\tfrac{\left(68 \chi ^4-184 \chi ^3+55 \chi ^2+16 \chi -12\right) }{2 (\chi -1)^3}\log \chi \\
&+\tfrac{2 \left(40 \chi ^4-80 \chi ^3+84 \chi ^2-44 \chi +9\right)}{(\chi -1)^4 \chi ^2}\,\zeta(3)+\tfrac{535 \chi ^2-535 \chi +88}{4 (\chi -1)^2}\,,
\end{split}
\end{align}
\begin{align}
\begin{split}
f_2^{(3)}(\chi)&=-\tfrac{2 (\chi -2) \left(9 \chi ^6-27 \chi ^5+43 \chi ^4-23 \chi ^3-11 \chi ^2+27 \chi -9\right)}{(\chi -1)^4 \chi }L_3(\chi)\\
&-\tfrac{2 \left(9 \chi ^7-35 \chi ^6+49 \chi ^5-31 \chi ^4+\chi ^3+\chi ^2+15 \chi -8\right)}{(\chi -1)^3 \chi ^2}L_3(1-\chi)\\
&-\tfrac{2 \chi ^2 \left(105 \chi ^6-648 \chi ^5+1693 \chi ^4-2414 \chi ^3+2017 \chi ^2-972 \chi +324\right)}{3 (\chi -1)^5} \log ^3\chi\\
& +\tfrac{2 \left(315 \chi ^8-1614 \chi ^7+3360 \chi ^6-3589 \chi ^5+2020 \chi ^4-491 \chi ^3-61 \chi ^2+63 \chi -18\right)}{3 (\chi -1)^4 \chi }  \log ^2\chi \log (1-\chi )\\
&-\tfrac{\left(2 \left(315 \chi ^8-1284 \chi ^7+1987 \chi ^6-1412 \chi ^5+422 \chi ^4-59 \chi ^3+100 \chi ^2-111 \chi +43\right)\right) }{3 (\chi -1)^3 \chi ^2}\log \chi  \log ^2(1-\chi )\\
&+\tfrac{2 \left(105 \chi ^6-108 \chi ^5-\chi ^4-10 \chi ^3-4 \chi ^2-11 \chi -21\right)}{3 \chi ^3} \log ^3(1-\chi )\\
&+\tfrac{\chi ^2 \left(123 \chi ^4-594 \chi ^3+1162 \chi ^2-1136 \chi +568\right)}{(\chi -1)^4} \log ^2\chi\\
& -\tfrac{\left(246 \chi ^6-939 \chi ^5+1335 \chi ^4-837 \chi ^3+306 \chi ^2-190 \chi +76\right) }{(\chi -1)^3 \chi } \log \chi \log (1-\chi )\\
&+\tfrac{\left(123 \chi ^4-99 \chi ^3-22 \chi ^2-38 \chi -78\right)}{\chi ^2} \log ^2(1-\chi )\\
&-\tfrac{\left(57 \chi ^4-199 \chi ^3+227 \chi ^2-56 \chi +28\right)}{2 (\chi -1)^3} \log \chi +\tfrac{\left(57 \chi ^4-154 \chi ^3+89 \chi ^2+88 \chi -68\right)}{2 (\chi -1)^2 \chi } \log (1-\chi )\\
&-\tfrac{2 \left(8 \chi ^4-14 \chi ^2+23 \chi -8\right)}{(\chi -1)^4 \chi ^2}\zeta(3)+\tfrac{1253 \chi ^2-2147 \chi +806}{4 (\chi -1)^2}\,,
\end{split}
\end{align}
\begin{align}
\begin{split}
f_3^{(3)}(\chi)&=-\tfrac{2 \left(2 \chi ^8-18 \chi ^7+64 \chi ^6-116 \chi ^5+126 \chi ^4-76 \chi ^3+70 \chi ^2-36 \chi +9\right)}{(\chi -1)^6 \chi }L_3(\chi)\\
&-\tfrac{2 (\chi +1) \left(2 \chi ^8-18 \chi ^7+57 \chi ^6-98 \chi ^5+113 \chi ^4-98 \chi ^3+57 \chi ^2-18 \chi +2\right)}{(\chi -1)^5 \chi ^3}L_3(1-\chi)\\
&+\tfrac{2 \chi ^2 \left(30 \chi ^7-166 \chi ^6+370 \chi ^5-387 \chi ^4+119 \chi ^3+167 \chi ^2-7 \chi +324\right)}{3 (\chi -1)^7} \log ^3\chi\\
& -\tfrac{2 \left(90 \chi ^9-403 \chi ^8+720 \chi ^7-647 \chi ^6+346 \chi ^5-171 \chi ^4+38 \chi ^3-20 \chi ^2-12 \chi +9\right) }{3 (\chi -1)^6 \chi } \log ^2\chi \log (1-\chi )\\
& +\tfrac{2 \left(90 \chi ^{10}-308 \chi ^9+430 \chi ^8-396 \chi ^7+386 \chi ^6-375 \chi ^5+333 \chi ^4-277 \chi ^3+147 \chi ^2-23 \chi -5\right)}{3 (\chi -1)^5 \chi ^3} \log \chi \log ^2(1-\chi ) \\
&-\tfrac{2 \left(30 \chi ^8-11 \chi ^7+28 \chi ^6-34 \chi ^5+24 \chi ^4-34 \chi ^3+28 \chi ^2-11 \chi +30\right) }{3 (\chi -1)^2 \chi ^4}\log ^3(1-\chi )\\
&-\tfrac{\chi  \left(50 \chi ^6-138 \chi ^5+7 \chi ^4+420 \chi ^3-558 \chi ^2+848 \chi +16\right)}{(\chi -1)^6} \log ^2\chi\\
& +\tfrac{\left(100 \chi ^8-168 \chi ^7-108 \chi ^6+340 \chi ^5+7 \chi ^4-477 \chi ^3+338 \chi ^2-54 \chi -8\right)}{(\chi -1)^5 \chi ^2}  \log \chi \log (1-\chi )\\
&-\tfrac{50 \chi ^6+70 \chi ^5+33 \chi ^4-60 \chi ^3+33 \chi ^2+70 \chi +50}{(\chi -1)^2 \chi ^3}\log ^2(1-\chi )\\
&+\tfrac{\left(88 \chi ^6-46 \chi ^5-276 \chi ^4+619 \chi ^3+375 \chi ^2-52 \chi -8\right) }{2 (\chi -1)^5 \chi }\log (\chi )\\
&-\tfrac{\left(44 \chi ^6+25 \chi ^5-87 \chi ^4+35 \chi ^3-87 \chi ^2+25 \chi +44\right)}{(\chi -1)^4 \chi ^2} \log (1-\chi )\\
&+\tfrac{2 \left(40 \chi ^5-56 \chi ^4+80 \chi ^3-55 \chi ^2+18 \chi -2\right)}{(\chi -1)^6 \chi ^3}\zeta(3)-\tfrac{26 \chi ^4-4 \chi ^3-17 \chi ^2-4 \chi +26}{2 (\chi -1)^4 \chi }\,.
\end{split}
\end{align}
We would like to make a comment on a subtlety concerning this result, which it would be interesting to clarify in the future. As usual, the bootstrap algorithm leaves a number of free parameters corresponding to higher-derivative contact terms, and we can constrain these parameters by analyzing the Regge limit\footnote{See also \cite{Bliard:2023zpe} for similar observations in a related context.}. In particular, we require that 
\begin{align}\label{2222largeDeltagamma3}
\langle\gamma^{(3)}\rangle_{\Delta}\sim \#\Delta^4\,,\quad (\Delta\to \infty)\,,
\end{align}
for all three exchanged representations ($\{0,[0,0]\}$, $\{0,[0,2]\}$ and $\{0,[2,0]\}$), consistently with the result obtained for $\langle\mathcal{D}_1\mathcal{D}_1\mathcal{D}_1\mathcal{D}_1\rangle^{(3)}$. However, simply requiring that $\Delta^5$ and higher are absent still leaves us with two unfixed parameters. It turns out that requiring the coefficient of $\Delta^4$ in \eqref{2222largeDeltagamma3} to be the same for any pair of exchanged representations fixes an additional parameter, and moreover gives
\begin{align}
\langle\gamma^{(3)}_{0,[0,0]}\rangle_{\Delta}\sim\langle\gamma^{(3)}_{0,[0,2]}\rangle_{\Delta}\sim\langle\gamma^{(3)}_{0,[2,0]}\rangle_{\Delta}\sim\frac{\pi^2-30}{48} \Delta^4\,,\quad (\Delta\to \infty)\,,
\end{align}
that is only setting two of them equal makes all three representations equal, and moreover the coefficient of $\Delta^4$ is the same obtained in the $\langle\gamma^{(3)}\rangle_{\Delta}$ obtained from $\langle\mathcal{D}_1\mathcal{D}_1\mathcal{D}_1\mathcal{D}_1\rangle^{(3)}$. We take this as a strong indication that this constraint is indeed a property of the theory, although it would be interesting to better understand its origin. At this point we are left with only one parameter, which we can also fix using a somewhat speculative procedure. If one focuses on the terms in $f^{(3)}_i(\chi)$ of transcendentality one or zero, which are the ones affected by the ambiguity, it is possible to observe that there is a special choice of the remaining free parameters such that, for all three functions: 1) the coefficient of $\log \chi$ is less singular for $\chi\to 1$, 2) the coefficient of $\log(1-\chi)$ is less singular for $\chi \to 0$, and 3) the rational term is less singular for both $\chi \to 0$ and $\chi \to 1$. By ``less singular'' we mean that such a choice for the free parameter reduces the order of the pole of the rational functions by at least one unit. While this would of course require a detailed justification, it is highly non trivial that just by fixing one parameter one obtains these many independent simplifications. We then take this as our final constraint and obtain the results quoted above. It is interesting to observe that we have a non trivial check for our result: from the correlator above it is possible to extract the average $\langle a^{(0)}\gamma^{(3)}+a^{(1)}\gamma^{(2)}\rangle_{\Delta=4}$, which we use at the end of Section \ref{sec:mixing} in order to study the mixing problem at third order for singlet operators of free theory dimension $\Delta=4$. The upshot of that analysis is the third order anomalous dimension for the exact eigenstates of the dilatation operator, which can then be compared with a fit on the results obtain from integrability around $\lambda=\infty$ \cite{Cavaglia:2021bnz,Cavaglia:2022qpg}. We find perfect agreement.

\section{The mixing problem}\label{sec:mixing}

Let us now turn to the main technical novelty of this paper: the study of the mixing problem at second order (one loop). We remind the reader that in Section \ref{sec:1111} we have bootstrapped the four-point function of the displacement operator using the strategy of Section \ref{sec:bootstrap}, which uses the knowledge of CFT data at previous perturbative orders to compute the coefficients of the highest logarithmic singularities of a certain correlator (the equivalent of the double discontinuity in higher dimensions). However, as it is often the case in perturbative CFTs, extracting CFT data from a single correlator turns out to be insufficient, due to the presence of degeneracy between operators in the free theory. We have anticipated some general considerations on the mixing problem in Section \ref{sec:mixing_bootstrap}, while in this section we will consider the specific problem of interest: that of computing the average
\begin{align}\label{gamma2^2_mixing}
\left.\langle a^{(0)}\,(\gamma^{(2)})^2\rangle_{\D}\right|_{\langle \mathcal{D}_1\mathcal{D}_1\mathcal{D}_1\mathcal{D}_1\rangle}\equiv \sum_{\mathcal{O}\,|\,\h^{(0)}_{\mathcal{O}}=\D}(\mu^{(0)}_{\mathcal{O}})^2\,(\gamma^{(2)}_{\mathcal{O}})^2\,,
\end{align}
for singlet long multiplets, where $\mathcal{O}$ are exact eigenstates of the dilatation operator whose dimension at $\lambda=\infty$ is $\D$, while $\mu^{(0)}_{\mathcal{O}}$ are the OPE coefficients in the $\mathcal{D}_1\times \mathcal{D}_1$ OPE in the free theory. This gives the quantity \eqref{gamma2squared}, which we have already used in Section \ref{sec:1111} and that we will be derived in this section.

\subsection{The length basis}

While \eqref{gamma2^2_mixing} is the most intuitive way of expressing the quantity of interest, as it involves eigenstates of the dilatation operator (the operators $\mathcal{O}$) and their eigenvalues ($\gamma^{(2)}_{\mathcal{O}}$), it is not necessarily the most useful. In particular, at this stage we do not have much information on the dilatation operator at one loop, and certainly we do not know how to diagonalize it. It is therefore convenient to choose a different basis, which we can construct explicitly, and since we are working perturbatively in $1/\sqrt{\lambda}$ we construct such a basis at $\lambda=\infty$, where the theory is free. We remind the reader that the operators we are interested in are the superconformal primaries of long, singlet superconformal multiplets, and in \cite{Ferrero:2023znz} we discuss precisely how such operators can be constructed from the fundamental fields in the AdS Lagrangian description. In particular we will choose a basis of operators with fixed length $L$: as we discuss later, this will greatly simplify our task. 

To be more precise, for each fixed $\D$ the operators $\mathcal{O}$ that appear in \eqref{gamma2^2_mixing} belong to the degeneracy space $\mathtt{d}(\mathcal{L}^{\D}_{0,[0,0]})$ of long singlet supermultiplets whose free theory conformal dimension is $\h^{(0)}=\D$. For brevity, let us from now on denote $ \mathtt{d}(\D)\equiv\mathtt{d}(\mathcal{L}^{\D}_{0,[0,0]})$, since we will only deal with singlets in what follows. As discussed in \ref{sec:setup} and more at length in \cite{Ferrero:2023znz}, it is convenient to organize such space as
\begin{align}
\mathtt{d}(\D)=\bigoplus_{L=2}^{\infty}\mathtt{d}_L(\D)\,,
\end{align}
where each $\mathtt{d}_L(\D)$ is a vector space whose dimension for large $\D$ grows as \cite{Ferrero:2023znz}
\begin{align}
\text{dim}\left[\mathtt{d}_L(\D)\right]\sim \D^{L-2}\,,\qquad (\D\to \infty)\,.
\end{align}
We then choose a basis of operators $\widehat{\mathcal{O}}$ of fixed $L$, namely each of them belongs to a given $\mathtt{d}_L(\D)$, while the choice of basis within $\mathtt{d}_L(\D)$ is completely arbitrary. Such basis is related to the basis of eigenstates $\mathcal{O}$ of \eqref{gamma2^2_mixing} by 
\begin{align}
\widehat{\mathcal{O}}_{\alpha}=M_{\alpha}^{\,\,\,\beta}\,\mathcal{O}_{\beta}\,,
\end{align}
where the indices $\alpha,\beta=1,\ldots \text{dim}\left[\mathtt{d}(\D)\right]$ span the degeneracy space $\mathtt{d}(\D)$ and $M\in GL( \text{dim}\left[\mathtt{d}(\D)\right])$. Note that $\widehat{\mathcal{O}}$ need not be normalized and one can introduce a metric tensor 
\begin{align}
\mathtt{g}:\,\,\mathtt{d}(\D)\times \mathtt{d}(\D)\to \mathbb{R}\,,\qquad \mathtt{g}_{\alpha\beta}\equiv \langle \widehat{\mathcal{O}}_{\alpha}\,|\,\widehat{\mathcal{O}}_{\beta}\rangle^{(0)}\,,
\end{align}
containing the two-point functions computed in the free theory. Since obviously operators of different length are orthogonal in the free theory, one also has a block-diagonal decomposition
\begin{align}
\mathtt{g}_L=\bigoplus_L \mathtt{g}_L\,,\qquad \mathtt{g}_L:\,\,\mathtt{d}_L(\D)\times \mathtt{d}_L(\D)\to \mathbb{R}\,.
\end{align}
The metric tensors $\mathtt{g}_L$ can be easily computed by Wick contractions in the free theory, once a basis of operators is found with the methods described in \cite{Ferrero:2023znz}. Following completely analogous steps to those of Section \ref{sec:mixing_bootstrap}, one can then express the various averages of interest in terms of an arbitrary basis of operators. For an arbitrary four-point function $\langle S_1S_2S_3S_4\rangle$, focusing on the exchange of singlets, we have
\begin{align}
\begin{split}
\left.\langle a^{(0)}\rangle_{\D}\right|_{\langle S_1S_2S_3S_4\rangle}&=\mu^{(0)}_{12\alpha}\,(\mathtt{g}_{\D}^{-1})^{\alpha\beta}\,\mu^{(0)}_{12\alpha}\,,\\
\left.\langle a^{(0)}\,\gamma^{(2)}\rangle_{\D}\right|_{\langle S_1S_2S_3S_4\rangle}&=\mu^{(0)}_{12\alpha}\,(\Gamma^{(2)}_{\D})^{\alpha\beta}\,\mu^{(0)}_{12\alpha}\,,\\
\left.\langle a^{(0)}\,(\gamma^{(2)})^2\rangle_{\D}\right|_{\langle S_1S_2S_3S_4\rangle}&=\mu^{(0)}_{12\alpha}\,(\Gamma^{(2)}_{\D})^{\alpha\beta}\,(\mathtt{g}_{\D})_{\beta\gamma}(\Gamma^{(2)}_{\D})^{\gamma\delta}\,\,\mu^{(0)}_{12\delta}\,.
\end{split}
\end{align}
$\Gamma^{(2)}$ is a representation of the one-loop dilatation operator in the chosen basis, restricted to the space of exchanged operators: we shall refer to it as the anomalous dimensions matrix and its eigenvectors and eigenvalues are the pairs $\{\mathcal{O},\gamma^{(2)}_{\mathcal{O}}\}$ of \eqref{gamma2^2_mixing}. For arbitrary operators $S_i$ in the representation $\mathcal{R}_i$ of $\mathfrak{osp}(4^*|4)$ we have defined the OPE coefficients in the free theory
\begin{align}
\mu^{(0)}_{S_iS_jS_k}:\,\,\mathtt{d}(\mathcal{R}_i)\times \mathtt{d}(\mathcal{R}_j)\times \mathtt{d}(\mathcal{R}_k)\to \mathbb{C}^{\#}\,,\qquad \mu^{(0)}_{S_iS_jS_k}=\langle S_i S_j S_k\rangle\,,
\end{align}
where $\#$ denotes the number of invariant structures of the type $\langle\mathcal{R}_i \mathcal{R}_j\mathcal{R}_k\rangle$: only $\#=1$ is relevant in this work, such that each $\mu^{(0)}$ is simply a number. 

As anticipated, while one could in principle work with any arbitrary basis for $\mathtt{d}(\D)$, we find it convenient to work in a basis of operators with definite length, which we will consider from now on. This is due to the observations of Section \ref{sec:dilatationoperator} on the structure of the dilatation operator in this model: we have proved that at one loop the dilatation operator only connects operators whose length differs at most by two units. In terms of the anomalous dimensions matrix $\Gamma^{(2)}$, this implies the structure 
\begin{align}\label{Gamma2blocks}
\Gamma^{(2)}_{\D}=
\begin{pmatrix}
\Gamma^{(2)}_{\D,2\to 2} & \Gamma^{(2)}_{\D,2\to 4} & 0 & 0 &\,\,\, \ldots & \quad 0\\
(\Gamma^{(2)}_{\D,2\to 4})^T & \Gamma^{(2)}_{\D,4\to 4} & \Gamma^{(2)}_{\D,4\to 6} & 0 & \,\,\, \ldots & \quad 0\\
0 & (\Gamma^{(2)}_{\D,4\to 6})^T & \Gamma^{(2)}_{\D,6\to 6} & \Gamma^{(2)}_{\D,6\to 8} &  \,\,\,\ldots & \quad 0\\
0 & 0 & (\Gamma^{(2)}_{\D,6\to 8})^T & \Gamma^{(2)}_{\D,8\to 8}&  \,\,\,\ldots & \quad 0\\
\ldots & \ldots& \ldots& \ldots&\,\,\, \ldots &\quad \ldots
\end{pmatrix}\,,
\end{align}
where
\begin{align}
\Gamma^{(2)}_{\D,L_1\to L_2}:\,\,\,\mathtt{d}_{L_1}(\D)\to \mathtt{d}_{L_2}(\D) \,.
\end{align}
Now that we have understood the general structure, let us go back at the object of interest, namely \eqref{gamma2squared}, expressing it in this basis. A crucial fact is that in the $\mathcal{D}_1\times \mathcal{D}_1$ OPE in the free theory only operators with $L=2$ can appear, or in other words
\begin{align}\label{11exchangesL=2}
\mu^{(0)}_{\mathcal{D}_1\mathcal{D}_1\widehat{\mathcal{O}}^{\D}_{L,\alpha}}\propto \delta_{L,2}\,.
\end{align}
Moreover, we learn from \cite{Ferrero:2023znz} that 
\begin{align}
\text{dim}[\mathtt{d}_{L=2}(\D)]=1\,,
\end{align}
that is the length two operators are non-degenerate. Since these are the only operators exchanged in $\langle \mathcal{D}_1\mathcal{D}_1\mathcal{D}_1\mathcal{D}_1\rangle^{(0)}$, we also have that
\begin{align}
(\mu^{(0)}_{\mathcal{D}_1\mathcal{D}_1\widehat{\mathcal{O}}_{L=2}^{\D}})^2=\left. \langle a^{(0)}\rangle_{\D}\right|_{\langle \mathcal{D}_1\mathcal{D}_1\mathcal{D}_1\mathcal{D}_1\rangle}=\frac{\Gamma[3+\D]\Gamma[1+\D](\D-1)}{\Gamma[2+2\D]}\,,
\end{align}
which is just the quantity \eqref{a0free1111} (note that in Section \ref{sec:1111} all the averages are taken over $\langle \mathcal{D}_1\mathcal{D}_1\mathcal{D}_1\mathcal{D}_1\rangle$).We then have
\begin{align}\label{gamma2^2_mixing_L24}
\left.\langle (\gamma^{(2)})^2\rangle\right|_{\langle \mathcal{D}_1\mathcal{D}_1\mathcal{D}_1\mathcal{D}_1\rangle}=\frac{\left.\langle a^{(0)}\,(\gamma^{(2)})^2\rangle_{\D}\right|_{\langle \mathcal{D}_1\mathcal{D}_1\mathcal{D}_1\mathcal{D}_1\rangle}}{\left. \langle a^{(0)}\rangle_{\D}\right|_{\langle \mathcal{D}_1\mathcal{D}_1\mathcal{D}_1\mathcal{D}_1\rangle}}
=\left(\Gamma^{(2)}_{\D,2\to 2}\right)^2+\,\delta \Gamma^{(2)}_{\mathtt{sq}}(\D)\,,
\end{align}
where
\begin{align}
\delta \Gamma^{(2)}_{\mathtt{sq}}(\D)=\Gamma^{(2)}_{\D,2\to 4}\cdot \mathtt{g}_4\cdot \Gamma^{(2)}_{\D,2\to 4}\,.
\end{align}
Moreover, still using the fact that only operators with $L=2$ are exchanged in the $\mathcal{D}_1\times \mathcal{D}_1$ OPE, we also have that 
\begin{align}
\left.\langle \gamma^{(2)}\rangle\right|_{\langle \mathcal{D}_1\mathcal{D}_1\mathcal{D}_1\mathcal{D}_1\rangle}=\frac{\left.\langle a^{(0)}\,\gamma^{(2)}\rangle_{\D}\right|_{\langle \mathcal{D}_1\mathcal{D}_1\mathcal{D}_1\mathcal{D}_1\rangle}}{\left. \langle a^{(0)}\rangle_{\D}\right|_{\langle \mathcal{D}_1\mathcal{D}_1\mathcal{D}_1\mathcal{D}_1\rangle}}
=\Gamma^{(2)}_{\D,2\to 2}\,,
\end{align}
so that from \eqref{1loopCFTdata1111} we immediately read off
\begin{align}
\Gamma^{(2)}_{\D,2\to 2}=\gamma^{(1)}_{\D}\,\partial_{\D}\gamma^{(1)}_{\D}+\frac{\JJ}{8}\left(-11-\frac{6}{\JJ+2}+4\,H_{1+\D}\right)\,,
\end{align}
where $\gamma^{(1)}_{\D}=-\tfrac{1}{2}\JJ=-\tfrac{1}{2}\D(\D+3)$. We have therefore achieved a great simplification compared to the initial problem: in order to extract the quantity \eqref{gamma2^2_mixing} of interest, we only have to compute the vectors $\Gamma^{(2)}_{\D,2\to 4}$. Note that only even values of $\D$ appear in the OPE, so that the length of the vectors $\Gamma^{(2)}_{\D,2\to 4}$ is determined by \cite{Ferrero:2023znz}
\begin{align}
\text{dim}[\mathtt{d}_{L=4}(\D)]=\lfloor (\Delta/4)^2\rfloor\,,\qquad (\Delta\,\,\,\text{even})\,.
\end{align}
Thus, the number of operators that contributes to the mixing problem grows {\it quadratically} with their conformal dimension: this can be contrasted with other cases where mixing problems have been successfully solved, such as those arising in the study of holographic correlators\footnote{In the case of the $\epsilon$-expansion \cite{Alday:2017zzv} one makes an ansatz of pure transcendentality, with only a handful of parameters to fix, so that it is enough to consider the mixing problem only for low dimension of the exchanged operators.}. In all those cases \cite{Aprile:2017bgs,Alday:2017xua,Alday:2020tgi,Alday:2021ajh,Behan:2022uqr,Alday:2022rly} one is interested in the mixing between ``double-trace'' operators, and the growth of the degeneracy is only linear in the dimension. This makes our problem considerably harder. 

\subsection{The strategy}

Having identified the quantity of interest, let us discuss how it can be computed. The common strategy that is adopted in the literature is that of bootstrapping/computing families of correlators, extracting averaged CFT data from each one, and computing the relevant quantities once enough averages are available. More precisely, from a one-loop correlator $\langle S_1S_2S_3S_4\rangle$ such that singlet supermultiplets appear both in $S_1\times S_2$ and in $S_3\times S_4$, one extracts the average
\begin{align}\label{a0gamma2_general}
\left.\langle a^{(0)}\,\gamma^{(2)}\rangle_{\D}\right|_{\langle S_1S_2S_3S_4\rangle}&=\mu^{(0)}_{12\alpha}\,(\Gamma^{(2)}_{\D})^{\alpha\beta}\,\mu^{(0)}_{12\alpha}\,.
\end{align}
The OPE coefficients can be computed in the free theory once an explicit basis of operators is chosen. By varying the external operators one computes, through the quantities \eqref{a0gamma2_general}, linear combinations of the entries of $\Gamma^{(2)}_{\D}$: so long as these are linearly independent from each other, for each $\D$ a finite number of averages \eqref{a0gamma2_general} will be enough to extract all the entries of the matrix. The question of linear independence is not an irrelevant one, and is related to whether the free theory OPE coefficients span, when varying the external operators within the chosen family, all the directions in the degeneracy space $\mathtt{d}(\Delta)$.

This work will be no exception from this point of view, but we would like to stress again an important difference with the double-trace mixing encountered in the study of holographic correlators. In that context, it is enough to consider four-point functions between half-BPS operators or arbitrary weight, which are dual to KK modes of fundamental fields on the internal space. This would correspond to studying $\langle \mathcal{D}_p\mathcal{D}_p\mathcal{D}_q\mathcal{D}_q\rangle$ correlators at one loop\footnote{Note that singlet supermultiplets appear in the OPE $\mathcal{D}_{k_1}\times \mathcal{D}_{k_2}$ only for $k_1=k_2$, so our restriction to pairwise equal weights still gives access to the same amount of data as $\langle \mathcal{D}_{k_1}\mathcal{D}_{k_2}\mathcal{D}_{k_3}\mathcal{D}_{k_4}\rangle$, at least so long as singlets are concerned.}, which was done in the previous section and was our initial motivation to study that specific set of observables. However, as we discussed in Section \ref{sec:ppqq}, the situation here is very different: as opposed to KK modes, the operators $\mathcal{D}_p$ are not independent from each other but rather obtained as powers of $\mathcal{D}_1$. This reflects into the considerations around \eqref{C_pplong_Wickproperty}: consider the three-point function $\langle \mathcal{D}_{p}\mathcal{D}_{p} \widehat{\mathcal{O}}_{L}\rangle^{(0)}$, where $\widehat{\mathcal{O}}_{L}$ is an element of $\mathtt{d}_L(\D)$ (hence we can think of it as a vector). This vanishes for $p<L/2$, as one can easily see from Wick contractions,while for $p\ge L/2$, regardless of the precise definition of $\widehat{\mathcal{O}}_{L}$, one has
\begin{align}
\langle \mathcal{D}_{p}\mathcal{D}_{p} \widehat{\mathcal{O}}_{L}\rangle^{(0)}=\binom{p}{L/2}\,\langle \mathcal{D}_{L/2}\mathcal{D}_{L/2} \widehat{\mathcal{O}}_{L}\rangle^{(0)}\,.
\end{align}
One can then choose a basis in each $\mathtt{d}_L(\D)$ such that $\langle \mathcal{D}_{L/2}\mathcal{D}_{L/2} \widehat{\mathcal{O}}_{L}\rangle^{(0)}$ is only non-zero for one specific operator, and the same will be true for all $\langle \mathcal{D}_{p}\mathcal{D}_{p} \widehat{\mathcal{O}}_{L}\rangle^{(0)}$. Hence, $\langle \mathcal{D}_p\mathcal{D}_p\mathcal{D}_q\mathcal{D}_q\rangle$ correlators only give access to a one-dimensional subspace of $\mathtt{d}_L(\D)$ for each $L$, and therefore cannot be used to extract the whole vector $\Gamma^{(2)}_{\D,2\to 4}$: one can use them to extract exactly one entry, which is only enough for $\D=4$ but already insufficient at $\D=6$.

To overcome this problem, one has to consider a different family of correlators where at least one of the external operators belongs to a long multiplet: since there are no semi-short representations at strong coupling, this is the only alternative to half-BPS multiplets. In particular, we focus on the following class of observables:
\begin{align}\label{112L}
\langle \mathcal{D}_1\mathcal{D}_1\mathcal{D}_2\mathcal{L}_{\ext}\rangle\equiv\langle \mathcal{D}_1\mathcal{D}_1\mathcal{D}_2\mathcal{L}^{\hExt}_{0,[0,0]}\rangle\,,
\end{align}
which we bootstrap up to one loop still using an adaptation of the method of Section \ref{sec:bootstrap}, as we discuss in the next subsection. Let us first explain how we use \eqref{112L} to compute $\Gamma^{(2)}_{\D,2\to 4}$. First, note that in the direct channel OPE one has 
\begin{align}
\mathcal{D}_1\times \mathcal{D}_1=\mathcal{I}\oplus \mathcal{D}_2\oplus \sum_\h \mathcal{L}^{\h}_{0,[0,0]}\,,\qquad 
\mathcal{D}_2\times \mathcal{L}^{\hExt}_{0,[0,0]}=\mathcal{D}_2\oplus \sum_\h \mathcal{L}^{\h}_{0,[0,0]}\oplus \ldots\,,
\end{align}
where in the second expression we have suppressed terms that do not contribute to \eqref{112L}, so that the exchanged operators in \eqref{112L} in the direct channel are the same as those exchanged in $\langle \mathcal{D}_1\mathcal{D}_1\mathcal{D}_1\mathcal{D}_1\rangle$. The only caveat here is that the exchanged long operator in $\mathcal{D}_2\times \mathcal{L}^{\hExt}_{0,[0,0]}$ is not the superprimary of $\mathcal{L}^{\h}_{0,[0,0]}$, but rather its (unique!) descendant in the representation $[0,2]^{\h}_0$, see \cite{Ferrero:2023znz}. We also note that since the three-point function $\langle \mathcal{D}_1\mathcal{D}_1\mathcal{L}^{\h}_{0,[0,0]}\rangle$ is completely fixed in terms of a single number by superconformal symmetry, we have (see \cite{Ferrero:2023znz} for a derivation and more comments)
\begin{align}
\mu^{(0)}_{\mathcal{D}_1\mathcal{D}_1\widetilde{\mathcal{O}}^{\h}}=\h(\h+1)\,\mu^{(0)}_{\mathcal{D}_1\mathcal{D}_1\mathcal{O}^{\h}}\,,
\end{align}
where following the notation of \cite{Ferrero:2023znz} we have denoted by $\mathcal{O}^{\h}$ the superconformal primary of the relevant $\mathcal{L}^{\h}_{0,[0,0]}$ supermultiplet and by $\widetilde{\mathcal{O}}^{\h}$ its descendant in the representation $[0,2]^{\h}_0$. Then, say we are able to bootstrap \eqref{112L} at one loop for a certain choice of $\mathcal{L}_{\ext}$: this allows to extract averaged anomalous dimensions which, using again \eqref{11exchangesL=2} and the structure \eqref{Gamma2blocks} can be expressed as
\begin{align}\label{a0gamma2_112L}
\left.\langle a^{(0)}\,\gamma^{(2)}\rangle_{\D}\right|_{\langle \mathcal{D}_1\mathcal{D}_1\mathcal{D}_2\mathcal{L}_{\ext}\rangle}=\left.\langle a^{(0)}\rangle_{\D}\right|_{\langle \mathcal{D}_1\mathcal{D}_1\mathcal{D}_2\mathcal{L}_{\ext}\rangle}\,\Gamma^{(2)}_{\D,2\to 2}+\mu^{(0)}_{\mathcal{D}_1\mathcal{D}_1\widehat{\mathcal{O}}_{L=2}^{\D}}\,\mathsf{X}_{\D,\mathcal{L}_\ext}\,,
\end{align}
where $\mu^{(0)}_{\mathcal{D}_1\mathcal{D}_1\widehat{\mathcal{O}}_{L=2}^{\D}}$ is the three-point function coefficient between two $\mathcal{D}_1$ insertions and the {\it unique} operator of dimension $\D$ and $L=2$, while
\begin{align}\label{X_DeltaExt}
\mathsf{X}_{\D,\mathcal{L}_\ext}=\left(\Gamma^{(2)}_{\D,2\to 4}\right)\cdot \mu^{(0)}_{\mathcal{D}_2\mathcal{L}_\ext\widehat{\mathcal{O}}_{L=4}^{\D}}\,,
\end{align}
where $\mu^{(0)}_{\mathcal{D}_2\mathcal{L}_\ext\widehat{\mathcal{O}}_{L=4}^{\D}}$ should be thought of as a vector of OPE coefficients, whose entries are determined by all possible choices of $\widehat{\mathcal{O}}_{L=4}^{\D}\in \mathtt{d}_{L=4}(\D)$. Thus, for each choice of $\mathcal{L}_\ext$ one has access to a given linear combination of the entries of $\Gamma^{(2)}_{\D,2\to 4}$, defined by \eqref{X_DeltaExt}. The coefficients in the linear combination are determined by the OPE coefficients $\mu^{(0)}_{\mathcal{D}_2\mathcal{L}_\ext\widehat{\mathcal{O}}_{L=4}^{\D}}$ which, upon varying $\mathcal{L}_\ext$, determine linearly independent vectors, as it can be seen by computing such objects directly in the free theory using Wick contractions. Hence, as opposed to $\langle \mathcal{D}_p\mathcal{D}_p\mathcal{D}_q\mathcal{D}_q\rangle$, the correlators $\langle \mathcal{D}_1\mathcal{D}_1\mathcal{D}_2\mathcal{L}_{\ext}\rangle$ do provide enough information to compute the entries of $\Gamma^{(2)}_{\D,2\to 4}$.

As a concluding remark for this subsection, we would like to emphasize that the choice of $\langle \mathcal{D}_1\mathcal{D}_1\mathcal{D}_2\mathcal{L}^{\hExt}_{0,[0,0]}\rangle$ is arbitrary, as one could have chosen for instance $\langle \mathcal{D}_p\mathcal{D}_p\mathcal{D}_q\mathcal{L}^{\hExt}_{0,[0,q]}\rangle$, where moreover the component of the singlet exchanged supermultiplet is the superconformal primary and not one of its descendants. However, a crucial ingredient of the procedure streamlined above is the knowledge of the explicit form on the external and exchanged long operators in terms of free fields in AdS. This is what allows to compute the free theory correlators that provide the starting point of the bootstrap analysis of the next subsection, as well as allowing to compute all free theory OPE coefficients from Wick contractions. The construction of the superconformal primaries of $\mathcal{L}^{\h}_{0,[0,0]}$ multiplets and their $[0,2]^{\h}_0$ descendants, relevant for $\langle \mathcal{D}_1\mathcal{D}_1\mathcal{D}_2\mathcal{L}^{\hExt}_{0,[0,0]}\rangle$ is discussed in \cite{Ferrero:2023znz}. In particular, we are only interested in exchanged operators with $L=2,4$ and we will only consider $\mathcal{L}_\ext$ of the same length, which turns out to be sufficient for our purposes (note that  in the free theory $\langle \mathcal{D}_1\mathcal{D}_1\mathcal{D}_2\mathcal{L}_\ext\rangle$ vanishes when $L[\mathcal{L}_\ext]>4$). Similarly, we will only consider $\mathcal{L}_\ext$ of even dimension (in the free theory), since this is the relevant case for intermediate operators. On the other hand, the application of the procedure discussed above to $\langle \mathcal{D}_p\mathcal{D}_p\mathcal{D}_q\mathcal{L}^{\hExt}_{0,[0,q]}\rangle$ would require to construct independently both singlet and $\{0,[0,q]\}$ superconformal primaries.

\subsection{Bootstrapping $\langle \mathcal{D}_1\mathcal{D}_1\mathcal{D}_2\mathcal{L}_{\ext}\rangle$}

Let us now describe how the bootstrap algorithm presented in Section \ref{sec:bootstrap} is adapted to the case of $\langle \mathcal{D}_1\mathcal{D}_1\mathcal{D}_2\mathcal{L}_{\mathtt{ext}}\rangle$ correlators. The superconformal kinematics and blocks are discussed in \cite{Ferrero:2023znz}, here we will simply use the results. The correlator is determined in terms of a unique function $F(\chi)$ as
\begin{align}
\langle \mathcal{D}_1\mathcal{D}_1\mathcal{D}_2\mathcal{L}_{\mathtt{ext}}\rangle=\frac{(Y_1\cdot Y_3)(Y_2\cdot Y_3)}{t_{13}^2t_{23}^2}\left(\frac{t_{12}}{t_{14}t_{24}}\right)^{\hExt}F(\chi)\,,
\end{align} 
where $Y_i$ are polarization vectors for the operators $\mathcal{D}_1$ and $\mathcal{D}_2$, see \cite{Ferrero:2023znz} for more details. Since the four-point function is not crossing-symmetric, we defined two inequivalent OPE channels which again we refer to as direct ($d$) and crossed ($c$) channel, with associated behavior of the cross ratio and exchanged operators given by 
\begin{align}\label{112L_OPEchannels}
\begin{split}
\text{direct}\,\, (\chi\to 0)\,&:\quad \mathcal{D}_2\,\oplus\,\sum_{\h} \mathcal{L}^{\h}_{0,[0,0]}\,,\\
\text{crossed}\,\, (\chi\to 1)\,&:\quad \mathcal{D}_1\,\oplus\,\sum_{\h} \mathcal{L}^{\h}_{0,[0,1]}\,,
\end{split}
\end{align}
where recall that, as discussed above and in \cite{Ferrero:2023znz}, the exchanged unprotected operator in the direct channel is actually the $[0,2]^{\h}_0$ descendant of $\mathcal{L}^{\h}_{0,[0,0]}$. The associated superconformal blocks are derived in \cite{Ferrero:2023znz} and we will not repeat them here. We denote them with $\mathfrak{F}_{d,\mathcal{O}}(\chi)$ $(\mathfrak{F}_{c,\mathcal{O}}(\chi)$) for the exchange of an operator $\mathcal{O}$ in the direct (crossed) channel. The main observation regarding the structure of the conformal blocks is that they not only depend on the dimension of the exchanged operators, but also on the external long dimension $\hExt$. When expanding around $\lambda=\infty$, the appearance of derivatives of blocks and logarithms in the OPE will therefore have two origins: the expansion of the CFT data of {\it exchanged} operators around their free theory value, as well as the expansion
\begin{align}
\hExt=\DExt+\frac{1}{\sqrt{\lambda}}\,\gamma^{(1)}_{\ext}+\frac{1}{\lambda}\,\gamma^{(2)}_{\ext}+\ldots\,,
\end{align}
of the dimension of $\mathcal{L}_\ext$, where we have denoted with $\DExt$ its dimension in the free theory, and we will only consider the case of even $\DExt$. As usual, it will be useful to organize the analytic structure of the correlator highlighting the explicit powers of $\log$ generated by the OPE in both channels, as
\begin{align}\label{logpieces112L}
F^{(\ell)}(\chi)=\sum_{k=0}^{\ell}F^{(\ell)}_{d,\log^k}(\chi)\,\log^k(\chi)=\sum_{k=0}^{\ell}F^{(\ell)}_{c,\log^k}(\chi)\,\log^k(1-\chi)\,.
\end{align}
With these ingredients, let us consider the bootstrap problem order by order.

\paragraph{Free theory.} Consider a specific choice of $\mathcal{L}_\ext$: then free theory correlators are simply computed by Wick contractions. As discussed, this requires an explicit construction for $\mathcal{L}_\ext$: for operators of length $L=2,4$ this is given in \cite{Ferrero:2023znz} (note that there are no singlet long multiplets with $L=1,3$), while these correlators vanish identically for $L>4$. The OPE in the free theory reads
\begin{align}\label{freeOPE112L}
\begin{split}
F^{(0)}(\chi)&=a^{(0)}_{d,\mathcal{D}_2}\,\mathfrak{F}_{d,\mathcal{D}_2}(\chi)+\sum_{\Delta}\langle a^{(0)}_d\rangle_{\Delta}\,\mathfrak{F}_{d,\Delta}(\chi)\\
&=a^{(0)}_{c,\mathcal{D}_1}\,\mathfrak{F}_{c,\mathcal{D}_1}(\chi)+\sum_{\Delta}\langle a^{(0)}_c\rangle_{\Delta}\,\mathfrak{F}_{c,\Delta}(\chi)\,,
\end{split}
\end{align}
where regardless of the specific choice for $\mathcal{L}_\ext$, the sum in the direct channel runs over {\it even} values of $\Delta\ge 2$,  while in the crossed channel all $\Delta\ge 3$ contribute. This is related to the emergent braiding symmetry at strong coupling, where as opposed to a generic 1d CFT all permutations of the four external operators lead to symmetries of the associated four-point functions. In this case, the first two operators ($\mathcal{D}_1$) are identical and this leads to the invariance 
\begin{align}
F^{(0)}(\chi)=F^{(0)}\big(\tfrac{\chi}{\chi-1}\big)\,.
\end{align}
The presence of only even values of $\D$ in the direct channel OPE is as usual related to this fact and the property of superconformal blocks
\begin{align}
\mathfrak{F}_{d,\mathcal{O}}\big(\tfrac{\chi}{\chi-1}\big)=(-1)^{\h_{\mathcal{O}}+\hExt}\mathfrak{F}_{d,\mathcal{O}}(\chi)\,, 
\end{align}
which is a symmetry when $\hExt+\h_{\mathcal{O}}$ is an even integer, which will be always the case for the correlators considered here (in the free theory limit). As in the previous cases, although the braiding symmetry is naively only present in the free theory, we will always be able to use it as a constraint order by order in perturbation theory, in the spirit of \eqref{braiding1d}, and we will always be able to define an extension of the correlators beyond $\chi\in (0,1)$ such that braiding is an exact symmetry.

In terms of explicit results, there is an obvious difference between the case $L[\mathcal{L}_\ext]=2$ and $L[\mathcal{L}_\ext]=4$: in the former, there is exactly one operator for each dimension, of the schematic form $\mathcal{O}_{L=2}^{(k)}\sim \varphi \partial^{k}\varphi$, with dimension $\DExt=2+k$ in the free theory for even $k$ (see \cite{Ferrero:2023znz} for explicit expressions). It will be easy to give closed form expressions as functions of $k$ in this sector. In particular, for free theory correlators (where we normalize $\mathcal{O}_{L=2}^{(k)}$ such that $\mathtt{g}_2=1$) one has
\begin{align}
F_{(L=2)}^{(0)}(\chi)=\frac{\mathcal{N}(k)^{1/2}(1+k)(2+k)(4+k)!}{\sqrt{2}(5+2k)!}\frac{1+(1-\chi)^{2+k}}{\chi^{2+k}}\,,
\end{align}
which using \eqref{freeOPE112L} gives the OPE data
\begin{align}
\begin{split}
a^{(0)}_{d,\mathcal{D}_2}&=\frac{\sqrt{2}\mathcal{N}(k)^{1/2}(1+k)(2+k)(4+k)!}{(5+2k)!}\,,\\
\langle a^{(0)}_{d}\rangle_{\D}&=\frac{2\sqrt{2}(2+\D)!(1+k+\D)!}{(1+2\D)!\mathcal{N}(k)^{1/2}}\,,
\end{split}
\end{align}
in the direct channel,  
where
\begin{align}
\mathcal{N}(k)=\frac{16(3+2k)(5+2k)(2k+1)!}{(2+k)(3+k)(4+k)}\,,
\end{align}
and
\begin{align}
\begin{split}
a^{(0)}_{c,\mathcal{D}_1}&=\frac{a^{(0)}_{d,\mathcal{D}_2}}{2}\,,\\
\langle a^{(0)}_{c}\rangle_{\D}&=
\begin{cases}
0\,, \quad \D\le 2+k\,,\\
(-1)^{\D+1}\,
\frac{(\Delta-2)!(\Delta+k+5)!(\Delta-k-2)}{(k+1)!(2\Delta+2)!}\, a^{(0)}_{d,\mathcal{D}_2}
\,, \quad \D >k+2\,,
\end{cases}
\end{split}
\end{align}
in the crossed channel. On the other hand, the large degeneracy present for $L[\mathcal{L}_\ext]=4$ makes the result heavily dependent on the choice of basis, which is completely arbitrary, and we have not found an organizing principle that selects a particularly convenient basis. For this reason, we will not give explicit results in that sector but the interested reader can contact the authors. The main structure that we have observed in the results is that one can choose a basis for $\mathtt{d}_{L=4}(\DExt)$ such that
\begin{align}\label{free112L_L=4}
F^{(0)}(\chi)=\left\{
\frac{(1-\chi)^{\DExt/2}}{\chi^{\DExt}}\,, \quad
\frac{(1-\chi)^{\DExt/2-1}}{\chi^{\DExt-2}}\,,\quad\dots \,\quad
\frac{(1-\chi)^2}{\chi^4}\,,0\,,\dots\,,0
\right\}
\end{align}
where we notice that out of a total of $\text{dim}[\mathtt{d}_{L=4}(\DExt)]=\lfloor( \DExt/4)^2\rfloor$ operators,  only the first $\frac{\DExt-2}{2}$ give non-vanishing correlators (with a very simple structure),  while the remaining ones give zero. This fact can be explained by noticing that the number of $L=4$ operators $\mathcal{O}^{\DExt}_{L=4}$ that give a non-vanishing four-point function $\langle \mathcal{D}_1\mathcal{D}_1\mathcal{D}_2\mathcal{O}^{\DExt}_{L=4}\rangle^{(0)}$ is the same as the number of long operators of dimension $\DExt$ appearing in the $(\mathcal{D}_1\times \mathcal{D}_1)\times \mathcal{D}_2$ triple OPE. In turn, this can be obtained from the OPE $\mathcal{O}^{\Delta}_{L=2}\times \mathcal{D}_2$ by varying $\Delta$, and this number grows linearly with $\DExt$.

We have no closed-form expression for the squared OPE coefficients appearing in the blocks decomposition \eqref{112L_OPEchannels} in this case,  we limit to mention the fact that in the basis \eqref{free112L_L=4} the $\langle a^{(0)}_{d}\rangle_\D$ are all zero for $\Delta\ge \DExt-2$ and $\langle a^{(0)}_{c}\rangle_\D$ all vanish for $\D\ge \DExt$,  for all cases in \eqref{free112L_L=4}\footnote{A similar behavior was also observed in a related context in \cite{Bliard:2023zpe,Barrat:2024nod}.}

\paragraph{Tree level.} Moving to the first perturbative order, we make as usual an ansatz of transcendentality one
\begin{align}\label{Ftreeansatz}
F^{(1)}(\chi)=r_1(\chi)+r_2(\chi)\,\log(1-\chi)+r_3(\chi)\,\log\chi\,,
\end{align}
It is convenient to introduce the following combinations
\begin{align}
\begin{split}
\mathfrak{F}_{d,\D}^{\left(\ell_{1}, \ell_{2}\right)}(\chi)&=\chi^{\D-\DExt}\left(\partial_{\D}\right)^{\ell_{1}}\left(\partial_{\DExt}\right)^{\ell_{2}} \chi^{-\D+\DExt} \mathfrak{F}_{d,\mathcal{O}}(\chi)\,,\\
\mathfrak{F}_{c,\D}^{\left(\ell_{1}, \ell_{2}\right)}(\chi)&=\chi^{1-\D}\left(\partial_{\D}\right)^{\ell_{1}}\left(\partial_{\DExt}\right)^{\ell_{2}}\chi^{\D-1}\mathfrak{F}_{c,\mathcal{O}}(\chi)\,,
\end{split}
\end{align}
which allow to express the OPE at this order as
\begin{align}\label{treeOPE112L}
\begin{split}
F^{(1)}_{d,\log^1}(\chi)=&-\gamma^{(1)}_{\mathtt{ext}}\,F^{(0)}(\chi)+\sum_{\Delta}\langle a^{(0)}_{d}\gamma^{(1)}_{d}\rangle_{\Delta} \mathfrak{F}_{d,\Delta}(\chi)\,,\\
F^{(1)}_{c,\log^1}(\chi)=&\sum_{\Delta}\langle a^{(0)}_{c}\gamma^{(1)}_{c}\rangle_{\Delta} \mathfrak{F}_{c,\Delta}(\chi)\,\\
F^{(1)}_{d,\log^0}(\chi)=&\gamma^{(1)}_{\mathtt{ext}}\big[a^{(0)}_{d,\mathcal{D}_2}\mathfrak{F}^{(0,1)}_{d,\mathcal{D}_2}(\chi)+\sum_{\Delta}\langle a^{(0)}_{d}\rangle_{\Delta} \mathfrak{F}^{(0,1)}_{d,\Delta}(\chi)\big]+\sum_{\Delta}\langle a^{(0)}_{d}\gamma^{(1)}_{d}\rangle_{\Delta} \mathfrak{F}^{(1,0)}_{d,\Delta}(\chi)\\
&+\sum_{\Delta}a^{(1)}_{d,\mathcal{D}_2}\mathfrak{F}_{d,\mathcal{D}_2}(\chi)+\sum_{\Delta}\langle a^{(1)}_{d}\rangle_{\Delta} \mathfrak{F}_{d,\Delta}(\chi)\,,\\
F^{(1)}_{c,\log^0}(\chi)=&\gamma^{(1)}_{\mathtt{ext}}\big[a^{(0)}_{c,\mathcal{D}_1}\mathfrak{F}^{(0,1)}_{c,\mathcal{D}_1}(\chi)+\sum_{\Delta}\langle a^{(0)}_{c}\rangle_{\Delta} \mathfrak{F}^{(0,1)}_{c,\Delta}(\chi)\big]+\sum_{\Delta}\langle a^{(0)}_{c}\gamma^{(1)}_{c}\rangle_{\Delta} \mathfrak{F}^{(1,0)}_{c,\Delta}(\chi)\\
&+\sum_{\Delta}a^{(1)}_{c,\mathcal{D}_2}\mathfrak{F}_{c,\mathcal{D}_1}(\chi)+\sum_{\Delta}\langle a^{(1)}_{c}\rangle_{\Delta} \mathfrak{F}_{c,\Delta}(\chi)\,.
\end{split}
\end{align}
Given the discussion of the previous sections, we have by now proved that the anomalous dimensions at first order are proportional to the quadratic superconformal Casimir and in particular given by \eqref{treegammas}. We can therefore use such result, and in particular
\begin{align}
\gamma^{(1)}_{\mathtt{ext}}=-\frac{1}{2}\Delta_{\mathtt{ext}}(\Delta_{\mathtt{ext}}+3)\,, \quad 
\gamma^{(1)}_{d,\Delta}=-\frac{1}{2}\Delta(\Delta+3)\,, \qquad
\gamma^{(1)}_{c,\Delta}=-\frac{1}{2}\Delta(\Delta+3)+2\,,
\end{align}
to resum the first two lines in \eqref{treeOPE112L} and obtain the two functions $r_2(\chi)$ and $r_3(\chi)$ in \eqref{Ftreeansatz}. This determines the full correlator only up to the rational function $r_1(\chi)$, which is constrained by braiding symmetry:
\begin{align}
r_1(\chi)=r_1\big(\tfrac{\chi}{\chi-1}\big)\,,
\end{align}
as well as by the boundary conditions dictated by the OPE
\begin{align}\label{OPElimit112L}
\begin{split}
F^{(1)}(\chi)&\sim  \chi^{-\DExt}+\mathcal{O}(\chi^{-\DExt+1})\,, \quad (\chi\to 0)\,,\\
F^{(1)}(\chi)&\sim 1+\mathcal{O}(1-\chi)\,, \quad (\chi\to 1)\,.
\end{split}
\end{align}
Put together,  these constraints turn out to fix $F^{(1)}(\chi)$ for each $\mathcal{L}_{\ext}$ of dimension $\DExt$ up to $\DExt/2$ parameters, which are related to the necessity of resolving mixing for the tree-level OPE coefficients. However, it will turn out to that it is still possible to compute the averages $\left.\langle a^{(0)}\,\gamma^{(2)}\rangle_{\D}\right|_{\langle \mathcal{D}_1\mathcal{D}_1\mathcal{D}_2\mathcal{L}_{\ext}\rangle}$ without fixing this ambiguity, so we disregard it for the time being.

As for the free theory, results for $L[\mathcal{L}_\ext]=4$ have to be discussed case by case and we shall not give them here, while for $L[\mathcal{L}_\ext]=2$ we find, for example,
\begin{align}
\begin{split}
F^{(1)}_{d,\log^1}(\chi)&=\frac{(2+k)a^{(0)}_{d,\mathcal{D}_2}}{4(1-\chi)\chi^{k+2}}(5+k-5\chi-(2+k)\chi-(\chi-1)^{k+3}(5+k+2\chi))\,,\\
F^{(1)}_{c,\log^1}(\chi)&=-\frac{(2+k)(5+k+2\chi)a^{(0)}_{d,\mathcal{D}_2}}{4}\left(\frac{\chi-1}{\chi}\right)^{2+k}\,.
\end{split}
\end{align}
Moreover, let us stress that the explicit form of tree level correlator is not necessary nor particularly illuminating: for our purposes, they are just a necessary intermediate step since (as we discuss in the next paragraph) one needs to compute $\langle a^{(1)}_d\rangle_\D$ (up to the ambiguities discussed above) in order to extract the averages $\left.\langle a^{(0)}\,\gamma^{(2)}\rangle_{\D}\right|_{\langle \mathcal{D}_1\mathcal{D}_1\mathcal{D}_2\mathcal{L}_{\ext}\rangle}$ from one loop correlators.

\paragraph{One loop.} The highest perturbative order we are interested in is one loop, where we make an ansatz of transcendentality two, this time {\it assuming} the absence of $\Li_2(\chi)$ from the final answer. This seems natural given the fact that it never appears in the other results discussed so far and moreover as stressed in Section \ref{sec:1111/remarks} it cannot be obtained from the diagonal limit of single-valued HPLs in higher dimensions, which is the case for all other functions appearing in our results. We then set
\begin{align}\label{ansatz1loop112L}
\begin{split}
F^{(2)}(\chi)=&r_1(\chi)+[r_2(\chi)+r_3(\chi)\log(1-\chi)]\,\log(1-\chi)\\
&+[r_4(\chi)+r_5(\chi)\log(1-\chi)+r_6(\chi)\,\log\chi]\,\log\chi\,.
\end{split}
\end{align}
Note, however, that we are not interested in bootstrapping the full correlator at this order, since our motivation for considering $\langle \mathcal{D}_1\mathcal{D}_1\mathcal{D}_2\mathcal{L}_{\ext}\rangle$ is really just related to the necessity of computing the averages \eqref{a0gamma2_112L}, which can be extracted from the $\log\chi$ singularity in the direct channel:
\begin{align}
F^{(2)}_{d,\log^1}=[r_4(\chi)+r_5(\chi)\log(1-\chi)]\,\log\chi\,.
\end{align}
Let us then discuss how this quantity can be computed. The OPE for the relevant parts of the correlator reads
\begin{align}\label{1loopOPE112L}
\begin{split}
F^{(2)}_{d,\log^2}(\chi)=&\frac{1}{2}(\gamma^{(1)}_{\mathtt{ext}})^2 F^{(0)}(\chi)+\frac{1}{2}\sum_{\Delta}\langle a^{(0)}_{d}\gamma^{(1)}_{d}(\gamma^{(1)}_{d}-2\gamma^{(1)}_{\mathtt{ext}})\rangle_{\Delta} \mathfrak{F}_{d,\Delta}(\chi)\,,\\
F^{(2)}_{c,\log^2}(\chi)=&\frac{1}{2}\sum_{\Delta}\langle a^{(0)}_{c}(\gamma^{(1)}_{c})^2\rangle_{\Delta} \mathfrak{F}_{c,\Delta}(\chi)\,,\\
F^{(2)}_{d,\log^1}(\chi)=&-\gamma^{(2)}_{\mathtt{ext}}\,F^{(0)}(\chi)-(\gamma^{(1)}_{\mathtt{ext}})^2\big[ a^{(0)}_{d,\mathcal{D}_2}\,\mathfrak{F}^{(1,0)}_{d,\mathcal{D}_2}(\chi)+\langle a^{(0)}_{d}\rangle_{\Delta}\,\mathfrak{F}^{(1,0)}_{d,\Delta}(\chi) \big]\\
&-\gamma^{(1)}_{\mathtt{ext}}\,\big[a^{(1)}_{d,\mathcal{D}_2}\,\mathfrak{F}_{d,\mathcal{D}_2}(\chi)+\sum_{\Delta}\big(\langle a^{(1)}_{d}\rangle_{\Delta}\,\mathfrak{F}_{d,\Delta}(\chi)+\langle a^{(0)}_{d}\,\gamma^{(1)}_{d}\rangle_{\Delta}(\mathfrak{F}^{(1,0)}_{d,\Delta}(\chi)-\mathfrak{F}^{(0,1)}_{d,\Delta}(\chi)) \big)\big]\\
&+\sum_{\Delta}\big(\langle a^{(0)}_{d}\,\gamma^{(2)}_{d}+a^{(1)}_{d}\,\gamma^{(1)}_{d}\rangle_{\Delta} \mathfrak{F}_{d,\Delta}(\chi)+\langle a^{(0)}_{d}\,(\gamma^{(1)}_{d})^2\rangle_{\Delta} \mathfrak{F}^{(1,0)}_{d,\Delta}(\chi)\big)\,,
\end{split}
\end{align}
from which it is clear that the highest logarithmic singularities, $F^{(2)}_{d,\log^2}(\chi)$ and $F^{(2)}_{c,\log^2}(\chi)$, can be computed from CFT data at previous orders using the fact that the degeneracy is not lifted at tree level. One then obtains
\begin{align}
r_3(\chi)=F^{(2)}_{c,\log^2}(\chi)\,,\qquad r_6(\chi)=F^{(2)}_{d,\log^2}(\chi)\,,
\end{align}
while braiding symmetry fixes the remaining term of transcendentality two to be
\begin{align}
r_5(\chi)=-r_3(\chi)+r_3\big(\tfrac{\chi}{\chi-1}\big)\,,
\end{align}
the other function appearing in $F^{(2)}_{d,\log^1}$ is constrained to be braiding symmetric
\begin{align}\label{r4braiding}
r_4(\chi)=r_4\big(\tfrac{\chi}{\chi-1}\big)\,,
\end{align}
and the other constraints from braiding will not be relevant here. Compatibility with the OPE also requires that
\begin{align}
F^{(2)}_{d,\log}(\chi)\sim \chi^{-\DExt}\,,\quad (\chi \to 0)\,,
\end{align}
and some experimenting with the ansatz shows that the correct denominator for $r_4(\chi)$ seems to be $\chi^{\DExt}\,(1-\chi)$, although this can be shown in practice by looking for solutions with other denominators and imposing the constraints. All together,  these constraints fix $F^{(2)}_{d,\log^1}(\chi)$ up to a finite number of coefficients,  in analogy with what happens for $F^{(1)}_{d,\log^0}(\chi)$.  However,  in this case $F^{(2)}_{d,\log^1}(\chi)$ is what allows us to compute the averages \eqref{a0gamma2_112L}, so it is important to understand how to fix such coefficients.

Before explaining how this is done, an important remark is in order. In \eqref{1loopOPE112L}, the first term in the expansion of $F^{(2)}_{\log^1}(\chi)$ contains $\gamma^{(2)}_\ext$ and we should clarify what we mean by that, since the external operators $\mathcal{L}_\ext$ we are working with are {\it not} eigenstates of the dilatation operator, and therefore their conformal dimension is not well-defined. One should really interpret that part of the OPE as follows. For an external operator $\mathcal{L}_\ext$ with $\hExt^{(0)}=\DExt$, fix an arbitrary basis $\widehat{\mathcal{O}}_{\alpha}$ in $\mathtt{d}(\Delta_{\mathtt{ext}})$, with $\alpha=1,\ldots,\text{dim}[\mathtt{d}(\Delta_{\mathtt{ext}})]$. Then, the term we are focusing on should really be interpreted as
\begin{align}\label{F2fromF0}
\left[F^{(2)}_{d,\log^1}(\chi)\right]_\alpha = \left[\Gamma^{(2)}_{\DExt}\right]^{\alpha\beta}\,\left[F^{(0)}(\chi)\right]_\beta+\ldots\,,
\end{align} 
where note that $\Gamma^{(2)}_{\DExt}$ above is expressed in the basis chosen for the {\it external} operators, which needs not be the same as that used for the {\it exchanged} operators. In principle, one should then either choose the same basis for the two sets of operators or work out the explicit change of basis relating the two. In practice, however, it turns out to be sufficient to treat the entries of the matrix $\Gamma^{(2)}_{\DExt}$ appearing in \eqref{F2fromF0} as unknown parameters, with no need to input their correct values: the constraints that we discuss below are still sufficient to fix the result uniquely, and this allows us to treat the external and exchanged operators independently.

Let us now go back to the problem of fixing the undetermined parameters in $r_4(\chi)$. We use the recursive procedure outlined in \cite{Ferrero:2021bsb}: imagine to know the averages $\langle a^{(0)}\gamma^{(2)}\rangle_{\D}$ computed on $\langle \mathcal{D}_1\mathcal{D}_1\mathcal{D}_2\mathcal{L}_\ext\rangle$ correlators for {\it all} external operators $\mathcal{L}_\ext$ of length $L=2,4$ and free theory dimension $2\le \DExt\le \widehat{\D}-2$, for a certain even integer $\widehat{\D}$. Combined with the knowledge of the free theory OPE coefficients, that we can compute with Wick contractions, such averages are enough to compute the components of $\Gamma^{(2)}_{\D,2\to4}$ for $\D=2,...,\widehat{\D}+2$, using \eqref{a0gamma2_112L}. This information can be used to compute $\langle a^{(0)}\gamma^{(2)}\rangle_{\D}$ for $\langle \mathcal{D}_1\mathcal{D}_1\mathcal{D}_2\mathcal{L}_\ext\rangle$ where now we increase the dimension of the external operator by two units: $\DExt= \widehat{\Delta}$, for all values of $\D \le \widehat{\D}+2$. These averages can be used as constraints on the small $\chi$ expansion of $F^{(2)}_{d,\log^1}$ using the OPE \eqref{1loopOPE112L}, and they turn out to provide exactly enough information to fix all the undetermined parameters in $r_4(\chi)$, even if we have been agnostic about the entries of $\Gamma^{(2)}_{\DExt}$ in \eqref{F2fromF0} and we have not fixed the ambiguity in the determination of $\langle a^{(1)}_d\rangle_\D$. By studying the OPE for operators with $\D>\widehat{\D}+2$ we can then obtain all the averages $\langle a^{(0)}\gamma^{(2)}\rangle_\D$ for all $\D$ and all $\mathcal{L}_\ext$ with $\DExt=\widehat{\D}$. This completes our recursive step and one can then move to the next by repeating the above with $\widehat{\D}\to \widehat{\D}+2$. Note that for the procedure to work it is important that we always allow $\mathcal{L}_\ext$ to range over {\it all} operators with $L=2,4$ and a certain dimension $\DExt$, as only this way one can explore enough directions in the degeneracy space $\mathtt{d}_{L=4}(\D)$. The starting point of the recursion is $\widehat{\D}=2$, so we should explain how we computed $\Gamma^{(2)}_{\D,2 \to 4}$ for $\D=2,4$. The answer is simple: for $\D=2$ there is a unique operator with $L=2$ and no operators of $L=4$, so $\Gamma^{(2)}_{\D=2,2 \to 4}$ is empty; for $\D=4$ there is a unique operator with $L=4$ so $\Gamma^{(2)}_{\D=4,2 \to 4}$ is a real number which can be extracted from, {\it e.g.}, $\langle\mathcal{D}_1\mathcal{D}_1\mathcal{D}_2\mathcal{D}_2\rangle^{(2)}$.

The explicit results for $\Gamma^{(2)}_{\D=4,2 \to 4}$ depend on the choice of basis, so they are generally not particularly illuminating, although we shall give some examples in the next subsection. What we are really interested in, on the other hand, is the average \eqref{gamma2^2_mixing_L24}, which does not depend on the basis. In practice, we compute all entries of $\Gamma^{(2)}_{\D=4,2 \to 4}$ in an arbitrary basis up to a certain $\D$, obtaining enough data as to allow us to guess a closed-form expression for \eqref{gamma2^2_mixing_L24}. In particular, considering all $\mathcal{L}_\ext$ with $L=2,4$ and $2\le\DExt\le 14$ allows us to access entries of $\Gamma^{(2)}_{\D=4,2 \to 4}$ for $2\le \D\le 16$. With an ansatz based on the expected transcendentality of the result (in terms of harmonic sums) and the reciprocity principle\footnote{Note that in \eqref{gamma2^2_mixing_L24} we are expressing $\langle(\gamma^{(2)})^2\rangle_\D=\langle\gamma^{(2)}\rangle_\D^2+\delta \Gamma^{(2)}_{\mathtt{sq}}(\D)$. The whole quantity is not expected to have an expansion in powers of $\JJ$, due to the corrections discussed in Appendix \ref{app:perturbativeOPE}, but the half-integer powers are fully captured by $\langle\gamma^{(2)}\rangle_\D^2$, so that $\delta \Gamma^{(2)}_{\mathtt{sq}}(\D)$ should have an expansion in integer powers of $\JJ$ for large $\D$. All terms in \eqref{deltaGamma2sq}, including the harmonic sums, have this property.}, this is enough to guess
\begin{align}\label{deltaGamma2sq}
\begin{split}
\delta \Gamma^{(2)}_{\mathtt{sq}}=&\JJ\,\left(-\frac{\JJ-2}{2}(S_{-2}(\D+1)+\zeta(2))+\frac{3\JJ-4}{8}H^2_{\D+1}\right)\\
&-\frac{4(\JJ)^3+8(\JJ)^4-15(\JJ)^2-36}{4(\JJ+2)}H_{\D+1}+\frac{29(\JJ)^3+108(\JJ)^4-56(\JJ)^2-288}{32(\JJ+2)}\,.
\end{split}
\end{align}
We then performed a consistency check by pushing our procedure to $\DExt=16$ and comparing the average thus obtained for $\D=18$ with the prediction of \eqref{deltaGamma2sq}, finding perfect agreement. Moreover, it is important to highlight that at every step of the recursive procedure described above the total number of equations that determine the entries $\Gamma^{(2)}_{\D,2\to 4}$ is more than the number of unknowns, so that we are always dealing with over-constrained systems of equations: the fact that they admit a solution represents a non-trivial check of the validity of this procedure and of our assumptions.

In the simplest case where $\mathcal{L}_\ext$ is the unique operator with $L=2$ and $\DExt=2$, the correlator does not depend on any free parameter and we can bootstrap it completely, obtaining
\begin{align}
\begin{split}
F^{(0)}(\chi)=&\frac{2}{\sqrt{5}}\frac{1+(1-\chi)^2}{\chi^2}\,,\\
F^{(1)}(\chi)=&\frac{2}{\sqrt{5}\chi^2}\Big[\frac{5-7\chi+(1-\chi)^3(5+2\chi)}{1-\chi}\log\chi-(1-\chi)^2(5+2\chi)\log(1-\chi)\\&-\frac{194(1-\chi)+109\chi^2}{24}\Big]\,,\\
F^{(2)}(\chi)=&\frac{1}{\sqrt{5}\chi^2}\Big[\frac{50.150\chi+145\chi^2-40\chi^3+15\chi^4-20\chi^5+9\chi^6}{(1-\chi)^2}\log^2\chi\\
&-\frac{50-100\chi+45\chi^2+9\chi^3+18\chi^4-18\chi^5}{1-\chi}\log\chi\log(1-\chi)\\
&+(1-\chi)^2(20+20\chi+9\chi^2)\log^2(1-\chi)+\frac{-130+260\chi-117\chi^2-13\chi^3+28\chi^4}{(1-\chi)}\log\chi\\
&+\frac{6+56\chi-113\chi^2+19\chi^3+28\chi^4}{\chi}\log(1-\chi)+\frac{17278(1-\chi)+11519\chi^2}{576}
\Big]\,.
\end{split}
\end{align}
Another interesting result that can be expressed in a simple way is the average $\langle\gamma^{(2)}\rangle_\D$ computed for external operators of length two:
\begin{align}\label{gamma2_112L_L=2}
\begin{split}
 \left.\langle \gamma^{(2)}\rangle_{\D}\right|_{\langle\mathcal{D}_1\mathcal{D}_1\mathcal{D}_2\mathcal{O}^{(k)}_{L=2}\rangle}=\,&\gamma^{(1)}_{\D}\partial_{\Delta}\gamma^{(1)}_{\D}+(j^2_k/2+\JJ)H_{1+\D}+(j^2_k/2+\JJ/2)H_{3+k}\\
&+\frac{2j^2_k(10+3j^2_k)+4\JJ(-19-7j^2_k+j^4_k)-(\JJ)^2(28+15j^2_k)}{8(2+\JJ)(2+j^2_k)}\,,
\end{split}
\end{align}
where $\gamma^{(1)}_{\Delta}=-\JJ/2$ is the dimension of operators $\mathcal{L}^{\Delta}_{0,[0,0]}$ at tree level,  with $\JJ=\Delta(\Delta+3)$ and $j^2_k=\DExt(\DExt+3)=(2+k)(5+k)$.

\subsection{CFT data, unmixed }

Let us collect here some of the results that we have obtained from the study of the mixing problem at strong coupling. 
\vspace{0.2cm}
\paragraph{Singlet operators with $\D=4$.} We start from the $\mathtt{d}(\mathcal{L}^{\D=4}_{0,[0,0]})$, that is the sector of singlet long multiplets whose conformal dimension is four at strong coupling. There are two such operators, with $L=2$ and $L=4$ respectively, so there is no ambiguity in the choice of basis and the two operators have the schematic form (note we use $\widehat{\mathcal{O}}$ to stress that operators are in the basis of fixed length, as opposed to the eigenstates basis)
\begin{align}
\widehat{\mathcal{O}}^{(\D=4)}_{L=2}\sim \varphi\,\partial^2\varphi+\ldots\,,\qquad 
\widehat{\mathcal{O}}^{(\D=4)}_{L=4}\sim\varphi^4\,,
\end{align} 
where the derivatives act in such a way as to give a superconformal primary and the dots denote terms with superconformal descendants of $\varphi$, see \cite{Ferrero:2023znz} for more details. On the other hand, the second operator only contains terms in $\varphi$ and we are only being schematic with its normalization. Given that there is just one operator for each length, the mixing problem in this sector can actually be solved by looking just at $\langle \mathcal{D}_p\mathcal{D}_p\mathcal{D}_q\mathcal{D}_q\rangle$, and given that the maximum length is four it would in principle be enough to consider $p=1,2$ for some $q\ge p$. One would normally resolve the degeneracy for the free theory OPE coefficients together with the first order anomalous dimensions. However, as mentioned several times $\Gamma^{(1)}_{\D}$ is proportional to the identity and therefore cannot be used for this purpose. We then consider the two averages $\langle a^{(0)}\rangle_{\D=4}$ and $\langle a^{(0)}\,\gamma^{(2)}\rangle_{\D=4}$ for this purpose. In particular, we have\footnote{We are implicitly normalizing the operators. Moreover, since they have different lengths, they are of course orthogonal in the sense of the free-theory two point functions. }
\begin{align}\label{mixedsystemDelta4}
\begin{split}
\left. \langle a^{(0)}\rangle_{\D=4}\right|_{\langle \mathcal{D}_p\mathcal{D}_p\mathcal{D}_q\mathcal{D}_q\rangle}&=\mu^{(0)}_{pp\widehat{\mathcal{O}}^{(\D=4)}}\cdot \mu^{(0)}_{qq\widehat{\mathcal{O}}^{(\D=4)}}=\frac{p\,q}{7}\,\left[1+\tfrac{2}{5}(p-1)(q-1)\right]\,,\\
\left. \langle a^{(0)}\,\gamma^{(2)}\rangle_{\D=4}\right|_{\langle \mathcal{D}_p\mathcal{D}_p\mathcal{D}_q\mathcal{D}_q\rangle}&=(\mu^{(0)}_{pp\widehat{\mathcal{O}}^{(\D=4)}})^T\,\Gamma^{(2)}_{\D=4}\,\mu^{(0)}_{qq\widehat{\mathcal{O}}^{(\D=4)}}=p^2\,q^2+\frac{p\,q}{30}\left[269+57(p-1)(q-1)\right]\,,
\end{split}
\end{align}
where we have defined the vector
\begin{align}\label{C0ppDelta4}
\mu^{(0)}_{pp\widehat{\mathcal{O}}^{(\D=4)}}=\left(\mu^{(0)}_{pp\widehat{\mathcal{O}}^{(\D=4)}_{L=2}},\quad \mu^{(0)}_{pp\widehat{\mathcal{O}}^{(\D=4)}_{L=4}}\right)\,.
\end{align}
It is clear from the diagrammatics that the free theory OPE coefficients \eqref{C0ppDelta4} can be at most quadratic in $p$, and moreover since the second operator has length four the OPE coefficient must vanish when $p=1$\footnote{Note that this input is crucial, as it distinguishes between the two operators. Without this, one could only unmix this sector up to a $O(2)$ rotation parameter.}. We could of course (and we have) compute \eqref{C0ppDelta4} in the free theory, but it turns out that these observations on their dependence on $p$ together with the constraints \eqref{mixedsystemDelta4} are enough to find both the OPE coefficients \eqref{C0ppDelta4} and the entries of $\Gamma^{(2)}_{\D=4}$ (which, of course, must be independent of $p$ and $q$!). The result is
\begin{align}\label{C0ppDelta4_result}
\mu^{(0)}_{pp\widehat{\mathcal{O}}^{(\D=4)}}=\left(\frac{p}{\sqrt{7}}\,,\quad p(p-1)\,\sqrt{\frac{2}{35}}\right)\,,
\end{align}
as well as
\begin{align}\label{Gamma2_Delta4}
\Gamma^{(2)}_{\D=4}=
\begin{pmatrix}
\frac{2093}{30} & 7\sqrt{\frac{5}{2}}\\
7\sqrt{\frac{5}{2}} & \frac{203}{4}
\end{pmatrix}\,.
\end{align}
Note that all entries are non-vanishing and in particular this implies that the eigenstates of the dilatation operator mix different lengths at second order. We find that the matrix \eqref{Gamma2_Delta4} is diagonalized by the two states
\begin{align}
\left(\mathcal{O}^{(\D=4)}_{1}\,,\,\, \mathcal{O}^{(\D=4)}_{2}\right)=M^{(2)}_{\D=4}\,\left(\widehat{\mathcal{O}}^{(\D=4)}_{L=2}\,,\,\, \widehat{\mathcal{O}}^{(\D=4)}_{L=4}\right)\,,\quad 
M^{(2)}_{\D=4}=
\begin{pmatrix}
\sqrt{\tfrac{1}{2}-\tfrac{163}{2\sqrt{62569}}} &\,\,\, -\sqrt{\tfrac{1}{2}+\tfrac{163}{2\sqrt{62569}}}\\
\sqrt{\tfrac{1}{2}+\tfrac{163}{2\sqrt{62569}}} &\,\,\,  \sqrt{\tfrac{1}{2}-\tfrac{163}{2\sqrt{62569}}}
\end{pmatrix}\,.
\end{align}
The corresponding OPE coefficients are
\begin{align}\label{OPEeigenstatesDelta4Order2}
\begin{split}
\mu^{(0)}_{pp{\mathcal{O}}^{(\D=4)}}&=\tfrac{p}{60\sqrt{70}}\left(\sqrt{\tfrac{1}{2}-\tfrac{163}{2\sqrt{62569}}}(-43+\sqrt{62659}-120p)\,,\,\,\sqrt{\tfrac{1}{2}+\tfrac{163}{2\sqrt{62569}}}(43+\sqrt{62659}+120p)\right)\\
& \simeq \big(0.0998 p(p-1.7277)\,,\,\, 0.2172p(p+2.4443)\big)\,,
\end{split}
\end{align}
while the eigenvalues of $\Gamma^{(2)}_{\D=4}$ read
\begin{align}
\gamma^{(2)}_{\mathcal{O}^{(\D=4)}_1}=\frac{7}{120}(1033-\sqrt{62659})\simeq 45.6565\,,\quad 
\gamma^{(2)}_{\mathcal{O}^{(\D=4)}_2}=\frac{7}{120}(1033+\sqrt{62659})\simeq 74.8602\,,
\end{align}
and note that we are fixing a sign ambiguity in determining the eigenstates (and associated OPE coefficients) by requiring that \eqref{OPEeigenstatesDelta4Order2} are both positive for $p=1$. Other choices are possible and equally valid. One could have actually derived the same results by starting directly with the basis of eigenstates, and thinking of the averages \eqref{mixedsystemDelta4} as
\begin{align}\label{mixedsystemDelta4_eigenstates}
\begin{split}
\left. \langle a^{(0)}\rangle_{\D=4}\right|_{\langle \mathcal{D}_p\mathcal{D}_p\mathcal{D}_q\mathcal{D}_q\rangle}&=\mu^{(0)}_{pp{\mathcal{O}}^{(\D=4)}}\cdot \mu^{(0)}_{qq{\mathcal{O}}^{(\D=4)}}\,,\\
\left. \langle a^{(0)}\,\gamma^{(2)}\rangle_{\D=4}\right|_{\langle \mathcal{D}_p\mathcal{D}_p\mathcal{D}_q\mathcal{D}_q\rangle}&=(\mu^{(0)}_{pp{\mathcal{O}}^{(\D=4)}})^T\,
\begin{pmatrix}
\gamma^{(2)}_{\mathcal{O}^{(\D=4)}_1} & 0\\
0 & \gamma^{(2)}_{\mathcal{O}^{(\D=4)}_2}
\end{pmatrix}
\,\mu^{(0)}_{qq{\mathcal{O}}^{(\D=4)}}\,.
\end{split}
\end{align}
Now we have one less unknown, as we have assumed a diagonal form for the anomalous dimension matrix. On the other hand, in this basis both OPE coefficients are ({\it a priori}) arbitrary polynomials of degree two in $p$: we have simply traded the information on one of them vanishing for $p=1$ with the vanishing of the off-diagonal entries of $\Gamma^{(2)}_{\D=4}$ in the eigenstates basis. Requiring the OPE coefficients to be positive for $p=1$ and choosing, as above, $\gamma^{(2)}_{\mathcal{O}^{(\D=4)}_1}<\gamma^{(2)}_{\mathcal{O}^{(\D=4)}_2}$, one recovers the same results.

It turns out that in the $\D=4$ sector we can actually go further, exploiting the fact that in Section \ref{sec:ppqq} we have derived the two-loop correlators $\langle \mathcal{D}_1\mathcal{D}_1\mathcal{D}_q\mathcal{D}_q\rangle$ and $\langle \mathcal{D}_2\mathcal{D}_2\mathcal{D}_2\mathcal{D}_2\rangle$. This will allow us to unmix the first order OPE coefficients together with the third order anomalous dimensions. We have the averages
\begin{align}\label{mixedsystemDelta4_2loops}
\begin{split}
\left. \langle a^{(1)}\rangle_{\D=4}\right|_{\langle \mathcal{D}_1\mathcal{D}_1\mathcal{D}_q\mathcal{D}_q\rangle}&=q\,\left(\tfrac{91}{90}-\tfrac{3}{14}q\right)\,,\\ 
\left. \langle a^{(1)}\rangle_{\D=4}\right|_{\langle \mathcal{D}_2\mathcal{D}_2\mathcal{D}_2\mathcal{D}_2\rangle}&=-\tfrac{388}{315}\,,\\
\left. \langle a^{(0)}\,\gamma^{(3)}+a^{(1)}\,\gamma^{(2)}\rangle_{\D=4}\right|_{\langle \mathcal{D}_1\mathcal{D}_1\mathcal{D}_q\mathcal{D}_q\rangle}&=\tfrac{q}{2700}(90563-46125q-8100q^2)\,,\\
\left. \langle a^{(0)}\,\gamma^{(3)}+a^{(1)}\,\gamma^{(2)}\rangle_{\D=4}\right|_{\langle \mathcal{D}_2\mathcal{D}_2\mathcal{D}_2\mathcal{D}_2\rangle}&=-\tfrac{224377}{675}\,.
\end{split}
\end{align}
We can interpret these in the length basis, where
\begin{align}\label{a1gamma3ppqq_system}
\begin{split}
\left. \langle a^{(1)}\rangle_{\D=4}\right|_{\langle \mathcal{D}_1\mathcal{D}_1\mathcal{D}_q\mathcal{D}_q\rangle}&=\mu^{(0)}_{pp\widehat{\mathcal{O}}^{(\D=4)}}\cdot \mu^{(1)}_{qq\widehat{\mathcal{O}}^{(\D=4)}}+\mu^{(1)}_{pp\widehat{\mathcal{O}}^{(\D=4)}}\cdot \mu^{(0)}_{qq\widehat{\mathcal{O}}^{(\D=4)}}\,,\\
\left. \langle a^{(0)}\,\gamma^{(3)}+a^{(1)}\,\gamma^{(2)}\rangle_{\D=4}\right|_{\langle \mathcal{D}_1\mathcal{D}_1\mathcal{D}_q\mathcal{D}_q\rangle}&=
(\mu^{(0)}_{pp\widehat{\mathcal{O}}^{(\D=4)}})^T\,\Gamma^{(3)}_{\D=4}\, \mu^{(0)}_{qq\widehat{\mathcal{O}}^{(\D=4)}}\\
&+(\mu^{(0)}_{pp\widehat{\mathcal{O}}^{(\D=4)}})^T\,\Gamma^{(2)}_{\D=4}\, \mu^{(1)}_{qq\widehat{\mathcal{O}}^{(\D=4)}}\\
&+(\mu^{(1)}_{pp\widehat{\mathcal{O}}^{(\D=4)}})^T\,\Gamma^{(2)}_{\D=4}\, \mu^{(0)}_{qq\widehat{\mathcal{O}}^{(\D=4)}}\,.
\end{split}
\end{align}
We now can solve \eqref{mixedsystemDelta4_2loops} for the first order OPE coefficients and third order anomalous dimensions. The study of tree level correlators from Section \ref{sec:ppqq} suggests that $\mu^{(1)}_{pp\widehat{\mathcal{O}}^{(\D=4)}}$ should be of degree three in $p$, but this time we lack any basic consideration based on diagrammatics that can impose further constraints like we had in the free theory. Thus, although just using the data in \eqref{mixedsystemDelta4_2loops} we can solve for $\mu^{(1)}_{pp\widehat{\mathcal{O}}^{(\D=4)}}$ for generic $p$, as well as for $\Gamma^{(3)}_{\D=4}$, we are only able to to so up to one free parameter, that we label $\alpha$. The results read
\begin{align}\label{C1_Delta4}
\mu^{(1)}_{pp\widehat{\mathcal{O}}^{(\D=4)}}=
\left(\tfrac{p(386-135p)}{90\sqrt{7}},\,\,\tfrac{p(p-1)(1-36p)}{6\sqrt{70}}\right)+\alpha\,\epsilon\,\mu^{(0)}_{pp\widehat{\mathcal{O}}^{(\D=4)}}\,,\quad 
\epsilon=
\begin{pmatrix}
0 & 1\\
-1 & 0
\end{pmatrix}\,,
\end{align}
and
\begin{align}\label{Gamma3_Delta4}
\Gamma^{(3)}_{\D=4}=
\begin{pmatrix}
-\tfrac{4424}{15} & -\tfrac{10087}{36\sqrt{10}}\\
-\tfrac{10087}{36\sqrt{10}} & -\tfrac{4459}{36}
\end{pmatrix}
+\alpha\,\left(\epsilon\,\Gamma^{(2)}_{\D=4}-\Gamma^{(2)}_{\D=4}\,\epsilon\right)\,,
\end{align}
where $\Gamma^{(2)}_{\D=4}$ is the second order anomalous dimension matrix in the length basis, as appearing in \eqref{Gamma2_Delta4}. The undetermined parameter $\alpha$ is clearly related to the ambiguity of performing infinitesimal rotations in the free theory. We will soon resolve such ambiguity in two distinct ways, but first let us note that its presence is not related to the fact that we have considered only few correlators from which to extract the averages. To make this more explicit, note that plugging in \eqref{a1gamma3ppqq_system} the solution \eqref{C1_Delta4} and \eqref{Gamma3_Delta4} we can compute the averages
\begin{align}\label{a1gamma3ppqq_solution}
\begin{split}
\left. \langle a^{(1)}\rangle_{\D=4}\right|_{\langle \mathcal{D}_1\mathcal{D}_1\mathcal{D}_q\mathcal{D}_q\rangle}&=\tfrac{p q}{315}\left(54\left(p^2(1-q)+q^2(1-p)\right)-\tfrac{3}{2}(83(p+q)-74 p q)+389\right)\,,\\
\left. \langle a^{(0)}\,\gamma^{(3)}+a^{(1)}\,\gamma^{(2)}\rangle_{\D=4}\right|_{\langle \mathcal{D}_1\mathcal{D}_1\mathcal{D}_q\mathcal{D}_q\rangle}&=-\tfrac{p\,q}{2700}\,\left(23490\,p\,q\,(p+q)-15390(p^2+q^2)\right.\\
&\left.\,\,\,\,+43710\,(p+q)-21075\,p\,q-118883\right)\,,
\end{split}
\end{align}
which are independent of the undetermined parameter $\alpha$. Note, moreover, that while we do not have access to other correlators at two loops, we can cross-check the prediction for $\langle a^{(1)}\rangle$ against the results of Section \ref{sec:ppqq}, finding perfect agreement. A similar logic applies to averages extracted to other four-point functions.

There are two ways to fix the coefficient $\alpha$. The first is a bit indirect and consists in realizing that there has to be a unique value of $\alpha$ and physical quantities should be independent of it. Now, suppose that we knew the product of the eigenvalues of $\Gamma^{(3)}_{\Delta}$, which we will call $\gamma^{(3)}_{\mathcal{O}^{(\D=4)}_1}$ and $\gamma^{(3)}_{\mathcal{O}^{(\D=4)}_2}$. Then we could solve the equation
\begin{align}
\det\,\Gamma^{(3)}_{\D}=\gamma^{(3)}_{\mathcal{O}_1}\,\gamma^{(3)}_{\mathcal{O}_2}=\tfrac{49}{64800}(-1126242 \alpha ^2+174342 \sqrt{10} \alpha +37927675)\,,
\end{align} 
which in general has two solutions. A unique solution is present if and only if 
\begin{align}
\alpha=\frac{29057 }{187707}\sqrt{\frac{5}{2}}\,,
\end{align}
which gives
\begin{align}
\mu^{(1)}_{pp\widehat{\mathcal{O}}^{(\D=4)}}=
\left(\tfrac{p(23279924-7575105p)}{5631210\sqrt{7}}\,,\,\,\tfrac{p (-62569 (p-1) (36 p-1)-290570)}{375414 \sqrt{70}}\right)\,,
\end{align}
and
\begin{align}\label{Gamma3_Delta4_sol}
\Gamma^{(3)}_{\D=4}=
\begin{pmatrix}
 -\frac{90573427}{312845} & -\frac{11071459 }{187707} \sqrt{\frac{5}{2}}\\
 -\frac{11071459 }{187707} \sqrt{\frac{5}{2}}& -\frac{291199111}{2252484} \\
\end{pmatrix}\,.
\end{align}
Note that, {\it a posteriori}, the criterion that fixes $\alpha$ seems to be that $\mu^{(1)}_{pp\widehat{\mathcal{O}}^{(\D=4)}_{L=2}}$ is actually one degree less than $\mu^{(1)}_{pp\widehat{\mathcal{O}}^{(\D=4)}_{L=4}}$ as a polynomial in $p$. The same holds for the free theory result \label{C0ppDelta4_result} and presumably can be used as a criterion at higher orders as well. There is, however, a more instructive way to fix $\alpha$. Note that in \eqref{a1gamma3ppqq_system} we have insisted on using the length basis, but we could have used the basis of eigenstates instead: the solution for $\mu^{(0)}$ and $\gamma^{(2)}$ is known from the analysis above, while $\mu^{(1)}$ and $\gamma^{(3)}$ are unknowns, but since we are diagonalizing the anomalous dimensions matrix we have one less unknown than previously, where the analysis was done in the length basis.  The anomalous dimensions $\gamma^{(3)}_{\mathcal{O}^{(\D=4)}_1}$ and $\gamma^{(3)}_{\mathcal{O}^{(\D=4)}_2}$ found with this method match precisely the eigenvalues of \eqref{Gamma3_Delta4_sol}, thus confirming the validity of the previous analysis. We are then ready to put everything together, and give the result for the dimension of $\mathcal{O}^{(\D=4)}_1$ and $\mathcal{O}^{(\D=4)}_2$
\begin{align}
\begin{split}
h[{\mathcal{O}}^{(\D=4)}_{1}]&=4-\frac{14}{\lambda^{1/2}}+\frac{7\,(1033-\sqrt{62569})}{120\,\lambda}-\frac{7\,(673805561-1581637 \sqrt{62569})}{22524840\,\lambda^{3/2}}+\ldots\,,\\
h[{\mathcal{O}}^{(\D=4)}_{2}]&=4-\frac{14}{\lambda^{1/2}}+\frac{7\,(1033+\sqrt{62569})}{120\,\lambda}-\frac{7\,(673805561+1581637 \sqrt{62569})}{22524840\,\lambda^{3/2}}+\ldots\,,\\
\end{split}
\end{align}
and their OPE coefficients
\begin{align}\label{OPEs_Delta=4sector_intro}
\begin{split}
\mu^2_{pp{\mathcal{O}}^{(\D=4)}_{1}}=&\frac{(43+120p-\sqrt{62569})^2\,p^2}{875966-2282 \sqrt{62569}}+p^2\Big[\frac{389-249p+219p^2-108p^3}{630}\\
&-\frac{510526661+5856154953 p-5169773793 p^2+1101464676 p^3}{39418470 \sqrt{62569}}\Big]\frac{1}{\lambda^{1/2}}+\ldots\,,\\
\mu^2_{pp{\mathcal{O}}^{(\D=4)}_{2}}=&\frac{(43+120p+\sqrt{62569})^2\,p^2}{875966+2282 \sqrt{62569}}+p^2\Big[\frac{389-249p+219p^2-108p^3}{630}\\
&+\frac{510526661+5856154953 p-5169773793 p^2+1101464676 p^3}{39418470 \sqrt{62569}}\Big]\frac{1}{\lambda^{1/2}}+\ldots\,.
\end{split}
\end{align}
These results are found to agree perfectly with the numerical analysis of \cite{Cavaglia:2021bnz,Cavaglia:2022qpg} after a numerical fit at strong coupling (for $p=1$). We thank the authors of those papers for sharing the relevant data with us.
\vspace{0.2cm}
\paragraph{Singlet operators with $\D=6$.} Moving to $\D=6$, we have a dimension four vector space $\mathtt{d}(\mathcal{L}^{\D=6}_{0,[0,0]})$, containing one operator of length two, two operators of length four and one operator of length six. We choose our basis to be\footnote{More precisely, we have a derivative structure of the type
\begin{align}
\widehat{\mathcal{O}}^{\D=6}_{L=4,a}=(\varphi\partial^2\varphi-\tfrac{15}{8}(\partial\varphi)^2)\,\varphi^2+\tfrac{3}{8}(\varphi\partial\varphi)^2+\ldots\,,\quad 
\widehat{\mathcal{O}}^{\D=6}_{L=4,b}=(\varphi\partial^2\varphi+\tfrac{3}{2}(\partial\varphi)^2)\,\varphi^2-3(\varphi\partial\varphi)^2+\ldots\,,
\end{align}
see \cite{Ferrero:2023znz} for details of the explicit construction.}
\begin{align}
\widehat{\mathcal{O}}^{\D=6}_{L=2}\sim \partial^4\varphi^2+\ldots\,,\quad 
\widehat{\mathcal{O}}^{\D=6}_{L=4,a}\sim \partial^2\varphi^4+\ldots\,,\quad 
\widehat{\mathcal{O}}^{\D=6}_{L=4,b}\sim \partial^2\varphi^4+\ldots\,,\quad 
\widehat{\mathcal{O}}^{\D=6}_{L=6}\sim \varphi^6\,,
\end{align}
where again the dots denote terms with $\psi$ and $f$. Moreover, although we have not been careful in spelling out the structure and the normalization, we are considering all operators to be unit-normalized in the sense of the two-point functions. Note that since there are two operators for $L=4$ there is arbitrariness in the choice of basis within $\mathtt{d}_{L=4}(\mathcal{L}^{\D=6}_{0,[0,0]})$: we have chosen an orthonormal basis for simplicity, but this is in general not necessary. We need to choose four correlators that can explore this four-dimensional degeneracy space. We can use $\langle\mathcal{D}_p\mathcal{D}_p\mathcal{D}_q\mathcal{D}_q\rangle$ correlators to span one direction in $\mathtt{d}_L(\D)$ for each $L$, so we need one more independent piece of information since there are two operators with length four. To this end, we can use $\langle\mathcal{D}_1\mathcal{D}_1\mathcal{D}_2\mathcal{L}_\ext \rangle$ choosing the simplest $\mathcal{L}_\ext$, namely the unique operator with $\D=2$ and $L=2$ in the free theory. Note that the $\langle \gamma^{(2)}\rangle$ extracted from this can be read off from \eqref{gamma2_112L_L=2}. We will not repeat a detailed analysis here, but following the same logic as above we can obtain
\begin{align}
\Gamma^{(2)}_{\D=6}=
\begin{pmatrix}
\frac{110619}{560} & \frac{9 \sqrt{165}}{8} & \frac{3 \sqrt{429}}{2} & 0 \\
\frac{9 \sqrt{165}}{8} & \frac{3541}{24} & \frac{28}{3} \sqrt{\frac{13}{5}} & \sqrt{\frac{1365}{2}} \\
\frac{3 \sqrt{429}}{2} & \frac{28}{3} \sqrt{\frac{13}{5}} & \frac{20323}{120} & \sqrt{42} \\
0 & \sqrt{\frac{1365}{2}} & \sqrt{42} & \frac{1035}{8}
\end{pmatrix}\,,
\end{align}
from which the eigenvalues can be easily read off. An interesting observation is that despite the structure \eqref{Gamma2blocks}, that is respected by this result, the eigenstates of $\Gamma^{(2)}_{\D=6}$ mix operators of all lengths: the simplification in the structure of the dilatation operator seems to be present only in the length basis but is completely lost once we go to the basis of eigenstates.
\vspace{0.2cm}
\paragraph{Higher $\D$.} Moving to higher $\D$ one has operators of length up to $L=\D$ and moreover the degeneracy spaces $\mathtt{d}_L(\D)$ are non-trivial for $L\ge 6$ starting from $\D=8$. Thus, $\langle\mathcal{D}_p\mathcal{D}_p\mathcal{D}_q\mathcal{D}_q\rangle$ are no longer enough to resolve the degeneracy for $L\ge 6$ and $\langle\mathcal{D}_1\mathcal{D}_1\mathcal{D}_2\mathcal{L}_\ext \rangle$ can only be used for $L=4$. Therefore, starting with $\D=8$ the system of correlators that we have considered is no longer enough to fully resolve the degeneracy and determine the eigenstates and eigenvalues of the dilatation operator. Instead, for $\D\ge 8$ we limit to extracting $\Gamma^{(2)}_{\D,2\to 4}$ and computing its norm, which is what allows us to guess \eqref{deltaGamma2sq}.
\vspace{0.2cm}
\paragraph{Operators in the $\{0,[0,1]\}$ representation.} While our main goal in addressing the mixing problem was to compute the average \eqref{gamma2^2_mixing}, which concerns the sector of singlet long multiplets, as a byproduct of our analysis one can extract plenty of information regarding the spectrum of the theory. In particular, so long as there is only one degenerate operator for each length the $\langle\mathcal{D}_p\mathcal{D}_p\mathcal{D}_q\mathcal{D}_q\rangle$ correlators are enough to determine the exact eigenstates and eigenvalues of the dilatation operator. As an example, let us consider long multiplets $\mathcal{L}^{\D}_{0,[0,1]}$. From the analysis of \cite{Ferrero:2023znz}, we can argue the degeneracies listed in table \ref{tab:[0,1]degeneracies} for the first few values of $\D$.
\begin{table}[h!]
\centering
\begin{tabular}{|c||c c c c|} 
\hline
$\D$ & 3 & 4 & 5 & 6\\
\hline \hline
$L=3$ & 1 & 1 & 1 & 1\\
\hline 
$L=5$& 0 & 0 & 1 & 1 \\
\hline 
\end{tabular}
\caption{Degeneracies of long multiplets $\mathcal{L}^{\D}_{0,[0,1]}$ at strong coupling, organized by length.}
\label{tab:[0,1]degeneracies}
\end{table}
We thus see that for $\D=3,4$ there is no degeneracies and indeed the one-loop anomalous dimension of such operators follows from the general formula \eqref{gamma2nondegenerate} for non-degenerate operators. On the other hand, for $\D=5,6$ the degeneracy is twofold but the two operators have different lengths in both cases, so that from $\langle\mathcal{D}_p\mathcal{D}_p\mathcal{D}_q\mathcal{D}_q\rangle$ correlators one can extract the matrices
\begin{align}
\Gamma^{(2)}(\mathcal{L}^{\D=5}_{0,[0,1]})=
\left(
\begin{array}{cc}
 \frac{979}{10} & \sqrt{231} \\
 \sqrt{231} & \frac{303}{4} \\
\end{array}
\right)\,,\qquad 
\Gamma^{(2)}(\mathcal{L}^{\D=5}_{0,[0,1]})=
\left(
\begin{array}{cc}
 \frac{3455}{24} & \frac{5 \sqrt{91}}{3} \\
 \frac{5 \sqrt{91}}{3} & \frac{2495}{24} \\
\end{array}
\right)\,,
\end{align}
where in both cases we have used a basis where the first operator has $L=3$ and the second has $L=5$.

\section{Discussion}\label{sec:discussion}

In this paper we have presented several results for perturbative four-point in the 1d defect CFT defined by the half-BPS Wilson loop in planar $\mathcal{N}=4$ SYM, focusing on the case of the fundamental Wilson loop at strong 't~Hooft coupling $\lambda$. Such results are obtained through the application of an analytic bootstrap method that consists in formulating an ansatz in terms of HPLs, which has proven particularly useful in the context of 1d holographic CFTs. Our results include correlation functions between half-BPS operators as well as mixed correlators between short and long multiplets, which we bootstrapped to second order in perturbation theory. This was instrumental to addressing a certain mixing problem due to the degeneracy of operators at $\lambda=\infty$, which in turn allowed to bootstrap the four-point function of the displacement supermultiplet up to fourth perturbative order, corresponding to three-loop Witten diagrams.

There are a few aspects that we would like to emphasize and that are worth of future investigations. The first concerns the explicit expression of our perturbative results, which have turned out to be considerably simpler (especially at high loop order) than the ansatz that we have started with. More precisely, in the final expressions the HPLs combine to form certain functions known as Lewin polylogarithms, in such a way that the results can be expressed purely in terms of ordinary logs and the weight-three Lewin polylog. It is then natural to investigate the origin of such simplicity and try to exploit it to obtain new results. In the main text we have discussed a relation between the functions appearing in our results and the single-valued HPLs that naturally occur in higher-dimensional four-point functions: it would certainly be interesting to understand whether such connection persists at higher perturbative orders and to develop a precise understanding of the type of functions that can appear in the 1d correlators we are interested in.\footnote{An understanding of the structure of four-point functions in this model can also be developed considering the large charge approach of \cite{Giombi:2022anm}, where perturbative results in terms of HPLs are also obtained.} This would allow, for example, to formulate a more stringent ansatz for the displacement four-point function at higher orders, possibly allowing to bypass (at least to a certain extent) the study of the mixing problem at higher orders. This would be in the spirit of \cite{Huang:2021xws,Drummond:2022dxw}, where a suitably constrained ansatz was formulated for the spacetime expression of the two-loop graviton four-point amplitude in AdS$_5\times S^5$, allowing to obtain the result without addressing the mixing problem.\footnote{See also \cite{Huang:2023oxf} for an analogous gluon scattering amplitude in AdS$_5\times S^3$. In that case, however, it is not completely clear that the two-loop gluon amplitude can be separated from the contribution of gravitons, which appear at the same order in the large central charge expansion.}

The mixing problem itself is certainly one of the central points of our work. The more involved nature of this problem compared to those usually encountered at first order in holography has led us to conduct a careful investigation of the structure of the dilatation operator in perturbation theory, reminiscent of what has been done for gauge theories at weak coupling \cite{Beisert:2004ry}. Moreover, the structure of the theory has forced us to include in our analysis mixed correlators between short and long multiplets, which has not been considered so far in the study of holographic correlators. We would like to propose the idea that the methods presented here should be of universal applicability in other perturbative CFTs: in the mixing problems considered so far in the literature one was always able to resort to simpler methods, but enlarging the class of observables and understanding the structure of the dilatation operator will presumably be instrumental to tackle mixing at higher orders. A possible application of this could be the study of mixing between triple-trace operators in the holographic regime of $\mathcal{N}=4$ SYM, where in \cite{Bissi:2021hjk} it was suggested that external $\tfrac{1}{4}$-BPS operators could play a role.\footnote{See \cite{Harris:2024nmr} for a recent investigation of triple twist operators in the $\epsilon$-expansion using higher-point correlation functions.} Moreover, the study of mixed short-long correlators provides an entirely new class of observables in the study of holographic correlators, providing access to a vast array of new CFT data. This could potentially be relevant in the study of gluon scattering in AdS$_5\times S^3$, which after a series of recent works \cite{Alday:2021odx,Alday:2021ajh,Drummond:2022dxd,Huang:2023oxf,Behan:2023fqq,Glew:2023wik} is now on a similar footing to graviton scattering in AdS$_5\times S^5$. In this case, gravitons propagating in the bulk reduce to a combination of both short and long multiplets in AdS$_5$, so that a better understanding of mixed correlators is necessary for the study of mixed scattering amplitudes between gluons and gravitons in that context.

The challenge presented by the mixing problem has led us to a detailed analysis of the spectrum of the theory at strong coupling, as well as to an investigation of the structure of the dilatation operator in perturbation theory. While at tree level we have been able to obtain the complete result for the spectrum of anomalous dimensions of long operators, which takes the simple form of a quadratic Casimir, already at second order we have focused on a particular subsector of operators. On the other hand, given the compactness of the first order result and the simplifications observed in the general structure of the dilatation operator at higher orders, it would be interesting to take a more systematic approach to understand whether higher order corrections also admit analogous expressions which could shed light on the result at all orders. This type of analysis would also be relevant for other holographic theories, for which so far investigations of the dilatation operator have been limited to the double-trace spectrum, with a thorough analysis only available for $\mathcal{N}=4$ SYM \cite{Aprile:2017xsp,Aprile:2018efk,Aprile:2019rep}. We hope that our work will provide useful input in this direction.

It is also worth mentioning that the same model considered here has recently been subject of thorough investigations using a combination of integrability and numerical bootstrap methods, a technique that has been dubbed ``bootstrability'' \cite{Grabner:2020nis,Cavaglia:2021bnz,Cavaglia:2022qpg,Cavaglia:2022yvv}. This exploits the knowledge of the spectrum from integrability as an input for a numerical bootstrap algorithm, which combined with the integrated correlator constraints found in \cite{Cavaglia:2022qpg,Cavaglia:2022yvv,Drukker:2022pxk} has been shown to put stringent bounds on the OPE coefficients of certain long multiplets. All of these developments make the half-BPS Wilson line defect an extremely interesting playground where one can learn how to combine different approaches in an attempt to completely solve the model. In this context, it is going to be interesting to work out the constraints arising from the study of mixed correlators (with short and long multiplets) in the bootstrability approach. The methods and results developed in the companion paper \cite{Ferrero:2023znz}, as well as the analytic results obtained here at strong coupling, provide essential input and cross-checks to explore this new territory.

Another interesting direction would be that of considering generalizations of the current setup, which in the context of the half-BPS Wilson line in $\mathcal{N}=4$ SYM could include the study of $1/N$ corrections, Wilson lines in other representations than the fundamental, or higher point functions. We expect the bootstrap techniques used here to apply to those cases as well, and variations on the current setting would improve our understanding of the perturbative structure of correlation functions. This is true in particular for higher-point functions, which have only recently become subject of investigations in the context of the analytic bootstrap \cite{Goncalves:2019znr,Alday:2022lkk,Barrat:2021tpn,Barrat:2022eim,Goncalves:2023oyx,Alday:2023kfm}.\footnote{See also \cite{Giombi:2023zte} for a recent computation of a six-point function on the half-BPS Wilson line in $\mathcal{N}=4$ SYM from direct computation using a novel method.} Moreover, the integrated correlator constraints of \cite{Cavaglia:2022qpg,Cavaglia:2022yvv,Drukker:2022pxk} have not played a role in our analysis\footnote{One can check, however, that they are satisfied, which provides a further consistency check of the validity of our results.}, but could be instrumental to the study of these generalizations, and it would be interesting to study their extension to higher-point functions or four-point functions between half-BPS operators of higher rank.

Besides the half-BPS Wilson line, there are other interesting defect SCFTs that one could study with similar methods. This is certainly true for the half-BPS Wilson line in ABJM, which has been studied at tree level with analogous techniques to those used here \cite{Bianchi:2020hsz}, and where the study of perturbative corrections would probably have a similar structure to that presented in our paper. On the other hand, the half-BPS surface defect in 6d $\mathcal{N}=(2,0)$ CFTs initiated in \cite{Drukker:2020swu}\footnote{See also \cite{Meneghelli:2022gps} for a study of bulk correlators in the presence of this defect.} provides an even more interesting extension, since one can use Mellin space methods which have the potential to drastically simplify the expression of the results, in a model where we expect the structure of the dilatation operator and the type of mixing problem to be analogous to the ones that we have considered.

Finally, we would also like to mention that having obtained a large class of explicit results for four-point functions one could use them to further investigate analytic tools for 1d CFTs. In particular, an inherently one-dimensional formulation of Mellin space was proposed in \cite{Bianchi:2021piu}, and it would certainly be interesting to understand whether it allows to reformulate the bootstrap problem directly in Mellin space, with similar simplifications to those occurring in higher dimensions. This would be also relevant for the study of the structure of Mellin amplitudes for holographic correlators beyond one loop, which is currently unknown. In a similar spirit, a 1d version of the Lorentzian inversion formula was proposed in \cite{Mazac:2018qmi}, but it was never applied to a concrete 1d model. It is natural to ask whether the use of such an inversion formula allows for a simpler derivation of our results, or a more systematic reformulation, or even an explanation of why only certain combinations of HPLs seem to appear in our results.

\section{Acknowledgments}

We thank Fernando Alday and Pedro Liendo for collaboration at early stages of this work and for many fruitful discussions. We thank Fernando Alday for sharing with us unpublished notes and are grateful to Alex Gimenez-Grau and Johan Henriksson for sharing with us some Mathematica code and some unpublished results respectively. P.F. would like to thank Leonardo Rastelli, Gabriel Cuomo and Adar Sharon for useful discussions. The work of C.M. has received funding from the European Union’s Horizon 2020 research and innovation program under the Marie Sklodowska Curie grant agreement No 754496. The work of P.F. has received funding from the European Research Council (ERC) under the European Union’s Horizon 2020 research and innovation program (grant agreement No 787185).

\appendix
\addtocontents{toc}{\protect\setcounter{tocdepth}{1}}

\setcounter{tocdepth}{1}
\section{Harmonic polylogarithms}\label{app:polylogs}

In this appendix we discuss HPLs more in detail, giving an explicit characterization of the basis of functions used in this paper, as well as its properties. In particular, we shall discuss the transformation of the HPLs of interest under cyclic and braiding transformations, as well as their properties as functions of a complex variable.

\subsection{A basis of harmonic polylogarithms}

Multiple polylogarithms were introduced in the mathematical literature long ago \cite{MultiplePolylogs},  and their structure and properties was more recently explored by Goncharov \cite{Goncharov:1998kja,Goncharov:2001iea}, to the extent that they are sometimes referred to as Goncharov polylogarithms in the physics literature.  They can be defined alternatively as nested sums or as iterated integrals (see \cite{Duhr:2011zq} for a physics-oriented review).  Here we shall take the latter approach and define,  for $n\ge 0$,  multiple polylogarithms recursively via
\begin{align}
G\left(a_{1}, \ldots, a_n ; \chi\right)=\int_{0}^{\chi} \frac{\text{d} t}{t-a_{1}} G\left(a_{2}, \ldots, a_n ; t\right)\,,
\end{align}
with $G(\chi)=G(;\chi)=1$,  for some constants $a_i\in\mathbb{C}$.  Following \cite{Duhr:2011zq},  we shall refer to the vector $\vec{a}$ as the {\it vector of singularities},  whose length $n$ we call the {\it weight},  or {\it transcendentality} (denoted with $\mathtt{t}$ in the main text) of the functions $G(\vec{a};\chi)$.  

In this paper we are interested in a special class of multiple polylogarithms,  which we refer to as {\it harmonic polylogarithms} (HPLs) following, {\it e.g.}, \cite{Drummond:2013nda}\footnote{HPLs were introduced in the physics literature in \cite{Remiddi:1999ew},  where the entries of the vector of singularities $\vec{a}$ were allowed to take values in $\{-1,0,1\}$.  Here we use the term HPL in a restricted sense.}. We then obtain the definition given in Section \ref{sec:bootstrap} of HPLs as
\begin{align}\label{defHPLsappendix}
H(a_1,\dots,a_n;\chi)=\int_{0}^{\chi}\text{d}t\,f_{a_1}(t)\,H(a_2,\dots,a_n;t)\,, \qquad a_i\in\{0,1\}\,,
\end{align}
with 
\begin{align}\label{f01HPLs}
f_0(t)=\frac{1}{t}\,, \qquad f_1(t)=\frac{1}{1-t}\,,
\end{align}
and 
\begin{align}
H(;\chi)=1\,, \quad H(\vec{0}_n;\chi)=\frac{1}{n!}\log^n\chi\,.
\end{align}
From the definition \eqref{defHPLsappendix} two basic properties of HPLs should be clear. First, for fixed transcendentality $\mathtt{t}$ there are $2^{\mathtt{t}}$ independent HPLs, corresponding to the number of ways of assigning the entries of the vector of singularities from the set $\{0,1\}$. Second, when $\chi$ is taken to be a complex variable the HPLs are holomorphic functions for $\chi\in\mathbb{C}\setminus (-\infty,0]\cup [1,+\infty)$, with branch points located at $\chi=0,1,\infty$, due to the poles of the functions $f_a(t)$ in \eqref{f01HPLs}. This reflects the analytic structure of 1d correlators,  as discussed in \ref{sec:bootstrap}. In particular,  while the branch point at $\chi=\infty$ is always present,  if $a_n=0$ ($a_n=1$),  then there is also a branch point at $\chi=0$ ($\chi=1$). Such branch points are related to logarithmic singularities: if for some $k$ one has $a_{n-k-1}=1$ and $a_{n-k}=...=a_n=0$, then in the expansion around $\chi=0$ one encounters powers of $\log\chi$ up to and including $\log^k\chi$. The same holds interchanging 0 and 1, in which case one has powers of $\log(1-\chi)$ in the expansion around $\chi=1$.

A perhaps less obvious, but extremely useful for our purposes, property of HPLs is that they form a closed set under the crossing symmetry group $S_3$ that is relevant for four-point functions \cite{Drummond:2013nda}.  Let us recall that the action of $S_3$ on the unique cross-ratio $\chi$ of one-dimensional CFTs is given by
\begin{align}\label{S3onchi}
\chi \,\,\to\,\,\left\{ \chi\,,1-\chi\,, \frac{\chi}{\chi-1}\,,\frac{1}{\chi}\,,\frac{1}{1-\chi}\,,\frac{\chi-1}{\chi}\right\}\,,
\end{align}
thus permuting the three singularities of HPLs located at $\chi=\{0,1,\infty\}$. This means that when evaluating a certain HPL at one of the arguments given in \eqref{S3onchi}, this can always be expressed in terms of sums and products of HPLs of the same weight or lower, evaluated at $\chi$. More details on this procedure can be found in section 6 of \cite{Remiddi:1999ew}. This gives the precise sense in which we can use HPLs as a {\it basis} of functions in the bootstrap problem: when requiring invariance of a certain correlator under the transformations \eqref{S3onchi}, one can always map HPLs with argument in \eqref{S3onchi} to argument $\chi$, and then require that the coefficient of each one of this vanishes, given that they provide the basis as a vector space, in the sense of \cite{AlgebrePolylogs,UlanskiiHPLs,BROWN2004527}. The identities satisfied by HPLs after the transformations \eqref{S3onchi} also involve certain constants, known as {\it multiple $\zeta$-values} (MZVs), which for our purposes can be defined to be the result of evaluating HPLs at $\chi=1$, when the result is finite. In particular, since we will consider HPLs of transcendentality $\mathtt{t}\le 4$, we will only be interested in ordinary $\zeta$-values, which arise from 
\begin{align}
H(\vec{0}_{n-1},1;1)=\zeta(n)=\sum_{k=1}^{\infty}\frac{1}{k^n}\,,
\end{align}
which is finite for $n\ge 2$. Notice that MZVs are naturally assigned a transcendentality, which is that of the HPLs that they originate from.

The definition \eqref{defHPLsappendix} provides a complete characterization of HPLs, as well as a uniform way to label them which is uniform in their transcendentality. However, for the purposes of this paper, they do not provide an ideal basis, mainly for the following reason. The natural decomposition \eqref{Gpowerslog} of perturbative correlators, which is inherited by the structure of the OPE, one would like a basis of functions where all powers of $\log\chi$ can be easily read off: ideally, either one sees an explicit power of $\log\chi$, or the function is regular in its expansion around $\chi=0$. While it is clear which HPLs are regular at $\chi=0$ and which are not, each of the singular one can contain all powers of $\log\chi$ up to a maximum value, so that in \eqref{Gpowerslog} each HPL could in principle contribute to various $G^{(\ell)}_{\log^k}$, while we want to be able to make an ansatz for each $G^{(\ell)}_{\log^k}$ independently. It is then convenient to make a change of basis and switch, for each transcendentality, to a set of functions where the logarithmic singularities are completely transparent. A systematic way to achieve this can be defined using the shuffle algebra (see, {\it e.g.}, \cite{Remiddi:1999ew}) of HPLs and is described for general weight in section 5 of \cite{Maitre:2005uu}. Fortunately, since we are ultimately interested in $\mathtt{t}\le 4$, all HPLs we are interested in can be expressed in terms of sums and products of $\log\chi$ and the simpler class of Nielsen's generalized polylogarithms, defined by \cite{NielsenPolys}
\begin{align}\label{defNielsen}
S_{n,p}(\chi)=\frac{(-1)^{n+p-1}}{(n-1)!p!}\int_0^1\frac{\mathrm{d} t}{t}\,\log^{n-1}t\,\log^p(1-\chi\,t)=\sum_{k=0}^{\infty}\frac{(-1)^k\, \mathcal{S}^{(p)}_{p+k}}{(p+k)!(p+k)^n}\,\chi^{p+k}=H(\vec{0}_n,\vec{1}_p;\chi)\,,
\end{align}
where we have stressed that Nielsen's polylogs admit a regular expansion around $\chi=0$, given in terms of Stirling numbers of the first kind,  defined for $k\ge m$
\begin{align}\label{defStirling}
\mathcal{S}^{(m)}_k=\sum_{i=0}^{k-m}\frac{1}{i!}\dbinom{k-1+i}{k-m+i}\dbinom{2k-m}{k-m-i}\sum_{j=0}^i(-1)^{i+j}\dbinom{i}{j}(i-j)^{k-m+i}\,,
\end{align}
with $\mathcal{S}^{(k)}_k=1$.  A special case of Nielsen's polylogs is that of ordinary polylogs, which arise for $p=1$
\begin{align}\label{defPolylogs}
\Li_n(\chi)=S_{n-1,1}(\chi)=H(\vec{0}_{n-1},1;\chi)=\sum_{k=1}^{\infty}\frac{\chi^k}{k^n}\,,
\end{align}
with the logarithm corresponding to the special case $n=1$: $\Li_1(\chi)=-\log(1-\chi)$.  For a detailed discussion of the properties of ordinary polylogarithms,  see \cite{lewin1981polylogarithms,lewin1991structural}. These functions allow us, for each fixed transcendentality $\mathtt{t}$, to define the basis $\mathcal{B}_{\mathtt{t}}$ of polylogs used in the main text,  for $0\le \mathtt{t}\le 4$. All HPLs of transcendentality $\mathtt{t}$ can be written as linear combinations of the elements of the set $\mathcal{B}_{\mathtt{t}}$, as shown explicitly in \eqref{HPLs_to_Nielsen}, so using the functions below to formulate our ansatz is equivalent to using HPLs. Our choice is
\begin{align}\label{basis3loops}
\begin{split}
\mathcal{B}_0=&\,\big\{1\big\}\,,\\
\mathcal{B}_1=&\,\big\{\log(1-\chi),\,\log \chi\big\}\,,\\
\mathcal{B}_2=&\,\big\{\Li_2(\chi),\,\log^2(1-\chi),\,\log(1-\chi)\,\log \chi,\,\log^2\chi\big\}\,,\\
\mathcal{B}_3=&\,\big\{\Li_3(\chi),\,S_{1,2}(\chi)\,,\log^3(1-\chi)\,,\Li_2(\chi)\,\log(1-\chi),\,\Li_2(\chi)\log \chi,\\
&\,\,\log^2(1-\chi)\,\log \chi,\,\log(1-\chi)\log^2 \chi,\,\log^3\chi\big\}\,,\\
\mathcal{B}_4=&\,\big\{\Li_4(\chi)\,,S_{2,2}(\chi),\,S_{1,3}(\chi),\,\Li_3(\chi)\,\log(1-\chi)\,,S_{1,2}(\chi)\,\log(1-\chi)\,,\Li_2^2(\chi),\\
&\,\,\Li_2(\chi)\,\log(1-\chi),\,\log^4(1-\chi),\,\Li_3(\chi)\,\log \chi,\,S_{1,2}(\chi)\,\log \chi\,,\log^3(1-\chi)\,\log \chi,\\
&\,\,\Li_2(\chi)\,\log(1-\chi)\,\log \chi,\,\Li_2(\chi)\log^2 \chi,\,\log^2(1-\chi)\,\log^2 \chi,\,\log(1-\chi)\log^3 \chi,\,\log^4\chi\big\}\,.
\end{split}
\end{align}
By construction all functions involved,  except for {\it explicit} powers of $\log\chi$,  are analytic at $\chi=0$,  and actually they are so for $\chi\in \mathbb{C}\setminus[1\,+\infty)$.  Moreover, we notice that each $\mathcal{B}_{\mathtt{t}}$ can be written as the disjoint union of two sets of the same size. The first contains only functions that are regular at $\chi=0$ while the second, which contains functions with logarithmic singularities, can be obtained multiplying each function in $\mathcal{B}_{\mathtt{t}-1}$ by $\log\chi$. In other words, we have
\begin{align}
\mathcal{B}_{\mathtt{t}}=\mathcal{B}^{(0)}_{\mathtt{t}}\cup\left[\mathcal{B}_{\mathtt{t}-1}\,\times\,\log \chi\right]\,,
\end{align}
where $\mathcal{B}^{(0)}_{\mathtt{t}}$ contains functions that are regular at $\chi=0$ (the first $2^{\mathtt{t}-1}$ in each set given in \eqref{basis3loops}). Then, each function $G^{(\ell)}_{\log^k}(\chi)$ in \eqref{Gpowerslog} can be written as a linear combination of functions in $\mathcal{B}^{(0)}_{\ell-k}$, with coefficients that are rational functions. Note that,  in principle,  we could have used only ordinary polylogarithms,  using the identities
\begin{align}
\begin{split}
S_{1,2}(\chi)=\,&\Li_3(\chi)+\Li_3\left(\frac{\chi}{\chi-1}\right)-\Li_2(\chi)\,\log(1-\chi)-\frac{1}{6}\log^3(1-\chi)\,,\\
S_{2,2}(\chi)=\,&\zeta(4)+\zeta(3)\,\log(1-\chi)+\frac{1}{2}\zeta(2)\log^2(1-\chi)+\frac{1}{24}\log^4(1-\chi)-\frac{1}{6}\log^3(1-\chi)\log \chi\\
&-\Li_3(\chi)\,\log(1-\chi)-\Li_4(1-\chi)+\Li_4(\chi)+\Li_4\left(\frac{\chi}{\chi-1}\right)\,,\\
S_{1,3}(\chi)=\,&\zeta(4)+\zeta(3)\,\log(1-\chi)+\frac{1}{2}\zeta(2)\log^2(1-\chi)+\frac{1}{6}\log^4(1-\chi)-\frac{1}{6}\log^3(1-\chi)\,\log \chi\\
&+\frac{1}{2}\Li_2(\chi)\,\log^2(1-\chi)-\Li_3(\chi)\,\log(1-\chi)-\Li_3\left(\frac{\chi}{\chi-1}\right)\,\log(1-\chi)-\Li_4(1-\chi)\,,
\end{split}
\end{align}
to replace $S_{1,2}(\chi)$,  $S_{2,2}(\chi)$ and $S_{1,3}(\chi)$ in \eqref{basis3loops} with $\Li_3\left(\frac{\chi}{\chi-1}\right)$,  $\Li_4(1-\chi)$ and $\Li_4\left(\frac{\chi}{\chi-1}\right)$.  However,  this necessarily introduces $\Li_4(1-\chi)$ into the game,  which has a branch point at $\chi=0$,  spoiling the property that we discussed before.  One could still avoid the use of $S_{1,2}(\chi)$ and $S_{1,3}(\chi)$ at the expense of dealing with polylogarithms of argument $\frac{\chi}{\chi-1}$,  but we chose to work with the basis \eqref{basis3loops} as it also has the property that all functions have an expansion around $\chi=0$ which can be given in closed form,  using the definitions \eqref{defPolylogs} and \eqref{defNielsen}.

Finally, we can give the expressions of all HPLs of $\mathtt{t}\le 4$ in terms of our basis \eqref{basis3loops}.
\begin{align}\label{HPLs_to_Nielsen}
\begin{split}
\mathtt{t}=0:\,\,&H(;\chi)=1\,,\\
\mathtt{t}=1:\,\,&H(0;\chi)=\log\chi\,, \quad
H(1;\chi)=-\log(1-\chi)\,,\\
\mathtt{t}=2:\,\,&H(0,0;\chi)=\frac{1}{2}\log^2\chi\,, \quad
H(0,1;\chi)=\Li_2(\chi)\,, \\
&H(1,0;\chi)=-\Li_2(\chi)-\log(1-\chi)\log\chi\,, \quad
H(1,1;\chi)=\frac{1}{2}\log^2(1-\chi)\,,\\
\mathtt{t}=3:\,\,&H(0,0,0;\chi)=\frac{1}{3!}\log^3\chi\,, \quad H(0,0,1;\chi)=\Li_3(\chi)\,,\quad H(0,1,0;\chi)=-2\Li_3(\chi)+\Li_2(\chi)\,\log\chi\,, \\
&H(1,0,0;\chi)=\Li_3(\chi)-\Li_2(\chi)\,\log\chi-\frac{1}{2}\log(1-\chi)\,\log^2\chi\,, \quad
H(0,1,1;\chi)=S_{1,2}(\chi)\,,\\
&H(1,0,1;\chi)=-2S_{1,2}(\chi)-\Li_2(\chi)\,\log(1-\chi)\,,\quad
H(1,1,1;\chi)=-\frac{1}{3!}\log^3(1-\chi)\,,\\
&H(1,1,0;\chi)=S_{1,2}(\chi)+\Li_2(\chi)\,\log(1-\chi)+\frac{1}{2}\log^2(1-\chi)\,\log\chi\,,\\
\mathtt{t}=4:\,\,&H(0,0,0,0;\chi)=\frac{1}{4!}\log^4\chi\,, \quad
H(0,0,0,1;\chi)=\Li_4(\chi)\,, \\
&H(0,0,1,0;\chi)=-3\,\Li_4(\chi)+\Li_3(\chi)\,\log\chi\,,\\
&H(0,1,0,0;\chi)=3\Li_4(\chi)-2\Li_3(\chi)\,\log\chi+\frac{1}{2}\Li_2(\chi)\,\log^2\chi\,,\\
&H(1,0,0,0;\chi)=-\Li_4(\chi)+\Li_3(\chi)\,\log\chi-\frac{1}{2}\Li_2(\chi)\,\log^2\chi-\frac{1}{6}\log(1-\chi)\,\log^3\chi\,,\\
&H(0,0,1,1;\chi)=S_{2,2}(\chi)\,, \quad
H(0,1,0,1;\chi)=-2S_{2,2}(\chi)+\frac{1}{2}\Li_2^2(\chi)\,,\\
&H(1,0,0,1;\chi)=-\Li_3(\chi)\,\log(1-\chi)-\frac{1}{2}\Li_2^2(\chi)\, \quad 
H(0,1,1,0;\chi)=S_{1,2}(\chi)\,\log\chi-\frac{1}{2}\Li_2^2(\chi)\,\\
&H(1,1,0,0;\chi)=-S_{2,2}(\chi)+S_{1,2}(\chi)\,\log\chi-\Li_3(\chi)\,\log(1-\chi)\\
&\hspace{2.7cm}+\Li_2(\chi)\,\log(1-\chi)\,\log\chi+\frac{1}{4}\log^2(1-\chi)\,\log^2\chi\,,\\
&H(1,0,1,0;\chi)=2S_{2,2}(\chi)-2S_{1,2}(\chi)\,\log\chi+2\Li_3(\chi)\,\log(1-\chi)+\frac{1}{2}\Li_2^2(\chi)\\
&\hspace{2.7cm}-\Li_2(\chi)\,\log(1-\chi)\,\log\chi\,,\quad H(0,1,1,1;\chi)=S_{1,3}(\chi)\,,\\
&H(1,0,1,1;\chi)=-S_{1,3}(\chi)-S_{1,2}(\chi)\,\log(1-\chi)\,,\\
&H(1,1,0,1;\chi)=3 S_{1,3}(\chi)+2S_{1,2}(\chi)\,\log(1-\chi)+\frac{1}{2}\Li_2(\chi)\,\log^2(1-\chi)\,,\\
&H(1,1,1,0;\chi)=-S_{1,3}(\chi)-S_{1,2}(\chi)\,\log(1-\chi)-\frac{1}{2}\Li_2(\chi)\,\log^2(1-\chi)\\
&\hspace{2.7cm}-\frac{1}{6}\log^2(1-\chi)\,\log^2\chi\,,\quad
H(1,1,1,1;\chi)=\frac{1}{4!}\log^4(1-\chi)\,.
\end{split}
\end{align}

The basis of Nielsen's polylogs is also particularly convenient because it allows us to express the transformation properties of the basis functions \eqref{basis3loops} under the $S_3$ transformations \eqref{S3onchi} in a uniform way. The whole group is generated by the two basic transformations
\begin{align}\label{appBcrossingbraiding}
\begin{split}
\text{cyclic}:&\quad \chi \to 1-\chi\,,\\
\text{braiding}:&\quad \chi\to \frac{\chi}{\chi-1}\,,
\end{split}
\end{align}
which are ultimately all we are interested in, as detailed in Section \ref{sec:bootstrap}. From \cite{NielsenPolys} we read 
\begin{align}\label{S3NielsenPolylogs}
\begin{split}
S_{n,p}(1-\chi)&=\frac{(-1)^p}{n!p!}\log^n(1-\chi)\log^p\chi+\sum_{j=0}^{n-1}\frac{\log^j}{j!}\left(S_{n-j,p}(1)-\sum_{k=0}^{p-1}(-1)^k\frac{\log^k\chi}{k!}S_{p-k,n-j}(\chi)\right)\,,\\
S_{n,p}\big(\tfrac{\chi}{\chi-1}\big)&=(-1)^n\frac{\log^{n+p}(1-\chi)}{(n+p)!}+\sum_{j=0}^{n-1}\sum_{k=0}^{n-j-1}(-1)^{p+j+k}\frac{\log^j(1-\chi)}{j!}\binom{p+k-1}{k}S_{n-j-k,p+k}(\chi)\,,
\end{split}
\end{align}
which are sufficient to work out functional identities for all HPLs in our basis \eqref{basis3loops}. 

\subsection{Single-valued polylogarithms}

One of the main upshots of Section \ref{sec:1111} is that the result for $\langle \mathcal{D}_1\mathcal{D}_1\mathcal{D}_1\mathcal{D}_1 \rangle$ at three loops is much simpler than it could have been given the ansatz \eqref{3loopansatz1111}. Such simplicity manifests itself in two, related, ways: first, the number of HPLs appearing in the result is less than the expected one; second, such HPLs appear in special combinations that originate from single-valued HPLs. 

A well-known property of HPLs is that, while it is possible to extend their definition for complex values of their argument, giving rise to holomorphic functions, they are not single-valued. The problem of defining a generalization of HPLs that gives single-valued functions was solved in general in \cite{BROWN2004527}, by combining in a suitable manner holomorphic and anti-holomorphic HPLs. More precisely, for each HPL $H(z)$ it is possible to define a function $\tilde{H}(z,\bar{z})$ such that it is real-analytic in the punctured complex plane $\mathbb{C}\setminus\{0,1\}$ and, when $z$ and $\bar{z}$ are treated as independent complex variables, its discontinuities around $z=0$ and $z=1$ are related
(see \cite{Drummond:2012bg})
\begin{align}
\text{disc}_{z=0}\tilde{H}=\text{disc}_{\bar{z}=0}\tilde{H}\,, \qquad 
\text{disc}_{z=1}\tilde{H}=\text{disc}_{\bar{z}=1}\tilde{H}\,.
\end{align}
Besides their mathematical interest, single-valued HPLs appear in several instances of perturbative computations in physics \cite{Drummond:2012bg, Dixon:2012yy, Schnetz:2013hqa,Drummond:2013nda, Chicherin:2015edu,  DelDuca:2016lad,Aprile:2017bgs,Huang:2021xws}, so their relevance to the present context should be no surprise. In particular, CFT four-point functions are required to be single-valued in the Euclidean configuration $\bar{z}=z^*$, so that if they can be expressed in terms of HPLs, these necessarily have to be single valued. Given that it is possible to express all of our results using only the functions $\log$ and $\Li_n$ with arguments $\chi$ and $1-\chi$, we are mostly interested in single-valued generalizations of these two functions. For $\log$ this is of course straightforward, since naturally the single-valued extension of $\log z$ is $\log(z\bar{z})$, as already discussed around \eqref{D1111_1d}. For the ordinary polylogs $\Li_n(z)$, various single-valued extensions are possible, due to various authors\cite{Lewin1985TheOO,Zi,Wo1,Zag2,Zag3}, see section 2 of \cite{2021arXiv210501543P} for a nice review. We are particularly interested in the definition given by Wojtkowiak, who in \cite{Wo1} introduced the functions
\begin{align}
\tilde{L}_n(z,\bar{z})=\mathfrak{R}_n\left(\sum_{k=0}^{n-1}\frac{(-1)^k}{k!}\Li_{n-k}(z)\,\log^k|z|-\frac{(-1)^n}{n!}\log|1-z|\,\log^{n-1}|z|\right)\,,
\end{align}
where $\mathfrak{R}_n=\mathfrak{R}$ for odd $n$ and $\mathfrak{R}_n=\mathfrak{I}$ for even $n$, which are real-analytic and single-valued for $z\in\mathbb{C}\setminus\{0,1\}$. These are interesting for us because of their connection with the functions $L_n(\chi)$ introduced by Lewin \cite{Lewin1985TheOO} for real $\chi$, given by\footnote{It is interesting to notice that these can be defined for $\chi\in(0,1)$ as those functions satisfying
\begin{align}
\frac{\mathrm{d}}{\mathrm{d}\chi}L_n(\chi)=\frac{(-1)^{n-1}}{n\,(n-2)!}\,\log^{n-2}\chi\,\left(\frac{\log\chi}{1-\chi}+\frac{\log(1-\chi)}{\chi}\right)\,.
\end{align}}
\begin{align}\label{generalLewinLn}
L_n(\chi)=\sum_{k=0}^{n-1}\frac{(-1)^k}{k!}\Li_{n-k}(\chi)\,\log^k|\chi|-\frac{(-1)^n}{n!}\log|1-\chi|\,\log^{n-1}|\chi|\,,
\end{align}
naturally defined for $\chi\in (-1,1)$ but whose definition can be extended to $\chi\in\mathbb{R}\setminus\{0,1\}$ via $L_n(1/\chi)=(-1)^{n-1}L_n(\chi)$. We notice that these are related to Wojtkowiak's function by what we could refer to as ``diagonal limit'' in CFT language, {\it i.e.} by taking $z=\bar{z}=\chi\in\mathbb{R}$. In particular, we have
\begin{align}
\lim_{z\to\bar{z}\equiv\chi}\tilde{L}_n(z,\bar{z})=
\begin{cases}
0\,, \quad n\,\,\text{even}\,,\\
L_n(\chi)\,,\quad n\,\,\text{odd}\,.
\end{cases}
\end{align}
We can see this as a justification of the fact that, if we replace all $\Li_n$ with $L_n$ in our results, we only find $L_3$ but no $L_2$ or $L_4$: while there exists a single-valued extension of ordinary polylogs with even weight, its ``diagonal limit'' vanishes and does not define a function of a real variable. 

To conclude, let us highlight an interesting connection between Lewin's polylogs $L_n(\chi)$ and another interesting class of single-valued polylogs that is often encountered in the physics literature: ladder integrals \cite{USSYUKINA1993363,Usyukina:1993ch,Isaev:2003tk,Aprile:2017bgs} 
\begin{align}\label{defLadder}
\Phi^{(L)}(z,\bar{z})=-\frac{(1-z)(1-\bar{z})}{z-\bar{z}}\sum_{k=0}^L(-1)^r\frac{(2L-k)!}{k!L!(L-k)!}\,\log^{k}(z\bar{z})\,\left(\Li_{2L-k}(z)-\Li_{2L-k}(\bar{z})\right)\,,
\end{align}
that are related to each other by the differential relation
\begin{align}
\partial_z\,\partial_{\bar{z}}\,\left[\frac{z-\bar{z}}{(1-z)(1-\bar{z})}\Phi^{(L)}(z,\bar{z})\right]=-\frac{z-\bar{z}}{z\bar{z}(1-z)(1-\bar{z})}\Phi^{(L-1)}(z,\bar{z})\,.
\end{align}
It is interesting to notice that these are related to Lewin's polylogarithms by
\begin{align}
\lim_{z\to\bar{z}\equiv\chi}\Phi^{(L)}(z,\bar{z})=\frac{(1-\chi)^2}{\chi}\,\sum_{k=1}^{L-1}\frac{(-1)^{L-k}\,(2k+2)!}{(k+1)!L!(L-k-1)!}\,\log^{2L-2k-2}|\chi|\,L_{2k+1}(\chi)\,,
\end{align}
which, to the best of our knowledge, was not previously observed in the literature. In agreement with our discussion above, only $L_n$ of odd weight appear in this expression.

\section{OPE in perturbative 1d CFTs}\label{app:perturbativeOPE}

In this appendix we discuss in more detail some aspects of the expansion of perturbative correlators over conformal blocks and their derivatives,  and its connection to the role of braiding transformations $\chi\to \frac{\chi}{\chi-1}$.  This complements the discussion of Section \ref{sec:bootstrap}.

\subsection{Perturbative OPE}

As in Section \ref{sec:bootstrap} we discuss here a non-supersymmetric model with $\mathfrak{sl}(2)$ invariance. With little adaptation, everything that we are going to discuss also applies to the Wilson line defect theory. For simplicity, let us focus on the case of a four-point function of identical operators $\varphi$ of dimension $h_{\varphi}$
\begin{align}\label{generic4pt}
\langle \varphi(t_1)\,\varphi(t_2)\,\varphi(t_3)\,\varphi(t_4)\rangle=\frac{1}{t_{12}^{2h_{\varphi}}\,t_{34}^{2h_{\varphi}}}\,G(\chi)\,,
\end{align}
where the function $G(\chi)$ can be expanded over 1d conformal blocks $g_{h}(\chi)$ in the channel where $t_1\to t_2$:
\begin{align}\label{fullOPE}
G(\chi)=\sum_{h}a_h \, g_{h}(\chi)\,,
\end{align}
where the sum runs over all operators $\mathcal{O}_h$ appearing in the $\varphi\times\varphi$ OPE,  whose (non-perturbative) dimension is denoted with $h$,  and $a_h=\mu_{12h}\,\mu_{34h}$ is the product of the OPE coefficients of pairs of external operators with $\mathcal{O}_h$.  Let us also recall that the $\mathfrak{sl}(2)$ blocks for this configuration are given by
\begin{align}\label{1dblocks}
g_h(\chi)=\chi^h\,_2F_1(h,h,2h;\chi)\,.
\end{align}
We now consider the following scenario,  which is a model for most cases of interest in this paper\footnote{In Section \ref{sec:mixing} we also consider a case in which one of the external operators is not protected,  so it also acquires an anomalous dimension.  This amounts to a small modification of what we discuss here, where one also expands the dimension of external operators perturbatively}: the external dimensions $h_{\varphi}$ are frozen, while we expand the correlator,  the dimensions of the exchanged operators and their OPE coefficients in a small parameter $\g$.  For the purposes of this appendix we will neglect the issue of mixing, which is discuss in detail in Section \ref{sec:mixing}. When a degeneracy of operators is present, one should simply replaced the products of CFT data appearing below with their average value over each family of degenerate operators: {\it e.g.} $a^{(0)}_{\Delta} \,\gamma^{(1)}_{\Delta} \rightarrow \langle a^{(0)} \,\gamma^{(1)} \rangle_{\Delta}$.  With this understanding, we can write
\begin{align}\label{perturbativeexpansions}
\begin{split}
G(\chi)=\sum_{\ell=0}^{\infty}&\g^{\ell}\,G^{(\ell)}(\chi)\,,\\
h=\Delta+\sum_{\ell=1}^{\infty}\g^{\ell}\,\gamma_{\Delta}^{(\ell)}\,,\quad &\quad
a_h=\sum_{\ell=0}^{\infty}\g^{\ell}\,a_{\Delta}^{(\ell)}\,,
\end{split}
\end{align}
where throughout all this paper we denote with $\Delta$ the dimensions of the exchanged operators in the free theory ($\ell=0$), and we are only interested in cases where $\Delta$ is a positive integer.

An elementary,  but crucial fact about the 1d conformal blocks is that while the hypergeometric function (and its derivatives with respect to $\h$) appearing in \eqref{1dblocks} is analytic at $\chi=0$,  the factor of $\chi^h$ produced powers of $\log\chi$ when the dimension $\h$ is expanded using \eqref{perturbativeexpansions}.  To highlight these logarithms,  it is then useful to introduce 
\begin{align}\label{derblocks}
g^{(n)}_h(\chi)=\chi^h\,\left(\frac{\partial}{\partial h}\right)^{n}\,g_h(\chi)\,,
\end{align}
which are analytic at $\chi=0$ when evaluated at $h=\Delta$,  for integer $\Delta$.  It is then useful to collect each power of $\log\chi$ appearing in the expansion of $G^{(\ell)}(\chi)$,  via
\begin{align}\label{logexpappendix}
G^{(\ell)}(\chi)=\sum_{k=0}^{\ell}G^{(\ell)}_{\log^k}(\chi)\,(\log\chi)^k\,,
\end{align}
where each $G^{(\ell)}_{\log^k}(\chi)$ is analytic at $\chi=0$.  One can then rearrange the OPE collecting the powers of $\log\chi$ produced by derivatives of blocks.  This translates into an expansion for each $G^{(\ell)}_{\log^k}(\chi)$ in terms of the functions in \eqref{derblocks}:
\begin{align}\label{blocklog}
G^{(\ell)}_{\log^k}(\chi)=\sum_{\Delta} H_{\Delta}^{(\ell,k)}(\chi)\,,
\end{align}
where for each $\Delta$ we have
\begin{align}\label{Hblocks}
\begin{split}
H_{\Delta}^{(\ell,k)}(\chi)=
\frac{1}{k!}\sum_{p=0}^{\ell-k}
\frac{1}{p!}\left(\gamma_{\Delta}^{(1)}\right)^{k+p}\,
\Big[a_{\Delta}^{(\ell-k-p)}+\sum_{q=1}^{\ell-k-p}\left(\gamma_{\Delta}^{(1)}\right)^{-q}\,\binom{k+p}{q}\,\alpha^{(\ell,k)}_{\Delta}(p,q)\Big]\,g^{(p)}_{\Delta}(\chi)\,.
\end{split}
\end{align}
The coefficients $\alpha^{(\ell,k)}_{\Delta}(p,q)$ are given by sums of products of one power of $a^{(\#)}_{\Delta}$ and $q$ powers of $\gamma^{(\#)}_{\Delta}$:
\begin{align}\label{alphacoeff}
\begin{split}
\alpha^{(\ell,k)}_{\Delta}(p,q)=\sum_{m=p}^{\ell-k-q}\sum_{k_{q-2}=0}^{\ell-k-q-m}\sum_{k_{q-3}=0}^{k_{q-2}}\cdot\cdot\cdot \sum_{k_{0}=0}^{k_{1}} &a_{\Delta}^{(m-p)}\,\gamma_{\Delta}^{(2+k_0)}\,\gamma_{\Delta}^{(2+k_1-k_0)}\,\\
&\times\cdot\cdot\cdot\times\,\gamma_{\Delta}^{(2+k_{q-2}-k_{q-3})}\,\gamma_{\Delta}^{(\ell-k-q+2-m-k_{q-2})}\,.
\end{split}
\end{align}
The most prominent feature of the $H_{\Delta}^{(\ell,k)}(\chi)$ is that for $2\le k\le \ell$ they only depend on CFT data that can be extracted from previous perturbative orders,  the simplest instance being the case $k=\ell$:
\begin{align}
G^{(\ell)}_{\log^{\ell}}(\chi)=\sum_{\Delta}\frac{1}{\ell!}a^{(0)}_{\Delta}\,\left(\gamma^{(1)}_{\Delta}\right)^{\ell}\,g_{\Delta}(\chi)\,,
\end{align}
which only depends on free theory and tree level data.  This justifies the claim made in Section \ref{sec:bootstrap} that,  for $\ell\ge 2$,  one can reconstruct all the functions $G^{(\ell)}_{\log^k}(\chi)$ for $k\ge 2$ from CFT data at previous orders, modulo resolving mixing,  if present. The first time new CFT data appear is for $k=1$,  where only $\gamma^{(\ell)}_{\Delta}$ is present,  while $a^{(\ell)}_{\Delta}$ makes its first appearance for $k=0$.

\subsection{Braiding}\label{app:braiding}

Let us now turn to the issue of braiding,  that is how the transformation $\chi \to \frac{\chi}{\chi-1}$ can be used to constrain 1d four-point functions in perturbation theory. This issue was already considered from different perspectives in \cite{Liendo:2018ukf,Ferrero:2019luz,Bianchi:2020hsz} and is closely related to the fact that in 1d there is no continuous group of rotation.\footnote{There is only a residual $\mathbb{Z}_2$ parity invariance, sometimes referred to as S-parity -- see \cite{Billo:2013jda}.} However, as discussed in Section \ref{sec:bootstrap}, the holographic origin of the model that we study implies that braiding symmetry must be somehow restored, and can therefore be used to put constraints on correlation functions. In this section we discuss in detail how this can be done.

Once again, we will take as a reference a four-point function of identical operators in a theory with $\mathfrak{sl}(2)$ invariance, as in \eqref{generic4pt}. In that case, we notice that the conformal blocks \eqref{1dblocks} have a simple transformation property under braiding, 
\begin{align}\label{braidingblocksappendix}
g_h(\chi)=(-1)^h\,g_h\big(\tfrac{\chi}{\chi-1}\big)\,.
\end{align}
Now, in (generalized) free theories it is not uncommon that the conformal dimensions of the operators are integers and in certain cases (often related to holography) the dimensions of the exchanged operators in a certain correlator are either all even or all odd integers. A simple example is the case of a generalized free theory, as discussed around \eqref{GFT}, where if the external operators have integer dimension then all the exchanged ones have even integer dimension. In that case, we can say that in \eqref{perturbativeexpansions} for $\g=0$ one has $\Delta\in 2\mathbb{N}$ and therefore
\begin{align}\label{braidingfree}
G^{(0)}(\chi)=G^{(0)}\big(\tfrac{\chi}{\chi-1}\big)\,, 
\end{align}
since it is true for each individual conformal block appearing in the OPE. While this might seem a pretty stringent requirement, it is a property shared by all free theory correlators of interest for this paper. Note, however, that not all 1d free CFTs share this property, even when the external operators have integer dimension. For instance, even the simplest $\langle \mathcal{D}_1\mathcal{D}_1\mathcal{D}_1\mathcal{D}_1\rangle$ correlator on the Wilson line at weak coupling ($\lambda\to 0$) has no well-defined transformation properties under braiding -- see, {\it e.g.}, \cite{Kiryu:2018phb}.

Clearly, turning on perturbative corrections breaks the symmetry \eqref{braidingfree} as the exchanged operators acquire anomalous dimensions, but the crucial point that we wish to make in this section is that the symmetry is only broken in a weak and controlled sense, which still allows to put constraints on our ansatz in perturbation theory. To see how this works, notice that if we map $\chi \to \tfrac{\chi}{\chi-1}$ in \eqref{logexpappendix} we obtain, at each order,
\begin{align}\label{logexpbraiding1}
G^{(\ell)}\big(\tfrac{\chi}{\chi-1}\big)=\sum_{k=0}^{\ell}G^{(\ell)}_{\log^k}\big(\tfrac{\chi}{\chi-1}\big)\left(\log(-\chi)-\log(1-\chi)\right)^k\,,
\end{align}
but at the same time $G\big(\tfrac{\chi}{\chi-1}\big)$ can be expanded as a sum over the transformed conformal blocks
\begin{align}\label{hatblock}
\widehat{g}_h(\chi)=g_h\big(\tfrac{\chi}{\chi-1}\big)=(-\chi)^h\,{}_2F_1(h,h,2h;\chi)\,,
\end{align}
as
\begin{align}
G\big(\tfrac{\chi}{\chi-1}\big)=\sum_h a_h\,\widehat{g}_h(\chi)\,,
\end{align}
where the OPE coefficients $a_h$ are the same as those appearing in \eqref{fullOPE}. Following the same logic as in the previous subsection, the perturbative expansions \eqref{perturbativeexpansions} produce powers of $\log$ in the OPE and at each order one obtains
\begin{align}\label{logexpbraiding2}
G^{(\ell)}\big(\tfrac{\chi}{\chi-1}\big)=\sum_{k=0}^{\ell}G^{(\ell)}_{\log^k}(\chi)\log^k(-\chi)\,,
\end{align}
where the functions $G^{(\ell)}_{\log^k}(\chi)$ are exactly those appearing in \eqref{logexpappendix}, for two reasons: first, the hypergeometric function in \eqref{hatblock} is the same as that in \eqref{1dblocks}, and moreover in \eqref{blocklog} one always sums over the free theory dimensions $\Delta$, which by assumption are even integers and therefore $(-\chi)^{\Delta}=\chi^{\Delta}$. One should then equate \eqref{logexpbraiding1} and \eqref{logexpbraiding2}, where the equality should be interpreted as follows. Since as discussed each $G^{(\ell)}_{\log^k}(\chi)$ (and therefore also $G^{(\ell)}_{\log^k}\big(\tfrac{\chi}{\chi-1}\big)$) is holomorphic at $\chi=0$, and so is $\log(1-\chi)$, one should interpret the equality as holding for the coefficient of each power of $\log(-\chi)$, independently. From this, one can easily derive
\begin{align}\label{braidinglogpiecesappendix}
G^{(\ell)}_{\log^k}(\chi)=\sum_{m=0}^{\ell-k}\binom{k+m}{m}(-1)^m\,\log^m(1-\chi)\,G^{(\ell)}_{\log^{k+m}}\big(\tfrac{\chi}{\chi-1}\big)\,,
\end{align}
which as discussed in Section \ref{sec:bootstrap} is the constraint that we enforce on our ansatz at each perturbative order.

An alternative way to see how this can be derived is to work with the expansion \eqref{blocklog}, over the blocks $H^{(\ell,k)}_{\Delta}(\chi)$ defined in \eqref{Hblocks}. Each of these depends on conformal blocks and their derivatives \eqref{derblocks}, where note that the derivatives really only act on the hypergeometric function appearing in the blocks, all of them evaluated on the free theory dimension $\h=\Delta$, which is an even integer. We have already discussed how the blocks transform under braiding in this case in \eqref{braidingblocksappendix}, where one should set $(-1)^h=(-1)^{\Delta}=1$. It is not hard to see that the ``derivative blocks'' \eqref{derblocks} transform with
\begin{align}\label{braidingderblocks}
g^{(n)}_{\Delta}\big(\tfrac{\chi}{\chi-1}\big)=\sum_{k=0}^{n}\binom{n}{k}\,\log^{n-k}(1-\chi)\,g^{(k)}_{\Delta}(\chi)\,.
\end{align}
Now, at each order, one can rephrase \eqref{logexpappendix} together with \eqref{blocklog} as
\begin{align}\label{blockexplogsymm}
G^{(\ell)}(\chi)=\sum_{\Delta}\sum_{k,m=0}^{\ell}\beta^{(\ell,k)}_{\Delta}\,\binom{k}{m}\,\log^{m}\chi\,g^{(k-m)}_{\Delta}(\chi)\,,
\end{align}
for certain coefficients $\beta^{(\ell,k)}_{\Delta}$ (that we stress are independent of $m$) that depend on the CFT data\footnote{The explicit expression, which is not important for the argument, can be given in terms of the $\alpha^{(\ell,k)}_{\Delta}(p,q)$ defined in \eqref{alphacoeff},  as
\begin{align}
\beta^{(\ell,k)}_{\Delta}=\frac{1}{k!}\,\left(\gamma_{\Delta}^{(1)}\right)^k\,\left[a^{(\ell-k)}_{\Delta}+\sum_{q=1}^{\ell-k}\left(\gamma_{\Delta}^{(1)}\right)^{-q} \, \binom{k}{q}\,\alpha^{(\ell,k)}_{\Delta}(0,q)\right]\,.
\end{align} }.
One can then combine \eqref{braidingderblocks} and \eqref{blockexplogsymm} to compute
\begin{align}
\begin{split}
G^{(\ell)}\big(\tfrac{\chi}{\chi-1}\big)&=\sum_{\Delta}\sum_{k,m=0}^{\ell}\beta^{(\ell,k)}_{\Delta}\,\binom{k}{m}\sum_{a=0}^{k-m}\sum_{b=0}^m(-1)^b \binom{k-m}{a}\binom{m}{b}\\
&\hspace{5cm}\log^{k-m-a+b}(1-\chi)\,\log^{m-b}(-\chi)\,g^{(a)}_{\Delta}(\chi)\\
&=\sum_{\Delta}\sum_{a,c,k,m=0}^{\ell}\beta^{(\ell,k)}_{\Delta}\,\binom{k}{m}(-1)^{m-c}\,\binom{m}{m-c}\,\binom{k-m}{a}\\
&\hspace{5cm}\log^{k-a-c}(1-\chi)\,\log^c(-\chi)\,g^{(a)}_{\Delta}(\chi)\,,
\end{split}
\end{align}
where we have used that, given the presence of the binomials,  all sums can be actually computed up to $\ell$,  and we have changed the summation index $b$ to $c=m-b$.  At this point we note that neither the exponents of the logarithms nor the derivative index of the blocks depend on $m$,  so we can perform the sum over $m$ explicitly.  We find
\begin{align}
\sum_{m=0}^{\ell}(-1)^{m-c}\,\binom{m}{m-c}\,\binom{k-m}{a}=\binom{k}{a}\,\delta_{c,k-a}\,,
\end{align}
that is the sum over $m$ localizes the sum over $c$ on $c=k-a$,  and replacing we get
\begin{align}\label{almostfinalbraiding}
G^{(\ell)}\big(\tfrac{\chi}{\chi-1}\big)= \sum_{\Delta}\sum_{k,a=0}^{\ell}\beta^{(\ell,k)}_{\Delta}\,\binom{k}{a}\,\log^{k-a}(-\chi)\,g^{(a)}_{\Delta}(\chi)\,,
\end{align}
which is just the same as \eqref{blockexplogsymm} with each $\log\chi$ replaced with $\log(-\chi)$. This gives us another way to interpret how perturbative correlators behave under braiding transformations: if we take $\chi\in(0,1)$ and $\log(-\chi)=\log\chi+i\,\pi$, then comparing with \eqref{blockexplogsymm} we can read \eqref{almostfinalbraiding} as
\begin{align}\label{finalbraiding}
G^{(\ell)}\big(\tfrac{\chi}{\chi-1}\big)=G^{(\ell)}(\chi)+ \sum_{\Delta}\sum_{k,a=0}^{\ell}(i\,\pi)^{k-a}\,\beta^{(\ell,k)}_{\Delta}\,\binom{k}{a}\,g^{(a)}_{\Delta}(\chi)\,,
\end{align}
where the second term is entirely due to the branch point of $\log\chi$ at $\chi=0$ and provides what we previously described as a ``weak'' breaking of the braiding symmetry of the free theory. 

In practical computations, one can make use of this last point in an efficient way. In particular, thanks to the fact that the basis of HPLs chosen in Appendix \ref{app:polylogs} only contains functions that are either holomorphic at $\chi=0$, or have explicit powers of $\log\chi$, all HPLs in the ansatz for the blocks $G^{(\ell)}_{\log^k}(\chi)$ in \eqref{logexpappendix} are holomorphic at $\chi=0$. Then, all branch points in $G^{(\ell)}(\chi)$ are due to the explicit powers of $\log\chi$ displayed in \eqref{logexpappendix}. Then, when applying a braiding transformation, one can simply ``forget'' the powers of $i\,\pi$ arising from $\log(-\chi)-\log\chi=i\,\pi$, which corresponds to neglecting the second term on the right hand side of \eqref{finalbraiding}. The result is an equality between $G^{(\ell)}\big(\tfrac{\chi}{\chi-1}\big)$ and $G^{(\ell)}(\chi)$, which is completely equivalent to \eqref{braidinglogpiecesappendix}. Equivalently, one could replace each $\log\chi$ appearing in \eqref{logexpappendix} with $\log|\chi|$, to define 
\begin{align}
\bar{G}^{(\ell)}(\chi)=\sum_{k=0}^{(\ell)}G^{(\ell)}_{\log^k}(\chi)\,\log^k|\chi|\,,
\end{align}
so that for this function one obtains, from \eqref{finalbraiding}, 
\begin{align}\label{braidingbarG}
\bar{G}^{(\ell)}\big(\tfrac{\chi}{\chi-1}\big)=\bar{G}^{(\ell)}(\chi)\,.
\end{align}

Finally, let us observe that repeating the same exercise for the OPE in the other channel, which leads to an expansion around $\chi=1$, gives exactly the same results with powers of $\log(1-\chi)$ and one would obtain analogous statements, with the conclusion that to use the full crossing symmetry group $S_3$ as a constraint on the correlator, one would need to replace $\log(1-\chi)$ with $\log|1-\chi|$, in this case making the transformation $\chi\to 1/\chi$ an exact symmetry. This also justifies the definition of the extensions $\bar{f}^{(\ell)}(\chi)$ of each $f^{(\ell)}(\chi)$ to the whose real axis, in Section \ref{sec:1111}: each of those functions satisfies exact identities under crossing symmetry, as in \eqref{braidingbarG}. Similarly, such prescription provides a justification of the absolute values appearing in the Wilson line correlators discussed in \cite{Giombi:2017cqn}.

\subsection{Reciprocity principle}

One aspect of the results that we have presented that we would like to briefly review and interpret is the fact that all anomalous dimensions satisfy the 1d version of the reciprocity principle. This was first introduced in  \cite{Dokshitzer:2005bf,  Basso:2006nk} and concerns certain properties of the expansion of anomalous dimensions in higher-dimensional CFTs for large values of the spin. It was then proved for general CFTs in \cite{Alday:2015eya}, following analogous observations on properties of the anomalous dimensions in $\mathcal{N}=4$ SYM \cite{Beccaria:2008fi,Fiamberti:2009jw,Velizhanin:2010cm}. The one-dimensional analogue of the higher-dimensional reciprocity principle can be obtain studying the Regge limit of four-point functions, where operators with large conformal dimension dominate the OPE \cite{Ferrero:2019luz}. From that, it is possible to prove that anomalous dimensions $\gamma_{\Delta}\equiv h-\Delta$ in 1d CFTs are functions of the combination $\Delta+2\gamma_{\Delta}$, 
\begin{align}\label{1dreciprocityappendix}
\gamma_{\Delta}=F(\Delta+2\gamma_{\Delta})\,.
\end{align}
This statement is equivalent to the requirement that, when expanded in (inverse) powers of the ``bare'' conformal Casimir $\JJ$, $F(\Delta)$ contains only even powers of $j_{\Delta}$\footnote{Note that here we are discussing this in the context of a theory with only $\mathfrak{sl}(2)$ invariance, but the same statements apply to superconformal theories where the expression of the conformal Casimir is replaced with the one relevant for that case.} -- see \cite{Alday:2015eya} for more details in the higher-dimensional case. One should think of \eqref{1dreciprocityappendix} as a nested equation, whose full content can be appreciated expanding the anomalous dimensions $\gamma_{\Delta}$ in perturbation theory and comparing the two sides. This procedure reveals which combination of perturbative anomalous dimensions $\gamma^{(\ell)}_{\Delta}$ (and derivatives thereof), at each order, should admit such expansion in even (inverse) powers of $j_{\Delta}$. More precisely, expanding perturbatively
\begin{align}
F(\Delta+2\gamma_{\Delta})=\sum_{\ell=0}^{\infty}\frac{1}{\lambda^{\ell/2}}F^{(\ell)}_{\Delta}\,,
\end{align}
we find that the combinations
\begin{align}
F^{(\ell)}_{\Delta}=\sum_{k=0}^{\ell-1}\frac{(-1)^k}{(k+1)!}(\partial_{\Delta})^k\,\eta^{(k,\ell)}_{\Delta}\,,
\end{align}
admit an expansion in even powers of $j_{\Delta}$ for large $\Delta$ at each order, where we have introduced the combinations
\begin{align}
\eta^{(k,\ell)}_{\Delta}=\sum_{m_{k}=1}^{\ell-1}\sum_{m_{k-1}=m_{k}}^{\ell-1}\cdots \sum_{m_1=m_2+1}^{\ell-1}\gamma^{(\ell-m_1)}_{\Delta}\,\gamma^{(m_1-m_2)}_{\Delta}\,\cdots\,\gamma^{(m_k)}_{\Delta}\,.
\end{align}
The first few orders read
\begin{align}
\begin{split}
F^{(1)}_{\Delta}&=\gamma^{(1)}_{\Delta}\,,\\
F^{(2)}_{\Delta}&=\gamma^{(2)}_{\Delta}-\tfrac{1}{2!} \partial_{\Delta}(\gamma^{(1)}_{\Delta})^2\,,\\
F^{(3)}_{\Delta}&=\gamma^{(3)}_{\Delta}-\partial_{\Delta}(\gamma^{(1)}_{\Delta}\gamma^{(2)}_{\Delta})+\tfrac{1}{3!} \partial_{\Delta}^2(\gamma^{(1)}_{\Delta})^3\,,\\
F^{(4)}_{\Delta}&=\gamma^{(4)}_{\Delta}-\partial_{\Delta}(\gamma^{(1)}_{\Delta}\gamma^{(3)}_{\Delta}+\tfrac{1}{2}(\gamma^{(2)}_{\Delta})^2)+\tfrac{1}{2}\partial_{\Delta}^2((\gamma^{(1)}_{\Delta})^2\gamma^{(2)}_{\Delta})-\tfrac{1}{4!} \partial_{\Delta}^3(\gamma^{(1)}_{\Delta})^4\,.
\end{split}
\end{align}

In terms of the type of functions appearing in the anomalous dimensions found in Section \ref{sec:1111}, it is convenient to consider the case $j^2_{\Delta}=\Delta(\Delta+3)$ of interest for that section. One can observe that all the harmonic sums encountered there, namely $H_{1+\Delta}$ and $S_{-2}(1+\Delta)$, admit an expansion in even powers of $j_{\Delta}$ (and powers of $\log j^2_{\Delta}$) for large $\Delta$, while other harmonic sums such as $H^{(2)}_{1+\Delta}$, which do not share such property, do not appear. Finally, let us observe that in the main text we are always dealing with averaged quantities over degeneracy spaces, but the reciprocity principle seems to apply to those quantities as well.

\section{$\langle \mathcal{D}_1\mathcal{D}_1\mathcal{L}^{\D=2}_{0,[0,0]}\mathcal{L}^{\D=2}_{0,[0,0]}\rangle$}\label{app:bootstrapresults}

Although strictly speaking it did not prove necessary for the purpose of obtaining the three-loops result for $\langle\mathcal{D}_1\mathcal{D}_1\mathcal{D}_1\mathcal{D}_1\rangle$, in the course of our investigations we have also considered correlation functions involving two $\mathcal{D}_1$ multiplets and two singlet long multiplets. The superconformal blocks for this four-point function have been found in \cite{Ferrero:2023znz} and here we would like to present the bootstrap results up to one loop for the simplest correlator of this kind: $\langle \mathcal{D}_1\mathcal{D}_1\mathcal{L}^{\D=2}_{0,[0,0]}\mathcal{L}^{\D=2}_{0,[0,0]}\rangle$, where $\mathcal{L}^{\D=2}_{0,[0,0]}$ is the {\it unique} supermultiplet whose superconformal primary is a singlet of dimension two at strong coupling. Our bootstrap algorithm proceeds in the same way as usual, with the main difference that here the four-point function of the superconformal primaries is not enough to fix the whole correlator in superspace. Rather, as explained in \cite{Ferrero:2023znz} we should consider three independent four-point functions. Let $\mathcal{O}_{\D=2}$ be the superconformal primary of $\mathcal{L}^{\D=2}_{0,[0,0]}$. The three functions correspond to the correlators between the components of $\mathcal{D}_1$ and $\mathcal{O}_{\D=2}$:
\begin{align}
\begin{split}
\langle\varphi(t_1,y_1)\varphi(t_2,y_2)\mathcal{O}_{\D=2}(t_3)\mathcal{O}_{\D=2}(t_4)\rangle&=\frac{y_{12}^2}{t_{12}^2t_{34}^4}\,H^{(\varphi)}(\chi)\,,\\
\langle\Psi_{\alpha a}(t_1,y_1)\Psi_{\beta b}(t_2,y_2)\mathcal{O}_{\D=2}(t_3)\mathcal{O}_{\D=2}(t_4)\rangle&=\frac{\epsilon_{\alpha\beta}\epsilon_{ac}\epsilon_{bd}y_{12}^{cd}}{t_{12}^3t_{34}^4}\,H^{(\Psi)}(\chi)\,,\\
\langle f(t_1)f(t_2)\mathcal{O}_{\D=2}(t_3)\mathcal{O}_{\D=2}(t_4)\rangle&=\frac{\epsilon_{\alpha\gamma}\epsilon_{\beta\delta}+\epsilon_{\alpha\delta}\epsilon_{\beta\gamma}}{8\,t_{12}^4t_{34}^4}\,H^{(f)}(\chi)\,.
\end{split}
\end{align}
Note that all three correlators are invariant under braiding (in the usual sense) since the first two operators are always identical.

The free-theory results are determined by Wick contractions and read
\begin{align}
H^{(0,\varphi)}(\chi)=1+\frac{2(1+(1-\chi)^2)\,\chi^2}{5(1-\chi)^2}\,, \quad
H^{(0,\Psi)}(\chi)=-2\,, \quad
H^{(0,f)}(\chi)=-6\,.
\end{align}
Moving to tree-level, one can implement the usual bootstrap strategy using the first-order anomalous dimensions as an input. Braiding invariance and the input of certain low-$\D$ CFT data that are known from other correlators are then enough to completely fix the result. Alternatively, it is relatively straightforward to compute the six-point function of $\varphi$ and take a suitable limit where two pairs of points are taken to coincide in a suitable way as to create two $\mathcal{O}_{\D=2}$ operators. The results (for $H^{(1,\varphi)}$) are the same. The three correlators at tree level read
\begin{align}
\begin{split}
H^{(1,\varphi)}(\chi)=&\tfrac{2\chi^2(10-20\chi+21\chi^2-11\chi^3+2\chi^4)}{5(\chi-1)^3}\log\chi-\tfrac{120-240\chi+302\chi^2-182\chi^3+67\chi^4}{30(\chi-1)^2}\\
&-\tfrac{2(10-25\chi+20\chi^2-10\chi^3+2\chi^4-4\chi^5+2\chi^6)}{5\chi(\chi-1)^2}\log(1-\chi)\,,\\
H^{(1,\Psi)}(\chi)=&\tfrac{2\chi^4(3-3\chi+\chi^2)}{(\chi-1)^3}\log\chi-\tfrac{-12+24\chi-9\chi^2-3\chi^3+2\chi^4}{(\chi-1)^2}-\tfrac{2(-6+3\chi+\chi^4)}{\chi}\log(1-\chi)\,,\\
H^{(1,f)}(\chi)=&48-\tfrac{24(\chi-2)}{\chi}\log(1-\chi)\,,
\end{split}
\end{align}
The agreement of the Witten diagrams computation and the bootstrap result is a non-trivial check of the validity of the results discussed in this paper. In particular, this latter result provides additional evidence that {\it all} tree-level correlators between {\it any} external operators have transcendentality one, as they can be always factorized as the product of a contact diagram and a certain number of free theory propagators. In AdS$_2$, contact diagrams between external operators of integer dimension have transcendendality one for any number of external legs \cite{Bliard:2022xsm}. This should be contrasted with what happens for analogous results in higher dimensions, at least in certain cases, where the tree-level four-point function of two single-particle operators and two bound states has the analytic structure of one-loop correlators between single particle states in Mellin space \cite{Ma:2022ihn}.

Moving to one loop, once again one could in principle proceed computing six-point Witten diagrams or with the bootstrap. However, we lack the knowledge of six-point Witten diagrams at order $1/\lambda$ at this stage, so we proceed with the bootstrap and it turns out that the usual constraints are enough to fully fix the result. We find
\begin{align}
\begin{split}
H^{(2,\varphi)}=&\tfrac{\chi^2(50-150\chi+295\chi^2-340\chi^3+215\chi^4-70\chi^5+9\chi^6)}{5(\chi-1)^4}\log^2\chi\\
&-\tfrac{-10+35\chi-45\chi^2-25\chi^3+75\chi^4-165\chi^5+191\chi^6-98\chi^7+18\chi^8}{5\chi(\chi-1)^3}\log\chi \log(1-\chi)\\
&+\tfrac{75-220\chi+220\chi^2-80\chi^3+35\chi^4-20\chi^5+34\chi^6-28\chi^7+9\chi^8}{5\chi^2(\chi-1)^2}\log^2(1-\chi)\\
&-\tfrac{-30+90\chi+485\chi^2-1120\chi^3+1404\chi^4-829\chi^5+189\chi^6}{15(\chi-1)^3}\log\chi\\
&+\tfrac{1970-4865\chi+3232\chi^2-1193\chi^3+634\chi^4-1547\chi^5+756\chi^6}{60\chi(\chi-1)^2}\log(1-\chi)\\
&+\tfrac{535-1070\chi+1568\chi^2-1033\chi^3+349\chi^4}{30(\chi-1)^2}\,,
\end{split}
\end{align}
\begin{align}
\begin{split}
H^{(2,\Psi)}(\chi)=&\tfrac{\chi^4(51-102\chi+98\chi^2-47\chi^3+9\chi^4)}{(\chi-1)^4}\log^2\chi+
\tfrac{-120+114\chi-17\chi^2-3\chi^5+18\chi^6}{2\chi^2}\log^2(1-\chi)\\
&-\tfrac{12-42\chi+54\chi^2-30\chi^3+12\chi^4-111\chi^5+183\chi^6-133\chi^7+36\chi^8}{2\chi(\chi-1)^3}\\
&+\tfrac{24-96\chi+226\chi^2-342\chi^3+796\chi^4-1134\chi^5+751\chi^6-225\chi^7+20\chi^8}{4(\chi-1)^4}\log\chi\\
&-\tfrac{522-1293\chi+1002\chi^2-181\chi^3-170\chi^4+323\chi^5-185\chi^6+20\chi^7}{4\chi(\chi-1)^2}\log(1-\chi)\\
&-\tfrac{-564+1692\chi-1395\chi^2-30\chi^3+575\chi^4-278\chi^5+40\chi^6}{8(1-\chi)^3}\,,
\end{split}
\end{align}
\begin{align}
\begin{split}
H^{(2,f)}(\chi)=&-\tfrac{18\chi^4(1+(\chi-1)^4)}{(\chi-1)^4}\log^2\chi-\frac{6(50-48\chi+9\chi^2+3\chi^6)}{\chi^2}\log^2(1-\chi)\\
&+\tfrac{12(1+\chi)(2-3\chi+3\chi^2-3\chi^3+3\chi^4)}{\chi}\log\chi\log(1-\chi)\\
&+\tfrac{-24+120\chi-262\chi^2+328\chi^3-144\chi^4
-212\chi^5+412\chi^6-328\chi^7+165\chi^8
-55\chi^9+10\chi^{10}}{(\chi-1)^5}\log\chi\\
&-\tfrac{648-312\chi+22\chi^2+22\chi^3+40\chi^4-5\chi^5+10\chi^6}{\chi}\log(1-\chi)\\
&-\tfrac{(696-2784\chi+4221\chi^2-2919\chi^3+920\chi^4-223\chi^5
+169\chi^6-80\chi^7+20\chi^8)}{2(\chi-1)^4}\,.
\end{split}
\end{align}


\providecommand{\href}[2]{#2}\begingroup\raggedright\endgroup


\begin{thebibliography}{100}

\bibitem{Ferrero:2023znz}
P.~Ferrero and C.~Meneghelli, {\it {Unmixing the Wilson line defect CFT. Part
  I: spectrum and kinematics}},  \href{http://arxiv.org/abs/2312.12550}{{\tt
  arXiv:2312.12550}}.

\bibitem{Ferrero:2021bsb}
P.~Ferrero and C.~Meneghelli, {\it {Bootstrapping the half-BPS line defect CFT
  in $\mathcal{N}=4$ SYM at strong coupling}},
  \href{http://arxiv.org/abs/2103.10440}{{\tt arXiv:2103.10440}}.

\bibitem{Minahan:2002ve}
J.~A. Minahan and K.~Zarembo, {\it {The Bethe ansatz for N=4 superYang-Mills}},
   {\em JHEP} {\bf 03} (2003) 013,
  [\href{http://arxiv.org/abs/hep-th/0212208}{{\tt hep-th/0212208}}].

\bibitem{Pestun:2007rz}
V.~Pestun, {\it {Localization of gauge theory on a four-sphere and
  supersymmetric Wilson loops}},  {\em Commun.Math.Phys.} {\bf 313} (2012)
  71--129, [\href{http://arxiv.org/abs/0712.2824}{{\tt arXiv:0712.2824}}].

\bibitem{Maldacena:1997re}
J.~M. Maldacena, {\it {The Large N limit of superconformal field theories and
  supergravity}},  {\em Adv. Theor. Math. Phys.} {\bf 2} (1998) 231--252,
  [\href{http://arxiv.org/abs/hep-th/9711200}{{\tt hep-th/9711200}}].

\bibitem{Rattazzi:2008pe}
R.~Rattazzi, V.~S. Rychkov, E.~Tonni, and A.~Vichi, {\it {Bounding scalar
  operator dimensions in 4D CFT}},  {\em JHEP} {\bf 0812} (2008) 031,
  [\href{http://arxiv.org/abs/0807.0004}{{\tt arXiv:0807.0004}}].

\bibitem{Maldacena:1998im}
J.~M. Maldacena, {\it {Wilson loops in large N field theories}},  {\em Phys.
  Rev. Lett.} {\bf 80} (1998) 4859--4862,
  [\href{http://arxiv.org/abs/hep-th/9803002}{{\tt hep-th/9803002}}].

\bibitem{Cooke:2017qgm}
M.~Cooke, A.~Dekel, and N.~Drukker, {\it {The Wilson loop CFT: Insertion
  dimensions and structure constants from wavy lines}},  {\em J. Phys.} {\bf
  A50} (2017), no.~33 335401, [\href{http://arxiv.org/abs/1703.03812}{{\tt
  arXiv:1703.03812}}].

\bibitem{Cooke:2018obg}
M.~Cooke, A.~Dekel, N.~Drukker, D.~Trancanelli, and E.~Vescovi, {\it
  {Deformations of the circular Wilson loop and spectral (in)dependence}},
  {\em JHEP} {\bf 01} (2019) 076, [\href{http://arxiv.org/abs/1811.09638}{{\tt
  arXiv:1811.09638}}].

\bibitem{Giombi:2017cqn}
S.~Giombi, R.~Roiban, and A.~A. Tseytlin, {\it {Half-BPS Wilson loop and
  AdS$_2$/CFT$_1$}},  {\em Nucl. Phys.} {\bf B922} (2017) 499--527,
  [\href{http://arxiv.org/abs/1706.00756}{{\tt arXiv:1706.00756}}].

\bibitem{Kiryu:2018phb}
N.~Kiryu and S.~Komatsu, {\it {Correlation Functions on the Half-BPS Wilson
  Loop: Perturbation and Hexagonalization}},  {\em JHEP} {\bf 02} (2019) 090,
  [\href{http://arxiv.org/abs/1812.04593}{{\tt arXiv:1812.04593}}].

\bibitem{Grabner:2020nis}
D.~Grabner, N.~Gromov, and J.~Julius, {\it {Excited States of One-Dimensional
  Defect CFTs from the Quantum Spectral Curve}},  {\em JHEP} {\bf 07} (2020)
  042, [\href{http://arxiv.org/abs/2001.11039}{{\tt arXiv:2001.11039}}].

\bibitem{Cavaglia:2021bnz}
A.~Cavagli\`a, N.~Gromov, J.~Julius, and M.~Preti, {\it {Integrability and
  conformal bootstrap: One dimensional defect conformal field theory}},  {\em
  Phys. Rev. D} {\bf 105} (2022), no.~2 L021902,
  [\href{http://arxiv.org/abs/2107.08510}{{\tt arXiv:2107.08510}}].

\bibitem{Cavaglia:2022qpg}
A.~Cavagli\`a, N.~Gromov, J.~Julius, and M.~Preti, {\it {Bootstrability in
  defect CFT: integrated correlators and sharper bounds}},  {\em JHEP} {\bf 05}
  (2022) 164, [\href{http://arxiv.org/abs/2203.09556}{{\tt arXiv:2203.09556}}].

\bibitem{Giombi:2018qox}
S.~Giombi and S.~Komatsu, {\it {Exact Correlators on the Wilson Loop in
  $\mathcal{N}=4$ SYM: Localization, Defect CFT, and Integrability}},  {\em
  JHEP} {\bf 05} (2018) 109, [\href{http://arxiv.org/abs/1802.05201}{{\tt
  arXiv:1802.05201}}]. [Erratum: JHEP 11, 123 (2018)].

\bibitem{Giombi:2018hsx}
S.~Giombi and S.~Komatsu, {\it {More Exact Results in the Wilson Loop Defect
  CFT: Bulk-Defect OPE, Nonplanar Corrections and Quantum Spectral Curve}},
  {\em J. Phys. A} {\bf 52} (2019), no.~12 125401,
  [\href{http://arxiv.org/abs/1811.02369}{{\tt arXiv:1811.02369}}].

\bibitem{Giombi:2020amn}
S.~Giombi, J.~Jiang, and S.~Komatsu, {\it {Giant Wilson loops and
  AdS$_{2}$/dCFT$_{1}$}},  {\em JHEP} {\bf 11} (2020) 064,
  [\href{http://arxiv.org/abs/2005.08890}{{\tt arXiv:2005.08890}}].

\bibitem{Drukker:2005kx}
N.~Drukker and B.~Fiol, {\it {All-genus calculation of Wilson loops using
  D-branes}},  {\em JHEP} {\bf 02} (2005) 010,
  [\href{http://arxiv.org/abs/hep-th/0501109}{{\tt hep-th/0501109}}].

\bibitem{Gomis:2006sb}
J.~Gomis and F.~Passerini, {\it {Holographic Wilson Loops}},  {\em JHEP} {\bf
  08} (2006) 074, [\href{http://arxiv.org/abs/hep-th/0604007}{{\tt
  hep-th/0604007}}].

\bibitem{Gomis:2006im}
J.~Gomis and F.~Passerini, {\it {Wilson Loops as D3-Branes}},  {\em JHEP} {\bf
  01} (2007) 097, [\href{http://arxiv.org/abs/hep-th/0612022}{{\tt
  hep-th/0612022}}].

\bibitem{Giombi:2020kvo}
S.~Giombi and B.~Offertaler, {\it {Wilson loops in $\mathcal{N}=4$ $SO(N)$ SYM
  and D-Branes in $AdS_5\times \mathbb{RP}^5$}},
  \href{http://arxiv.org/abs/2006.10852}{{\tt arXiv:2006.10852}}.

\bibitem{Liendo:2016ymz}
P.~Liendo and C.~Meneghelli, {\it {Bootstrap equations for $ \mathcal{N} $ = 4
  SYM with defects}},  {\em JHEP} {\bf 01} (2017) 122,
  [\href{http://arxiv.org/abs/1608.05126}{{\tt arXiv:1608.05126}}].

\bibitem{Liendo:2018ukf}
P.~Liendo, C.~Meneghelli, and V.~Mitev, {\it {Bootstrapping the half-BPS line
  defect}},  {\em JHEP} {\bf 10} (2018) 077,
  [\href{http://arxiv.org/abs/1806.01862}{{\tt arXiv:1806.01862}}].

\bibitem{Gromov:2009tv}
N.~Gromov, V.~Kazakov, and P.~Vieira, {\it {Exact Spectrum of Anomalous
  Dimensions of Planar N=4 Supersymmetric Yang-Mills Theory}},  {\em Phys. Rev.
  Lett.} {\bf 103} (2009) 131601, [\href{http://arxiv.org/abs/0901.3753}{{\tt
  arXiv:0901.3753}}].

\bibitem{Gromov:2009bc}
N.~Gromov, V.~Kazakov, A.~Kozak, and P.~Vieira, {\it {Exact Spectrum of
  Anomalous Dimensions of Planar N = 4 Supersymmetric Yang-Mills Theory: TBA
  and excited states}},  {\em Lett. Math. Phys.} {\bf 91} (2010) 265--287,
  [\href{http://arxiv.org/abs/0902.4458}{{\tt arXiv:0902.4458}}].

\bibitem{Gromov:2013pga}
N.~Gromov, V.~Kazakov, S.~Leurent, and D.~Volin, {\it {Quantum Spectral Curve
  for Planar $\mathcal{N} = 4$ Super-Yang-Mills Theory}},  {\em Phys. Rev.
  Lett.} {\bf 112} (2014), no.~1 011602,
  [\href{http://arxiv.org/abs/1305.1939}{{\tt arXiv:1305.1939}}].

\bibitem{AlessandroThesis}
A.~Trenta, ``{Reinforcement Learning for Conformal Field Theories}.''
  \url{https://etd.adm.unipi.it/t/etd-04122023-120235/}, Master's thesis,
  University of Pisa, 2023.

\bibitem{Niarchos:2023lot}
V.~Niarchos, C.~Papageorgakis, P.~Richmond, A.~G. Stapleton, and M.~Woolley,
  {\it {Bootstrability in Line-Defect CFT with Improved Truncation Methods}},
  \href{http://arxiv.org/abs/2306.15730}{{\tt arXiv:2306.15730}}.

\bibitem{Ferrero:2019luz}
P.~Ferrero, K.~Ghosh, A.~Sinha, and A.~Zahed, {\it {Crossing symmetry,
  transcendentality and the Regge behaviour of 1d CFTs}},  {\em JHEP} {\bf 07}
  (2020) 170, [\href{http://arxiv.org/abs/1911.12388}{{\tt arXiv:1911.12388}}].

\bibitem{Rastelli:2016nze}
L.~Rastelli and X.~Zhou, {\it {Mellin amplitudes for $AdS_5\times S^5$}},  {\em
  Phys. Rev. Lett.} {\bf 118} (2017), no.~9 091602,
  [\href{http://arxiv.org/abs/1608.06624}{{\tt arXiv:1608.06624}}].

\bibitem{Rastelli:2017udc}
L.~Rastelli and X.~Zhou, {\it {How to Succeed at Holographic Correlators
  Without Really Trying}},  {\em JHEP} {\bf 04} (2018) 014,
  [\href{http://arxiv.org/abs/1710.05923}{{\tt arXiv:1710.05923}}].

\bibitem{Alday:2020lbp}
L.~F. Alday and X.~Zhou, {\it {All Tree-Level Correlators for M-theory on
  $AdS_7 \times S^4$}},  {\em Phys. Rev. Lett.} {\bf 125} (2020), no.~13
  131604, [\href{http://arxiv.org/abs/2006.06653}{{\tt arXiv:2006.06653}}].

\bibitem{Alday:2020dtb}
L.~F. Alday and X.~Zhou, {\it {All Holographic Four-Point Functions in All
  Maximally Supersymmetric CFTs}},  {\em Phys. Rev. X} {\bf 11} (2021), no.~1
  011056, [\href{http://arxiv.org/abs/2006.12505}{{\tt arXiv:2006.12505}}].

\bibitem{Alday:2021odx}
L.~F. Alday, C.~Behan, P.~Ferrero, and X.~Zhou, {\it {Gluon Scattering in AdS
  from CFT}},  {\em JHEP} {\bf 06} (2021) 020,
  [\href{http://arxiv.org/abs/2103.15830}{{\tt arXiv:2103.15830}}].

\bibitem{Aprile:2017bgs}
F.~Aprile, J.~M. Drummond, P.~Heslop, and H.~Paul, {\it {Quantum Gravity from
  Conformal Field Theory}},  {\em JHEP} {\bf 01} (2018) 035,
  [\href{http://arxiv.org/abs/1706.02822}{{\tt arXiv:1706.02822}}].

\bibitem{Alday:2017xua}
L.~F. Alday and A.~Bissi, {\it {Loop Corrections to Supergravity on $AdS_5
  \times S^5$}},  {\em Phys. Rev. Lett.} {\bf 119} (2017), no.~17 171601,
  [\href{http://arxiv.org/abs/1706.02388}{{\tt arXiv:1706.02388}}].

\bibitem{Alday:2020tgi}
L.~F. Alday, S.~M. Chester, and H.~Raj, {\it {6d (2,0) and M-theory at
  1-loop}},  {\em JHEP} {\bf 01} (2021) 133,
  [\href{http://arxiv.org/abs/2005.07175}{{\tt arXiv:2005.07175}}].

\bibitem{Alday:2021ajh}
L.~F. Alday, A.~Bissi, and X.~Zhou, {\it {One-loop gluon amplitudes in AdS}},
  {\em JHEP} {\bf 02} (2022) 105, [\href{http://arxiv.org/abs/2110.09861}{{\tt
  arXiv:2110.09861}}].

\bibitem{Behan:2022uqr}
C.~Behan, {\it {Holographic S-fold theories at one loop}},
  \href{http://arxiv.org/abs/2202.05261}{{\tt arXiv:2202.05261}}.

\bibitem{Alday:2022rly}
L.~F. Alday, S.~M. Chester, and H.~Raj, {\it {M-theory on AdS$_{4}$\texttimes{}
  S$^{7}$ at 1-loop and beyond}},  {\em JHEP} {\bf 11} (2022) 091,
  [\href{http://arxiv.org/abs/2207.11138}{{\tt arXiv:2207.11138}}].

\bibitem{Huang:2021xws}
Z.~Huang and E.~Y. Yuan, {\it {Graviton scattering in AdS$_{5}$\texttimes{}
  S$^{5}$ at two loops}},  {\em JHEP} {\bf 04} (2023) 064,
  [\href{http://arxiv.org/abs/2112.15174}{{\tt arXiv:2112.15174}}].

\bibitem{Drummond:2022dxw}
J.~M. Drummond and H.~Paul, {\it {Two-loop supergravity on
  AdS$_{5}$\texttimes{}S$^{5}$ from CFT}},  {\em JHEP} {\bf 08} (2022) 275,
  [\href{http://arxiv.org/abs/2204.01829}{{\tt arXiv:2204.01829}}].

\bibitem{Beisert:2004ry}
N.~Beisert, {\it {The Dilatation operator of N=4 super Yang-Mills theory and
  integrability}},  {\em Phys. Rept.} {\bf 405} (2004) 1--202,
  [\href{http://arxiv.org/abs/hep-th/0407277}{{\tt hep-th/0407277}}].

\bibitem{Penedones:2010ue}
J.~Penedones, {\it {Writing CFT correlation functions as AdS scattering
  amplitudes}},  {\em JHEP} {\bf 03} (2011) 025,
  [\href{http://arxiv.org/abs/1011.1485}{{\tt arXiv:1011.1485}}].

\bibitem{Fitzpatrick:2011ia}
A.~L. Fitzpatrick, J.~Kaplan, J.~Penedones, S.~Raju, and B.~C. van Rees, {\it
  {A Natural Language for AdS/CFT Correlators}},  {\em JHEP} {\bf 11} (2011)
  095, [\href{http://arxiv.org/abs/1107.1499}{{\tt arXiv:1107.1499}}].

\bibitem{Bianchi:2021piu}
L.~Bianchi, G.~Bliard, V.~Forini, and G.~Peveri, {\it {Mellin amplitudes for 1d
  CFT}},  {\em JHEP} {\bf 10} (2021) 095,
  [\href{http://arxiv.org/abs/2106.00689}{{\tt arXiv:2106.00689}}].

\bibitem{Gimenez-Grau:2019hez}
A.~Gimenez-Grau and P.~Liendo, {\it {Bootstrapping line defects in
  $\mathcal{N}=2$ theories}},  {\em JHEP} {\bf 03} (2020) 121,
  [\href{http://arxiv.org/abs/1907.04345}{{\tt arXiv:1907.04345}}].

\bibitem{Bianchi:2020hsz}
L.~Bianchi, G.~Bliard, V.~Forini, L.~Griguolo, and D.~Seminara, {\it {Analytic
  bootstrap and Witten diagrams for the ABJM Wilson line as defect CFT$_{1}$}},
   {\em JHEP} {\bf 08} (2020) 143, [\href{http://arxiv.org/abs/2004.07849}{{\tt
  arXiv:2004.07849}}].

\bibitem{FernandoNotes}
L.~F. Alday, {\it {Unpublished notes} (2017)}, .

\bibitem{Abl:2021mxo}
T.~Abl, P.~Heslop, and A.~E. Lipstein, {\it {Higher-dimensional symmetry of
  AdS$_{2}$\texttimes{}S$^{2}$ correlators}},  {\em JHEP} {\bf 03} (2022) 076,
  [\href{http://arxiv.org/abs/2112.09597}{{\tt arXiv:2112.09597}}].

\bibitem{Mazac:2018mdx}
D.~Mazac and M.~F. Paulos, {\it {The Analytic Functional Bootstrap I: 1D CFTs
  and 2D S-Matrices}},  \href{http://arxiv.org/abs/1803.10233}{{\tt
  arXiv:1803.10233}}.

\bibitem{Mazac:2018ycv}
D.~Mazac and M.~F. Paulos, {\it {The analytic functional bootstrap. Part II.
  Natural bases for the crossing equation}},  {\em JHEP} {\bf 02} (2019) 163,
  [\href{http://arxiv.org/abs/1811.10646}{{\tt arXiv:1811.10646}}].

\bibitem{Giombi:2009ds}
S.~Giombi and V.~Pestun, {\it {Correlators of local operators and 1/8 BPS
  Wilson loops on S**2 from 2d YM and matrix models}},  {\em JHEP} {\bf 10}
  (2010) 033, [\href{http://arxiv.org/abs/0906.1572}{{\tt arXiv:0906.1572}}].

\bibitem{Caron-Huot:2017vep}
S.~Caron-Huot, {\it {Analyticity in Spin in Conformal Theories}},  {\em JHEP}
  {\bf 09} (2017) 078, [\href{http://arxiv.org/abs/1703.00278}{{\tt
  arXiv:1703.00278}}].

\bibitem{Aharony:2016dwx}
O.~Aharony, L.~F. Alday, A.~Bissi, and E.~Perlmutter, {\it {Loops in AdS from
  Conformal Field Theory}},  {\em JHEP} {\bf 07} (2017) 036,
  [\href{http://arxiv.org/abs/1612.03891}{{\tt arXiv:1612.03891}}].

\bibitem{Alday:2017zzv}
L.~F. Alday, J.~Henriksson, and M.~van Loon, {\it {Taming the
  $\epsilon$-expansion with large spin perturbation theory}},  {\em JHEP} {\bf
  07} (2018) 131, [\href{http://arxiv.org/abs/1712.02314}{{\tt
  arXiv:1712.02314}}].

\bibitem{Carmi:2020ekr}
D.~Carmi, J.~Penedones, J.~A. Silva, and A.~Zhiboedov, {\it {Applications of
  dispersive sum rules: $\epsilon$-expansion and holography}},
  \href{http://arxiv.org/abs/2009.13506}{{\tt arXiv:2009.13506}}.

\bibitem{Drukker:2020swu}
N.~Drukker, S.~Giombi, A.~A. Tseytlin, and X.~Zhou, {\it {Defect CFT in the 6d
  (2,0) theory from M2 brane dynamics in AdS$_7 \times$S$^4$}},  {\em JHEP}
  {\bf 07} (2020) 101, [\href{http://arxiv.org/abs/2004.04562}{{\tt
  arXiv:2004.04562}}].

\bibitem{Drukker:2005af}
N.~Drukker and S.~Kawamoto, {\it {Circular loop operators in conformal field
  theories}},  {\em Phys. Rev. D} {\bf 74} (2006) 046002,
  [\href{http://arxiv.org/abs/hep-th/0512150}{{\tt hep-th/0512150}}].

\bibitem{Drukker:2006xg}
N.~Drukker and S.~Kawamoto, {\it {Small deformations of supersymmetric Wilson
  loops and open spin-chains}},  {\em JHEP} {\bf 07} (2006) 024,
  [\href{http://arxiv.org/abs/hep-th/0604124}{{\tt hep-th/0604124}}].

\bibitem{Barrat:2021tpn}
J.~Barrat, P.~Liendo, G.~Peveri, and J.~Plefka, {\it {Multipoint correlators on
  the supersymmetric Wilson line defect CFT}},
  \href{http://arxiv.org/abs/2112.10780}{{\tt arXiv:2112.10780}}.

\bibitem{Barrat:2022eim}
J.~Barrat, P.~Liendo, and G.~Peveri, {\it {Multipoint correlators on the
  supersymmetric Wilson line defect CFT II: Unprotected operators}},
  \href{http://arxiv.org/abs/2210.14916}{{\tt arXiv:2210.14916}}.

\bibitem{Zwiebel:2011bx}
B.~I. Zwiebel, {\it {From Scattering Amplitudes to the Dilatation Generator in
  N=4 SYM}},  {\em J. Phys. A} {\bf 45} (2012) 115401,
  [\href{http://arxiv.org/abs/1111.0083}{{\tt arXiv:1111.0083}}].

\bibitem{Qiao:2017xif}
J.~Qiao and S.~Rychkov, {\it {A tauberian theorem for the conformal
  bootstrap}},  {\em JHEP} {\bf 12} (2017) 119,
  [\href{http://arxiv.org/abs/1709.00008}{{\tt arXiv:1709.00008}}].

\bibitem{Hogervorst:2013kva}
M.~Hogervorst, H.~Osborn, and S.~Rychkov, {\it {Diagonal Limit for Conformal
  Blocks in $d$ Dimensions}},  {\em JHEP} {\bf 1308} (2013) 014,
  [\href{http://arxiv.org/abs/1305.1321}{{\tt arXiv:1305.1321}}].

\bibitem{Sen:2019lec}
K.~Sen, A.~Sinha, and A.~Zahed, {\it {Positive geometry in the diagonal limit
  of the conformal bootstrap}},  {\em JHEP} {\bf 11} (2019) 059,
  [\href{http://arxiv.org/abs/1906.07202}{{\tt arXiv:1906.07202}}].

\bibitem{Hogervorst:2013sma}
M.~Hogervorst and S.~Rychkov, {\it {Radial Coordinates for Conformal Blocks}},
  {\em Phys.Rev.} {\bf D87} (2013) 106004,
  [\href{http://arxiv.org/abs/1303.1111}{{\tt arXiv:1303.1111}}].

\bibitem{Qiao:2017lkv}
J.~Qiao and S.~Rychkov, {\it {Cut-touching linear functionals in the conformal
  bootstrap}},  {\em JHEP} {\bf 06} (2017) 076,
  [\href{http://arxiv.org/abs/1705.01357}{{\tt arXiv:1705.01357}}].

\bibitem{Mazac:2018qmi}
D.~Maz\'a\v{c}, {\it {A Crossing-Symmetric OPE Inversion Formula}},  {\em JHEP}
  {\bf 06} (2019) 082, [\href{http://arxiv.org/abs/1812.02254}{{\tt
  arXiv:1812.02254}}].

\bibitem{Bissi:2020wtv}
A.~Bissi, G.~Fardelli, and A.~Georgoudis, {\it {Towards all loop supergravity
  amplitudes on AdS5\texttimes{}S5}},  {\em Phys. Rev. D} {\bf 104} (2021),
  no.~4 L041901, [\href{http://arxiv.org/abs/2002.04604}{{\tt
  arXiv:2002.04604}}].

\bibitem{Bissi:2020woe}
A.~Bissi, G.~Fardelli, and A.~Georgoudis, {\it {All loop structures in
  supergravity amplitudes on AdS5 \texttimes{} S5 from CFT}},  {\em J. Phys. A}
  {\bf 54} (2021), no.~32 324002, [\href{http://arxiv.org/abs/2010.12557}{{\tt
  arXiv:2010.12557}}].

\bibitem{Cavaglia:2023mmu}
A.~Cavagli\`a, N.~Gromov, and M.~Preti, {\it {Computing Four-Point Functions
  with Integrability, Bootstrap and Parity Symmetry}},
  \href{http://arxiv.org/abs/2312.11604}{{\tt arXiv:2312.11604}}.

\bibitem{Dixon:2012yy}
L.~J. Dixon, C.~Duhr, and J.~Pennington, {\it {Single-valued harmonic
  polylogarithms and the multi-Regge limit}},  {\em JHEP} {\bf 10} (2012) 074,
  [\href{http://arxiv.org/abs/1207.0186}{{\tt arXiv:1207.0186}}].

\bibitem{Drummond:2013nda}
J.~Drummond, C.~Duhr, B.~Eden, P.~Heslop, J.~Pennington, and V.~A. Smirnov,
  {\it {Leading singularities and off-shell conformal integrals}},  {\em JHEP}
  {\bf 08} (2013) 133, [\href{http://arxiv.org/abs/1303.6909}{{\tt
  arXiv:1303.6909}}].

\bibitem{Duhr:2011zq}
C.~Duhr, H.~Gangl, and J.~R. Rhodes, {\it {From polygons and symbols to
  polylogarithmic functions}},  {\em JHEP} {\bf 10} (2012) 075,
  [\href{http://arxiv.org/abs/1110.0458}{{\tt arXiv:1110.0458}}].

\bibitem{Gehrmann:2000zt}
T.~Gehrmann and E.~Remiddi, {\it {Two loop master integrals for gamma*
  ---\ensuremath{>} 3 jets: The Planar topologies}},  {\em Nucl. Phys. B} {\bf
  601} (2001) 248--286, [\href{http://arxiv.org/abs/hep-ph/0008287}{{\tt
  hep-ph/0008287}}].

\bibitem{Vogt:2004mw}
A.~Vogt, S.~Moch, and J.~A.~M. Vermaseren, {\it {The Three-loop splitting
  functions in QCD: The Singlet case}},  {\em Nucl. Phys. B} {\bf 691} (2004)
  129--181, [\href{http://arxiv.org/abs/hep-ph/0404111}{{\tt hep-ph/0404111}}].

\bibitem{Moch:2004pa}
S.~Moch, J.~A.~M. Vermaseren, and A.~Vogt, {\it {The Three loop splitting
  functions in QCD: The Nonsinglet case}},  {\em Nucl. Phys. B} {\bf 688}
  (2004) 101--134, [\href{http://arxiv.org/abs/hep-ph/0403192}{{\tt
  hep-ph/0403192}}].

\bibitem{Vermaseren:2005qc}
J.~A.~M. Vermaseren, A.~Vogt, and S.~Moch, {\it {The Third-order QCD
  corrections to deep-inelastic scattering by photon exchange}},  {\em Nucl.
  Phys. B} {\bf 724} (2005) 3--182,
  [\href{http://arxiv.org/abs/hep-ph/0504242}{{\tt hep-ph/0504242}}].

\bibitem{Duhr:2012fh}
C.~Duhr, {\it {Hopf algebras, coproducts and symbols: an application to Higgs
  boson amplitudes}},  {\em JHEP} {\bf 08} (2012) 043,
  [\href{http://arxiv.org/abs/1203.0454}{{\tt arXiv:1203.0454}}].

\bibitem{Broedel:2013tta}
J.~Broedel, O.~Schlotterer, and S.~Stieberger, {\it {Polylogarithms, Multiple
  Zeta Values and Superstring Amplitudes}},  {\em Fortsch. Phys.} {\bf 61}
  (2013) 812--870, [\href{http://arxiv.org/abs/1304.7267}{{\tt
  arXiv:1304.7267}}].

\bibitem{Broedel:2014vla}
J.~Broedel, C.~R. Mafra, N.~Matthes, and O.~Schlotterer, {\it {Elliptic
  multiple zeta values and one-loop superstring amplitudes}},  {\em JHEP} {\bf
  07} (2015) 112, [\href{http://arxiv.org/abs/1412.5535}{{\tt
  arXiv:1412.5535}}].

\bibitem{Bern:2005iz}
Z.~Bern, L.~J. Dixon, and V.~A. Smirnov, {\it {Iteration of planar amplitudes
  in maximally supersymmetric Yang-Mills theory at three loops and beyond}},
  {\em Phys. Rev. D} {\bf 72} (2005) 085001,
  [\href{http://arxiv.org/abs/hep-th/0505205}{{\tt hep-th/0505205}}].

\bibitem{Bern:2006ew}
Z.~Bern, M.~Czakon, L.~J. Dixon, D.~A. Kosower, and V.~A. Smirnov, {\it {The
  Four-Loop Planar Amplitude and Cusp Anomalous Dimension in Maximally
  Supersymmetric Yang-Mills Theory}},  {\em Phys. Rev. D} {\bf 75} (2007)
  085010, [\href{http://arxiv.org/abs/hep-th/0610248}{{\tt hep-th/0610248}}].

\bibitem{DelDuca:2009au}
V.~Del~Duca, C.~Duhr, and V.~A. Smirnov, {\it {An Analytic Result for the
  Two-Loop Hexagon Wilson Loop in N = 4 SYM}},  {\em JHEP} {\bf 03} (2010) 099,
  [\href{http://arxiv.org/abs/0911.5332}{{\tt arXiv:0911.5332}}].

\bibitem{DelDuca:2010zg}
V.~Del~Duca, C.~Duhr, and V.~A. Smirnov, {\it {The Two-Loop Hexagon Wilson Loop
  in N = 4 SYM}},  {\em JHEP} {\bf 05} (2010) 084,
  [\href{http://arxiv.org/abs/1003.1702}{{\tt arXiv:1003.1702}}].

\bibitem{Goncharov:2010jf}
A.~B. Goncharov, M.~Spradlin, C.~Vergu, and A.~Volovich, {\it {Classical
  Polylogarithms for Amplitudes and Wilson Loops}},  {\em Phys. Rev. Lett.}
  {\bf 105} (2010) 151605, [\href{http://arxiv.org/abs/1006.5703}{{\tt
  arXiv:1006.5703}}].

\bibitem{Dixon:2011pw}
L.~J. Dixon, J.~M. Drummond, and J.~M. Henn, {\it {Bootstrapping the three-loop
  hexagon}},  {\em JHEP} {\bf 11} (2011) 023,
  [\href{http://arxiv.org/abs/1108.4461}{{\tt arXiv:1108.4461}}].

\bibitem{Golden:2013xva}
J.~Golden, A.~B. Goncharov, M.~Spradlin, C.~Vergu, and A.~Volovich, {\it
  {Motivic Amplitudes and Cluster Coordinates}},  {\em JHEP} {\bf 01} (2014)
  091, [\href{http://arxiv.org/abs/1305.1617}{{\tt arXiv:1305.1617}}].

\bibitem{Dixon:2013eka}
L.~J. Dixon, J.~M. Drummond, M.~von Hippel, and J.~Pennington, {\it {Hexagon
  functions and the three-loop remainder function}},  {\em JHEP} {\bf 12}
  (2013) 049, [\href{http://arxiv.org/abs/1308.2276}{{\tt arXiv:1308.2276}}].

\bibitem{Dixon:2014voa}
L.~J. Dixon, J.~M. Drummond, C.~Duhr, and J.~Pennington, {\it {The four-loop
  remainder function and multi-Regge behavior at NNLLA in planar N = 4
  super-Yang-Mills theory}},  {\em JHEP} {\bf 06} (2014) 116,
  [\href{http://arxiv.org/abs/1402.3300}{{\tt arXiv:1402.3300}}].

\bibitem{Dixon:2014iba}
L.~J. Dixon and M.~von Hippel, {\it {Bootstrapping an NMHV amplitude through
  three loops}},  {\em JHEP} {\bf 10} (2014) 065,
  [\href{http://arxiv.org/abs/1408.1505}{{\tt arXiv:1408.1505}}].

\bibitem{Drummond:2014ffa}
J.~M. Drummond, G.~Papathanasiou, and M.~Spradlin, {\it {A Symbol of
  Uniqueness: The Cluster Bootstrap for the 3-Loop MHV Heptagon}},  {\em JHEP}
  {\bf 03} (2015) 072, [\href{http://arxiv.org/abs/1412.3763}{{\tt
  arXiv:1412.3763}}].

\bibitem{Dixon:2015iva}
L.~J. Dixon, M.~von Hippel, and A.~J. McLeod, {\it {The four-loop six-gluon
  NMHV ratio function}},  {\em JHEP} {\bf 01} (2016) 053,
  [\href{http://arxiv.org/abs/1509.08127}{{\tt arXiv:1509.08127}}].

\bibitem{Dixon:2016nkn}
L.~J. Dixon, J.~Drummond, T.~Harrington, A.~J. McLeod, G.~Papathanasiou, and
  M.~Spradlin, {\it {Heptagons from the Steinmann Cluster Bootstrap}},  {\em
  JHEP} {\bf 02} (2017) 137, [\href{http://arxiv.org/abs/1612.08976}{{\tt
  arXiv:1612.08976}}].

\bibitem{Caron-Huot:2020bkp}
S.~Caron-Huot, L.~J. Dixon, J.~M. Drummond, F.~Dulat, J.~Foster,
  O.~G\"urdo\u{g}an, M.~von Hippel, A.~J. McLeod, and G.~Papathanasiou, {\it
  {The Steinmann Cluster Bootstrap for $N$ = 4 Super Yang-Mills Amplitudes}},
  {\em PoS} {\bf CORFU2019} (2020) 003,
  [\href{http://arxiv.org/abs/2005.06735}{{\tt arXiv:2005.06735}}].

\bibitem{Henriksson:2018myn}
J.~Henriksson and M.~Van~Loon, {\it {Critical O(N) model to order $\epsilon^4$
  from analytic bootstrap}},  {\em J. Phys. A} {\bf 52} (2019), no.~2 025401,
  [\href{http://arxiv.org/abs/1801.03512}{{\tt arXiv:1801.03512}}].

\bibitem{Guha:2019ipe}
S.~Guha and K.~Sen, {\it {Relating diagrammatic expansion with conformal
  correlator expansion}},  \href{http://arxiv.org/abs/1911.11188}{{\tt
  arXiv:1911.11188}}.

\bibitem{Eden:2000mv}
B.~Eden, C.~Schubert, and E.~Sokatchev, {\it {Three loop four point correlator
  in N=4 SYM}},  {\em Phys. Lett. B} {\bf 482} (2000) 309--314,
  [\href{http://arxiv.org/abs/hep-th/0003096}{{\tt hep-th/0003096}}].

\bibitem{Eden:2011we}
B.~Eden, P.~Heslop, G.~P. Korchemsky, and E.~Sokatchev, {\it {Hidden symmetry
  of four-point correlation functions and amplitudes in N=4 SYM}},  {\em Nucl.
  Phys. B} {\bf 862} (2012) 193--231,
  [\href{http://arxiv.org/abs/1108.3557}{{\tt arXiv:1108.3557}}].

\bibitem{Chicherin:2015edu}
D.~Chicherin, J.~Drummond, P.~Heslop, and E.~Sokatchev, {\it {All three-loop
  four-point correlators of half-BPS operators in planar $ \mathcal{N} $ = 4
  SYM}},  {\em JHEP} {\bf 08} (2016) 053,
  [\href{http://arxiv.org/abs/1512.02926}{{\tt arXiv:1512.02926}}].

\bibitem{Carmi:2019ocp}
D.~Carmi, {\it {Loops in AdS: From the Spectral Representation to Position
  Space}},  {\em JHEP} {\bf 06} (2020) 049,
  [\href{http://arxiv.org/abs/1910.14340}{{\tt arXiv:1910.14340}}].

\bibitem{Carmi:2021dsn}
D.~Carmi, {\it {Loops in AdS: from the spectral representation to position
  space. Part II}},  {\em JHEP} {\bf 07} (2021) 186,
  [\href{http://arxiv.org/abs/2104.10500}{{\tt arXiv:2104.10500}}].

\bibitem{Heckelbacher:2022fbx}
T.~Heckelbacher, I.~Sachs, E.~Skvortsov, and P.~Vanhove, {\it {Analytical
  evaluation of AdS${}_4$ Witten diagrams as flat space multi-loop Feynman
  integrals}},  \href{http://arxiv.org/abs/2201.09626}{{\tt arXiv:2201.09626}}.

\bibitem{Giombi:2022anm}
S.~Giombi, S.~Komatsu, and B.~Offertaler, {\it {Large Charges on the Wilson
  Loop in $\mathcal{N}=4$ SYM: II. Quantum Fluctuations, OPE, and Spectral
  Curve}},  \href{http://arxiv.org/abs/2202.07627}{{\tt arXiv:2202.07627}}.

\bibitem{Cutkosky:1960sp}
R.~E. Cutkosky, {\it {Singularities and discontinuities of Feynman
  amplitudes}},  {\em J. Math. Phys.} {\bf 1} (1960) 429--433.

\bibitem{Meltzer:2019nbs}
D.~Meltzer, E.~Perlmutter, and A.~Sivaramakrishnan, {\it {Unitarity Methods in
  AdS/CFT}},  {\em JHEP} {\bf 03} (2020) 061,
  [\href{http://arxiv.org/abs/1912.09521}{{\tt arXiv:1912.09521}}].

\bibitem{Meltzer:2020qbr}
D.~Meltzer and A.~Sivaramakrishnan, {\it {CFT unitarity and the AdS Cutkosky
  rules}},  {\em JHEP} {\bf 11} (2020) 073,
  [\href{http://arxiv.org/abs/2008.11730}{{\tt arXiv:2008.11730}}].

\bibitem{Alday:2019nin}
L.~F. Alday and X.~Zhou, {\it {Simplicity of AdS Supergravity at One Loop}},
  {\em JHEP} {\bf 09} (2020) 008, [\href{http://arxiv.org/abs/1912.02663}{{\tt
  arXiv:1912.02663}}].

\bibitem{Alday:2021ymb}
L.~F. Alday, S.~M. Chester, and H.~Raj, {\it {ABJM at strong coupling from
  M-theory, localization, and Lorentzian inversion}},  {\em JHEP} {\bf 02}
  (2022) 005, [\href{http://arxiv.org/abs/2107.10274}{{\tt arXiv:2107.10274}}].

\bibitem{Drukker:2009sf}
N.~Drukker and J.~Plefka, {\it {Superprotected n-point correlation functions of
  local operators in N=4 super Yang-Mills}},  {\em JHEP} {\bf 0904} (2009) 052,
  [\href{http://arxiv.org/abs/0901.3653}{{\tt arXiv:0901.3653}}].

\bibitem{Heemskerk:2009pn}
I.~Heemskerk, J.~Penedones, J.~Polchinski, and J.~Sully, {\it {Holography from
  Conformal Field Theory}},  {\em JHEP} {\bf 10} (2009) 079,
  [\href{http://arxiv.org/abs/0907.0151}{{\tt arXiv:0907.0151}}].

\bibitem{Bissi:2022mrs}
A.~Bissi, A.~Sinha, and X.~Zhou, {\it {Selected Topics in Analytic Conformal
  Bootstrap: A Guided Journey}},  \href{http://arxiv.org/abs/2202.08475}{{\tt
  arXiv:2202.08475}}.

\bibitem{Lewin1985TheOO}
L.~Lewin, {\it The order-independence of the polylogarithmic ladder
  structure—implications for a new category of functional equations},  {\em
  aequationes mathematicae} {\bf 30} (1985) 1--20.

\bibitem{Aprile:2017xsp}
F.~Aprile, J.~M. Drummond, P.~Heslop, and H.~Paul, {\it {Unmixing
  Supergravity}},  {\em JHEP} {\bf 02} (2018) 133,
  [\href{http://arxiv.org/abs/1706.08456}{{\tt arXiv:1706.08456}}].

\bibitem{USSYUKINA1993363}
N.~Ussyukina and A.~Davydychev, {\it An approach to the evaluation of three-
  and four-point ladder diagrams},  {\em Physics Letters B} {\bf 298} (1993),
  no.~3 363--370.

\bibitem{Usyukina:1993ch}
N.~I. Usyukina and A.~I. Davydychev, {\it {Exact results for three and four
  point ladder diagrams with an arbitrary number of rungs}},  {\em Phys. Lett.
  B} {\bf 305} (1993) 136--143.

\bibitem{Isaev:2003tk}
A.~P. Isaev, {\it {Multiloop Feynman integrals and conformal quantum
  mechanics}},  {\em Nucl. Phys. B} {\bf 662} (2003) 461--475,
  [\href{http://arxiv.org/abs/hep-th/0303056}{{\tt hep-th/0303056}}].

\bibitem{Alday:2016njk}
L.~F. Alday, {\it {Large Spin Perturbation Theory for Conformal Field
  Theories}},  {\em Phys. Rev. Lett.} {\bf 119} (2017), no.~11 111601,
  [\href{http://arxiv.org/abs/1611.01500}{{\tt arXiv:1611.01500}}].

\bibitem{Giombi:2023zte}
S.~Giombi, S.~Komatsu, B.~Offertaler, and J.~Shan, {\it {Boundary
  reparametrizations and six-point functions on the AdS$_2$ string}},
  \href{http://arxiv.org/abs/2308.10775}{{\tt arXiv:2308.10775}}.

\bibitem{Bliard:2023zpe}
G.~J.~S. Bliard, {\em {Perturbative and non-perturbative analysis of defect
  correlators in AdS/CFT}}.
\newblock PhD thesis, Humboldt U., Berlin, 2023.
\newblock \href{http://arxiv.org/abs/2310.18137}{{\tt arXiv:2310.18137}}.

\bibitem{Barrat:2024nod}
J.~Barrat, {\em {Line defects in conformal field theory}}.
\newblock PhD thesis, Humboldt U., Berlin, Humboldt-Universit\"at zu Berlin,
  2024.
\newblock \href{http://arxiv.org/abs/2401.10336}{{\tt arXiv:2401.10336}}.

\bibitem{Huang:2023oxf}
Z.~Huang, B.~Wang, E.~Y. Yuan, and X.~Zhou, {\it {AdS super gluon scattering up
  to two loops: A position space approach}},
  \href{http://arxiv.org/abs/2301.13240}{{\tt arXiv:2301.13240}}.

\bibitem{Bissi:2021hjk}
A.~Bissi, G.~Fardelli, and A.~Manenti, {\it {Rebooting quarter-BPS operators in
  $ \mathcal{N} $ = 4 super Yang-Mills}},  {\em JHEP} {\bf 04} (2022) 016,
  [\href{http://arxiv.org/abs/2111.06857}{{\tt arXiv:2111.06857}}].

\bibitem{Harris:2024nmr}
S.~Harris, A.~Kaviraj, J.~A. Mann, L.~Quintavalle, and V.~Schomerus, {\it {Comb
  Channel Lightcone Bootstrap II: Triple-Twist Anomalous Dimensions}},
  \href{http://arxiv.org/abs/2401.10986}{{\tt arXiv:2401.10986}}.

\bibitem{Drummond:2022dxd}
J.~M. Drummond, R.~Glew, and M.~Santagata, {\it {Bern-Carrasco-Johansson
  relations in AdS5\texttimes{}S3 and the double-trace spectrum of super
  gluons}},  {\em Phys. Rev. D} {\bf 107} (2023), no.~8 L081901,
  [\href{http://arxiv.org/abs/2202.09837}{{\tt arXiv:2202.09837}}].

\bibitem{Behan:2023fqq}
C.~Behan, S.~M. Chester, and P.~Ferrero, {\it {Gluon scattering in AdS at
  finite string coupling from localization}},
  \href{http://arxiv.org/abs/2305.01016}{{\tt arXiv:2305.01016}}.

\bibitem{Glew:2023wik}
R.~Glew and M.~Santagata, {\it {The Veneziano amplitude in AdS$_5 \times$S$^3$
  from an 8-dimensional effective action}},
  \href{http://arxiv.org/abs/2305.01013}{{\tt arXiv:2305.01013}}.

\bibitem{Aprile:2018efk}
F.~Aprile, J.~Drummond, P.~Heslop, and H.~Paul, {\it {Double-trace spectrum of
  $N=4$ supersymmetric Yang-Mills theory at strong coupling}},  {\em Phys. Rev.
  D} {\bf 98} (2018), no.~12 126008,
  [\href{http://arxiv.org/abs/1802.06889}{{\tt arXiv:1802.06889}}].

\bibitem{Aprile:2019rep}
F.~Aprile, J.~Drummond, P.~Heslop, and H.~Paul, {\it {One-loop amplitudes in
  AdS$_{5}$\texttimes{}S$^{5}$ supergravity from $ \mathcal{N} $ = 4 SYM at
  strong coupling}},  {\em JHEP} {\bf 03} (2020) 190,
  [\href{http://arxiv.org/abs/1912.01047}{{\tt arXiv:1912.01047}}].

\bibitem{Cavaglia:2022yvv}
A.~Cavagli\`a, N.~Gromov, J.~Julius, and M.~Preti, {\it {Integrated correlators
  from integrability: Maldacena-Wilson line in $ \mathcal{N} $ = 4 SYM}},  {\em
  JHEP} {\bf 04} (2023) 026, [\href{http://arxiv.org/abs/2211.03203}{{\tt
  arXiv:2211.03203}}].

\bibitem{Drukker:2022pxk}
N.~Drukker, Z.~Kong, and G.~Sakkas, {\it {Broken Global Symmetries and Defect
  Conformal Manifolds}},  {\em Phys. Rev. Lett.} {\bf 129} (2022), no.~20
  201603, [\href{http://arxiv.org/abs/2203.17157}{{\tt arXiv:2203.17157}}].

\bibitem{Goncalves:2019znr}
V.~Gon\c{c}alves, R.~Pereira, and X.~Zhou, {\it {$20'$ Five-Point Function from
  $AdS_5\times S^5$ Supergravity}},  {\em JHEP} {\bf 10} (2019) 247,
  [\href{http://arxiv.org/abs/1906.05305}{{\tt arXiv:1906.05305}}].

\bibitem{Alday:2022lkk}
L.~F. Alday, V.~Gon\c{c}alves, and X.~Zhou, {\it {Supersymmetric Five-Point
  Gluon Amplitudes in AdS Space}},  {\em Phys. Rev. Lett.} {\bf 128} (2022),
  no.~16 161601, [\href{http://arxiv.org/abs/2201.04422}{{\tt
  arXiv:2201.04422}}].

\bibitem{Goncalves:2023oyx}
V.~Gon\c{c}alves, C.~Meneghelli, R.~Pereira, J.~Vilas~Boas, and X.~Zhou, {\it
  {Kaluza-Klein Five-Point Functions from $\textrm{AdS}_5\times S_5$
  Supergravity}},  \href{http://arxiv.org/abs/2302.01896}{{\tt
  arXiv:2302.01896}}.

\bibitem{Alday:2023kfm}
L.~F. Alday, V.~Gon\c{c}alves, M.~Nocchi, and X.~Zhou, {\it {Six-Point AdS
  Gluon Amplitudes from Flat Space and Factorization}},
  \href{http://arxiv.org/abs/2307.06884}{{\tt arXiv:2307.06884}}.

\bibitem{Meneghelli:2022gps}
C.~Meneghelli and M.~Tr\'epanier, {\it {Bootstrapping string dynamics in the 6d
  \ensuremath{\mathscr{N}} = (2, 0) theories}},  {\em JHEP} {\bf 07} (2023)
  165, [\href{http://arxiv.org/abs/2212.05020}{{\tt arXiv:2212.05020}}].

\bibitem{MultiplePolylogs}
J.~Lappo-Danilevskij, {\it {Memoires sur la theorie des systemes des equations
  differentielles lineaires. Vol. I}},  {\em Travaux Inst. Physico-Math.
  Stekloff} {\bf 6} (1934) 1232--1258.

\bibitem{Goncharov:1998kja}
A.~B. Goncharov, {\it {Multiple polylogarithms, cyclotomy and modular
  complexes}},  {\em Math. Res. Lett.} {\bf 5} (1998) 497--516,
  [\href{http://arxiv.org/abs/1105.2076}{{\tt arXiv:1105.2076}}].

\bibitem{Goncharov:2001iea}
A.~B. Goncharov, {\it {Multiple polylogarithms and mixed Tate motives}},
  \href{http://arxiv.org/abs/math/0103059}{{\tt math/0103059}}.

\bibitem{Remiddi:1999ew}
E.~Remiddi and J.~A.~M. Vermaseren, {\it {Harmonic polylogarithms}},  {\em Int.
  J. Mod. Phys. A} {\bf 15} (2000) 725--754,
  [\href{http://arxiv.org/abs/hep-ph/9905237}{{\tt hep-ph/9905237}}].

\bibitem{AlgebrePolylogs}
J.~Hoeven, M.~Petitot, and .~Hoang Ngoc~Minh, {\it L'algèbre des
  polylogarithmes par les séries génératrices}, .

\bibitem{UlanskiiHPLs}
E.~Ulanskii, {\it Identities for generalized polylogarithms},  {\em
  Mathematical Notes} {\bf 73} (2003) 571--581.

\bibitem{BROWN2004527}
F.~C. Brown, {\it Polylogarithmes multiples uniformes en une variable},  {\em
  Comptes Rendus Mathematique} {\bf 338} (2004), no.~7 527--532.

\bibitem{Maitre:2005uu}
D.~Maitre, {\it {HPL, a mathematica implementation of the harmonic
  polylogarithms}},  {\em Comput. Phys. Commun.} {\bf 174} (2006) 222--240,
  [\href{http://arxiv.org/abs/hep-ph/0507152}{{\tt hep-ph/0507152}}].

\bibitem{NielsenPolys}
K.~Kolbig, {\it {Nielsen’s Generalized Polylogarithms}},  {\em SIAM J. Math.
  Anal.} {\bf 17 (5)} (1986) 1232--1258.

\bibitem{lewin1981polylogarithms}
L.~Lewin, {\em Polylogarithms and Associated Functions}.
\newblock North Holland, 1981.

\bibitem{lewin1991structural}
L.~Lewin, {\em Structural Properties of Polylogarithms}.
\newblock Mathematical surveys and monographs. American Mathematical Society,
  1991.

\bibitem{Drummond:2012bg}
J.~M. Drummond, {\it {Generalised ladders and single-valued polylogarithms}},
  {\em JHEP} {\bf 02} (2013) 092, [\href{http://arxiv.org/abs/1207.3824}{{\tt
  arXiv:1207.3824}}].

\bibitem{Schnetz:2013hqa}
O.~Schnetz, {\it {Graphical functions and single-valued multiple
  polylogarithms}},  {\em Commun. Num. Theor. Phys.} {\bf 08} (2014) 589--675,
  [\href{http://arxiv.org/abs/1302.6445}{{\tt arXiv:1302.6445}}].

\bibitem{DelDuca:2016lad}
V.~Del~Duca, S.~Druc, J.~Drummond, C.~Duhr, F.~Dulat, R.~Marzucca,
  G.~Papathanasiou, and B.~Verbeek, {\it {Multi-Regge kinematics and the moduli
  space of Riemann spheres with marked points}},  {\em JHEP} {\bf 08} (2016)
  152, [\href{http://arxiv.org/abs/1606.08807}{{\tt arXiv:1606.08807}}].

\bibitem{Zi}
C.~K. {Zickert}, {\it {Holomorphic polylogarithms and Bloch complexes}},  {\em
  arXiv e-prints} (Feb., 2019) arXiv:1902.03971,
  [\href{http://arxiv.org/abs/1902.03971}{{\tt arXiv:1902.03971}}].

\bibitem{Wo1}
Z.~Wojtkowiak, {\it A construction of analogs of the bloch-wigner function.},
  {\em MATHEMATICA SCANDINAVICA} {\bf 65} (Dec., 1989) 140–142.

\bibitem{Zag2}
D.~Zagier, {\it {The Bloch-Wigner-Ramakrishnan polylogarithm function}},  {\em
  Math. Ann.} {\bf 286} (1990) 613--624.

\bibitem{Zag3}
D.~Zagier, {\em Polylogarithms, Dedekind Zeta Functions, and the Algebraic
  K-Theory of Fields}, pp.~391--430.
\newblock Birkh{\"a}user Boston, Boston, MA, 1991.

\bibitem{2021arXiv210501543P}
L.~{Pirio}, {\it {On webs, polylogarithms and cluster algebras}},  {\em arXiv
  e-prints} (May, 2021) arXiv:2105.01543,
  [\href{http://arxiv.org/abs/2105.01543}{{\tt arXiv:2105.01543}}].

\bibitem{Billo:2013jda}
M.~Billó, M.~Caselle, D.~Gaiotto, F.~Gliozzi, M.~Meineri, and R.~Pellegrini,
  {\it {Line defects in the 3d Ising model}},  {\em JHEP} {\bf 07} (2013) 055,
  [\href{http://arxiv.org/abs/1304.4110}{{\tt arXiv:1304.4110}}].

\bibitem{Dokshitzer:2005bf}
Y.~L. Dokshitzer, G.~Marchesini, and G.~P. Salam, {\it {Revisiting parton
  evolution and the large-x limit}},  {\em Phys. Lett. B} {\bf 634} (2006)
  504--507, [\href{http://arxiv.org/abs/hep-ph/0511302}{{\tt hep-ph/0511302}}].

\bibitem{Basso:2006nk}
B.~Basso and G.~P. Korchemsky, {\it {Anomalous dimensions of high-spin
  operators beyond the leading order}},  {\em Nucl. Phys. B} {\bf 775} (2007)
  1--30, [\href{http://arxiv.org/abs/hep-th/0612247}{{\tt hep-th/0612247}}].

\bibitem{Alday:2015eya}
L.~F. Alday, A.~Bissi, and T.~Lukowski, {\it {Large spin systematics in CFT}},
  {\em JHEP} {\bf 11} (2015) 101, [\href{http://arxiv.org/abs/1502.07707}{{\tt
  arXiv:1502.07707}}].

\bibitem{Beccaria:2008fi}
M.~Beccaria and V.~Forini, {\it {Reciprocity of gauge operators in N=4 SYM}},
  {\em JHEP} {\bf 06} (2008) 077, [\href{http://arxiv.org/abs/0803.3768}{{\tt
  arXiv:0803.3768}}].

\bibitem{Fiamberti:2009jw}
F.~Fiamberti, A.~Santambrogio, and C.~Sieg, {\it {Five-loop anomalous dimension
  at critical wrapping order in N=4 SYM}},  {\em JHEP} {\bf 03} (2010) 103,
  [\href{http://arxiv.org/abs/0908.0234}{{\tt arXiv:0908.0234}}].

\bibitem{Velizhanin:2010cm}
V.~N. Velizhanin, {\it {Six-Loop Anomalous Dimension of Twist-Three Operators
  in N=4 SYM}},  {\em JHEP} {\bf 11} (2010) 129,
  [\href{http://arxiv.org/abs/1003.4717}{{\tt arXiv:1003.4717}}].

\bibitem{Bliard:2022xsm}
G.~Bliard, {\it {Notes on n-point Witten diagrams in AdS$_{2}$}},  {\em J.
  Phys. A} {\bf 55} (2022), no.~32 325401,
  [\href{http://arxiv.org/abs/2204.01659}{{\tt arXiv:2204.01659}}].

\bibitem{Ma:2022ihn}
W.-J. Ma and X.~Zhou, {\it {Scattering bound states in AdS}},  {\em JHEP} {\bf
  08} (2022) 107, [\href{http://arxiv.org/abs/2204.13419}{{\tt
  arXiv:2204.13419}}].

\end{thebibliography}
\end{document}